\documentclass[12pt, a4paper, twoside]{report}
\pdfoutput=1
\usepackage{import}
\usepackage{Preamble}

\usepackage[style=phys, articletitle=true, biblabel=brackets, backend=bibtex,
chaptertitle=false, pageranges=false, eprint=true,%
natbib=true, maxbibnames=10, giveninits=true, sorting=none]{biblatex} 

\DeclareFieldFormat[article]{journaltitle}{\mkbibitalic{#1\isdot}} 
\makeatletter
\newrobustcmd{\mkbibfixedbrackets}[1]{%
	\begingroup
	\blx@blxinit
	\blx@setsfcodes
	\bibleftbracket#1\bibrightbracket
	\endgroup}

\renewbibmacro*{doi+eprint+url}{%
	\iftoggle{bbx:doi}
	{\printfield{doi}}
	{}%
	\ifboolexpr{test {\iffieldequalstr{eprinttype}{arxiv}} or test {\iffieldequalstr{eprinttype}{arXiv}}}
	{\setunit{\addspace}\newblock}
	{\newunit\newblock}%
	\iftoggle{bbx:eprint}
	{\usebibmacro{eprint}}
	{}%
	\newunit\newblock
	\iftoggle{bbx:url}
	{\usebibmacro{url+urldate}}
	{}}

\xpatchbibdriver{online}
{\newunit\newblock
	\iftoggle{bbx:eprint}
	{\usebibmacro{eprint}}
	{}}
{\ifboolexpr{test {\iffieldequalstr{eprinttype}{arxiv}} or test {\iffieldequalstr{eprinttype}{arXiv}}}
	{\setunit{\addspace}\newblock}
	{\newunit\newblock}%
	\iftoggle{bbx:eprint}
	{\usebibmacro{eprint}}
	{}}
{}{}

\DeclareFieldFormat{eprint:arxiv}{%
	\mkbibbrackets{%
		\ifhyperref
		{\href{http://arxiv.org/\abx@arxivpath/#1}{%
				arXiv\addcolon
				\nolinkurl{#1}%
				\iffieldundef{eprintclass}
				{}
				{\addspace\UrlFont{\mkbibfixedbrackets{\thefield{eprintclass}}}}}}
		{arXiv\addcolon
			\nolinkurl{#1}%
			\iffieldundef{eprintclass}
			{}
			{\addspace\UrlFont{\mkbibfixedbrackets{\thefield{eprintclass}}}}}}}
\makeatother

\counterwithout{footnote}{chapter}

\fancypagestyle{stylenormal}{
	\fancyhf{}
	\fancyhead[L]{Chapter \thechapter \hspace{0.5pt} -- \leftmark}
	\fancyhead[R]{\thepage}
}

\begin{document}

\pagestyle{fancy}
\pagenumbering{Roman}

\begin{titlepage}
\graphicspath{{Images/}} 

\begin{center}
     
\vspace*{2mm}
     
{\LARGE \bf Correlation functions of conserved currents \\[2.5mm] in (super)conformal field theory} 

\vspace{1cm}

{\Large{\textbf{Benjamin J. Stone}}}\\

\vspace{1cm}
{\large  
\begin{tabular}{  m{0.3\textwidth}  m{0.55\textwidth}  }
	\hfill Supervisor: & \hspace{1mm}A/Prof. Evgeny I. Buchbinder \hfill \\
	\hfill Co-supervisor: & \hspace{5.6mm} Prof. Sergei M. Kuzenko \hfill
\end{tabular}}


\vspace{1cm}

\includegraphics[width=0.4\textwidth]{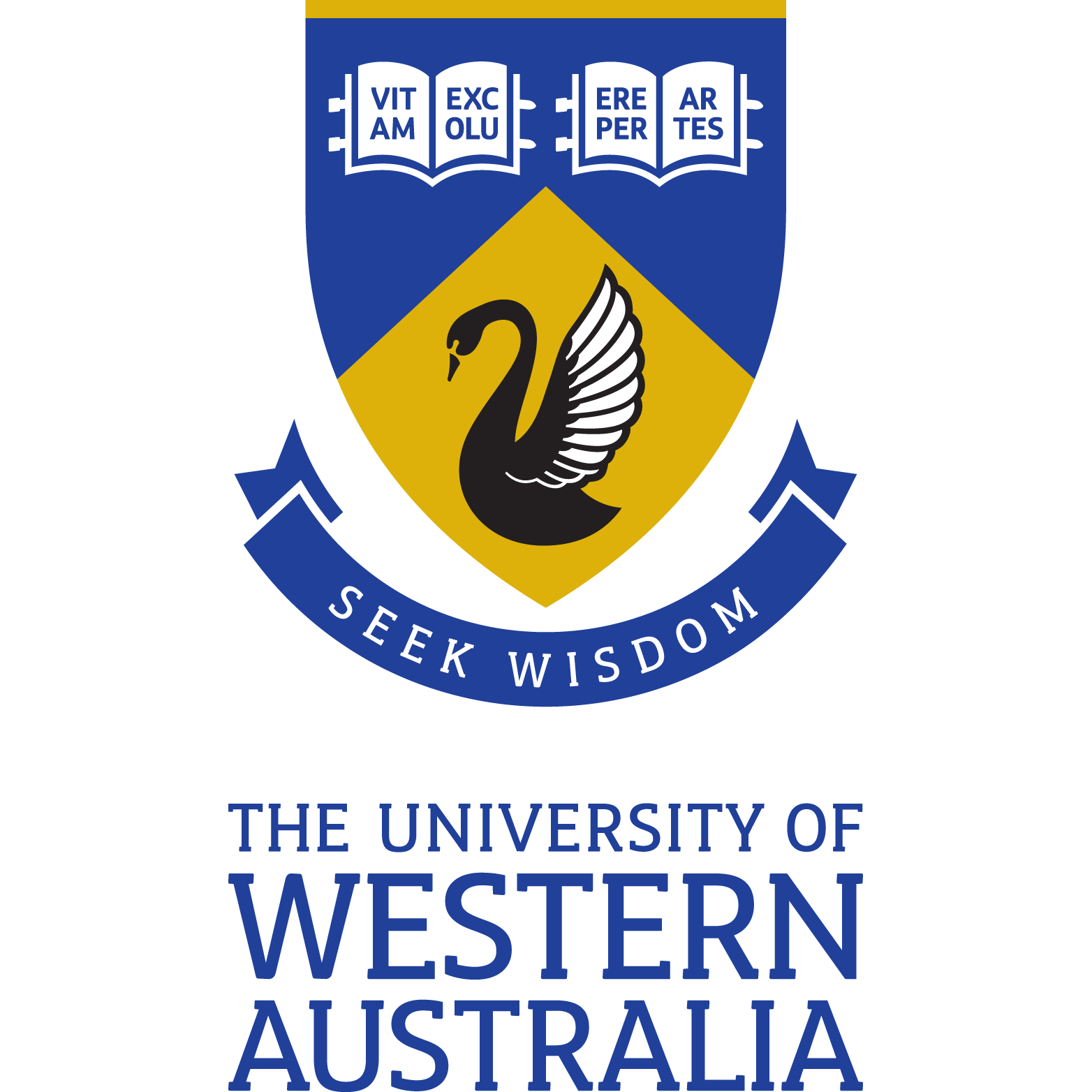}

\vspace{1cm}

{\large
This thesis is presented for the degree of Doctor of Philosophy\\
The University of Western Australia\\
Department of Physics\\[5mm]
Dec 2023

\vspace{1cm}
\begin{tabular}{  m{0.4\textwidth}  m{0.6\textwidth}  }
		Examiners: & \hfill \\
		Prof. Johanna Erdmenger \hfill & \hfill (University of W\"{u}rzburg, Germany) \\
		Prof. Jeong-Hyuck Park \hfill & \hfill (Sogang University, South Korea)
\end{tabular}}
\vspace{2cm}

\end{center}

\end{titlepage}


\chapter*{Thesis Declaration}
I, Benjamin J. Stone, certify that:\\[-6pt]

This thesis has been substantially accomplished during enrolment in this degree.
This thesis does not contain material which has been submitted for the award of any other degree or diploma in my name, in any university or other tertiary institution.\\[-6pt]

In the future, no part of this thesis will be used in a submission in my name, for any other degree or diploma in any university or other tertiary institution without the prior approval of The University of Western Australia, and where applicable, any partner institution responsible for the joint-award of this degree.\\[-6pt]

This thesis does not contain any material previously published or written by another person, except where due reference has been made in the text and, where relevant, in the Authorship Declaration that follows.\\[-6pt]

This thesis does not violate or infringe any copyright, trademark, patent, or other rights whatsoever of any person.\\[-6pt]

This thesis contains published work and/or work prepared for publication, some of which has been co-authored. 


\chapter*{} 

%

\vspace{-20mm}

\begin{center}
	{\Huge\textbf{Abstract}}
\end{center}

This thesis is dedicated to analysing the general structure of two- and three-point correlation functions of conserved currents of arbitrary integer or half-integer spins in three- and four-dimensional (super)conformal field theory. 





We begin by analysing the two- and three-point correlation functions of conserved currents in three-dimensional (3D) conformal field theory (CFT). A general formalism is developed to construct three-point functions of conserved currents of arbitrary integer or half-integer spins, consistent with the requirements of conformal covariance. In particular, we utilise an analytic computational approach to efficiently impose all constraints which arise due to conservation equations and symmetries under permutations of spacetime points. It is demonstrated that the three-point functions of conserved currents are generically fixed up to two parity-even structures and one parity-odd structure, with the existence of the latter subject to triangle inequalities in the spins. 

Next, we undertake the analysis of two- and three-point correlation functions of conserved currents belonging to an arbitrary Lorentz representation in four-dimensional (4D) conformal field theory. We elucidate how to impose all the relevant constraints arising due to conservation equations, reality conditions and symmetries under permutations of spacetime points. The general structure of three-point functions of conserved currents is classified. We show that for vector-like higher-spin currents, the number of conserved structures in the three-point function increases linearly with the minimum spin. For currents belonging to more general spin representations, it is shown that there are some special cases where the number of structures deviates from this result.

The final chapter is dedicated to analysing the two- and three-point functions of conserved supercurrents in three-dimensional $\cN=1$ superconformal field theory (SCFT), where it is shown that supersymmetry imposes additional constraints on the structure of three-point functions of conserved currents. By utilising a manifestly supersymmetric construction, it is demonstrated that the three-point functions of conserved supercurrents which are overall Grassmann-even in superspace are generically fixed up to one parity-even structure and one parity-odd structure. The existence of the parity-odd structure is subject to triangle inequalities in the superspins. For three-point functions which are overall Grassmann-odd in superspace, it is proven analytically that they are fixed up to a single parity-even structure for arbitrary superspins.


\clearpage
\fancyhead[LE,LO]{Authorship Declaration}
\chapter*{Authorship Declaration}

\noindent This thesis is based on seven published papers \cite{Buchbinder:2021gwu,Buchbinder:2021qlb,Buchbinder:2022cqp,Buchbinder:2022mys,Buchbinder:2023fqv,Buchbinder:2023ndg,Buchbinder:2023coi}. It is formatted as a series of papers, with most of the material taken from \cite{Buchbinder:2022mys,Buchbinder:2023fqv,Buchbinder:2023ndg,Buchbinder:2023coi}. The details of these works are as follows:
\begin{enumerate}
	\item E. I. Buchbinder and B. J. Stone, ``Mixed three-point functions of conserved currents in three-dimensional superconformal field theory", \textit{Phys. Rev. D} \textbf{103}, 086023 (2021) [\href{https://arxiv.org/abs/2102.04827}{arXiv:2102.04827 [hep-th]}].
	\begin{itemize}
		\item \textbf{Location in thesis:} Chapter \ref{Chapter5}.
	\end{itemize}
	\vspace{2mm}
	
	\item E. I. Buchbinder and B. J. Stone, ``Three-point functions of a superspin-2 current multiplet in 3D, $\cN=1$ superconformal theory", \textit{Phys. Rev. D} \textbf{104}, 106004 (2021) [\href{https://arxiv.org/abs/2108.01865}{arXiv:2108.01865 [hep-th]}].
	\begin{itemize}
		\item \textbf{Location in thesis:} Chapter \ref{Chapter5}.
	\end{itemize}
	\vspace{2mm}
	
	\item E. I. Buchbinder and B. J. Stone, ``Three-point functions of a fermionic higher-spin current in 4D conformal field theory",\textit{ Phys. Rev. D} \textbf{105}, 125004 (2022) [\href{https://arxiv.org/abs/2204.04899}{arXiv:2204.04899 [hep-th]}].
	\begin{itemize}
		\item \textbf{Location in thesis:} Chapter \ref{Chapter3}.
	\end{itemize}
	\vspace{2mm}
	
	\item E. I. Buchbinder and B. J. Stone, ``Three-point functions of conserved currents in 3D CFT: General formalism for arbitrary spins", \textit{Phys. Rev. D} \textbf{107}, 046007 (2023) [\href{https://arxiv.org/abs/2210.13135}{arXiv:2210.13135 [hep-th]}].
	\begin{itemize}
		\item \textbf{Location in thesis:} Chapter \ref{Chapter4}.
	\end{itemize}
	\vspace{2mm}
	
	\item E. I. Buchbinder and B. J. Stone, ``Three-point functions of conserved supercurrents in 3D $\cN=1$ SCFT: General formalism for arbitrary superspins", \textit{Phys. Rev. D} \textbf{107}, 106001 (2023) [\href{https://arxiv.org/abs/2302.00593}{arXiv:2302.00593 [hep-th]}].
	\begin{itemize}
		\item \textbf{Location in thesis:} Chapter \ref{Chapter5}.
	\end{itemize}
	\vspace{2mm}

	\item E. I. Buchbinder and B. J. Stone, ``Grassmann-odd three-point functions of conserved supercurrents in 3D $\cN=1$ SCFT", \textit{Phys. Rev. D} \textbf{108}, 046001 (2023) [\href{https://arxiv.org/abs/2305.02233}{arXiv:2305.02233 [hep-th]}].
	\begin{itemize}
		\item \textbf{Location in thesis:} Chapter \ref{Chapter5}.
	\end{itemize}
	\vspace{2mm}
	
	\item E. I. Buchbinder and B. J. Stone, ``Three-point functions of conserved currents in 4D CFT: general formalism for arbitrary spins", \textit{Phys. Rev. D} \textbf{108}, 086017 (2023) [\href{https://arxiv.org/abs/2307.11435}{arXiv:2307.11435 [hep-th]}].
	\begin{itemize}
		\item \textbf{Location in thesis:} Chapter \ref{Chapter4}.
	\end{itemize}
\end{enumerate}
\vspace{3mm}

\noindent Permission has been granted by all co-authors to use the above work.

%

\vspace{5mm}

The \textit{Mathematica} code utilised for the projects above and the results in this thesis can be found in the author's GitHub repository: \href{https://github.com/BSTONE97/CFTcorrelators}{github.com/BSTONE97/CFTcorrelators}.

The author would like to point out that development of the computational tools required to carry out the analysis presented in this thesis involved a significant amount of work which is difficult to convey within the scope of this thesis. It is hoped that the techniques developed in these Mathematica notebooks will be of value to those interested in conformal field theory and related research.

\chapter*{Acknowledgements}

I would like to begin by extending my sincere gratitude to my supervisor, Prof. Evgeny Buchbinder, who I have been working with for the past four years. Throughout my PhD, Evgeny has provided me with excellent guidance and allowed me the freedom to approach my project in a way that appeals to my strengths. His kind and encouraging approach to supervision gave me the confidence to succeed in my research, and I feel very fortunate to have been his PhD student. It has been an incredible learning experience.

I would also like to thank my co-supervisor, Prof. Sergei Kuzenko, for his support and encouragement throughout my PhD. His expertise in the field is inspiring; he set high standards for the quality of my research and instilled in me the importance of self discipline. I am grateful for our many discussions over the past few years and it has been a privilege to undertake my PhD with the Field Theory and Quantum Gravity group at the University of Western Australia.

I have also had the privilege of working alongside some incredibly talented researchers throughout my PhD: Daniel Hutchings, Emmanouil Raptakis, Nowar Koning, Jake Stirling, James La Fontaine, Jessica Hutomo, Joshua Pinelli, Kai Turner and Michael Ponds. I thank all of them for many insightful discussions and the great times spent both at the office and at conferences together. You have all inspired me to do my best work. 

Last but not least, I would like to thank my friends, family, and my partner Alissa for their unwavering support and encouragement while undertaking my studies. 

The research undertaken in this thesis was supported by the Bruce and Betty Green Postgraduate Research Scholarship, and the Australian Government Research Training Program (RTP) Scholarship.


\clearpage
\fancyhead[LE,LO]{Contents}
{\hypersetup{hidelinks}
\tableofcontents}

\clearpage
\pagenumbering{arabic}
\newpage

\pagestyle{stylenormal}

\chapter{Introduction}\label{Chapter1}


The concept of symmetry plays a pivotal role in modern theoretical physics, providing a powerful framework in which to understand the fundamental interactions in our universe.  
For example, our modern understanding of space and time is based on Einstein's theory of special relativity \cite{Einstein:1905ve}. In this theory, the principle of relativity requires that the symmetry group of any closed relativistic system in four spacetime dimensions must include the Poincar\'e group, which is generated by Lorentz transformations and spacetime translations. These transformations are the isometries of a four-dimensional manifold known as Minkowski space, $\mathbb{R}^{1,3}$, parameterised by the coordinates $x^{m} = (ct, x, y, z)$, and equipped with the metric $\text{d}s^{2} = \eta_{m n} \text{d}x^{m} \text{d}x^{n}$, where $\eta_{mn} = \text{diag}(-1,1,1,1)$ and $c$ is the speed of light. Symmetry principles have also been paramount to the development of Quantum Field Theory (QFT), the theoretical framework which combines the principles of special relativity and quantum mechanics.  A particularly influential development guided by symmetry principles was the discovery of the Dirac equation \cite{Dirac:1928hu}, a relativistic wave equation describing all spin-$1/2$ particles which is consistent with both Poincar\'e invariance and the principles of quantum mechanics. Soon after, Wigner \cite{Wigner:1939cj} demonstrated that symmetry principles are intrinsically interwoven with particle physics. Using an elegant group-theoretic framework, he demonstrated that free elementary particles are associated with the unitary irreducible representations of the Poincar\'e group. In particular, the mass and spin of elementary particles are associated with the two Casimir operators of the Poincar\'e group.

The Standard Model of particle physics, the crowning achievement of late 20th century physics, is an example of a renormalisable quantum field theory describing three of the four fundamental interactions in our universe. It is a non-Abelian gauge theory described in terms of Yang-Mills gauge fields coupled to a matter sector, and possesses Poincar\'e symmetry in addition to local gauge symmetries: $SU(2) \times U(1)$ for the electroweak sector, resulting in the $W$ and $Z$ bosons and the photon, and $SU(3)$ for the strong force, which gives rise to gluons. All elementary particles in the Standard Model possess spin $s \leq 1$. The symmetries above have incredible predictive power; they foretold the existence of particles before they were experimentally verified, such as the charm quark \cite{Bjorken:1964gz,Glashow:1970gm}. Moreover, the Higgs mechanism, which explains why some particles possess mass, arises from spontaneous symmetry breaking of the electroweak sector (see the textbooks \cite{Weinberg:1995mt,Weinberg:1996kr}). The evolution and understanding of the Standard Model is a testament to the power and significance of symmetries in theoretical physics; they are not just aesthetic or mathematical constructs, but are deeply ingrained into the structure of the universe.

Despite the enormous success of the Standard Model of particle physics, it is widely recognised as an incomplete theory. One of its main shortcomings is that it does not incorporate gravitational interactions. At the classical level, gravity is described by Einstein's theory of general relativity \cite{Einstein:1915ca,Einstein:1916vd}; the theory describes the dynamics of the gravitational field, and may be viewed as a non-Abelian gauge theory of a spin-2 field possessing diffeomorphism invariance \cite{Buchbinder:1998qv}. Although quantum gravitational effects are not relevant at the energy scale of the Standard Model of particle physics, it is expected that they must be accounted for at the Planck scale (of order $10^{19}$ GeV), and play an important role in early universe cosmology. One of the main challenges of modern theoretical physics is to construct a quantum theory of gravity which consistently combines both the principles of quantum field theory and general relativity. However, conventional approaches to quantise Einstein's theory of gravity have failed due to it being a non-renormalisable theory. Efforts to construct a quantum theory of gravity have naturally led theoretical physicists to explore hypothetical extensions to the symmetries of spacetime. However, in the 1960s an important theorem was established by Coleman and Mandula \cite{Coleman:1967ad}, which places strong restrictions on such extensions. In particular, the theorem states that the only possible spacetime symmetry algebra of the S-matrix in four-dimensional quantum field theory with a mass gap is a (semi) direct sum of the Poincar\'e algebra (the conformal algebra in the massless case) and some other Lie algebra of internal (gauge) symmetries. Hence, in addition to providing us a deeper understanding of the fundamental forces, symmetry principles also impose restrictions on possible extensions to the Standard Model of particle physics. These principles also underpin modern developments such as superstring theory \cite{Green:1987sp}, which aims to unify all known fundamental forces. \\[4mm]
\noindent\textbf{Supersymmetry}

Among the key developments in late 20th century theoretical physics is the concept of supersymmetry. Originally introduced by Golfand and Likhtman \cite{Golfand:1971iw}, Volkov and Akulov \cite{Volkov:1972jx}, and Wess and Zumino \cite{Wess:1974tw,Wess:1973kz} (see the textbooks \cite{Gates:1983nr,Buchbinder:1998qv} for some important references), supersymmetry is the only consistent and nontrivial extension of Poincar\'e symmetry which is compatible with the principles of quantum field theory. It is a symmetry of spacetime which transforms bosonic fields onto fermionic fields and vice-versa, and thus it leads to unification of matter particles and force carriers by arranging them into so-called supermultiplets. Supersymmetry transformations are generated by spinor operators called supersymmetry generators (supercharges) which are fermionic in nature, obeying anti-commutation relations. Let us recall that according to the Coleman-Mandula theorem \cite{Coleman:1967ad}, the only possible spacetime symmetry algebra of the S-matrix in four-dimensional quantum field theory is a direct sum of the Poincar\'e algebra (the conformal algebra in the massless case) and a Lie algebra of internal symmetries. In other words, there cannot be a non-trivial mixing between spacetime and internal symmetries. However, the Coleman-Mandula theorem assumes that the only symmetry algebras are Lie algebras; by generalising the theorem to incorporate Lie superalgebras, it was shown by Haag, Lopuszanski and Sohnius \cite{Haag:1974qh} that supersymmetry is the only possible extension of the symmetry algebra consistent with the principles of quantum field theory. The extension of the Poincar\'e algebra to incorporate supersymmetry is known as the $\cN$-extended super-Poincar\'e algebra, where $\cN$ is proportional to the number of supercharges that the theory possesses. Theories with $\cN$-extended supersymmetry can admit an additional Lie algebra symmetry, known as R-symmetry; this is an internal symmetry which transforms the different supercharges into eachother, effectively bypassing one of the main restrictions of the Coleman-Mandula theorem. 

From the supersymmetry algebra it is clear that local supersymmetry implies invariance under local spacetime translations (general covariance), and the latter requires gravity. Hence, gravity is an intrinsic part of the gauge theory of supersymmetry, otherwise known as supergravity. In this theory supersymmetry is gauged with the introduction of the gravitino (spin-3/2), while its superpartner, the graviton (spin-2), is associated with diffeomorphism invariance (see the textbooks \cite{Gates:1983nr,Wess:1992cp,Buchbinder:1998qv,Weinberg:2000cr}, and the review articles \cite{Aharony:1999ti,Ferrara:2017hed}). It is the principle of local supersymmetry which unifies gravity with lower spin fields describing elementary particles and their non-gravitational interactions. In four spacetime dimensions, there are different versions of supergravity depending on the number of local supersymmetries the theory possesses ($\cN = 1, \dots, 8)$. The simplest supergravity theory was constructed in 1976 \cite{Freedman:1976xh,Freedman:1976py,Deser:1976eh}, while the first extended supergravity ($\cN=2$) was formulated soon after \cite{Fradkin:1979cw,deWit:1979dzm}. The latter realised Einstein's dream of unifying gravity and electromagnetism, which was an incredible theoretical achievement. Supergravity also arises as the low energy limit of string theory, the leading candidate for unifying the four fundamental forces. The discovery of supersymmetry is one of the most important advancements in theoretical high-energy physics and remains an active field of research due to its potential phenomenological and theoretical applications.

The most natural arena in which to carry out the study of supersymmetric field theory is in so-called superspace. Introduced by Volkov and Akulov \cite{Akulov:1974xz} and Salam and Strathdee \cite{Salam:1974yz}, in essence, superspace appends the (commuting) ``bosonic" spacetime coordinates with new (anti-commuting) Grassmann coordinates. In analogy with ordinary fields which are functions of spacetime coordinates, one can introduce the notion of superfields in superspace which are now functions of the spacetime coordinates and the additional Grassmann coordinates. Due to the anti-commuting nature of the Grassmann coordinates, superfields are typically expanded in power series in the Grassmann variables, which truncates at an order determined by the degree of supersymmetry, $\cN$. Indeed, it was proposed by Salam and Strathdee \cite{Salam:1974yz} that superspace and superfields are the appropriate tools that one should use to construct and study supersymmetric field theories. We will employ the superspace approach in the latter parts of this thesis.\\[4mm]
\noindent\textbf{Conformal symmetry}

In accordance with the Coleman-Mandula theorem \cite{Coleman:1967ad}, the spacetime symmetry algebra for a quantum field theory with no mass gap may be extended to conformal symmetry, the maximal possible symmetry algebra of the S-matrix. Conformal symmetry extends Poincar\'e symmetry by the inclusion of additional spacetime transformations, namely scale transformations, special conformal transformations and inversions. These are transformations which at most scale the spacetime metric, and the set of all such transformations forms the conformal group of (Minkowski) spacetime. Since the 1970s, conformal symmetry has quickly established its importance in many areas of physics, encompassing topics such as quantum field theory and the description of critical phenomena in statistical mechanics. However, an important motivation for the study of conformal symmetry was due to Cunningham \cite{Cunningham:1910} and Bateman \cite{Bateman:1909, Bateman:1910a, Bateman:1910b} in the early 20th century, who showed that the spacetime symmetry group of the free Maxwell equations was actually larger than the Poincar\'e group, coinciding instead with the conformal group of Minkowski space. Maxwell's electrodynamics is but one example of a conformally invariant field theory. Since then it has been realised that many massless field theories possess conformal symmetry at the classical level. A particularly important model of recent interest is the unique conformal and duality-invariant non-linear extension of Maxwell's electrodynamics, known as ModMax \cite{Bandos:2020jsw}.

It was realised in the 1970s that conformal symmetry appears as an asymptotic symmetry in many models of quantum field theories at high energies, where the presence of dimensionful parameters such as particle masses become irrelevant. A quantum field theory possessing conformal symmetry is known as a conformal field theory (CFT). The study of CFT has been pioneered by authors such as Mack and Salam \cite{Mack:1969rr,Mack:1976pa}, and Polyakov \cite{Polyakov:1970xd, Belavin:1984vu} (see also \cite{Schreier:1971um, Migdal:1971xh, Mack:1973mq}), where it has been shown to have far reaching applications. As a result of their scale invariance properties, CFTs (particularly 2D CFTs) are uniquely equipped to handle problems at critical points, such as phase transitions in statistical mechanics. For example, Cardy \cite{Cardy:1984bb} demonstrated how CFT can be used effectively to describe critical phenomena on surfaces. More recently, CFT methods have been applied to solve the three-dimensional Ising model at the critical temperature \cite{El-Showk:2012cjh}.


At the quantum level, conformal symmetry is much more restrictive than in the classical regime. Essentially, conformal invariance requires quantum field theory to be free of ultraviolet (UV) divergences, and this requirement corresponds to fixed points of renormalisation group (RG) flows. The renormalization group method, developed by authors such as Stueckelberg and Petermann \cite{Stueckelberg:1951gg}, Gell-Mann and Low \cite{Gell-Mann:1954yli}, Bogoliubov and Shirkov \cite{Bogolyubov:1956gh} and later by Callan and Symanzik \cite{Callan:1970yg,Symanzik:1970rt}, is a framework that has deeply influenced our understanding of the behaviour of systems across different scales. In QFT, the renormalization group describes how the properties of a system (for example, the coupling constants in a field theory) change as we change the energy scale at which we probe the system. An important realization, largely credited to Wilson (see e.g. \cite{Wilson:1970ag,Wilson:1971bg,Wilson:1971dh,Wilson:1973jj,Wilson:1974mb, Wilson:1983xri}), is that many different microscopic theories, when ``coarse-grained" to larger scales, can ``flow" to the same long-distance behavior; this flow can be described by the RG equations. When a particular QFT flows under the RG transformations to a ``fixed point", where the theory becomes scale-invariant, the symmetry often extends to conformal invariance. It is for this reason that the RG has been instrumental in identifying conformal field theories.
%
%

Due to the high degree of symmetry that conformal field theories possess, they can often be analysed using an approach known as the ``conformal bootstrap". The conformal bootstrap is a method to constrain and solve conformal field theories based on a limited set of fundamental principles, namely the symmetry properties of the theory and a few basic consistency conditions. The technique was pioneered by authors such as Mack and Salam \cite{Mack:1969rr}, Polyakov \cite{Polyakov:1970xd}, Ferrara, Grillo and Gatto, Migdal \cite{,,Migdal:1971fof,Ferrara:1973yt} (see also \cite{El-Showk:2012cjh,Poland:2018epd} for some recent developments). The core of the conformal bootstrap lies in the study of correlation functions, which are understood as the expectation value of a product of fields, often taken at different spacetime points. Since conformal field theories possess no overall energy-scale, the correlation functions of local operators/fields are the main observables of interest. Among these, the three-point functions hold a particularly significant position as they are fundamental building blocks that capture the simplest non-trivial interactions in the theory, serving as a cornerstone to our understanding of CFT dynamics. 

Given the symmetries inherent to CFT, notably scale and conformal invariance, the correlation functions are highly constrained. Specifically, conformal symmetry determines the overall form of three-point functions up to a few arbitrary parameters \cite{Osborn:1993cr}. This is in contrast to two-point functions, which are entirely fixed by conformal symmetry up to an overall coefficient, and four-point functions, which introduce significantly more complexity and are generically fixed up to arbitrary functions of conformally invariant cross-ratios constructed from four points \cite{Dolan:2000ut}. The three-point functions therefore offer the simplest non-trivial window into the dynamics of a given conformal field theory. The importance of three-point functions is multifold; first, the constants that define them are related to the Operator Product Expansion (OPE) coefficients \cite{Osborn:1993cr}. These coefficients provide information about how fields interact at short distances. Therefore, determining three-point functions grants insight into the OPE coefficients, elucidating the short-distance behavior of the theory. Second, higher-order correlation functions can often be understood by decomposing them in terms of products of three-point functions and conformal blocks, making the knowledge of three-point functions pivotal for exploring more complicated interactions.

Superconformal field theories (SCFTs) bring additional richness to this landscape. These theories possess superconformal symmetry, which is a combination of supersymmetry and conformal symmetry. However, according to Nahm's classification superconformal groups \cite{Nahm:1977tg} exist only in dimensions $d \leq 6$. The motivation for studying (super)conformal field theory in four dimensions was stimulated by the discovery of non-trivial fixed points in a large class of $\cN=2$ supersymmetric theories \cite{Howe:1983wj, Seiberg:1994aj, Argyres:1995xn}, as well as the famous $\cN=4$ supersymmetric Yang-Mills theory \cite{Sohnius:1981sn}, which has a vanishing beta function. The fixed points of these theories are non-trivial as they arise for non-zero values of the couplings, such that the conformal field theories are interacting. These discoveries have stimulated the study of (S)CFTs in general dimensions. In superconformal field theories the structure of the OPEs becomes more intricate, however the bootstrap approach remains potent; one can constrain the general form of correlation functions of superconformal primaries, and the consistency conditions derived can be used to learn information about the (S)CFT in question. In recent years, the (super)conformal bootstrap has undergone a resurgence, particularly with the development of advanced computational techniques. These tools have allowed researchers to determine the allowed values for OPE coefficients and scaling dimensions in a variety of (S)CFTs, see e.g. the review \cite{Poland:2018epd}. \\[4mm]
\noindent\textbf{Higher-spin symmetry}

Another possible extension to the symmetries of spacetime which has received continued interest is higher-spin symmetry. Initiated by pioneers such as Dirac \cite{Dirac:1936tg}, Fierz and Pauli \cite{Fierz:1939ix}, and Rarita and Schwinger \cite{Rarita:1941mf}, the idea of higher-spin symmetry is centered around the proposed existence of fields with spin $s>2$ propagating on a background spacetime. The most well-understood fields in quantum field theory are of spin-0 (scalars), spin-1/2 (fermions), and spin-1 (gauge bosons, such as photons or gluons), while gravitational interactions would introduce a hypothetical spin-2 tensor boson, the graviton. However, in accordance with Wigner's classification \cite{Wigner:1939cj}, a particle's spin takes non-negative (half)-integer values, $s = 0, \tfrac{1}{2}, 1, \tfrac{3}{2}, \dots$, with higher-spin fields being identified with higher-rank tensor representations of the Poincar\'e group. These higher-spin fields possess redundant degrees of freedom in the form of gauge symmetries, hence, field theories involving higher-spin fields are known as higher-spin gauge theories. 

The problem of constructing field theories describing the consistent propagation and interactions of higher-spin fields ($s > 2$) has a long history. In particular, for massless fields, the possible interactions are highly constrained by the higher-spin gauge symmetry which is required to decouple the redundant degrees of freedom. In flat spacetime, there are also powerful no-go theorems such as the Weinberg low-energy theorem \cite{Weinberg:1964ew}, the Coleman and Mandula theorem \cite{Coleman:1967ad}, and the Weinberg-Witten theorem \cite{Weinberg:1980kq} (see also \cite{Bekaert:2010hw} for a review) which severely constrain interactions involving massless HS particles. One of the main difficulties is that the inclusion of massless higher-spin fields results in higher-order conservation equations which are too restrictive to allow, under general assumptions, for a non-trivial S-matrix. Moreover, the Coleman-Mandula theorem forbids the existence of higher-spin symmetry generators (equivalently, conserved charges). 

Despite the issues above, significant progress has been made towards constructing models of higher-spin fields. At the classical level, Lagrangian formulations for massive higher-spin fields propagating on flat spacetime (and also (Anti-)de Sitter backgrounds) were first constructed by Singh and Hagen \cite{Singh:1974qz,Singh:1974rc}. However, these models are not gauge invariant; by considering the massless limit, Fronsdal \cite{Fronsdal:1978rb,Fronsdal:1978vb} succeeded in constructing a gauge invariant formulation for massless fields of an arbitrary integer spin. This was later extended by Fang and Fronsdal \cite{Fang:1978wz,Fang:1979hq} to half-integer spins. In the massive case, gauge invariant actions for higher-spin theories of an arbitrary integer spin in four dimensional Minkowski space were constructed by Klishevich and Zinoviev in \cite{Klishevich:1997pd}. The formulation was later generalised to $d$-dimensional (anti-)de Sitter spacetimes by Zinoviev in \cite{Zinoviev:2001dt}. The covariant quantisation of this model in Minkowski space was recently carried out in \cite{fegebank2023quantisation}.

A prominent motivation for the continued study of higher-spin theories is the promise they hold for a deeper understanding of quantum gravity. This is because the flat spacetime no-go theorems mentioned above can be circumvented if one assumes a non-zero cosmological constant, for example in Anti-de Sitter (AdS) backgrounds, where S-matrix arguments do not apply. Fradkin and Vasiliev \cite{Fradkin:1987ks} explicitly constructed cubic vertices of massless higher-spin fields in (A)dS. Subsequently, Vasiliev's equations, formulated in the late 20th century \cite{Vasiliev:1990en}, offered a fully non-linear realization of higher-spin gauge theories in Anti-de Sitter spacetimes. However, many questions remain unanswered concerning the fully consistent coupling of higher-spin fields to gravity and matter (see e.g. \cite{Vasiliev:2001ur,Vasiliev:2003cph,Sorokin:2004ie,Bekaert:2004qos,Bekaert:2010hw} for some reviews of recent progress). As far as applications are concerned, it has been conjectured that higher-spin gauge theories emerge at the tensionless limit of string theory \cite{Sundborg:2000wp} (see also \cite{Gross:1988ue}). In addition, higher-spin gauge theories are also central to AdS/CFT correspondence. Initially proposed by Maldacena \cite{Maldacena:1997re} and formulated more precisely by Witten \cite{Witten:1998qj} and Gubser, Klebanov and Polyakov \cite{Gubser:1998bc}, the AdS/CFT correspondence is a conjectured equivalence between string theory in an Anti-de Sitter spacetime and a conformal field theory formulated on its boundary. The duality potentially allows for higher-spin theories in Anti-de Sitter spacetimes to be dual to certain strongly coupled conformal field theories in one dimension lower. Hence, it provides a way to study quantum gravity problems using tools from conformal field theory. The prominent examples of AdS/CFT duality include the higher-spin/vector model conjecture of Klebanov and Polyakov \cite{Klebanov:2002ja} (see also \cite{Sezgin:2003pt,Leigh:2003gk}), and more recently the extensions of these conjectures to Chern-Simons theories coupled to vector models \cite{Aharony:2011jz,Giombi:2011kc}. The duality could offer insights into the nature of quantum gravity and the unification of the fundamental forces.\\[4mm]
\noindent\textbf{Correlation functions of conserved currents}

It is well-known that Noether's theorem \cite{Noether:1918zz} elegantly links the symmetries of a physical system to conserved quantities. In particular, for any physical system described by an action principle, Noether's theorem provides a construction of a conserved current associated with any continuous symmetry. The fundamental examples of conserved currents in any conformal field theory are the energy-momentum tensor and vector currents. The energy-momentum tensor is associated with invariance under spacetime translations, while vector currents are generally associated with internal symmetries. In a Poincar\'e invariant theory, one can always construct an energy-momentum tensor (Belinfante tensor), $T_{mn}$, which is both covariantly conserved, $\pa^{m} T_{mn} = 0$, and symmetric, $T_{mn} = T_{nm}$. However, in theories with conformal symmetry the energy-momentum tensor can also be made traceless, i.e. $\eta^{m n} T_{m n} = 0$, with the addition of suitable improvement terms. This is a defining property of any conformally invariant field theory. However, for these symmetries to hold in the quantum theory, the correlation functions of local operators must satisfy Ward-Takahashi identities, which are quantum analogues of the classical properties of the energy-momentum tensor. It is the correlation functions of conserved currents that contain fundamental information about the symmetries and interactions of a given conformal field theory. Similar considerations also hold for conserved higher-spin currents, which are associated with higher-spin symmetries. Hence, the study of their correlation functions is of fundamental importance. Indeed, the study of CFT correlation functions (specifically the three-point functions of conserved higher-spin currents) is central to tests of the AdS/CFT correspondence \cite{Gubser:1998bc,Freedman:1998tz} (see also \cite{Freedman:1998bj,Liu:1998ty,Mikhailov:2002bp,Giombi:2009wh,Aharony:2012nh,Giombi:2016zwa}). Some interesting tests were carried out recently in \cite{Sleight:2016dba} (see also \cite{Bekaert:2015tva,Sleight:2016hyl}); by employing the method of holographic reconstruction, explicit expressions for cubic interaction vertices for the type A minimal bosonic higher-spin gauge theory on $\text{AdS}_{d+1}$ were obtained from the three-point functions of conserved currents in the $d$-dimensional free $O(N)$ vector model.

In this thesis we are particularly interested in analysing the structure of three-point correlation functions of conserved currents for arbitrary integer or half-integer spins. The three-point functions of the energy-momentum tensor and vector currents were analysed by Osborn and Petkou, and Erdmenger and Osborn in \cite{Osborn:1993cr, Erdmenger:1996yc}, who introduced the systematic approach to study correlation functions of primary operators in conformal field theory in general dimensions ($d>2$) (see also refs.~\cite{Polyakov:1970xd, Schreier:1971um, Migdal:1971xh, Migdal:1971fof, Ferrara:1972cq, Ferrara:1973yt, Koller:1974ut, Mack:1976pa, Fradkin:1978pp, Stanev:1988ft} for earlier works). The formalism makes use of the fact that the conformal group is generated by rotations, translations and inversions. The three-point functions of primary operators of arbitrary spin are then constructed in terms of geometric building blocks which are functions of two and three points. Osborn and Petkou applied their formalism to conserved vector current operators $V_{m}$ of scale dimension $d-1$ and to the energy-momentum tensor $T_{mn}$ of dimension $d$, where it was shown that the three-point function of vector currents is fixed up to two independent tensor structures, while the three-point function of the energy-momentum tensor is fixed up to three independent tensor structures for $d>3$. For $d=4$ these three coefficient are related to certain coefficients in the expansion of the trace of the energy-momentum tensor in a curved background \cite{Osborn:1993cr}. However, for $d=3$ it was shown that there are only two possible linearly independent structures for the three-point function of the energy-momentum tensor. 

A particularly interesting feature of three-dimensional conformal field theories is that the three-point functions of conserved currents can contain parity-violating structures, which have been shown to be associated with theories of a Chern-Simons gauge field interacting with parity-violating matter \cite{Aharony:2011jz, Giombi:2011kc, Maldacena:2012sf, Jain:2012qi, GurAri:2012is, Aharony:2012nh, Giombi:2016zwa, Chowdhury:2017vel, Sezgin:2017jgm, Skvortsov:2018uru, Inbasekar:2019wdw}. These structures were not accounted for in the work of Osborn and Petkou, and instead were discovered later by Giombi, Prakash and Yin in~\cite{Giombi:2011rz}, where the analysis of three-point functions involving higher-spin currents was also undertaken (see also \cite{Costa:2011dw,Costa:2011mg}). In this study it was shown that the three-point functions of higher-spin currents also contain parity-violating structures and exist for arbitrary spins, provided that the spins of the currents satisfy triangle inequalities. Soon after, it was proven by Maldacena and Zhiboedov in~\cite{Maldacena:2011jn} that under certain assumptions all correlation functions involving the energy-momentum tensor and higher-spin currents are equal to those of free theories. This is an extension of the Coleman-Mandula theorem to conformal field theory.

The analysis of three-point functions of conserved higher-spin bosonic currents in four dimensions was undertaken by Stanev~\cite{Stanev:2012nq, Stanev:2013eha, Stanev:2013qra} (see also~\cite{Elkhidir:2014woa,Buchbinder:2022cqp}), and later by Zhiboedov \cite{Zhiboedov:2012bm} in general dimensions. In four dimensions it was shown that the number of independent structures in the three-point function of conserved bosonic ``vector-like"
currents, $J_{m_1\dots m_s}$, increases linearly with the minimum spin. 
This is quite different to the results found in three dimensions, where it has been shown that there are only three possible
independent conserved structures (see e.g. \cite{Giombi:2011rz,Maldacena:2011jn,Giombi:2016zwa,Jain:2021whr,Jain:2021gwa,Jain:2021vrv,Buchbinder:2022mys}).
For the correlation functions involving currents belonging to an arbitrary Lorentz representation in 4D CFT, less is known about their general structure. An approach to this problem was outlined in \cite{Elkhidir:2014woa}, however, it did not study correlation functions when the operators are all conserved currents. The intent of Chapter \ref{Chapter4} is to provide an answer to this question by providing a complete classification of three-point functions of conserved currents belonging to an arbitrary Lorentz representation in four-dimensional conformal field theory, where we obtain qualitatively new results.

The study of correlation functions in superconformal theories has been carried out in diverse dimensions using the group-theoretic approach developed in the following 
publications \cite{Park:1997bq, Osborn:1998qu, Park:1998nra, Park:1999pd, Park:1999cw, Kuzenko:1999pi, Nizami:2013tpa, Buchbinder:2015qsa, Buchbinder:2015wia, Kuzenko:2016cmf, Buchbinder:2021gwu, Buchbinder:2021izb, Buchbinder:2021kjk, Buchbinder:2021qlb, Jain:2022izp,Buchbinder:2021qlb,Buchbinder:2022cqp,Buchbinder:2022kmj,Buchbinder:2022mys}, where it has been shown that superconformal symmetry imposes additional restrictions on the three-point functions of conserved currents compared to non-supersymmetric theories. For example, it was pointed out
in~\cite{Buchbinder:2021gwu} that there is an apparent tension between supersymmetry and the existence of parity-violating structures in three-point functions of conserved currents. In contrast with the non-supersymmetric case, parity-odd structures are not found in the three-point functions of the energy-momentum tensor and conserved 
vector currents \cite{Buchbinder:2015qsa,Buchbinder:2015wia,Kuzenko:2016cmf,Buchbinder:2021gwu}. In general, for three-point functions 
involving conserved higher-spin currents, the conditions under which parity-violating structures can arise in supersymmetric theories are not well understood. The main goal of Chapter \ref{Chapter5} of this thesis is to clarify issues above by providing a complete classification of the three-point functions of conserved supercurrents in three-dimensional $\cN=1$ superconformal field theory.\\[4mm]
\noindent\textbf{Thesis outline}

This thesis is structured as a series of papers based on \cite{Buchbinder:2022mys,Buchbinder:2023coi,Buchbinder:2023fqv,Buchbinder:2023ndg}. We begin the thesis by presenting some background material in Chapter \ref{Chapter2}, which is devoted to reviewing some of the fundamental aspects of conformal field theory. We provide a ``pedestrian" discussion of conformal symmetry in general dimensions, the transformation properties of conformal primary fields, and the implications of conformal symmetry on the structure of energy-momentum tensor. We then discuss some general properties of correlation functions and the Ward-Takahashi identities associated with conformal symmetry. Finally, we provide an overview of the formalism of Osborn and Petkou \cite{Osborn:1993cr} and describe how it is used to construct two- and three-point correlation functions of conserved currents.

In Chapter \ref{Chapter3} we analyse the three-point functions of conserved currents in three-dimensional conformal field theory, where we develop an approach which augments the formalism of Osborn and Petkou with auxiliary spinors; this allows for the treatment of currents with an arbitrary integer or half-integer spin. In particular, we utilise a computational approach which efficiently and analytically imposes all the constraints on the three-point function which arise due to conservation equations and symmetries under permutations of spacetime points. We demonstrate that the three-point functions of conserved currents are generically fixed up to two parity-even structures and one parity-odd structure, with the existence of the latter depending on triangle inequalities in the spins of the currents. This chapter is based on the original work \cite{Buchbinder:2022mys}.

Next, in Chapter \ref{Chapter4}, we undertake the analysis of three-point correlation functions of conserved currents belonging to an arbitrary Lorentz representation in four-dimensional conformal field theory. In analogy with the three-dimensional construction, we show how to utilise the auxiliary spinor approach to classify the general structure of three-point functions and impose all the relevant constraints. In particular, we show that for vector-like higher-spin currents, the number of conserved structures in the three-point function increases linearly with the minimum spin, as is well known. However, for currents belonging to more general spin representations, we demonstrate that there are some special cases where the number of structures deviates from this result, which is qualitatively new. This chapter is based on the original works \cite{Buchbinder:2022cqp,Buchbinder:2023coi}.

Finally, Chapter \ref{Chapter5} is dedicated to analysing three-point functions of conserved supercurrents in three-dimensional $\cN=1$ superconformal field theory, where it is shown that supersymmetry imposes additional constraints on the structure of three-point functions of conserved currents. In particular, we utilise the superspace approach developed by Park and Osborn \cite{Park:1997bq,Osborn:1998qu,Park:1999cw,Park:1999pd} to construct manifestly superconformally covariant three-point functions. In analogy with the three-dimensional analysis, we develop the formalism by augmenting it with auxiliary spinors, which allows for the analysis of three-point functions of higher-spin supercurrents. We demonstrate that the three-point functions of conserved supercurrents which are Grassmann-even in superspace are generically fixed up to one parity-even structure and one parity-odd structure, with the existence of the latter depending on triangle inequalities in the superspins. For the three-point functions which are Grassmann-odd in superspace we use a method of irreducible decomposition to prove that they are fixed up to a single parity-even structure, while the parity-odd structure must vanish for arbitrary superspins. In particular, the construction of the parity-even sector is shown to reduce to solving a system of linear homogeneous equations with a tri-diagonal matrix of co-rank one. An explicit (unique) solution to this system of 
equations is obtained for arbitrary superspins. This chapter is based on the original works \cite{Buchbinder:2023fqv,Buchbinder:2023ndg} (see also the studies carried out in \cite{Buchbinder:2021gwu,Buchbinder:2021qlb}). 

In Chapter \ref{Chapter6} we conclude by discussing some potential applications of the results in this thesis and directions for future research.


 
\chapter{Background material}\label{Chapter2}
This chapter is devoted to a review of some essential topics in conformal field theory in general dimensions. We begin by discussing conformal symmetry and the transformation properties of primary fields. Next, we discuss conserved currents in conformal field theory, focusing primarily on the properties of the energy-momentum tensor. Finally, to motivate the analysis presented in the subsequent chapters, we explore the implications of conformal symmetry on the structure of correlation functions. In particular, we review the formalism of Osborn and Petkou \cite{Osborn:1993cr}, the group theoretic approach to constructing covariant two- and three-point correlation functions. The content in this chapter is not intended to be a complete description of these topics and we refer the reader to e.g. \cite{OsbornCFTNotes,Rychkov:2016iqz,Simmons-Duffin:2016gjk} and the relevant references cited throughout this chapter. The appendices \ref{Appendix2A} and \ref{Appendix2B} are devoted to the 3D and 4D conventions utilised throughout this thesis.

\section{Conformal symmetry in general dimensions}\label{section2.1}

\subsection{Conformal transformations}\label{subsection2.1.1}

Consider a $d$-dimensional Minkowski spacetime, $\mathbb{M}^{d} = \mathbb{R}^{1,d-1}$, with $d > 2$, parameterised by coordinates $x^{m} = (x^{0}, \vec{x})$ and metric $\eta_{m n} = \text{diag}(-1,1,...,1)$. A conformal transformation represented by a group element $g$ is a non-linear coordinate transformation which leaves the spacetime metric invariant up to a local scale factor \cite{Osborn:1993cr,DiFrancesco:1997nk,OsbornCFTNotes}
\begin{equation} \label{Ch02-Conformal transformations 1}
	x^{m}  \rightarrow x'^{m} = (g \cdot x)^{m} \, , \hspace{10mm} \text{d}x'^{2} = \Omega^{g}(x)^{2} \text{d}x^{2} \, , \hspace{10mm} \text{d} x^{2} = \eta_{m n} \text{d}x^{m} \text{d}x^{n} \, .
\end{equation}
For any conformal transformation we may define an associated local orthogonal transformation as follows:
\begin{equation} \label{Ch02-Conformal transformations 2}
	 \frac{\partial x'^{m}}{ \partial x^{n}} = \Omega^{g}(x) \, R^{m}{}_{ n}(x) \, , \hspace{10mm} R^{m}{}_{ n}(x) \in O(d-1,1) \, .
\end{equation}
When $\Omega(x) = 1$, we obtain conventional constant rotations and translations 
\begin{equation} \label{Ch02-Rotations and translations}
	x'^{m} = R^{m}{}_{n} \, x^{n} + a^{m} \, , \hspace{10mm} R^{m}{}_{n} \in O(d-1,1) \, , \hspace{10mm} \Omega^{(R,a)}(x) = 1 \, .
\end{equation}
where $a^{m}$ is an arbitrary constant four-vector. However, there are also constant scale transformations which form the dilatation group
\begin{equation} \label{Ch02-Dilatations}
	x'^{m} = \lambda  x^{m} \, , \hspace{10mm} \Omega^{\l}(x) = \lambda \, ,
\end{equation}
and the special conformal transformations given by
\begin{equation} \label{Ch02-Special conformal transformations}
	x'^{m} = \Omega^{b}(x) ( x^{m} + b^{m} x^{2} ) \, , \hspace{10mm} \Omega^{b}(x) = \frac{1}{1 + 2 b \cdot x + b^{2} x^{2}} \, .
\end{equation}
The set of coordinate transformations \eqref{Ch02-Conformal transformations 2} define the conformal group, which is isomorphic to $O(d,2)$ (see e.g. \cite{Kuzenko:2012tb} for a proof). 

An important role is also played by the conformal inversion, which is given by the transformation:
\begin{equation}
	x'^{m} = (i \cdot x)^{m} = \frac{x^{m}}{x^{2}} \hspace{5mm} \Rightarrow \hspace{5mm} R^{i}_{m n}(x) \equiv I_{m n}(x) = \eta_{m n} - \frac{2 x_{m} x_{n}}{x^{2}} \, , \hspace{5mm} \Omega^{i}(x) = x^{-2} \, ,
\end{equation}
where $I_{m n}(x)$ is known as the inversion tensor. It is essential to note that special conformal transformations \eqref{Ch02-Special conformal transformations} can be obtained by considering a composition of an inversion followed by a translation, followed by another inversion:
\begin{equation}
	x'^{m} = (( i \circ b \circ i ) \cdot x)^{m} = i \cdot \Big( \frac{x^{m}}{x^{2}} + b^{m} \Big) = \Omega^{b}(x) ( x^{m} + b^{m} x^{2} ) \, .
\end{equation}
Or, alternatively
\begin{equation}
	\frac{x'^{m}}{x'^{2}} = \frac{x^{m}}{x^{2}} + b^{m} \, , \hspace{10mm} x'^{2} = \frac{x^{2}}{1 + 2 b \cdot x + b^{2} x^{2}} \, .
\end{equation}
Hence, the full conformal group may be generated by combining rotations and translations, together with the inversion. 

It must be noted that the conformal inversion is not an element of the connected component of the conformal group, as $\text{det}(I) = -1$, and in general inversions are indeterminate for points on the lightcone $x^{2} = 0$ in Minkowski space. As a result, special conformal transformations are not globally defined in Minkowski space and the action of the conformal group must instead be realised on its conformal compactification (see e.g. \cite{Kuzenko:2012tb} for a construction). Hence, the discussion presented above for finite conformal transformations in Minkowski space is mostly formal and we note that only infinitesimal conformal transformations are well defined in general \cite{Buchbinder:1998qv}. 

To realise infinitesimal conformal transformations in Minkowski space we may instead consider infinitesimal transformations of the form
\begin{equation} \label{Ch02-Killing vectors}
	x'^{m} \approx x^{m} + \xi^{m}(x) \, , \hspace{10mm} \Omega^{g}(x) \approx 1 + \sigma(x) \, .
\end{equation}
From \eqref{Ch02-Conformal transformations 1}, we recall the transformation law for the Minkowski metric, $\eta_{m n}$, is
\begin{equation}
	\Omega^{g}(x)^{2} \eta_{m n} = \frac{\partial x'^{p}}{\partial x^{m}} \frac{\partial x'^{q}}{\partial x^{n}} \eta_{p q} \, .
\end{equation}
Substituting \eqref{Ch02-Killing vectors} into the above and keeping terms first order in $\xi$ and $\sigma$ we obtain the conformal Killing equation
\begin{align} \label{Ch02-Killing vector equation}
	2 \sigma \eta_{m n} &= \partial_{m} \xi_{n} + \partial_{n} \xi_{m} \, .
\end{align}
The solutions, $\xi^{m}(x)$, to the above equation are known as conformal Killing vectors, which span the conformal algebra. Eq. \eqref{Ch02-Killing vector equation} allows us to determine $\sigma$ as follows:
\begin{align} \label{Ch02-Killing vector divergence}
	\partial_{m} \xi^{m} = d \sigma \, .
\end{align}
Hence, $\s$ is determined in terms of $\xi$, and so we use the notation $\s(x) = \s_{\xi}(x)$. Now to find solutions to \eqref{Ch02-Killing vector equation} we compute the following sum:
\begin{align} \label{Ch02-Killing vector identity 1}
	& \frac{1}{2} \big( \partial_{p}(\partial_{m} \xi_{n} + \partial_{n} \xi_{m}) + \partial_{n}(\partial_{m} \xi_{p} + \partial_{p} \xi_{m}) - \partial_{m}(\partial_{p} \xi_{n} + \partial_{n} \xi_{p}) \big) \nonumber \\
	& \hspace{10mm} = \partial_{p} \partial_{n} \xi_{m} = \eta_{m p} \partial_{n} \sigma + \eta_{m n} \partial_{p} \sigma - \eta_{p n} \partial_{m} \sigma \, .
\end{align}
Now act on \eqref{Ch02-Killing vector equation} with $\partial_{m}$ to obtain:
\begin{equation}
	(d-2) \partial_{p} \partial_{n} \sigma = - \eta_{p n} \partial^{2} \sigma \, .
\end{equation}
We then take the trace of the expression above to obtain the result:
\begin{equation}
	(d-1) \partial^{2} \sigma = 0 \hspace{5mm} \Rightarrow \hspace{5mm} (d-1)(d-2) \partial_{m} \partial_{n} \sigma = 0 \, .
\end{equation}
This implies that, for $d>2$, $\sigma(x)$ is linear in $x$, i.e.
\begin{equation} \label{Ch02-Scaling function}
	\sigma(x) = \kappa - 2 b_{m} x^{m} \, .
\end{equation}
Substituting \eqref{Ch02-Scaling function} into \eqref{Ch02-Killing vector identity 1} we then obtain:
\begin{align}
	\partial_{p} \partial_{n} \xi_{m} = - 2 \eta_{m p} b_{n} -2  \eta_{m n} b_{p} + 2 \eta_{p n} b_{m} \, .
\end{align}
This may be solved consistently with \eqref{Ch02-Killing vector divergence} to arrive at the result:
\begin{equation} \label{Ch02-Conformal Killing vector}
	\xi^{m}(x) = a^{m} + \omega^{m}{}_{ n} x^{n} + \kappa x^{m} + b^{n} x^{2} I_{n}{}^{m}(x)\, , \hspace{10mm} \omega_{ m n} = - \omega_{ n m} \, ,
\end{equation}
where $a^{m}$, $\kappa$, $\omega^{m}{}_{n}$ and $b_{n}$ are real constant parameters and $I_{m}{}^{n}(x)$ is the aforementioned inversion tensor. The parameters $a^{m}$ and $\omega_{m n}$ correspond to Poincar\'e transformations, while the parameters $\kappa$ and $b^{m}$ induce scale and special conformal transformations. For understanding the infinitesimal transformations generated by $\xi^{m}$ it is also convenient to define $\hat{\omega}_{\xi,m n}(x) = \omega_{ m n} - 2 ( b_{m} x_{n} - b_{n} x_{m} )$, so that
\begin{equation} \label{Ch02-Conformal identity 3}
	\partial_{n} \xi_{m} = \sigma_{\xi} \eta_{m n} + \hat{\omega}_{\xi,m n} \, .
\end{equation}
As we will see later, the objects $\sigma_{\xi} $ and $ \hat{\omega}_{\xi,m n}$ are useful as they appear in the transformation formulae for conformal primary fields. We will often omit the subscript $\xi$ to streamline the notation.

Let us now count the number of independent components in the \eqref{Ch02-Conformal Killing vector}. The first term constitutes translations, for which there are $d$ parameters; the second term constitutes Lorentz transformations for which there are $\tfrac{1}{2}d(d-1)$ parameters; the third term corresponds to scale transformations, for which there is $1$ parameter, and the last two terms comprise the special conformal transformations, for which there are $d$ parameters. Therefore the total number of parameters defining conformal transformations is $\tfrac{1}{2}(d+1)(d+2)$. It is no coincidence that this is the same number of independent parameters used to describe $SO(d,2)$ rotations.

Now let $\xi_{(1)} = \xi_{(1)}^{m} \pa_{m}$, $\xi_{(2)} = \xi_{(2)}^{m} \pa_{m}$, be two conformal Killing vectors in Minkowski space. Their Lie bracket 
\begin{subequations}
\begin{gather}
	[ \xi_{(1)}, \xi_{(2)} ] = \xi_{(3)} = \xi_{(3)}^{m} \pa_{m} \, , \\
	\xi_{(3)}^{m} = \xi^{n}_{(1)} \pa_{n} \xi_{(2)}^{m} - \xi^{n}_{(2)} \pa_{n} \xi_{(1)}^{m}  \, ,
\end{gather}
\end{subequations}
results in another conformal Killing vector field $\xi_{(3)}$, which also satisfies \eqref{Ch02-Killing vector equation}. Therefore, the set of all conformal Killing vectors forms a Lie algebra, known as the conformal algebra. For any conformal Killing vector $\xi^{m}$ we may now introduce a basis $\{ P_{m}, L_{m n}, D, K_{m}\}$ for the conformal algebra as follows:
\begin{equation}
	\xi = \xi^{m}(x) \, \pa_{m} = \text{i} \,  \Big\{ a^{m} P_{m} - \sfrac{1}{2} \omega^{m n} L_{m n} + \kappa D + b^{m} K_{m} \Big\} \, ,
\end{equation}
where
\begin{subequations}
\begin{gather}
	P_{m} = - \text{i} \pa_{m} \, , \hspace{10mm} L_{mn} = - \text{i} ( x_{m} \pa_{n} - x_{n} \pa_{m} ) \, , \\
	D = - \text{i} x^{m} \pa_{m} \, , \hspace{10mm} K_{m} = - \text{i} x^{2} I_{m}{}^{n}(x) \, \pa_{n} \, .
\end{gather}
\end{subequations}
These differential operators, which are known as the conformal generators, satisfy the following commutation relations
\begin{subequations}
	\begin{gather} \label{Ch02-Conformal algebra}
		[D,P_{m}] = \text{i} P_{m} \, , \hspace{10mm} [D,K_{m}] = - \text{i} K_{m} \, , \\[2mm] 
		[K_{m},P_{n}] = 2 \text{i} (\eta_{m n} D - L_{m n}) \, , \\[2mm]
		[L_{a b}, P_{m}] = \text{i} (\eta_{m a} P_{b} - \eta_{m b} P_{a}) \, , \\[2mm]
		[L_{a b}, K_{m}] = \text{i} (\eta_{m a} K_{b} - \eta_{m b} K_{a}) \, , \\[2mm]
		[L_{m n},L_{p q}] = \text{i} (\eta_{m p} L_{n q} - \eta_{m q} L_{n p} + \eta_{n q} L_{m p} - \eta_{n p} L_{m q} ) \, .
	\end{gather}
\end{subequations}
Note that all other commutators vanish. We now postulate that the conformal algebra is an abstract Lie algebra defined by the commutation relations \eqref{Ch02-Conformal algebra} for a given basis of elements $\{ \boldsymbol{P}_{m}, \boldsymbol{L}_{m n}, \boldsymbol{D}, \boldsymbol{K}_{m}\}$. A general element $X$ of this Lie algebra is of the form 
\begin{equation} \label{Ch02-General conformal algebra element}
	X = - \text{i} \,  \Big\{ a^{m} \boldsymbol{P}_{m} - \sfrac{1}{2} \omega^{m n} \boldsymbol{L}_{m n} + \kappa \boldsymbol{D} + b^{m} \boldsymbol{K}_{m} \Big\} \, . 
\end{equation}
The conformal group, corresponding to the transformations \eqref{Ch02-Conformal transformations 1}, may then be (formally) obtained by exponentiation of \eqref{Ch02-General conformal algebra element}, with almost all elements (note that the inversion is excluded) of the group given by
\begin{equation}
	g = \exp\Big[ - \text{i} \,  \Big\{ a^{m} \boldsymbol{P}_{m} - \sfrac{1}{2} \omega^{m n} \boldsymbol{L}_{m n} + \kappa \boldsymbol{D} + b^{m} \boldsymbol{K}_{m} \Big\} \Big] \, .
\end{equation}
When restricted to infinitesimal group elements, exponentiation of each generator leads to the transformations \eqref{Ch02-Rotations and translations}, \eqref{Ch02-Dilatations}, \eqref{Ch02-Special conformal transformations} respectively.


\subsection{Primary fields}\label{Ch02-subsection2.1.2}

Consider a field $\phi_{\cA}(x)$, where $\cA$ denotes a collection of Lorentz/spinor indices, with scale dimension $\Delta$. A representation $T$ under conformal transformations is induced using the following rule: 
%
\begin{equation} \label{Ch02-Finite Field transformation}
	\phi \rightarrow \phi' = T_{g} \, \phi \, , \hspace{8mm} (T_{g} \, \phi)_{\cA}(x') = \Omega^{g}(x)^{-\Delta} \mathcal{D}[R(x)]_{\cA}{}^{\cB} \, \phi_{\cB}(x) \, ,
\end{equation}
where $\mathcal{R}_{\cA}{}^{\cB}(x) := \mathcal{D}[R(x)]_{\cA}{}^{\cB} \in O(d-1,1)$ is a representation matrix for $R_{m}{}^{n}(x)$ in the conformal transformation \eqref{Ch02-Conformal transformations 2}. Such a field is called primary with dimension $\Delta$. Infinitesimally we have
\begin{align}
	\mathcal{R}_{\cA}{}^{\cB}(x) = \big[ e^{ \frac{1}{2} \hat{\omega}^{m n}(x) \Sigma_{m n}} \big]_{\cA}{}^{\cB} \approx \mathbbm{1} + \sfrac{1}{2} \, \hat{\omega}^{m n}(x) ( \Sigma_{m n} )_{\cA}{}^{\cB} \, , \hspace{5mm} \Omega^{g}(x) \approx 1 + \sigma(x) \, ,
\end{align}
where the spin matrices, $\Sigma^{m n}$, are matrices which form a representation of the Lie algebra $\mathfrak{so}(d-1,1)$, satisfying the commutation relations
\begin{equation} \label{Ch02-Sigma matrix algebra}
	[ \Sigma_{mn}, \Sigma_{pq} ] = \eta_{mp} \Sigma_{nq} - \eta_{np} \Sigma_{mq} + \eta_{mq} \Sigma_{np} - \eta_{nq} \Sigma_{mp} \, .
\end{equation}
Hence, the field transforms as:
\begin{align}
	\phi'_{\cA}(x') &= (1+ \sigma(x))^{-\Delta} \big\{ \mathbbm{1} + \sfrac{1}{2} \, \hat{\omega}^{m n}(x) ( \Sigma_{m n} )_{\cA}{}^{\cB} \big\} \, \phi_{\cB}(x) \nonumber \\
	&= \phi_{\cA}(x) - \Delta \sigma(x) \phi_{\cA}(x) + \sfrac{ 1 }{2} \, \hat{\omega}^{m n}(x) ( \Sigma_{m n} )_{\cA}{}^{\cB} \phi_{\cB}(x) \, ,
\end{align}
and so, by defining $\delta_{\xi} \phi_{\cA}(x) = \phi'_{\cA}(x) - \phi_{\cA}(x)$, we obtain the infinitesimal transformation law for conformal primaries:
\begin{align} \label{Ch02-Field transformation}
	\delta_{\xi} \phi_{\cA}(x) &= - \xi^{m}(x) \, \partial_{m} \phi_{\cA}(x) - \Delta \sigma(x) \phi_{\cA}(x) + \sfrac{1}{2} \, \hat{\omega}^{m n}(x) ( \Sigma_{m n} )_{\cA}{}^{\cB} \phi_{\cB}(x) \, .
\end{align}
Here, the $x$-dependent parameters $\s(x)$ and $\hat{\omega}^{m n}(x)$ (expressed in terms of the Killing vector) represent local scale and Lorentz transformations combined with special conformal transformations.

Let us now provide a general (and somewhat informal) overview of some particular field representations. For conformal primary fields in the vector representation transforming according to \eqref{Ch02-Field transformation}, we consider the spin matrices
\begin{align}
	(\Sigma_{mn})_{a}{}^{b} = \d_{n}{}^{b} \eta_{m a} - \d_{m}{}^{b} \eta_{n a} \, ,
\end{align}
with similar results holding for higher-rank representations. For complex representations there also exists a conjugate spin matrix $\bar{\Sigma}_{mn}$ satisfying \eqref{Ch02-Sigma matrix algebra}. The conjugate conformal primary field $\bar{\phi}$ transforms as
\begin{align}
	\delta_{\xi} \bar{\phi}_{\bar{\cA}}(x) &= - \xi^{m}(x) \, \partial_{m} \bar{\phi}_{\bar{\cA}}(x) - \Delta \sigma(x) \bar{\phi}_{\bar{\cA}}(x) - \sfrac{1}{2} \, \hat{\omega}^{m n}(x) \, \bar{\phi}_{\bar{\cB}}(x) ( \bar{\Sigma}_{m n} )^{\bar{\cB}}{}_{\bar{\cA}}  \, .
\end{align}
To consider conformal primary chiral spinors $\psi$, $\bar{\psi}$, it is necessary to first define the appropriate gamma matrices. For even dimensions, $d=2n$, we may define inequivalent $2^{n-1} \times 2^{n-1}$ chiral gamma matrices $\gamma_{m}, \bar{\gamma}_{n}$ satisfying the Clifford algebra
\begin{equation}
	\g_{m} \bar{\g}_{n} + \g_{n} \bar{\g}_{m} = 2 \eta_{m n} \mathbbm{1} \, , \hspace{10mm} \bar{\g}_{m} \g_{n} + \bar{\g}_{n} \g_{m} = 2 \eta_{m n} \bar{\mathbbm{1}} \, ,
\end{equation}
where $\mathbbm{1}, \bar{\mathbbm{1}}$ are identity matrices for chiral and anti-chiral spinors respectively. We may then impose the Hermeticity conditions
\begin{equation}
	\g^{\dagger}_{m} = - \g_{m} \,, \hspace{10mm} \bar{\g}^{\dagger}_{m} = - \bar{\g}_{m} \, , \hspace{10mm} \bar{\psi} = \psi^{\dagger} \, ,
\end{equation}
where $\mathbbm{1}^{\dagger} = \bar{\mathbbm{1}}$. With these conventions, in even dimensions we may choose $\Sigma_{m n} = - \frac{1}{2} \g_{[m} , \bar{\g}_{n]} $, $\bar{\Sigma}_{m n} = - \frac{1}{2} \bar{\g}_{[m} , \g_{n]} $. In odd dimensions we have $\bar{\g}_{m} = \g_{m}$ and we use $\Sigma_{m n} = - \frac{1}{2} \g_{[m} , \g_{n]} $. The gamma matrices are typically chosen to be proportionate to the Pauli matrices. In three and four spacetime dimensions we use the conventions of \cite{Buchbinder:1998qv,Kuzenko:2010rp,Buchbinder:2015qsa}, which we outline in the appendices \ref{Appendix2A}, \ref{Appendix2B} respectively.

If we now restrict our attention to inversion transformations, then \eqref{Ch02-Finite Field transformation} becomes
\begin{equation} \label{Ch02-Inversion Field transformation}
	\phi'_{\cA}(i \cdot x)  = (x^{2})^{\Delta} \cI_{\cA}{}^{\cB}(x) \, \phi_{\cB}(x) \, ,
\end{equation}
where $\cI_{\cA}{}^{\cB}(x) = \mathcal{D}[ I(x)]_{\cA}{}^{\cB}$ is an appropriate representation matrix for the inversion tensor satisfying $\cI_{\cA}{}^{\cB}( \l x) = \cI_{\cA}{}^{\cB}(x)$ and $\cI_{\cA}{}^{\cA'}(x) \, \cI_{\cA'}{}^{\cB}(-x) = \delta_{\cA}{}^{\cB}$.
In general, the transformation $\phi \rightarrow \phi'$ for an inversion may not be possible, however, since $\text{det}(I) = -1$, inversions are connected to a reflection (parity transformation), denoted by $\mathfrak{R}_{0}$\footnote{Alternatively we could have defined the parity transformation to be a reflection of \textit{one} of the spatial coordinates. The analysis proceeds in much the same way.}
\begin{equation}
	x^{m} \rightarrow x'^{m} = \mathfrak{R}_{0} x^{m} \, , \hspace{5mm} \mathfrak{R}_{0} x^{0} = - x^{0} \, , \hspace{5mm} \mathfrak{R}_{0} \vec{x} = \vec{x} \, .
\end{equation}
To see this we may follow an argument by Osborn \cite{OsbornCFTNotes}; we consider the combination of a special conformal and scale transformation with a translation given by parameters $b^{m} = (\l, 0, \dots , 0 )$, $\kappa = \l^{2}$, $a^{m} = (-\l, 0, \dots, 0)$ respectively. We then have
\begin{align} \label{Ch02-Inversion transformation example}
	x'^{m}_{\l} = \frac{(1+\l^{2}) x^{m}}{1 + 2 \l x^{0} + \l^{2} x^{2}} \, , \hspace{3mm} m \neq 0 \, ,  \hspace{10mm} x'^{0}_{\l} = - \l + \frac{(1+\l^{2}) ( x^{0} + \l x^{2} )}{1 + 2 \l x^{0} + \l^{2} x^{2}} \, .
\end{align}
We note that $x'^{m}_{\l}|_{\l = 0} = x^{m}$, and as $\l \rightarrow \infty$ we have $x'^{m}_{\l} \rightarrow \mathfrak{R}_{0} x^{m}/x^{2}$. Hence, inversions are a symmetry provided that parity is also a symmetry.

For the action of inversion on vector fields and higher-rank symmetric and traceless tensors we use \eqref{Ch02-Inversion Field transformation}, where $\cI_{\cA}{}^{\cB}(x)$ is an appropriate symmetric and traceless tensor product of the fundamental inversion tensor, $I_{m}{}^{n}(x)$ \cite{Osborn:1993cr,DiFrancesco:1997nk}. Similar results also hold for infinitesimal inversion transformations. However, to describe the action of inversions on conformal primary chiral spinors $\psi$, $\bar{\psi}$, we may consider the infinitesimal form of the transformation \eqref{Ch02-Inversion transformation example} supplemented by the field transformation formula \eqref{Ch02-Field transformation}. After integrating the result we obtain the associated finite conformal transformation. In even dimensions we obtain, in the limit $\l \rightarrow \infty$
\begin{equation}
	\psi_{\l}(x) \rightarrow - (x^{2})^{-\D-\frac{1}{2}}  \gamma \cdot x \, \bar{\gamma}_{0} \psi( \mathfrak{R}_{0} \, x/x^{2}) \, .
\end{equation}
For parity-invariant theories, if we have spinors $\phi, \bar{\phi}$ of opposite chirality to $\psi, \bar{\psi}$ then the parity transformation is $\psi(x) \rightarrow \pm \gamma_{0} \, \phi(\mathfrak{R}_{0} \, x)$ (depending on convention), and so the action of inversion is
\begin{equation}
	\psi(x) \rightarrow (x^{2})^{-\D-\frac{1}{2}}  \gamma \cdot x \, \phi( x/x^{2}) \, .
\end{equation}
Hence, in the spinor representation the inversion tensor is of the form
\begin{align}
	\cI_{\cA}{}^{\bar{\cA}}(x) = (\gamma^{m})_{\cA}{}^{\bar{\cA}} \, \hat{x}_{m} \, , \hspace{10mm} \hat{x}_{m} = \frac{x^{m}}{\sqrt{x^{2}}} \, .
\end{align}
In general, spinors are mapped onto a complex conjugate representation by the inversion. Similar considerations also apply in odd dimensions, where the inversion maps spinors onto the same representation. For higher-rank representations of the spinor inversion, in particular for symmetric and traceless spin-tensors, one may simply take symmetric and traceless tensor products of the fundamental inversion. Hence, the action of inversions on fields is defined in any CFT.

\subsection{Field representations of the conformal algebra}\label{Ch02-subsection2.1.3}

The transformation law \eqref{Ch02-Field transformation} defines a representation of the conformal group on the space of fields. As a result, for a conformal transformation generated by a conformal Killing vector $\xi_{m}$, its action on fields can be expanded in terms of a basis of differential operators, $\{ \hat{P}_{m}, \hat{L}_{mn}, \hat{D}, \hat{K}_{m} \}$, as follows (field indices suppressed):
\begin{align}
	\delta_{\xi} \phi(x) = \text{i} \, \Big\{ - a^{m} \hat{P}_{m} + \sfrac{1}{2} \, \omega^{mn} \hat{L}_{m n} - \kappa \hat{D}  - b^{m} \hat{K}_{m} \Big\} \phi(x) \, ,
\end{align}
By using \eqref{Ch02-Field transformation} in addition to \eqref{Ch02-Conformal Killing vector}, \eqref{Ch02-Conformal identity 3}, the explicit form of the generators acting on fields may be read off
\begin{subequations}
	\begin{align} \label{Ch02-Conformal generators - differential operators}
		\hat{P}_{m} = - \text{i} \partial_{m} \hspace{38mm} & - \hspace{5mm} \textit{Translations} \\[2mm] 
		\hat{L}_{m n} = - \text{i} (x_{m} \partial_{n} - x_{n} \partial_{m}) - \text{i} \Sigma_{m n} \hspace{24mm} & - \hspace{5mm} \textit{Rotations} \\[2mm]
		\hat{D} = - \text{i} ( x^{m} \partial_{m} + \Delta ) \hspace{32mm} & - \hspace{5mm} \textit{Dilatations} \\[2mm]
		\hat{K}_{m} = - \text{i} x^{2} I_{m}{}^{n}(x) \, \partial_{n} + 2 \text{i} \Delta x_{m}  + 2 \text{i} x^{n} \, \Sigma_{m n} \hspace{5mm} & - \hspace{5mm} \textit{S.C.T}
	\end{align}
\end{subequations}
which are slightly modified versions of \eqref{Ch02-Conformal algebra}. The action of conformal transformations on a general field $\phi_{\cA}(x)$ may then be written in terms of commutators with each of the generators $\hat{P}_{m}$, $\hat{L}_{m n}$, $\hat{D}$ and $\hat{K}_{m}$ as follows:
\begin{subequations}
	\begin{align}
		\text{i} \, [ \hat{P}_{m} , \phi_{\cA}(x) ] &= \partial_{m} \phi_{\cA}(x) \, , \\[2mm]
		\text{i} \, [ \hat{L}_{m n} , \phi_{\cA}(x) ] &= (x_{m} \partial_{n} - x_{n} \partial_{m} ) \phi_{\cA}(x) + (\Sigma_{m n})_{\cA}{}^{\cB} \phi_{\cB}(x) \, , \\[2mm]
		\text{i} \, [ \hat{D} , \phi_{\cA}(x) ] &= (x^{m} \partial_{m} + \Delta ) \, \phi_{\cA}(x) \, , \\[2mm]
		\text{i} \, [ \hat{K}_{m} , \phi_{\cA}(x) ] &= x^{2} I_{m}{}^{n}(x) \, \partial_{n} \phi_{\cA}(x) - 2 x_{m}  \Delta  \phi_{\cA}(x) - 2 x^{n} ( \Sigma_{m n} )_{\cA}{}^{\cB} \phi_{\cB}(x)  \, .
	\end{align}
\end{subequations}
The generators satisfy the commutation relations \eqref{Ch02-Conformal algebra}, and hence they form a realisation of the conformal algebra acting on the space of fields. Note that the first two commutation relations in \eqref{Ch02-Conformal algebra} imply that $\hat{P}_{m}, \hat{K}_{m}$ act as raising and lowering operators for $\hat{D}$. In particular, for $\hat{K}_{m}$ we have
\begin{align}
	\hat{D} \hat{K}_{m} \phi(0) &= ( [ \hat{D}, \hat{K}_{m}] + \hat{K}_{m} \hat{D} ) \phi(0) \nonumber \\
	&= \text{i} (\D - 1) \hat{K}_{m} \phi(0) \, .
\end{align}
Thus, given an operator $\phi(0)$, we can repeatedly act with $\hat{K}_{m}$ to obtain new operators with arbitrarily low dimension.\footnote{The concept of radial quantization, described in many classic textbooks on CFT (e.g. \cite{DiFrancesco:1997nk}, see also the notes \cite{Simmons-Duffin:2016gjk}) allows one to define a Hilbert space of states in a CFT, which provides a mapping between local fields and operators. We will use these two terms interchangeably throughout this thesis.} On physical grounds, if we require the dimension to be bounded from below then this process will eventually terminate. Hence, there exists operators such that $[\hat{K}_{m}, \phi(0)] = 0$. Using \eqref{Ch02-Conformal algebra}, for fields at the origin we have:
\begin{subequations}
	\begin{align}
		[ \hat{L}_{mn}, \phi_{\cA}(0) ] &= - \text{i} (\Sigma_{m n})_{\cA}{}^{\cB} \phi_{\cB}(0) \, , \\
		[ \hat{D}, \phi_{\cA}(0) ] &= - \text{i} \Delta \phi_{\cA}(0) \, , \\
		[ \hat{K}_{m}, \phi_{\cA}(0) ] &= 0 \, ,
	\end{align}
\end{subequations}
which can be taken as the defining properties of a conformal primary field. Furthermore, given a conformal primary one can construct operators of higher-dimension called ``descendents", which are obtained by acting on a primary operator with momentum generators, which act as raising operators for dimension.

It is also essential to note that the commutation relations can be presented in a simpler form by defining the following generators:
\begin{subequations}
	\begin{align} \label{Conformal generators - combined}
		J_{m n} &= L_{m n} \, , & J_{-1,m} &= \frac{1}{2}(P_{m} - K_{m}) \, , \\[2mm] 
		J_{-1,0} &= D \, , & J_{0,m} &= \frac{1}{2}(P_{m} + K_{m}) \, .
	\end{align}
\end{subequations}
When defined in this way, the operators $J_{ab}$ with $a,b = -1,0,...,d$ form a representation of $\mathfrak{so}(d,2)$, as they satisfy
\begin{equation}
	[J_{a b},J_{c d}] = \text{i} (\eta_{b c} J_{a d} + \eta_{a d} J_{b c} - \eta_{a c} J_{b d} - \eta_{b d} J_{a c} ) \, .
\end{equation}
The fact that the conformal algebra is $\mathfrak{so}(d,2)$ is suggestive that it might be useful to think of conformal transformations as linear transformations in an ``embedding space" $\mathbb{R}^{d,2}$ instead of in $\mathbb{R}^{d-1,1}$, also known as the projective null-cone. The idea of the projective null-cone was introduced by Dirac \cite{Dirac:1936fq}, and has been studied by Mack and Salam \cite{Mack:1969rr}, Ferrara et. al. \cite{Ferrara:1972cq,Ferrara:1973yt}, Weinberg \cite{Weinberg:2010fx,Weinberg:2012mz} and many others. We will not discuss embedding space formalisms in this thesis, but these methods have been widely used in alternative approaches to the construction of correlation functions (see e.g. \cite{Costa:2011dw,Costa:2011mg,Costa:2014rya} for some recent work). The utility of the embedding space formalism depends on the specific problem under consideration.

\section{Conserved currents in conformal field theory}\label{Ch02-section2.2}

Suppose we have a classical field theory described by the action
\begin{equation}
	S[\phi, \partial_{m} \phi] = \int d^{d} x \, \mathcal{L} (\phi , \partial_{m} \phi ) \, .
\end{equation}
Given any continuous symmetry of the action, Noether's theorem \cite{Noether:1918zz} provides a construction of a conserved current. Now let us suppose that the field $\phi(x)$ transforms in the following way under a symmetry transformation (we utilise the construction presented in e.g. \cite{DiFrancesco:1997nk,Srednicki:2007qs}):
\begin{equation}
	x^{m} \rightarrow x'^{m}(x) \, , \hspace{10mm} \phi'(x') = \mathcal{F}(\phi(x)) \, .
\end{equation}
Accordingly, we have the associated infinitesimal transformation
\begin{subequations}  \label{Ch02-Infinitesimal field transformation}
	\begin{gather}
		 x^{m} \rightarrow x'^{m}(x) = x^{m} + \d_{\omega} x^{m} \, , \hspace{10mm} \delta_{\omega}x^{m} = \omega^{A} \frac{\delta x^{m} }{\delta \omega^{A}} \, , \nonumber \\[2mm]
		\phi(x) \rightarrow \phi'(x) = \phi(x) + \omega^{A} \frac{\d \mathcal{F}}{ \d \omega^{A}} \, ,
	\end{gather}
\end{subequations}
where $\omega^{A}$ are infinitesimal parameters (here $A$ denotes a collection of internal or spacetime indices). If this infinitesimal transformation is a symmetry of the action $S[\phi, \pa^{m} \phi]$, then we must have $\delta_{\omega} S = S' - S = 0$. The transformed action is given by
\begin{equation}
	S[\phi', \partial'_{m} \phi'] = \int d^{d} x \, \left\lvert \frac{\partial x'^{m}}{\partial x^{n}} \right\rvert \, \mathcal{L} \bigg(\phi + \omega^{A} \frac{\d \mathcal{F}}{ \d \omega^{A}} \, , \frac{\partial x^{n} }{ \partial x'^{m}} \partial_{n} \big( \phi + \omega^{A} \frac{\d \mathcal{F}}{ \d \omega^{A}} \big) \bigg) \, ,
\end{equation}
with
\begin{subequations}
	\begin{align}
		\frac{\partial x'^{m}}{ \partial x^{n}} &= \delta^{m}_{\; n} + \partial_{n} \bigg( \omega^{A} \frac{\d x^{m} }{\d \omega^{A}} \bigg) \, , \\[2mm]
		\text{det}(\mathbbm{1} + \epsilon) \approx 1 + \text{tr}(\epsilon) \hspace{5mm} &\Rightarrow \hspace{5mm} \left\lvert \frac{\partial x'^{m}}{\partial x^{n}} \right\rvert \approx 1 + \partial_{m} \bigg( \omega^{A} \frac{\d x^{m} }{\d \omega^{A}} \bigg) \, .
	\end{align}
\end{subequations}
Note that above, $\epsilon$ is an infinitesimal $d \times d$ matrix. After expanding the Lagrangian to first order in $\omega^{A}$, we then let $\omega^{A} \rightarrow \omega^{A}(x)$ and integrate by parts (with suitable boundary conditions) to obtain:
\begin{equation} \label{Ch02-Noether current - action variation}
	\delta _{\omega} S = - \int d^{d} x \, j^{m}_{A} \, \partial_{m} \omega^{A}(x) = \int d^{d} x \, \partial_{m} j^{m}_{A} \, \omega^{A}(x) \, ,
\end{equation}
where the current, $j^{m}_{A}(x)$, has the following form:
\begin{align} \label{Ch02-Noether current}
	j^{m}_{A}(x) = \bigg\{\frac{\partial \mathcal{L}}{\partial (\partial_{m} \phi) } \partial_{n} \phi  -\eta^{m n} \mathcal{L} \bigg\} \, \frac{\d x^{n}}{\d \omega^{A}} - \frac{\partial \mathcal{L}}{\partial (\partial_{m} \phi)} \frac{\d \mathcal{F}}{\d \omega^{A}} \, .
\end{align}
Noether's theorem states that if the field obeys the classical equations of motion
\begin{equation}
	\frac{\pa \cL}{\pa \phi} - \pa_{m} \bigg( \frac{\pa \cL}{\pa ( \pa_{m} \phi)} \bigg) = 0 \, ,
\end{equation}
then the action is invariant under an arbitrary variation of the fields, i.e., $\delta S$ must vanish for arbitrary parameters $\omega^{A}$. From \eqref{Ch02-Noether current - action variation} we then obtain the conservation equation
\begin{equation}
	\partial_{m} j^{m}_{A}(x) = 0 \, .
\end{equation}
In other words, a continuous symmetry of the action implies the existence of a conserved current given by \eqref{Ch02-Noether current}. Note that although we have restricted the discussion above to scalar fields, it may be readily extended to a multiplet of fields $\phi_{I} = ( \phi_{1}, \dots , \phi_{N})$ with the addition of appropriate indices.

\subsection{Energy-momentum tensor}\label{Ch02-subsection2.2.1}

Using the general form of the conserved current \eqref{Ch02-Noether current} and the conformal Killing vector \eqref{Ch02-Conformal Killing vector} for the infinitesimal transformations, a conformally invariant classical action $S[\phi, \pa^{m} \phi]$ gives rise to the following conserved currents:
\begin{subequations}
	\begin{align}
		T_{c}^{m n} = -\eta^{m n} \mathcal{L} + \frac{\partial \mathcal{L}}{\partial (\partial_{m} \phi)} \partial_{n} \phi \hspace{20mm} & - \hspace{5mm} \textit{Translation symmetry} \\[2mm]
		\mathcal{J}^{m n p} = T_{c}^{m n} \, x^{p} - T_{c}^{m p} \, x^{n} + \frac{\text{i}}{2} \frac{\partial \mathcal{L}}{\partial (\partial_{m} \phi)} \Sigma^{n p} \phi \hspace{5mm} & - \hspace{5mm} \textit{Lorentz symmetry} \\[2mm]
		j_{D}^{m} = T_{c \hspace{1mm} n}^{m} x^{n} + \frac{\partial \mathcal{L}}{\partial (\partial_{m} \phi)} \Delta \phi \hspace{20mm} & - \hspace{5mm} \textit{Dilatation symmetry}
	\end{align}
\end{subequations}
However, it must be noted that the canonical energy momentum tensor, $T_{c}^{m n}$, which arises as a result of the translational symmetry, is not manifestly symmetric, nor is it traceless in general. We can resolve this issue by exploiting the freedom of the energy momentum tensor and introducing improvement terms. Indeed if we define
\begin{equation} \label{Modified EM tensor}
	T^{m n} = T_{c}^{m n} + \partial_{p} B^{p m n} + \frac{1}{2} \partial_{p} \partial_{q} X^{p q m n} \, ,
\end{equation}
where the improvement terms are given by
\begin{subequations}
	\begin{align}
		B^{p m n} = \frac{\text{i}}{4} \bigg\{ \frac{\partial \mathcal{L}}{\partial (\partial_{p} \phi)} \Sigma^{m n} \phi - \frac{\partial \mathcal{L}}{\partial (\partial_{m} \phi)} \Sigma^{p n} \phi -  \frac{\partial \mathcal{L}}{\partial (\partial_{n} \phi)} \Sigma^{p m} \phi  \bigg\} \, , \\[4mm]
		X^{q p m n} = \frac{4}{d-2} \Big\{ \eta^{q [ p} \sigma_{+}^{m ] n} + \eta^{m [ n} \sigma_{+}^{q ] p} + \frac{1}{d-1} \, \eta^{q [p} \eta^{m ] n} \, \sigma^{a}_{+ \; a} \Big\} \, ,
	\end{align}
\end{subequations}
then it is possible to construct an energy-momentum tensor that is manifestly conserved, symmetric and traceless provided that $\sigma_{+}^{m n} = \sigma^{(m n)}$, where $\sigma^{m n}$ is constructed as a solution to the equation
\begin{equation}
	V^{m} = \partial_{n} \sigma^{n m} \, , \hspace{10mm} V^{m} = \frac{\partial \mathcal{L}}{\partial(\partial^{p} \phi)} \Big\{ \eta^{m p} \Delta + \frac{\text{i}}{2} \Sigma^{m p} \Big\} \phi \, .
\end{equation}
Here, $V^{m}$ is known as the Virial current. Note that the first two terms in \eqref{Modified EM tensor} constitute the Belinfante tensor, which can be constructed in any Poincar\'e invariant theory \cite{Weinberg:1995mt}. The last term is added to ensure overall tracelessness of the modified energy-momentum tensor. Furthermore, the symmetry properties of $B^{p m n}$ and $X^{p q m n}$ imply that conservation of the energy momentum tensor is preserved, as
\begin{equation}
	\partial_{m} \partial_{p} B^{p m n} = 0  \, , \hspace{10mm} \partial_{m} \partial_{p} \partial_{q} X^{p q m n} = 0 \, .
\end{equation}
In addition, the symmetry property $T_{m n} = T_{n m}$ of the energy momentum tensor is also preserved, as
\begin{equation}
	X^{q p [m n]} = \frac{2}{(d-1)(d-2)} \eta^{q [m} \eta^{n ] p} \, \sigma^{a}_{+ \; a} \, ,
\end{equation}
which vanishes upon contraction with partial derivatives. The trace of this term is
\begin{align}
	\frac{1}{2} \partial_{q} \partial_{p} X^{q p m}{}_{ m} &= \partial_{q} \partial_{p} \sigma_{+}^{q p} = \partial_{m} V^{m} \nonumber \\[2mm]
	&= \partial_{m} \bigg( \frac{\partial \mathcal{L}}{\partial(\partial_{m} \phi)} \Delta \phi \bigg) + \partial_{m} \bigg( \frac{\text{i}}{2} \frac{\partial \mathcal{L}}{\partial(\partial^{p} \phi)} \Sigma^{m p} \phi \bigg) \, .
\end{align}
Now, since
\begin{equation}
	\partial_{p} B^{p m}{}_{ m} = \frac{\text{i}}{2} \, \partial_{p} \bigg( \frac{\partial \mathcal{L}}{\partial(\partial^{m} \phi)} \Sigma^{m p} \phi \bigg) \, ,
\end{equation}
and, from conservation of $j_{D}$ we have
\begin{equation} \label{EM tensor symmetries and constraints}
	T_{c \; m}^{m} = - \partial_{m} \bigg( \frac{\partial \mathcal{L}}{\partial (\partial_{m} \phi)} \Delta \phi \bigg) \, ,
\end{equation}
we find that
\begin{align}
	T^{m}_{\; m} &= T_{c \; m}^{m} + \partial_{p} B^{p m}{}_{ m} +  \frac{1}{2} \partial_{q} \partial_{p} X^{q p m}{}_{ m} = 0 \, .
\end{align}
Therefore, the modified energy momentum tensor, $T^{m n}$, satisfies the following properties:
\begin{equation}
	\partial_{m} T^{m n} = 0 \, , \hspace{10mm} T^{[m n]} = 0 \, , \hspace{10mm} T^{m}_{\hspace{2mm} m} = 0 \, .
\end{equation}
Hence, we have shown that the scaling symmetry implies that one can construct an energy-momentum tensor that is manifestly traceless, provided that the Virial condition applies. A solution to the virial condition can be found for a broad class of field theories \cite{DiFrancesco:1997nk}.

In general, the question of whether scale invariance may be extended to conformal invariance has remained an important problem in conformal field theory (see e.g. \cite{Polchinski:1987dy}) and is intimately related to the renormalisation group. In two-dimensions the Zamolodchikov $c$-theorem \cite{Zamolodchikov:1986gt} demonstrated (using renormalisation group flow) that scale invariance implies conformal invariance at a renormalisation group fixed point. Essentially, Zamolodchikov proved that there is a quantity $C$ which decreases along renormalisation group flows from an ultraviolet to an infrared fixed point. At the fixed points, $C$ takes the value of the Virasoro central charge $c$. The quantity $C$ may be interpreted as a measure of the degrees of freedom, with its decrease corresponding to a reduction in the degrees of freedom in the large distance limit, where only massless fields are relevant. In four dimensions, attempts to construct a similar theorem were made by Cardy \cite{Cardy:1988cwa} and subsequently by Jack and Obsorn \cite{Osborn:1989td,Jack:1990eb}, where it is known as the $a$-theorem ($a$ is used as it relates to the trace anomaly of the energy-momentum tensor), however the results only apply perturbatively. More recently, Komargodski and Schwimmer \cite{Komargodski:2011vj} (see also \cite{Luty:2012ww}) managed to obtain a non-perturbative result, establishing the $a$-theorem in four-dimensions. The four dimensional analogue to the $c$-theorem might be an interesting tool for understanding the structure of QCD.

There is also an alternative approach to the construction of the energy-momentum tensor which uses a variation of Noether's theorem. The method utilises the constraints imposed on the conformal parameters $\xi_{m}$, $\sigma_{\xi}$ and $\hat{\omega}_{\xi,m n}$ defined in \eqref{Ch02-Conformal identity 3}. As is well known, if a CFT is derived from a conformally invariant action, then Noether's theorem provides a construction of the energy momentum tensor. Now suppose our theory is described by an action $S[\phi, \pa^{m} \phi]$ which is a functional of fundamental fields $\phi$ and its derivatives; because the action is conformally invariant, for Killing vectors $\xi_{m}$ there is an associated conformal transformation $\delta_{\xi}\phi$ such that $\delta_{\xi} S[\phi]=0$. Recall that conformal primary fields transform as in \eqref{Ch02-Field transformation}.
From the prior analysis of the conformal Killing vectors, we know that $\xi_{m}$, $\sigma_{\xi}(x)$ and $\hat{\omega}_{\xi,m n}(x)$ are constrained by the requirements of conformal invariance by the following four equations:
\begin{subequations}  \label{Ch02-Conformal parameter constraints}
	\begin{gather}
		\partial_{n} \xi_{m} - \hat{\omega}_{\xi,m n} - \sigma_{\xi} \eta_{m n} = 0 \, , \\
		\partial_{p} \hat{\omega}_{\xi, m n} -  \eta_{p n} \partial_{m} \sigma_{\xi} +  \eta_{p m} \partial_{n} \sigma_{\xi} = 0 \, , \\
		\partial_{p} \partial_{n} \xi_{m} - \eta_{m p} \partial_{n} \sigma_{\xi} - \eta_{m n} \partial_{p} \sigma_{\xi} + \eta_{p n} \partial_{m} \sigma_{\xi} = 0 \, , \\
		\partial_{m} \partial_{n} \sigma_{\xi} = 0 \, .
	\end{gather}
\end{subequations}
We now extend conformal transformations $\delta_{\xi} \phi$ so as to allow $\xi_{m}$ to be unconstrained, and promote $\hat{\omega}_{\xi , m n} \rightarrow \omega_{m n}$ and $\sigma_{\xi} \rightarrow \sigma$. We then have
\begin{align} \nonumber
	\delta_{\xi,\omega,\sigma} \phi_{\cA}(x) = - \xi^{m}(x) \partial_{m} \phi_{\cA}(x) - \Delta \sigma(x) \phi_{\cA}(x) + \sfrac{1}{2} \, \omega^{m n}(x) ( \Sigma_{m n} )_{\cA}{}^{\cB} \phi_{\cB}(x) \, .
\end{align}
The variation of the action may then be written in the form
\begin{align} \label{Ch02-EM tensor variation}
	\begin{split}
		\delta_{\xi,\omega,\sigma} S[\phi_{\cA}] &= \int d^{d} x \, \big\{ (\partial_{n} \xi_{m} - \omega_{m n} - \sigma \eta_{m n} ) T_{c}^{m n} + (\partial_{p} \omega_{m n} -  \eta_{p n} \partial_{m} \sigma +  \eta_{p m} \partial_{n} \sigma ) X^{p m n} \\
		& \hspace{20mm} + (\partial_{p} \partial_{n} \xi_{m} - \eta_{m p} \partial_{n} \sigma - \eta_{m n} \partial_{p} \sigma + \eta_{p n} \partial_{m} \sigma) Y^{m n p} + (\partial_{m} \partial_{n} \sigma ) Z^{m n}  \big\} \, ,
	\end{split}
\end{align}
where the fields $X$, $Y$ and $Z$ possess the following symmetry properties (consistent with the symmetries of the constraints \eqref{Ch02-Conformal parameter constraints}):
\begin{equation}
	X^{p m n} = - X^{p n m} \, , \hspace{10mm} Y^{m p n} = Y^{m n p} \, , \hspace{10mm} Z^{m n} = Z^{n m} \, .
\end{equation}
When restricted to infinitesimal conformal transformations, the variation vanishes and the equations of motion are found by requiring $\delta S[\phi_{\cA}] = 0$ for arbitrary $\delta \phi_{\cA}$. If we now vary $\xi_{m}$, $\omega_{ m n}$ and $\sigma$ independently we obtain:
\begin{subequations}
	\begin{align}
		\partial_{m} T_{c}^{m n} &= \partial_{p} \partial_{m} Y^{n p m} \, , \\[2mm]
		T_{c}^{[m n]} &= -\partial_{p} X^{p m n} \, , \\[2mm]
		\eta_{m n} T_{c}^{m n} &= 2 \eta_{p n} \partial_{m} X^{p m n} + 2 \eta_{p n} \partial_{m} Y^{p n m} - \eta_{p n} \partial_{m} Y^{m p n} + \partial_{m} \partial_{n} Z^{m n}  \, .
	\end{align}
\end{subequations}
We wish the right hand side of all of the above to vanish, which is non-trivial in general. Thus, we construct the modified energy momentum tensor $T^{m n}$ as follows:
\begin{equation}
	T^{m n} = T_{c}^{m n} + \partial_{p} ( X^{p m n} - X^{m p n} - X^{n p m} + Y^{p m n} - Y^{m n p} - Y^{n m p}) + \mathcal{D}^{m n \sigma p} Z_{\sigma p} \, ,
\end{equation}
with 
\begin{align}
	\begin{split}
		\mathcal{D}^{m n q p} &= \sfrac{1}{d-2} \big( \eta^{m (q} \partial^{p)} \partial^{n} + \eta^{n (q} \partial^{p)} \partial^{m} - \eta^{m (q} \eta^{p) n} \partial^{2} - \eta^{m n} \partial^{q} \partial^{p} \big) \\
		& \hspace{20mm} - \sfrac{1}{(d-1)(d-2)} \big( \partial^{m} \partial^{n} - \eta^{m n} \partial^{2} \big) \eta^{q p} \, ,
	\end{split}
\end{align}
which is constructed specifically to satisfy $\partial_{m} \mathcal{D}^{m n q p} = 0$, and $\eta_{m n} \mathcal{D}^{m n q p} = - \partial^{p} \partial^{q}$. As a result, $T^{m n}$ satisfies
\begin{equation} \nonumber
	\partial_{m} T^{m n} = 0 \, , \hspace{10mm} T^{[m n]} = 0 \, , \hspace{10mm} T^{m}{}_{ m} = 0 \, .
\end{equation}
As a consequence of the above analysis we now have
\begin{equation}
	\delta_{\xi,\omega,\sigma} S[\phi] = \int d^{d} x \, (\partial_{m} \xi_{n} - \omega_{m n} - \sigma \eta_{m n} ) \, T^{m n} \, .
\end{equation}

As a simple illustration of the construction above we may consider a free spinless scalar field with the action
\begin{align} \label{Ch02-Free scalar}
	S[\phi, \pa^{m} \phi] = - \tfrac{1}{2} \int d^{d}x \, \pa^{m} \phi \, \pa_{m} \phi \, , \hspace{10mm} \D_{\phi} = \tfrac{1}{2} (d-2) \, .
\end{align}
For a conformal transformation given by \eqref{Ch02-Field transformation}, the variation of $S$ may be expressed in the form of \eqref{Ch02-EM tensor variation} with 
\begin{align}
	T^{mn}_{c} = \pa^{m} \phi  \, \pa^{n} \phi - \sfrac{1}{2} \eta^{ m n} \pa^{p} \phi  \,\pa_{p} \phi \, , \hspace{5mm} Z^{m n} = - \tfrac{1}{2} (d-2) \eta^{m n} \phi^{2} \, .
\end{align}
We then find the general form of the improved energy-momentum tensor to be
\begin{align}
	T^{m n} = \pa^{m} \phi \, \pa^{n} \phi - \sfrac{1}{2} \eta^{ m n} \pa^{p} \phi \, \pa_{p} \phi  - \frac{d-2}{4(d-1)} ( \pa^{m} \pa^{n} - \eta^{m n} \pa^{2}) \phi^{2} \, .
\end{align}
As a result we find
\begin{align}
	\pa_{m}T^{m n} = \pa^{2} \phi \, \pa^{n} \phi \, , \hspace{10mm} \eta_{m n} T^{m n} = \tfrac{1}{2} (d-2) \phi \, \pa^{2} \phi \, ,
\end{align}
both of which vanish upon imposing the equations of motion $\pa^{2} \phi = 0$.


\subsection{Higher-spin currents}\label{Ch02-subsection2.2.2}

As we have established, a $d$-dimensional conformal field theory consists of local operators labeled by conformal dimension $\D$, and a representation $\cD[R\,]$ of $SO(d-1,1)$. It is also well known that in a unitary CFT, local operators satisfy unitarity bounds \cite{DiFrancesco:1997nk, OsbornCFTNotes}. For a spin-$s$ operator $J_{m_{1} \dots m_{s} }(x) = J_{(m_{1} \dots m_{s}) }(x)$ in the symmetric and traceless tensor representation of $SO(d-1,1)$, the unitarity bound is
\begin{align}
	\D \geq s + d - 2 \, , \hspace{10mm} s \geq 1 \, .
\end{align}
When the dimension saturates the unitary bound $(\D = s + d - 2)$, the field satisfies the conservation equation
\begin{equation} \label{Ch02-General conserved current}
	\pa^{m_{1}} J_{m_{1} \dots m_{s}} = 0 \, .
\end{equation}
Such a field is known as a conserved current of arbitrary integer spin $s$. Conserved currents are primary fields, as they possesses the following infinitesimal conformal transformation properties:
\begin{equation}
	\delta J_{m_{1} \dots m_{s}}(x) = - \xi J_{m_{1} \dots m_{s}}(x) - \Delta_{J} \, \s(x) J_{m_{1} \dots m_{s}}(x) + s \, \hat{\omega}_{( m_{1} }{}^{n}(x) \, J_{m_{2} \dots m_{s}) n}(x) \, , 
\end{equation}
where $\xi$ is a conformal Killing vector field, and $\s(x)$, $\omega_{m n}(x)$ are local parameters associated with local scale and combined Lorentz/special conformal transformations. Alternatively, the dimension $\Delta_{J}$ may be fixed by the conservation condition \eqref{Ch02-General conserved current}, which is conformally covariant provided that $\Delta_{J} = s + d - 2$. 
For currents of an arbitrary half-integer spin, one must fix the spacetime dimension and utilise the appropriate spin representation in addition to the transformation formula \eqref{Ch02-Field transformation}.

Higher-spin currents typically arise in free field theories. For example the free scalar field CFT \eqref{Ch02-Free scalar} possesses a larger symmetry, a ``higher-spin" extension of the conformal algebra. This may be seen by noting that one can construct, for each even spin $s$, a conserved spin-$s$ current. These currents are bilinears in the scalar field $\phi$ and involve higher derivatives. They are schematically of the form
\begin{align}
	J_{m_{1} \dots m_{s}} = \sum_{k = 0}^{s} c_{k}(s) \, \pa_{\{ m_{1}} \dots \pa_{m_{k}} \phi \, \pa_{m_{k+1}} \dots \pa_{m_{s} \}} \phi \, , \hspace{10mm} s = 2,4,6, \dots \, .
\end{align}
Here the brackets denote traceless symmetrisation, so that the current is overall totally symmetric and traceless, corresponding to an irreducible representation of $SO(d-1,1)$. These fields have dimension $\D = s + d - 2$, and by imposing the conservation condition
\begin{align}
	\pa^{m_{1}} J_{m_{1} \dots m_{s}} = 0 \, ,
\end{align}
and utilising the equation of motion $\pa^{2} \phi = 0$, it may be shown that the coefficients $c_{k}(s)$ are generated by Gegenbauer polynomials, see e.g. \cite{Craigie:1983fb} for details.\footnote{The explicit form of the higher-spin currents is more conveniently described after introducing auxiliary (null) vectors, $z^{m}$, satisfying $z^{m} z_{m} = 0$, and contracting the free tensor indices according to the rule $J_{s}(x;z) = J_{m_{1} \dots m_{s}}(x) z^{m_{1}} \dots z^{m_{s}}$. The explicit indices may then be restored by acting on this object with the ``Thomas derivative". We do not present this approach here, but see e.g. \cite{Costa:2011mg,Giombi:2016hkj}} 

The discussion above also applies to other free field models. An example which is of particular importance in the context of AdS/CFT is the free $O(N)$ vector model, described by the following action:
\begin{align}
	S[\phi, \pa^{m} \phi] = - \tfrac{1}{2} \int d^{d}x \, \pa^{m} \phi^{I} \, \pa_{m} \phi_{I} \, , \hspace{10mm} \D_{\phi} = \tfrac{1}{2} (d-2) \, ,
\end{align}
with equations of motion
\begin{align}
	\pa^{2} \phi^{I} = 0 \, , \hspace{10mm} I = 1, \dots, N \, .
\end{align}
The model has global $O(N)$ symmetry under which $\phi^{I}$ transforms in the vector representation. This model possesses conserved higher-spin currents carrying $O(N)$ indices
\begin{align}
	J^{IJ}_{m_{1} \dots m_{s}} = \sum_{k = 0}^{s} c_{k}(s) \, \pa_{\{ m_{1}} \dots \pa_{m_{k}} \phi^{I} \, \pa_{m_{k+1}} \dots \pa_{m_{s} \}} \phi^{J} \, ,
\end{align}
where the coefficients $c_{k}(s)$ are identical to those above. In this case the currents can be decomposed into irreducible representations of $O(N)$ 
\begin{align}
	J^{IJ}_{s} = \d^{IJ} J_{s} + J^{(IJ)}_{s} + J^{[IJ]}_{s} \, ,
\end{align}
and we obtain conserved currents of any integer spin $s = 1,2,3, \dots$.

The above models pertain to scalar CFTs, but one can also consider the theory of $N$ free massless fermions, which exists in general dimensions. For this we may consider the action of $N$ free fermions, $\psi^{I}$, $i = 1,\dots,N$ 
\begin{align}
	S[\psi, \bar{\psi}] = \int d^{d}x \, \bar{\psi}_{I} \gamma^{m} \pa_{m} \psi^{I} \, , \hspace{10mm} \D_{\psi} = \tfrac{1}{2} (d-1) \, .
\end{align}
This theory has $U(N)$ global symmetry under which $\psi^{I}$ transforms in the fundamental representation. Similar to the free scalar CFT, this model admits a tower of conserved higher-spin currents which are schematically of the form
\begin{align}
	J_{m_{1} \dots m_{s}} = \sum_{k=0}^{s-1} c_{k}(s) \bar{\psi}_{I} \g_{\{ m_{1}} \overset{\leftarrow}{\pa}_{m_{2}} \dots \overset{\leftarrow}{\pa}_{m_{k}} \overset{\rightarrow}{\pa}_{m_{k+1}} \dots \overset{\rightarrow}{\pa}_{m_{s} \}} \psi^{I} \, , \hspace{5mm} s = 1, 2, 3, \dots \, .
\end{align}
Analogous to the previous examples, the coefficients $c_{k}(s)$ are determined from conservation and the equations of motion.

Note that the above discussion is quite general and is intended to describe how higher-spin currents can be constructed for some well known free field theories. However, in this thesis we assume that conformal higher-spin currents of arbitrary integer or half-integer spin can exist without reference to a specific free field model. The conformal higher-spin currents built from free massless scalar and spinor fields in arbitrary dimensions may be found in \cite{Konstein:2000bi}.

The presence of exactly conserved currents of all spins implies that the CFT has an infinite
dimensional higher spin symmetry which includes the conformal symmetry as a subalgebra. Higher
spin symmetries turn out to be very constraining. In fact it has been proven that if a CFT possesses a spin-4 conserved current, then an infinite tower of conserved higher spin operators is present, and all correlation
functions of local operators coincide with those of a free CFT \cite{Maldacena:2011jn,Stanev:2012nq,Alba:2013yda,Alba:2015upa}. Furthermore, according to the AdS/CFT dictionary, exactly conserved currents of spin-$s$ in $\text{CFT}_{d}$ are dual to massless spin-$s$ gauge fields in $\text{AdS}_{d+1}$ \cite{Mikhailov:2002bp,Giombi:2009wh} (see also the lecture notes \cite{giombi2016abc,giombi2016tasi}).

\section{Correlation functions}\label{Ch02-section2.3}

\subsection{Generating functional}\label{Ch02-subsection2.3.1}
In quantum field theory we are typically interested in evaluating correlation functions of local operators/fields, which are computed using functional integrals. After performing a Wick rotation into Euclidean signature, the path-integral is convergent and the $n$-point correlation function $\langle \phi(x_{1}) \, \dots \, \phi(x_{n}) \rangle  \equiv \langle \Omega | \mathcal{T} \{ \phi(x_{1}) \,  \dots \, \phi(x_{n}) \} | \Omega \rangle $ (or Schwinger function), where $\cT$ denotes time-ordering, is computed as follows:
\begin{align}
	\langle \Omega | \mathcal{T} \{ \phi(x_{1}) \,  \dots \, \phi(x_{n}) \} | \Omega \rangle &= \frac{\int [\mathcal{D} \phi(x)] \, \phi(x_{1}) \dots \phi(x_{n}) \, e^{-S[\phi(x)]}}{\int [\mathcal{D} \phi(x)]  \, e^{-S[\phi(x)]}} \, .
\end{align}
Now consider the partition function
\begin{equation}
	\mathcal{Z}[J] = \int [\mathcal{D} \phi(x)] \, e^{ - S[\phi(x)] - \int d^{d}x J(x) \phi(x) } \, ,
\end{equation}
where $J(x)$ is some source term. This may be expanded as follows:
\begin{align}
	\hspace{10mm} \mathcal{Z}[J] &= \int [\mathcal{D} \phi(x)] \, e^{ - S[\phi(x)]} \Big( 1  - \int d^{d}x \, J(x) \phi(x) \nonumber \\[1mm]
	& \hspace{25mm} + \frac{(-1)^{2}}{2!} \int d^{d}x_{1} d^{d}x_{2} \, \phi(x_{1}) \phi(x_{2}) J(x_{1}) J(x_{2}) + \dots \Big) \nonumber \\[4mm]
	&= \int [\mathcal{D} \phi(x)] \, e^{ - S[\phi(x)]} -  \int d^{d}x \, J(x)  \int [\mathcal{D} \phi(x)] \, \phi(x) \, e^{ - S[\phi(x)]}  \\[1mm]  & \hspace{5mm} + \frac{(-1)^{2}}{2!} \int d^{d}x_{1} d^{d}x_{2} \, J(x_{1}) J(x_{2}) \int [\mathcal{D} \phi(x)] \, \phi(x_{1}) \phi(x_{2}) \, e^{ - S[\phi(x)]} + \dots \, . \nonumber
\end{align}
So we obtain
\begin{equation}
	\mathcal{Z}[J] = \sum_{n=0}^{\infty} \frac{1}{n!} \int \prod_{i=1}^{n} d^{d} x_{i} \, J(x_{1}) \, \dots \, J(x_{n}) \, \langle \phi(x_{1}) \,  \dots \, \phi(x_{n}) \rangle \, .
\end{equation}
$\mathcal{Z}[J]$ is the generating function for correlation functions, as by taking functional derivatives we obtain
\begin{equation}
	\langle \phi(x_{1}) \, \dots \, \phi(x_{n}) \rangle = \frac{(-1)^{n}}{\mathcal{Z}_{0}} \frac{\delta}{\delta J(x_{1})} \, \dots \frac{\delta}{\delta J(x_{n})} \mathcal{Z}[J] |_{J=0} \, .
\end{equation}
where $\mathcal{Z}_{0}$ is known as the vacuum functional, defined as
\begin{equation}
	\mathcal{Z}_{0} = \int [\mathcal{D} \phi(x)] \, e^{ - S[\phi(x)] } \, .
\end{equation}
Hence, the partition function may be computed as a sum over all $n$-point Green's functions, making them a fundamental object of study in quantum field theory. The conditions under which the Euclidean Green's functions can be analytically continued to Lorentzian signature are stated by the Osterwalder-Schrader theorem \cite{Osterwalder:1973dx,Osterwalder:1974tc}. Correlation functions are related to the S-matrix via the LSZ reduction formula \cite{Srednicki:2007qs}. 

For future reference we note that in a conformal field theory the conformal invariance of a general $n$-point correlation function is expressed through the condition
\begin{align}
	\langle \phi'_{\cA_{1}}(x_{1}) \, \dots \, \phi'_{\cA_{n}}(x_{n}) \rangle = \langle \phi_{\cA_{1}}(x_{1}) \, \dots \, \phi_{\cA_{n}}(x_{n}) \rangle \, ,
\end{align}
or equivalently, in accordance with \eqref{Ch02-Finite Field transformation}, we have 
\begin{align}
	\langle \phi'_{\cA_{1}}(x'_{1}) \, \dots \, \phi'_{\cA_{n}}(x'_{n}) \rangle &= \Omega(x_{1})^{-\Delta_{1}} \dots \Omega(x_{n})^{-\Delta_{n}} \nonumber \\
	& \times \mathcal{D}[R(x_{1})]_{\cA_{1}}{}^{\cB_{1}} \dots \mathcal{D}[R(x_{n})]_{\cA_{n}}{}^{\cB_{n}} \langle \phi_{\cB_{1}}(x_{1}) \, \dots \, \phi_{\cB_{n}}(x_{n}) \rangle  \, .
\end{align}

\subsection{Ward identities}\label{Ch02-subsection2.3.2}

At the classical level, invariance of the action under a continuous symmetry implies the existence of a conserved current. However, at the quantum level, classical symmetries result in constraints on the correlation functions, which are known as Ward-Takahashi identities. The symmetry is said to be anomalous if the functional measure within the path integral does not possess the symmetry of the action, i.e. $[\mathcal{D}\phi'] \neq [\mathcal{D}\phi]$. Now suppose that the classical action is invariant under the general transformation
\begin{equation}
	\phi'(x') = \mathcal{F}(\phi(x)) \, .
\end{equation}
Let us now examine the consequence of this symmetry on the correlation function:
\begin{align} \label{Transformation of correlator}
	\langle \phi(x'_{1}) \,  \dots \, \phi(x'_{n}) \rangle &= \mathcal{Z}_{0}^{-1} \int [\mathcal{D} \phi(x)] \, \phi(x'_{1}) \dots \phi(x'_{n}) \, e^{-S[\phi]} \nonumber \\
	&= \mathcal{Z}_{0}^{-1} \int [\mathcal{D} \phi'(x)] \, \phi'(x'_{1}) \dots \phi'(x'_{n}) \, e^{-S[\phi']} \nonumber \\
	&= \mathcal{Z}_{0}^{-1} \int [\mathcal{D} \phi(x)] \, \mathcal{F}(\phi(x_{1})) \dots \mathcal{F}(\phi(x_{n})) \, e^{-S[\phi(x)]} \nonumber \\
	&= \langle \mathcal{F}(\phi(x_{1})) \, \dots \, \mathcal{F}(\phi(x_{n})) \rangle \, ,
\end{align}
where we have made the non-trivial assumption that the functional measure is invariant under the transformation, i.e. $[\mathcal{D}\phi'] = [\mathcal{D}\phi]$ (i.e. the symmetry is non-anomalous). Now recall that an infinitesimal conformal transformation can be written in terms of its generator, $G_{A}$, as follows
\begin{align}
	\phi_{\cA}'(x) &= \phi_{\cA}(x) - \text{i} \omega^{A} (G_{A})_{\cA}{}^{\cB} \phi_{\cB}(x) \, ,
\end{align}
where $\omega^{A}$ are a set of constant infinitesimal parameters (here index $A$ denotes arbitrarily many indices). Letting $\omega^{A} \rightarrow \omega^{A}(x)$, the variation of the action $\delta_{\omega} S[\phi]$ is given by
\begin{equation}
	\delta_{\omega} S[\phi] = \int d^{d} x \, \partial_{m} j^{m}_{A}(x) \, \omega^{A}(x) \, .
\end{equation} 
Now let $X = \phi(x_{1}) \, ... \, \phi(x_{n})$, and its variation be denoted by $\delta_{\omega}X$. We have
\begin{equation}
	\langle X \rangle = \mathcal{Z}_{0}^{-1} \int [\mathcal{D} \phi'(x)] \, (X + \delta_{\omega} X) \, e^{-S[\phi(x)] - \int d^{d} x \, \partial_{m} j^{m}_{A}(x) \, \omega^{A}(x)} \, .
\end{equation}
Again, we assume the functional measure is invariant under the transformation and we expand the expression above to first order in $\omega^{A}$
\begin{align}
	\langle X \rangle &= \mathcal{Z}_{0}^{-1} \int [\mathcal{D} \phi'(x)] \, (X + \delta_{\omega} X) \, e^{-S[\phi(x)] - \int d^{d} x \, \partial_{m} j^{m}_{A}(x) \, \omega^{A}(x)} \nonumber \\
	&= \mathcal{Z}_{0}^{-1} \int [\mathcal{D} \phi'(x)] \, (X + \delta_{\omega} X) (1 - \int d^{d} x \, \partial_{m} j^{m}_{A}(x) \, \omega^{A}(x) + \dots ) \, e^{-S[\phi(x)] } \nonumber \\
	&= \langle X \rangle + \langle \delta_{\omega} X \rangle - \int d^{d} x \, \partial_{m} \Big( \mathcal{Z}_{0}^{-1} \int [\mathcal{D} \phi(x)] \, j^{m}_{A}(x) \, X \, e^{-S[\phi(x)] } \Big) \, \omega^{A}(x) \, .
\end{align}
Hence, the correlation function of the variation is given by:
\begin{equation}
	\Rightarrow \hspace{5mm} \langle \delta_{\omega} X \rangle = \int d^{d} x \, \partial_{m} \langle j^{m}_{A}(x) X \rangle \, \omega^{A}(x) \, .
\end{equation}
We now explicitly compute the variation:
\begin{align}
	\delta_{\omega}X &= -\text{i} \sum_{i=1}^{n} (\phi(x_{1}) \, \dots \, G_{A} \phi(x_{i}) \, \dots \, \phi(x_{n})) \, \omega^{A}(x_{i}) \nonumber \\
	&= -\text{i} \int d^{d} x \, \sum_{i=1}^{n} \delta^{(d)}(x-x_{i}) \, (\phi(x_{1}) \, \dots \, G_{A} \phi(x_{i}) \, \dots \, \phi(x_{n})) \, \omega^{A}(x) \\[2mm]
	\Rightarrow  \hspace{5mm}  & \langle \delta_{\omega} X \rangle = -\text{i} \int d^{d} x \,  \sum_{i=1}^{n} \delta^{(d)}(x-x_{i}) \, \langle \phi(x_{1}) \, \dots \, G_{A} \phi(x_{i}) \, \dots \, \phi(x_{n}) \rangle \, \omega^{A}(x) \, .
\end{align}
By combining the results above we arrive at the following local relation (as $\omega^{A}(x)$ are arbitrary) known as the Ward identity for the current $j^{m}_{A}(x)$:
\begin{equation} \label{Ward identity}
	\partial_{m} \langle j^{m}_{A}(x) \, \phi(x_{1}) \, \dots \, \phi(x_{n}) \rangle + \text{i} \, \sum_{i=1}^{n} \delta^{(d)}(x-x_{i}) \, \langle \phi(x_{1}) \, \dots \, G_{A} \phi(x_{i}) \, \dots \, \phi(x_{n}) \rangle = 0 \, .
\end{equation}
This is an infinitesimal version of \eqref{Transformation of correlator}. This can be seen by integrating the Ward identity over a region of spacetime that includes all $x_{i}$ while letting the boundary of this region extend to infinity, where the divergence $\partial_{m} \langle j^{m}_{A}(x) X \rangle$ vanishes by hypothesis. Again it must be stressed that the Ward identities apply only to symmetries which leave the integration measure invariant.


We will now present two different approaches to finding Ward identities in conformal field theories. The first approach uses the explicit form of the conserved currents for the given symmetries, while the second approach is faster and uses the constraints imposed by conformal invariance. There are three Ward identities associated with translation, Lorentz, and dilatation symmetry respectively. Associated with these symmetries are the following conserved currents:
\begin{subequations}
	\begin{align}
		T^{m n} = T_{c}^{m n} + \partial_{p} B^{p m n} + \sfrac{1}{2} \partial_{\lambda} \partial_{p} X^{\lambda p m n} \hspace{5mm} & - \hspace{5mm} \textit{Translation symmetry} \label{Noether current - translations}\\[2mm]
		\mathcal{\cJ}^{m n p} = T^{m n} \, x^{p} - T^{m p} \, x^{n}  \hspace{10mm} & - \hspace{5mm} \textit{Lorentz symmetry} \label{Noether current - rotations}\\[2mm]
		j_{D}^{m} = T^{m}{}_{ n} x^{n}  \hspace{25mm} & - \hspace{5mm} \textit{Dilatation symmetry} \label{Noether current - dilatations} \\[2mm]
		j_{K}^{mn} = - T^{m}{}_{p} I^{p n}(x) \hspace{20mm} & - \hspace{5mm} \textit{S.C.T}
	\end{align}
\end{subequations}
Here we have used the modified energy momentum tensor which we constructed in Section \ref{Ch02-subsection2.2.1}. Notice that the modified energy-momentum tensor is fundamental in the sense that the other conserved currents are constructed from it. We may now use \eqref{Ward identity} to establish Ward identities with each symmetry.\\[4mm]
\noindent\textbf{Translational symmetry:}\\
The generator of translations is the operator
\begin{equation}
	P_{m} = - \text{i} \partial_{m} \, ,
\end{equation}
We now substitute this and \eqref{Noether current - translations} into \eqref{Ward identity} to obtain
\begin{equation} \label{Ward identity - translations}
	\partial_{m} \langle T^{m n} \, \phi(x_{1}) \, \dots \, \phi(x_{n}) \rangle = - \sum_{i} \delta^{(d)}(x-x_{i}) \, \partial_{i}^{n} \langle \phi(x_{1}) \, \dots \, \phi(x_{i}) \, \dots \, \phi(x_{n}) \rangle \, .
\end{equation} 
This is the first Ward identity for the energy-momentum tensor.\\[4mm]
\noindent\textbf{Lorentz symmetry:}\\
The generator of Lorentz transformations acting on fields is given by
\begin{equation}
	L_{m n} = - \text{i} ( x_{m} \partial_{n} - x_{n} \partial_{m} ) - \text{i} \Sigma_{m n} \, .
\end{equation}
Inserting this into \eqref{Ward identity} and using \eqref{Noether current - rotations} gives
\begin{align}
	\partial_{m} \langle \mathcal{\cJ}^{m n p} \, \phi(x_{1}) \, \dots \, \phi(x_{n}) \rangle = \sum_{i} \delta^{(d)}(x-x_{i}) \, \big\{ x_{i}^{n} \partial_{i}^{p} - x_{i}^{p} \partial_{i}^{n} - \Sigma_{i}^{n p} \big\} \langle \phi(x_{1}) \, \dots \, \phi(x_{i}) \, \dots \, \phi(x_{n}) \rangle \, .
\end{align}
The partial derivative on the LHS acts on both $T^{m n}$ and $x^{p}$, and so we may use the Ward identity \eqref{Ward identity - translations} to simplify this result to
\begin{align} \label{Ward identity - rotations}
	\langle T^{[m n]} \, \phi(x_{1}) \, \dots \, \phi(x_{n}) \rangle = \frac{1}{2} \sum_{i} \delta^{(d)}(x-x_{i}) \, \Sigma_{i}^{m n} \langle \phi(x_{1}) \, \dots \, \phi(x_{i}) \, \dots \, \phi(x_{n}) \rangle \, ,
\end{align}
which is the second Ward identity.\\[4mm]
\noindent\textbf{Dilatation symmetry:}\\
The generator of dilatations is given by:
\begin{equation}
	D = - \text{i} ( x^{m} \partial_{m} + \Delta ) \, .
\end{equation}
After substitution into \eqref{Ward identity} and using \eqref{Noether current - dilatations} we obtain
\begin{equation}
	\partial_{m} \langle T^{m}{}_{ n} x^{n} \, \phi(x_{1}) \, \dots \, \phi(x_{n}) \rangle = - \sum_{i} \delta^{(d)}(x-x_{i}) \, [ x_{i}^{n} \partial_{i, \, n} + \Delta_{i}] \langle \phi(x_{1}) \, \dots \, \phi(x_{i}) \, \dots \, \phi(x_{n}) \rangle \, .
\end{equation}
We then use the first Ward identity to obtain
\begin{equation} \label{Ward identity - dilatations}
	\langle T^{m}{}_{ m} \, \phi(x_{1}) \, \dots \, \phi(x_{n}) \rangle = - \sum_{i} \delta^{(d)}(x-x_{i}) \, \Delta_{i} \langle \phi(x_{1}) \, \dots \, \phi(x_{i}) \, \dots \, \phi(x_{n}) \rangle \, .
\end{equation}
One can also compute the Ward identity associated with special conformal transformations, however, it may be shown that this does not give rise to any new constraints.

Notice that all three Ward identities are quantum analogues of the conditions in \eqref{EM tensor symmetries and constraints}. In particular, for distinct spacetime points, $x \neq x_{i}$, we obtain
\begin{subequations}
	\begin{align}
		\partial_{m} \langle T^{mn}(x) \, \phi(x_{1}) \, \dots \, \phi(x_{n}) \rangle &= 0 \, , \\
		\langle T^{[m n]}(x) \, \phi(x_{1}) \, \dots \, \phi(x_{n}) \rangle &= 0 \, , \\
		\langle T^{m}{}_{ m}(x) \, \phi(x_{1}) \, \dots \, \phi(x_{n}) \rangle &= 0 \, .
	\end{align}
\end{subequations}
The contributions for which $x = x_{i}$ are known as contact terms.

Another way to derive the Ward identities is by using the conformal invariance of the action supplemented by the modified Noether procedure presented in Subsection \ref{Ch02-subsection2.2.1}. Schematically we have:
\begin{equation}
	\langle \phi_{\cA}(x) \, \dots \, \rangle = \int [ \mathcal{D} \phi(x) ] \, e^{-S[\phi]} \phi_{\cA}(x) \, \dots \, .
\end{equation}
Now we assume that the functional measure is invariant under the variation $\delta_{\xi,\omega,\sigma} \phi$:
\begin{subequations}
	\begin{align}
		\delta_{\xi,\omega,\sigma} \phi_{\cA}(x) &= - \xi^{m}(x) \, \partial_{m} \phi_{\cA}(x) - \Delta \sigma_{\xi}(x) \phi_{\cA}(x) + \sfrac{1}{2} \, \omega_{\xi}^{m n}(x) ( \Sigma_{m n} )_{\cA}{}^{\cB} \phi_{\cB}(x) \, , \\[2mm]
		\delta_{\xi,\omega,\sigma} S[\phi] &= \int d^{d} x \, (\partial_{n} \xi_{m} - \omega_{m n} - \sigma \eta_{m n} ) \, T^{m n} \, .
	\end{align}
\end{subequations}
If we compute the variation of the correlator under an infinitesimal conformal transformation (in a similar manner to the E.M tensor in section 2.1.1), we obtain:
\begin{align}
	\langle \delta_{\xi,\omega,\sigma} \phi_{\cA}(x_{i}) \, \dots \, \rangle + \int d^{d}x \, (\partial_{n} \xi_{m} - \omega_{ m n} - \sigma \eta_{m n}) \, \langle T^{m n}(x) \, \phi_{\cA}(x_{i}) \, \dots \, \rangle = 0 \, .
\end{align}
Varying $\xi$, $\omega_{ m n}$ and $\sigma$ independently, we obtain the same three Ward identities:
\begin{subequations}
	\begin{align}
		\partial_{m} \langle T^{m n}(x) \, \phi_{\cA}(x_{i}) \, \dots \, \rangle &= - \delta^{(d)}(x-x_{i}) \, \partial_{i}^{n} \langle \phi_{\cA}(x_{i}) \, \dots \, \rangle \, , \\[2mm]
		\langle T^{[m n]}(x) \, \phi_{\cA}(x_{i}) \, \dots \, \rangle &= \frac{1}{2} \, \delta^{(d)}(x-x_{i}) \, \Sigma^{m n} \langle \phi_{\cA}(x_{i}) \, \dots \, \rangle \, , \\[2mm]
		\langle T^{m}{}_{m}(x) \, \phi_{\cA}(x_{i}) \, \dots \, \rangle &= - \delta^{(d)}(x-x_{i}) \, \Delta_{i} \langle \phi_{\cA}(x_{i}) \, \dots \, \rangle \, .
	\end{align}
\end{subequations}
The right hand side of the above expressions all vanish at distinct spacetime points. 

For correlation functions involving higher-spin currents, the Ward identities are assumed to reduce to their classical analogues at distinct spacetime points, i.e. within correlation functions the higher-spin currents are totally symmetric, traceless and transverse at distinct points as follows:
\begin{subequations}
	\begin{align}
		\partial^{m_{i}} \langle J_{m_{1} \dots m_{i} \dots m_{s}}(x) \, \phi_{\cA}(x_{i}) \, \dots \, \rangle &= 0 \, , \hspace{5mm} \forall i \, , \\[2mm]
		\langle J_{m_{1} \dots [m_{i} m_{j}] \dots m_{s}}(x) \, \phi_{\cA}(x_{i}) \, \dots \, \rangle &=  0 \, , \hspace{5mm} \forall i,j \, , \\[2mm]
		\eta^{m_{i} m_{j}}  \langle J_{m_{1} \dots m_{i} m_{j} \dots m_{s}}(x) \, \phi_{\cA}(x_{i}) \, \dots \, \rangle &= 0 \, , \hspace{5mm} \forall i,j \, .
	\end{align}
\end{subequations}
The main goal of this thesis is to determine the constraints imposed by the Ward identities on the structure of three-point functions of conserved currents, using the approach of \cite{Osborn:1993cr}.

\subsection{The conformal bootstrap for correlation functions}\label{Ch02-subsection2.3.3}

Perhaps one of the most important consequences of conformal symmetry is that it places stringent constraints on the general structure of correlation functions. In \cite{Osborn:1993cr} (see also \cite{Mack:1969rr,Schreier:1971um, Migdal:1971xh,Koller:1974ut,Mack:1976pa} for some earlier developments of the formalism), the group theoretic formalism to construct two- and three-point correlation functions was presented. Below we summarise the pertinent aspects of the formalism, which we will develop further in the remainder of this thesis.

\subsubsection{Two-point functions}
Given two points, $x_{1}$ and $x_{2}$, we define the covariant two-point function
\begin{equation} \label{Ch02-Two-point building blocks 1}
	x_{12}^{m} = (x_{1} - x_{2})^{m} \, , \hspace{10mm} x_{21}^{m} = - x_{12}^{m} \, . 
\end{equation}
Under a conformal transformation \eqref{Ch02-Conformal transformations 2}, the two-point functions have the following transformation property
\begin{equation}
	x'^{2}_{12} = \Omega(x_{1}) \, \Omega(x_{2}) \, x_{12}^{2} \, .
\end{equation}
If we now consider scalar fields $\phi, \psi$ of dimension $\Delta$ transforming according to \eqref{Ch02-Finite Field transformation}, we can construct a two-point function with the desired conformal transformation properties as follows:
\begin{align}
	\langle \phi(x_{1}) \psi(x_{2}) \rangle = \frac{C}{(x_{12}^{2})^{\Delta}} \, ,
\end{align}
where the constant parameter $C$ is fixed by normalisation. Typically it is conventional to rescale the operators entering the two-point function to obtain a canonical unit normalisation, i.e. $C=1$. 

To generalise this construction to compute two-point functions of operators with spin, we utilise the formalism of Osborn and Petkou \cite{Osborn:1993cr} and make use of the conformal inversion tensor
\begin{align} \label{Ch02-Inversion tensor}
	I_{mn}(x) = \eta_{mn} - 2 \, \hat{x}_{m} \hat{x}_{n} \, , \hspace{10mm} \hat{x}_{12}^{m} = \frac{x_{12}^{m}}{\sqrt{x_{12}^{2}}} \, , \hspace{10mm} I_{m a}(x) \, I^{a n}(x) = \d_{m}^{n} \, ,
\end{align}
where we have introduced the normalised two-point functions $\hat{x}_{12}$. Since the inversion transformation combines scale transformations and local conformal rotations, to construct a conformally invariant correlation function it suffices to demand invariance under translations and inversions. For two-point functions of operators with spin, it must possess the following conformal transformation properties consistent with \eqref{Ch02-Finite Field transformation}:
\begin{align}
	\langle \phi'_{\cA}(x'_{1}) \psi'_{\cB}(x'_{2}) \rangle &= \Omega(x_{1})^{-\D} \Omega(x_{2})^{-\D} \cR^{(1)}{}_{\cA}{}^{\cA'}(x_{1}) \, \cR^{(2)}{}_{\cB}{}^{\cB'}(x_{2}) \langle \phi_{\cA'}(x_{1}) \psi_{\cB'}(x_{2}) \rangle  \, ,
\end{align}
where $\cR^{(i)}{}_{\cA}{}^{\cA'}(x_{i}) = \cD_{(i)}[R(x_{i})]_{\cA}{}^{\cA'}$ is a representation matrix for the orthogonal transformation \eqref{Ch02-Conformal transformations 2} at the point $x_{i}$. As a starting point we may use the following ansatz
\begin{align}
	\langle \phi_{\cA}(x_{1}) \psi_{\cB}(x_{2}) \rangle = \frac{\cT_{\cA \cB}(x_{1}, x_{2})}{(x_{12}^{2})^{\Delta}} \, , \hspace{10mm} x_{1} \neq x_{2} \, ,
\end{align}
for some conformally covariant tensor $\cT_{\cA \cB}(x_{1}, x_{2})$ satisfying 
\begin{align} \label{Ch02-Tensor for two-point function}
	\cT_{\cA \cB}(x_{1}, x_{2}) = \cR^{(1)}{}_{\cA}{}^{\cA'}(x_{1}) \, \cR^{(2)}{}_{\cB}{}^{\cB'}(x_{2}) \, \cT_{\cA' \cB'}(x_{1}, x_{2}) \, .
\end{align}
The two-point correlation functions possesses scale and translational covariance provided that $\cT_{\cA \cB}(x_{1}, x_{2}) = \cT_{\cA \cB}(\hat{x}_{12})$. We now seek a solution for $\cT$ with the appropriate transformation properties under rotations and inversions. A solution to \eqref{Ch02-Tensor for two-point function} is
\begin{align}
	\cT_{\cA \cB}(x_{1}, x_{2}) = \cI^{(1)}{}_{\cA}{}^{\cA'}(x_{12}) \, g_{\cA' \cB} \, ,
\end{align}
where $\cI^{(1)}{}_{\cA}{}^{\cA'}(x) = \cD_{(1)}[I(x)]_{\cA}{}^{\cA'}$ is an appropriate representation of the inversion tensor and $g_{\cA \cB}$ is an invariant tensor for the representations $\cD_{(1)}$ and $\cD_{(2)}$, i.e. 
\begin{align} \label{Ch02-Invariant tensor for two-point function}
	g_{\cA \cB} = \cD_{(1)}[R \,]_{\cA}{}^{\cA'} \cD_{(2)}[R \,]_{\cB}{}^{\cB'} \, g_{\cA' \cB'} \, , \hspace{5mm} \forall R \, .
\end{align}
Indeed, the inversion tensor satisfies $\cD[R\,] \, \cD[I(x)] \, \cD[R\,]^{-1} = \cD[I(R x)]$, for a constant rotation $R$. However, for inversions we require $\cD[I(x_{1})] \, \cD[I(x_{12})] \, \cD[I(x_{2})] = \cD[I(x'_{12})]$, which automatically follows from the property
\begin{align} \label{Ch02-Inversion tensor transformation}
	I_{m}{}^{m'}(x_{1}) \, I_{m' n'}(x_{12}) \, I^{n'}{}_{n}(x_{2}) = I_{m n}(x'_{12}) \, , \hspace{10mm} x'_{12} = \frac{x_{1}}{x_{1}^{2}} - \frac{x_{2}}{x_{2}^{2}} \, .
\end{align}
In particular, \eqref{Ch02-Invariant tensor for two-point function} implies that a solution for the two-point function exists only for $\phi, \psi$ belonging to the same spin representation. 

Let us now present some examples for two-point functions. If we consider a vector field $V_{m}$, then the two-point function is fixed up to the following form:
\begin{align} \label{Ch02-Vector current two-point function}
	\langle V_{m}(x_{1}) V_{n}(x_{2}) \rangle = \frac{C_{V}}{(x_{12}^{2})^{\Delta}} I_{mn}(x_{12}) \, , \hspace{10mm} x_{1} \neq x_{2} \, .
\end{align}
Here $C_{V}$ may be fixed in terms of Ward-identites and/or the overall normalisation of the two-point functions of the fields comprising the currents. For a conserved vector field satisfying $\pa^{m} V_{m} = 0$, then the appropriate Ward identity becomes (at distinct spacetime points) 
\begin{align}
	\pa^{m}_{(1)} \langle V_{m}(x_{1}) V_{n}(x_{2}) \rangle = 0 \, ,
\end{align}
and similarly for conservation at $x_{2}$. Now due to the identity
\begin{equation} \label{Ch02-Inversion differential identity 1}
	\pa^{m} \bigg\{ \frac{I_{mn}(x)}{(x^{2})^{\D}} \bigg\} = 2(\D - d + 1) \frac{x_{n}}{(x^{2})^{\D+1}} \, ,
\end{equation}
the Ward identity is satisfied provided the dimension of the two-point function \eqref{Ch02-Vector current two-point function} is fixed to $\D = d - 1$, which is consistent with the dimension of a conserved vector field. Similar considerations also apply to two-point functions involving symmetric and traceless tensors, such as the energy-momentum tensor $T_{mn}$. In particular we have
\begin{align} \label{Ch02-EM tensor two-point function}
	\langle T_{m_{1} m_{2}}(x_{1}) T_{n_{1} n_{2}}(x_{2}) \rangle = \frac{C_{T}}{(x_{12}^{2})^{\Delta}} \, \cI_{m_{1} m_{2}, n_{1} n_{2}}(x_{12}) \, ,
\end{align}
where $\cI_{m_{1} m_{2}, n_{1} n_{2}}(x)$, which is separately symmetric and traceless in the indices $m_{i}$, $n_{j}$ respectively, is a representation of the inversion tensor acting on the space of symmetric and traceless tensors. It is constructed from $I_{mn}(x)$ as follows: 
\begin{align}
	\cI_{m_{1} m_{2}, n_{1} n_{2}}(x) = \frac{1}{2} \big\{ I_{m_{1} n_{1}}(x) I_{m_{2} n_{2}}(x) + I_{m_{1} n_{2}}(x) I_{m_{2} n_{1}}(x) \big\} - \frac{1}{d} \eta_{m_{1} m_{2}} \eta_{n_{1} n_{2} } \, .
\end{align}
This tensor satisfies the property
\begin{subequations}
	\begin{gather}
		\cI_{m_{1} m_{2}, m'_{1} m'_{2}}(x) \, \cI^{m'_{1} m'_{2}}{}_{n_{1} n_{2}}(x) = \mathcal{E}_{m_{1} m_{2}, n_{1} n_{2}} \, , \\
		\mathcal{E}_{m_{1} m_{2}, n_{1} n_{2}} = \frac{1}{2} ( \eta_{m_{1} n_{1}} \eta_{m_{2} n_{2}} + \eta_{m_{1} n_{2}} \eta_{m_{2} n_{1}}) - \frac{1}{d} \eta_{m_{1} m_{2}} \eta_{n_{1} n_{2} } \, ,
	\end{gather}
\end{subequations}
where $\mathcal{E}_{m_{1} m_{2}, n_{1} n_{2}}$ is a projection operator for symmetric and traceless tensors. If $T_{m_{1} m_{2}}$ is a conserved tensor satisfying $\pa^{m_{1}} T_{m_{1} m_{2}} = 0$, then, due to the identity
\begin{equation} \label{Ch02-Inversion differential identity 2}
	\pa^{m_{1}} \bigg\{ \frac{\cI_{m_{1} m_{2}, n_{1} n_{2}}(x)}{(x^{2})^{\D}} \bigg\} =  \frac{2(\D - d)}{(x^{2})^{\D+1}} \Big\{ \sfrac{1}{2} ( x_{n_{1}} I_{m_{2} n_{2}}(x) + x_{n_{2}} I_{m_{2} n_{1}}(x) ) + \frac{1}{d} x_{m_{2}} \eta_{n_{1} n_{2}} \Big\} \, ,
\end{equation}
the two-point function \eqref{Ch02-EM tensor two-point function} is compatible with conservation for $\D = d$, which is consistent with the dimension of a conserved symmetric and traceless tensor. 

Using the procedure above we can analogously construct two-point functions of symmetric and traceless tensor operators of arbitrary spin by considering appropriate symmetric and traceless tensor products of the inversion tensor, i.e. 
\begin{align} 
	\langle \phi_{m(s)}(x_{1}) \, \phi_{n(s)}(x_{2}) \rangle = \frac{C_{\phi}}{(x_{12}^{2})^{\Delta}} \left\{ I_{(m_{1} (n_{1}}(x_{12}) \dots  I_{m_{1}) n_{1})}(x_{12}) - \textit{Traces} \, \right\} \, .
\end{align}
In the case of conserved currents, the dimension $\Delta$ of the two-point function is fixed by imposing the conservation equation for the field $\phi$.


\subsubsection{Three-point functions}
Now given three distinct points in Minkowski space, $x_{i}$, with $i = 1,2,3$, we define conformally covariant three-point functions (or building blocks) in terms of the two-point functions as in \cite{Osborn:1993cr}
\begin{align}
	X_{ij} &= \frac{x_{ik}}{x_{ik}^{2}} - \frac{x_{jk}}{x_{jk}^{2}} \, , & X_{ji} &= - X_{ij} \, ,  & X_{ij}^{2} &= \frac{x_{ij}^{2}}{x_{ik}^{2} x_{jk}^{2} } \, , 
\end{align}
where $(i,j,k)$ is a cyclic permutation of $(1,2,3)$. For example we have
\begin{equation}
	X_{12}^{m} = \frac{x_{13}^{m}}{x_{13}^{2}} - \frac{x_{23}^{m}}{x_{23}^{2}} \, , \hspace{10mm} X_{12}^{2} = \frac{x_{12}^{2}}{x_{13}^{2} x_{23}^{2} } \, .
\end{equation}
Perhaps the most important property of the three-point functions is that, for an inversion transformation $x_{i} \rightarrow x'_{i}$, we have
\begin{equation} \label{Ch02-Three-point building block transformation}
	X'_{12 m} = x_{3}^{2} \, I_{m n}(x_{3}) \, X_{12}^{n} \, , \hspace{10mm} X'^{2}_{12} = (x_{3}^{2})^{2} X_{12}^{2} \, ,
\end{equation}
so that $X_{12}$ transforms as a conformally covariant vector at $x_{3}$. Similar properties also hold for $X_{13}$ and $X_{23}$. 

Using the inversion tensor and the two- and three-point building blocks, it is now possible to construct a conformally covariant ansatz for the three-point function of primary operators. Given three operators $\phi, \psi, \pi$ we may write
\begin{align} \label{Ch02-Three-point function ansatz}
	\langle \phi_{\cA_{1}}(x_{1}) \, \psi_{\cA_{2}}(x_{2}) \, \pi_{\cA_{3}}(x_{3}) \rangle = \frac{\cI^{(1)}{}_{\cA_{1}}{}^{\cA'_{1}}(x_{13}) \, \cI^{(2)}{}_{\cA_{2}}{}^{\cA'_{2}}(x_{23}) }{(x_{13}^{2})^{\D_{1}} (x_{23}^{2})^{\D_{2}}} \, \cT_{\cA'_{1} \cA'_{2} \cA_{3}}(x_{1}, x_{2}, x_{3}) \, ,
\end{align}
for some homogeneous tensor $\cT_{\cA_{1} \cA_{2} \cA_{3}}(x_{1}, x_{2}, x_{3})$ which is required to satisfy
\begin{align}
	\cT_{\cA_{1} \cA_{2} \cA_{3}}(Rx_{1}, Rx_{2}, Rx_{3}) &= \cD_{(1)}[R \,]_{\cA_{1}}{}^{\cA'_{1}} \cD_{(2)}[R \,]_{\cA_{2}}{}^{\cA'_{2}} \, \cD_{(3)}[R \,]_{\cA_{3}}{}^{\cA'_{3}} \nonumber \\
	& \hspace{35mm} \times \cT_{\cA'_{1} \cA'_{2} \cA'_{3}}(x_{1}, x_{2}, x_{3}) \, ,
\end{align}
for some constant orthogonal rotation $R$. Translational invariance is manifest provided that $\cT$ depends only on the two-point functions $x_{ij}$, and further, for an inversion $x_{i} \rightarrow x'_{i}$ it can be shown using \eqref{Ch02-Inversion tensor transformation} that $\cT$ is required to satisfy
\begin{align} \label{Ch02-H covariance}
	\cH_{\cA_{1} \cA_{2} \cA_{3}}(x'_{1}, x'_{2}, x'_{3}) &= (x_{3}^{2})^{\D_{3} - \D_{2}- \D_{1}} \cI^{(1)}{}_{\cA_{1}}{}^{\cA'_{1}}(x_{3}) \, \cI^{(2)}{}_{\cA_{2}}{}^{\cA'_{2}}(x_{3}) \nonumber \\
	&\hspace{30mm} \times \cI^{(3)}{}_{\cA_{3}}{}^{\cA'_{3}}(x_{3}) \, \cT_{\cA'_{1} \cA'_{2} \cA'_{3}}(x_{1}, x_{2}, x_{3}) \, .
\end{align}
Hence, due to the transformation property \eqref{Ch02-Three-point building block transformation}, a solution for $\cT$ is provided by 
\begin{align}
	\cT_{\cA_{1} \cA_{2} \cA_{3}}(x_{1}, x_{2}, x_{3}) = \cH_{\cA_{1} \cA_{2} \cA_{3}}(X_{12}) = X_{12}^{\D_{3} - \D_{2} - \D_{1}} \hat{\cH}_{\cA_{1} \cA_{2} \cA_{3}}(X_{12}) \, ,
\end{align}
where $\hat{\cH}_{\cA_{1} \cA_{2} \cA_{3}}(X)$ is homogeneous degree 0 in $X$. With this choice the ansatz \eqref{Ch02-Three-point function ansatz} possesses transformation properties consistent with that of the fields \eqref{Ch02-Finite Field transformation}. In \cite{Osborn:1993cr} it was shown that \eqref{Ch02-Three-point function ansatz} is consistent with the operator product expansion in such a way that the three-point function is completely determined by the leading singular operator product coefficient.

The significance of \eqref{Ch02-Three-point function ansatz} is that the structure of the three-point function is now completely determined by a tensor $\hat{\cH}_{\cA_{1} \cA_{2} \cA_{3}}(X)$ which depends on a single conformally covariant vector, $X^{m}$. The tensor $\hat{\cH}$ possesses the same tensorial symmetry properties as the fields in the three-point function and can be further constrained by differential equations in the case of conserved currents. In the vector representation, the solutions for $\hat{\cH}_{\cA_{1} \cA_{2} \cA_{3}}(X)$ may be realised as homogeneous polynomials in $X$, $\eta$. However, if we also allow $\hat{\cH}$ to be constructed from the Levi-Civita pseudo-tensor, $\epsilon$, then \eqref{Ch02-H covariance} is generically satisfied up to an overall sign. The structures involving $\epsilon$ are known as parity-violating (or parity-odd) structures, and were not accounted for in the analysis of \cite{Osborn:1993cr}. Note that by using the properties of the building blocks $\eta$, $X$ and $\epsilon$, the parity-odd structures in $\cH$ may be classified using the following formula which is analogous to \eqref{Ch02-H covariance}:
\begin{align} \label{Ch02-H covariance 2}
	\cH^{(\pm)}_{\cA_{1} \cA_{2} \cA_{3}}(X^{I}) &= \pm \, \cI^{(1)}{}_{\cA_{1}}{}^{\cA'_{1}}(X) \, \cI^{(2)}{}_{\cA_{2}}{}^{\cA'_{2}}(X) \, \cI^{(3)}{}_{\cA_{3}}{}^{\cA'_{3}}(X) \, \cH^{(\pm)}_{\cA'_{1} \cA'_{2} \cA'_{3}}(X) \, .
\end{align}
where we define $X^{I}_{m} = I_{mn}(X) \, X^{n} = - X_{m}$. Sometimes it is also useful to define the tensor $\cH^{I \, (\pm)}_{\cA_{1} \cA_{2} \cA_{3}}(X) = \cH^{(\pm)}_{\cA_{1} \cA_{2} \cA_{3}}(X^{I})$, which we will use in the subsequent chapters of the thesis. In the formula above the parity-even structures are denoted by $\cH^{(+)}$ while the parity-odd structures contain the Levi-Civita tensor $\epsilon$ and are denoted by $\cH^{(-)}$. Note that it is possible to construct parity-odd structures in only three and four spacetime dimensions. In the spinor representation, the classification of parity-even and parity-odd structures in three- and four-dimensional CFTs was carried out recently in \cite{Buchbinder:2022mys,Buchbinder:2023coi}.

There are several useful identities involving the two- and three-point functions and the conformal inversion tensor. For example we have the useful algebraic relations
\begin{subequations}
	\begin{align} \label{Ch02-Inversion tensor identities - vector case 1}
		I_{m}{}^{a}(x_{13}) \, I_{a n}(x_{23}) &= I_{m}{}^{a}(x_{12}) \, I_{a n}(X_{31}) \, ,  & I_{m n}(x_{23}) \, X_{12}^{n} &= \frac{x_{12}^{2}}{x_{13}^{2}} \, X_{31 \, m}^{I} \, ,
	\end{align} \vspace{-5mm}
	\begin{align} \label{Ch02-Inversion tensor identities - vector case 2}
		I_{m}{}^{a}(x_{23}) \, I_{a n}(x_{13}) &= I_{m}{}^{a}(x_{21}) \, I_{a n}(X_{23}) \, ,  & I_{m n}(x_{13}) \, X_{12}^{n} &= \frac{x_{12}^{2}}{x_{23}^{2}} \, X^{I}_{23 \, m} \, .
	\end{align}
\end{subequations}
These identities allow one to write analogous formula for the three-point function with $x_{1} \leftrightarrow x_{3}$ or $x_{2} \leftrightarrow x_{3}$ in terms of the building blocks $X_{23}$ or $X_{31}$ respectively. For example we may write
\begin{align} \label{Ch02-Three-point function ansatz 2}
	\langle \pi_{\cA_{3}}(x_{3}) \, \psi_{\cA_{2}}(x_{2}) \, \phi_{\cA_{1}}(x_{1}) \rangle = \frac{\cI^{(3)}{}_{\cA_{3}}{}^{\cA'_{3}}(x_{31}) \, \cI^{(2)}{}_{\cA_{2}}{}^{\cA'_{2}}(x_{21}) }{(x_{31}^{2})^{\D_{3}} (x_{21}^{2})^{\D_{2}}} \, \tilde{\cH}_{ \cA_{1} \cA'_{2} \cA'_{3} }(X_{23}) \, .
\end{align}
By equating \eqref{Ch02-Three-point function ansatz}, \eqref{Ch02-Three-point function ansatz 2} one can show, using the identities \eqref{Ch02-Inversion tensor identities - vector case 1}, \eqref{Ch02-Inversion tensor identities - vector case 1}, that 
\begin{align} \label{Ch02-Three-point function ansatz 2 - H Htilde relation}
	\tilde{\cH}_{\cA_{1} \cA_{2} \cA_{3}}(X) = (X^{2})^{\D_{1} - \D_{3}} \cI^{(2)}{}_{\cA_{2}}{}^{\cA'_{2}}(X) \, \cH_{\cA_{1} \cA'_{2} \cA_{3}}(-X) \, .
\end{align}
Hence, we see that the inversion tensor acts as an intertwining operator between the various representations of the three-point function. In particular, for correlation functions involving identical fields belonging to the same representation is necessary to impose
\begin{align}
	\cH_{\cA_{1} \cA_{2} \cA_{3}}(X) = \cH_{\cA_{2} \cA_{1} \cA_{3}}(-X) \, , \hspace{10mm} \tilde{\cH}_{\cA_{1} \cA_{2} \cA_{3}}(X) = \cH_{\cA_{3} \cA_{2} \cA_{1}}(-X) \, ,
\end{align}
with $\tilde{\cH}$ computed using \eqref{Ch02-Three-point function ansatz 2 - H Htilde relation}. 

For three conformal primary scalars $\phi_{1}$, $\phi_{2}$, $\phi_{3}$ we obtain the well known result
\begin{align}
		\langle \phi_{1}(x_{1}) \, \phi_{2}(x_{2}) \, \phi_{3}(x_{3}) \rangle &= \frac{C_{123}}{(x_{13}^{2})^{\D_{1}} (x_{23}^{2})^{\D_{2}}} \, X_{12}^{\D_{3} - \D_{2}- \D_{1}} \nonumber \\
		&=  \frac{C_{123}}{(x_{12}^{2})^{\frac{1}{2}(\D_{1} + \D_{2}- \D_{3})} (x_{23}^{2})^{\frac{1}{2}(\D_{2} + \D_{3}- \D_{1})}  (x_{13}^{2})^{\frac{1}{2}(\D_{1} + \D_{3} - \D_{2})} } \, .
\end{align}
where $C_{123}$ is a constant parameter. However it should be noted that although the general form above is consistent with conformal symmetry, in the case of identical operators, i.e. $\phi_{1} = \phi_{2} = \phi_{3}$, there is an additional constraint coming from Bose symmetry under exchange of any two operators in the correlation function. It follows that $C_{ijk}$ must be totally symmetric in its indices.

The construction is also useful for analysing three-point functions of conserved currents. As an illustrative example, let us consider the three-point function of a conserved vector current $V_{m}$ of dimension $\D_{V} = d - 1$ and two primary operators $\psi, \pi$ of arbitrary spin and dimension $\D_{2}$ and $\D_{3}$ respectively. Using \eqref{Ch02-Three-point function ansatz} the three-point function is fixed up to the following form:
\begin{align}
	\langle V_{m}(x_{1}) \, \psi_{\cA_{2}}(x_{2}) \, \pi_{\cA_{3}}(x_{3}) \rangle = \frac{I_{m}{}^{m'}(x_{13}) \, \cI^{(2)}{}_{\cA_{2}}{}^{\cA'_{2}}(x_{23}) }{(x_{13}^{2})^{\D_{V}} (x_{23}^{2})^{\D_{2}}} \, \cH_{m' \cA'_{2} \cA_{3}}(X_{12}) \, ,
\end{align}
For imposing the conservation condition on $V$, one can make use of \eqref{Ch02-Inversion differential identity 1} in addition to the following identities:
\begin{align} \label{Ch02-Inversion tensor identities - vector case 3}
	\pa_{(1) \, m} X_{12 \, n} = \frac{1}{x_{13}^{2}} I_{m n}(x_{13}) \, , \hspace{10mm} \pa_{(2) \, m} X_{12 \, n} = - \frac{1}{x_{23}^{2}} I_{m n}(x_{23}) \, .
\end{align}
As a result it may be shown that
\begin{align}
	\pa^{m}_{(1)} \langle V_{m}(x_{1}) \, \psi_{\cA_{2}}(x_{2}) \, \pi_{\cA_{3}}(x_{3}) \rangle = 0 \hspace{5mm} \Longrightarrow \hspace{5mm} \pa^{m}_{X} \cH_{m \cA_{2} \cA_{3}}(X) = 0 \, .
\end{align}
where $\pa_{X}$ denotes partial derivative with respect to the three-point covariant $X$. Hence, the conservation condition on $V$ results in a simple differential constraint on $\cH$. It is important to note that the construction does not treat the fields in the three-point function equally. In particular, for the ansatz \eqref{Ch02-Three-point function ansatz}, the field at $x_{3}$ is treated differently, and due to a lack of identities analogous to \eqref{Ch02-Inversion tensor identities - vector case 3} the conservation equation for a conserved operator at $x_{3}$ is difficult to impose. To rectify this issue one may use \eqref{Ch02-Three-point function ansatz 2} and \eqref{Ch02-Three-point function ansatz 2 - H Htilde relation} to construct an alternative representation for the three-point function in which conservation can be imposed in a straightforward manner. One may then use \eqref{Ch02-Inversion tensor identities - vector case 3} with $x_{1} \leftrightarrow x_{3}$.

Similar results are obtained for three-point functions where all operators are conserved currents. As a non-trivial example  let us consider the case of three conserved vector fields in $d=3$ for which we use the general ansatz
\begin{align}
	\langle V_{a}(x_{1}) \, V'_{b}(x_{2}) \,  V''_{c}(x_{3}) \rangle = \frac{I_{a}{}^{a'}(x_{13}) \, I_{b}{}^{b'}(x_{23})}{ (x_{13}^{2})^{2} (x_{23}^{2})^{2}} \, \mathcal{H}_{a'b'c}(X_{12}) \, .
\end{align}
In this case we must impose the conservation conditions
\begin{align}
	\pa^{a}_{X} \cH_{a b c}(X) = 0 \, , \hspace{8mm} \pa^{b}_{X} \cH_{a b c}(X) = 0 \, ,  \hspace{8mm} \pa^{c}_{X} \tilde{\cH}_{c b a}(X) = 0 \, . 
\end{align}
The general solution consistent with these constraints proves to be
\begin{align*}
	\mathcal{H}_{a b c}(X) &= \frac{a_{1}}{X^{5}} X_{a} X_{b} X_{c} + \frac{ a_{2} }{ X^{3}} (X_{a} \eta_{bc} + X_{b} \eta_{ac} - X_{c} \eta_{ab} ) \\
	& \hspace{15mm} + \frac{b}{X^{4}} ( - X_{a} X^{d} \epsilon_{bcd} + X_{b} X^{d} \epsilon_{acd} + X_{c} X^{d} \epsilon_{abd}) \, , 
\end{align*}
where $a_{1}$ and $a_{2}$ are free coefficients for the parity-even structures, while $b$ is a free coefficient for the parity-odd structure. When any of the vector fields in the correlation functions coincide it is also necessary to impose constraints arising from the symmetries under exchange of the spacetime points, which are of the form
\begin{align}
	\cH_{a b c}(X) = \cH_{b a c}(-X) \, , \hspace{10mm} \tilde{\cH}_{a b c}(X) = \cH_{c b a}(-X)  \, .
\end{align}
However, as shown in \cite{Osborn:1993cr} (see also \cite{Giombi:2011rz}), the remaining structures cannot satisfy these constraints unless the vector fields carry indices associated with a non-Abelian flavour group, i.e. $V_{a} = V_{a}^{\bar{a}} T^{\bar{a}}$, where $T^{\bar{a}}$ are appropriate generators. Since the conformal building blocks do not carry any flavour indices, the three-point function of these currents, $\langle V_{a}^{\bar{a}}(x_{1}) V_{b}^{\bar{b}}(x_{2}) V_{c}^{\bar{c}}(x_{3}) \rangle$, must factorise into a form which is proportional to an invariant tensor of the flavour group. Indeed, given generators $T^{\bar{a}}$ of the flavour group, we may define a completely anti-symmetric structure constant $f^{\bar{a} \bar{b} \bar{c}}$ as follows:
\begin{equation}
	[ T^{\bar{a}}, T^{\bar{b}} ] = \text{i} f^{\bar{a} \bar{b} \bar{c}} T^{\bar{c}} \, . 
\end{equation}
Due to the overall anti-symmetry of the structure constant, the three-point function of vector currents with flavour indices now possesses the correct properties under permutations of the fields.

In general, to solve for the explicit form of $\cH$ in vector representations one constructs an ansatz from all possible combinations of structures involving $X, \eta, \epsilon$ consistent with the tensorial symmetry properties of $\cH$. Schematically we have 
\begin{align}
	\cH_{\cA_{1} \cA_{2} \cA_{3}}(X) = X^{\D_{3} - \D_{2} - \D_{1}} \sum_{i=1}^{N} A_{i} \, \hat{\cH}_{i \, \cA_{1} \cA_{2} \cA_{3}}(X) \, ,
\end{align}
where $N$ denotes the number of possible structures and $A_{i}$ are coefficients for each possible structure. We then impose the conservation conditions, resulting in a linear system of equations in the coefficients $A_{i}$. The three-point functions of the energy-momentum tensor and conserved vector currents were first analysed using this approach in \cite{Osborn:1993cr}, however the formalism can also be applied to conserved currents of higher-rank. In general however, this procedure is complicated by issues such as: 1) how to identify all possible structures (which increases rapidly for increasing spins), and 2) linear dependence between the structures. Obtaining a general solution from ``first principles" for three-point functions of conserved currents of arbitrary spins remains an open problem.

The formalism outlined above generalises to three-point functions involving spinors and symmetric and traceless spin-tensors of arbitrary rank. For these cases one must fix the spacetime dimension and choose a particular spin representation. In such representations the higher-rank analogues of the inversion tensor are manifestly traceless and in some respects the analysis becomes more simple. The formalism has also been generalised to superconformal field theories in three and four spacetime dimensions by Park \cite{Park:1997bq,Park:1999pd,Park:1999cw} and Osborn \cite{Osborn:1998qu}. For more recent developments, see e.g. \cite{Nizami:2013tpa,Buchbinder:2015qsa,Buchbinder:2015wia,Kuzenko:2016cmf,Buchbinder:2021gwu,Buchbinder:2021izb,Buchbinder:2021kjk,Buchbinder:2021qlb,Buchbinder:2022kmj,Buchbinder:2023fqv,Buchbinder:2023ndg}.

The remainder of this thesis is dedicated to generalising the construction above to analyse the structure of two- and three-point correlation functions of conserved higher-spin currents of integer or half-integer spin in three and four spacetime dimensions. One of the primary goals of this thesis is to identify the conditions under which parity-odd solutions can exist in three-point functions of conserved currents. We also demonstrate how this approach can be generalised further to analyse the structure of three-point functions of conserved higher-spin supercurrents in three-dimensional $\cN=1$ superconformal field theory by developing the formalism presented in \cite{Park:1999cw}. In particular, we augment the approach of \cite{Osborn:1993cr, Park:1999cw} with auxiliary spinors, which simplifies the analysis considerably and allows us to propose a general classification for the structure of three-point functions of higher-spin currents in 3D $\cN=1$ SCFT.

\begin{subappendices}
	\section{4D conventions and notation}\label{Appendix2A}
Our conventions closely follow that of \cite{Buchbinder:1998qv}. For the Minkowski metric $\eta_{mn}$ we use the ``mostly plus'' convention: $\eta_{mn} = \text{diag}(-1,1,1,1)$. The Minkowski metric is used to raise and lower (flat) spacetime indices, i.e., given a contravariant four-vector $V^{a}$, $a = 0,1,2,3$ we may obtain the corresponding covariant vector $V_{a}$ by the rule $V_{a} = \eta_{a b} V^{b}$. We also make use of Planck units, so that $\hbar = c = G =1$.

For (anti-)symmetrisation of tensors (or spin-tensors) with an arbitrary number of indices, we use (brackets) parentheses according to the following conventions
\begin{subequations}
	\begin{align}
		(V_{S})_{a_{1} a_{2} \dots a_{n}} &\equiv V_{(a_{1} a_{2} \dots a_{n})} := \frac{1}{n!} \sum_{\pi \in S_{n}} V_{a_{\pi(1)} a_{\pi(2)} \dots a_{\pi(n)}} \, , \\[2mm]
		(V_{A})_{a_{1} a_{2} \dots a_{n}} &\equiv V_{[a_{1} a_{2} \dots a_{n}]} := \frac{1}{n!} \sum_{\pi \in S_{n}} \text{sgn}(\pi) V_{a_{\pi(1)} a_{\pi(2)} \dots a_{\pi(n)}} \, ,
	\end{align}
\end{subequations}
where $S_{n}$ is the symmetric group of $n$ elements. Sometimes we will also make use of the notation 
\begin{align}
	(V_{S.T})_{a_{1} a_{2} \dots a_{n}} \equiv V_{\{a_{1} a_{2} \dots a_{n} \}} := V_{(a_{1} a_{2} \dots a_{n})} - \frac{n}{n+d-1} \, \eta_{ ( a_{1} a_{2} } V_{a_{3} \dots a_{n}) k}{}^{k}  \, ,
\end{align}
which is to be understood as total symmetrisation with all possible traces removed, so that the tensor is overall symmetric and traceless.

The totally antisymmetric Levi-Civita tensor, $\e_{a b c d} = \e_{[a b c d]}$, is normalised such that $\e^{0123} = -\e_{0123} = 1$. Note that a product of Levi-Civita tensors may be written as
\begin{align}
	\e^{a b c d} \e_{a' b' c' d'} = - 4! \d^{a}_{[a'} \d^{b}_{b'} \d^{c}_{c'} \d^{d}_{d']} \, ,
\end{align}
with analogous results holding in general dimensions.

For the discussion of half-integer spin representations of the Lorentz group it is necessary to introduce the formalism of two-component Weyl spinors, which are representations of $\text{SL}(2,\mathbb{C})$, the universal covering group of the restricted Lorentz group $\text{SO}_{0}(3,1)$. We introduce the object $\psi_{\a}$, $\a = 1,2$, known as a left-handed Weyl spinor, which transforms under the fundamental representation of $\text{SL}(2,\mathbb{C})$ as follows:
\begin{align}
	\psi'_{\a} = N_{\a}{}^{\b} \psi_{\b} \, , \hspace{10mm} N_{\a}{}^{\b} \in \text{SL}(2,\mathbb{C}) \, .
\end{align}
This is known as the left-handed spinor representation of the Lorentz group, denoted by $(\frac{1}{2},0)$. Similarly, a right-handed Weyl spinor $\bar{\psi}_{\ad}$, $\ad = 1,2$ transforms in the complex conjugate representation
\begin{align}
		\bar{\psi}'_{\ad} = \bar{N}_{\ad}{}^{\bd} \bar{\psi}_{\bd} \, , \hspace{10mm} (N_{\a}{}^{\b})^{*} = \bar{N}_{\ad}{}^{\bd} \, .
\end{align}
This is known as the right-handed spinor representation of the Lorentz group, and is denoted by $(0,\frac{1}{2})$. The spinor indices on spin-tensors are raised and lowered using the $\text{SL}(2,\mathbb{C})$ invariant spinor metrics
\begin{subequations}
	\begin{align}
		\ve_{\a \b} = 
		\begingroup
		\setlength\arraycolsep{4pt}
		\begin{pmatrix}
			\, 0 & -1 \, \\
			\, 1 & 0 \,
		\end{pmatrix}
		\endgroup 
		\, , & \hspace{10mm}
		\ve^{\a \b} =
		\begingroup
		\setlength\arraycolsep{4pt}
		\begin{pmatrix}
			\, 0 & 1 \, \\
			\, -1 & 0 \,
		\end{pmatrix}
		\endgroup 
		\, , \hspace{10mm}
		\ve_{\a \g} \, \ve^{\g \b} = \d_{\a}{}^{\b} \, , \\[4mm]
		\ve_{\ad \bd} = 
		\begingroup
		\setlength\arraycolsep{4pt}
		\begin{pmatrix}
			\, 0 & -1 \, \\
			\, 1 & 0 \,
		\end{pmatrix}
		\endgroup 
		\, , & \hspace{10mm}
		\ve^{\ad \bd} =
		\begingroup
		\setlength\arraycolsep{4pt}
		\begin{pmatrix}
			\, 0 & 1 \, \\
			\, -1 & 0 \,
		\end{pmatrix}
		\endgroup 
		\, , \hspace{10mm}
		\ve_{\ad \gd} \, \ve^{\gd \bd} = \d_{\ad}{}^{\bd} \, .
	\end{align}
\end{subequations}
Given the spinor fields $\f_{\a}$, $\bar{\f}_{\ad}$, the spinor indices $\a = 1, 2$, $\ad = \bar{1}, \bar{2}$ are raised and lowered according to the following rules:
\begin{align}
	\f_{\a} &= \ve_{\a \b} \, \f^{\b} \, , & \f^{\a} &= \ve^{\a \b} \, \f_{\b} \, , & \bar{\f}_{\ad} &= \ve_{\ad \bd} \, \bar{\f}^{\b} \, , & \bar{\f}^{\ad} &= \ve^{\ad \bd} \, \bar{\f}_{\bd} \, .
\end{align}
We adopt the following conventions for contraction of spinor indices:
\begin{align}
	\psi \chi := \psi^{\a} \chi_{\a} = \chi \psi \, , \hspace{10mm} \bar{\psi} \bar{\chi} := \bar{\psi}_{\ad} \bar{\chi}^{\ad} = \bar{\chi} \bar{\psi} \, ,
\end{align}
in addition to $\psi^{2} = \psi^{\a} \psi_{\a}$, $\bar{\psi}^{2} = \bar{\psi}_{\ad} \bar{\psi}^{\ad}$. Conjugation of spinors should be understood as Hermitian conjugation
\begin{align}
	(\psi \chi)^{*} := (\psi^{\a} \chi_{\a})^{*} = (\chi_{\a})^{*} (\psi^{\a})^{*} = \bar{\chi}_{\ad} \bar{\psi}^{\ad} \, .
\end{align}

It is now useful to introduce the complex $2 \times 2$ $\s$-matrices, defined as follows:
\begin{align} \label{Ch02App01-sigma-matrices}
	\s_{0} &= 
	\begingroup
	\setlength\arraycolsep{4pt}
	\begin{pmatrix}
		\, 1 & 0 \, \\
		\, 0 & 1 \,
	\end{pmatrix}
	\endgroup 
	\, , & \hspace{5mm}
	\s_{1} &=
	\begingroup
	\setlength\arraycolsep{4pt}
	\begin{pmatrix}
		\, 0 & 1 \, \\
		\, 1 & 0 \,
	\end{pmatrix}
	\endgroup 
	\, , & \hspace{5mm}
	\s_{2} &=
	\begingroup
	\setlength\arraycolsep{4pt}
	\begin{pmatrix}
		\, 0 & -\text{i} \, \\
		\, \text{i} & 0 \,
	\end{pmatrix}
	\endgroup 
	\, , & \hspace{5mm}
	\s_{3} &=
	\begingroup
	\setlength\arraycolsep{4pt}
	\begin{pmatrix}
		\, 1 & 0 \, \\
		\, 0 & -1 \,
	\end{pmatrix}
	\endgroup 
	\, .
\end{align}
The $\s$-matrices span the Lie group $\text{SL}(2, \mathbb{C})$. Now let $\s_{m} = (\s_{0}, \vec{\s} )$, we denote the components of $\s_{m}$ as $(\s_{m})_{\a \ad}$, and define:
\begin{equation}
	(\tilde{\s}_{m})^{\ad \a} := \ve^{\ad \bd} \ve^{\a \b} (\s_{m})_{\b \bd} \, .
\end{equation}	
It can be shown that the $\s$-matrices possess the following useful properties:
\begin{subequations} \label{Ch02App01-sigma-matrix-relations}
	\begin{align}
		(\s_{m} \tilde{\s}_{n} + \s_{n} \tilde{\s}_{m}  )_{\a}{}^{\b} &= - 2 \eta_{m n} \, \d_{\a}^{\b} \, , \\
		(\tilde{\s}_{m } \s_{n} + \tilde{\s}_{n} \s_{m}  )^{\ad}{}_{\bd} &= - 2 \eta_{m n} \, \d^{\ad}_{\bd} \, , \\
		\text{Tr}(\s_{m} \tilde{\s}_{n} ) &= - 2 \eta_{m n} \, , \\
		(\s^{m})_{\a \ad} (\tilde{\s}_{m})^{\bd \b} &= - 2 \d_{\a}^{\b} \d_{\ad}^{\bd} \, .
	\end{align}
\end{subequations}
We may now introduce the antisymmetric and traceless matrices
\begin{subequations}
	\begin{align}
		(\s_{m n})_{\a}{}^{\b} = - \frac{1}{4} (\s_{m} \tilde{\s}_{n} - \s_{n} \tilde{\s}_{m})_{\a}{}^{\b} \, , \\
		(\tilde{\s}_{m n})^{\ad}{}_{\bd} = - \frac{1}{4} (\tilde{\s}_{m} \s_{n} - \tilde{\s}_{n} \s_{m})^{\ad}{}_{\bd} \, , 
	\end{align}
\end{subequations}
which are (anti-)self-dual due to the properties
\begin{align}
	\frac{1}{2} \e_{ a b c d } \s^{c d} = - \text{i} \s_{ab} \, , \hspace{10mm} \frac{1}{2} \e_{ a b c d } \tilde{\s}^{c d} = \text{i} \tilde{\s}_{ab} \, .
\end{align}
These objects also satisfy the Lorentz algebra
\begin{align} \label{App2B-Sigma matrix algebra}
	[ \s_{ab}, \s_{cd} ] = \eta_{a d} \s_{b c} - \eta_{a c} \s_{b d} + \eta_{b c} \s_{a d} - \eta_{b d} \s_{a c} \, .
\end{align}

The $\s$-matrices are used to convert spacetime indices into spinor ones (and vice-versa) according to the following rules:
\begin{equation}
	X_{\a \ad} = (\s^{m})_{\a \ad} X_{m} \, , \hspace{10mm} X_{m} = - \frac{1}{2} (\tilde{\s}_{m})^{\ad \a} X_{\a \ad} \, ,
\end{equation}
with similar results for partial derivatives. As an important example, we consider a real antisymmetric rank-2 tensor, $X_{m n} = - X_{n m}$; we may then convert $X_{m n}$ into spinor notation using the $\s$-matrices and decompose it into irreducible components as follows:
\begin{equation}
	X_{\a \ad, \b \bd} = (\s^{m})_{\a \ad} (\s^{n})_{\b \bd} X_{m n} = 2 \ve_{\a \b} \bar{X}_{\ad \bd} + 2 \ve_{\ad \bd} X_{\a \b} \, ,
\end{equation}
where we have defined the rank-2 spinors $X_{\a \b}$, $\bar{X}_{\ad \bd}$ according to the rules
\begin{align}
	X_{\a \b} = \frac{1}{2} (\s^{m n})_{\a \b} X_{m n} \, , \hspace{10mm}  	\bar{X}_{\ad \bd} = - \frac{1}{2} (\tilde{\s}^{m n})_{\ad \bd} X_{m n} \, .
\end{align}
This is particularly applicable to the Lorentz generators, $\Sigma_{m n} = - \Sigma_{n m}$, which are therefore equivalent to the pair of second rank spinors $(\Sigma_{\a \b}, \bar{\Sigma}_{\ad \bd})$. These generators satisfy the same algebra as the matrices $(\sigma_{m n}, \tilde{\sigma}_{m n})$, see  \eqref{App2B-Sigma matrix algebra}. The generators act on arbitrary spinors $\psi_{\g}, \bar{\psi}_{\gd}$ as follows:
\begin{subequations}
	\begin{align}
		\Sigma_{\a \b} \psi_{\g} = \ve_{\g (\a} \psi_{\b)} \, , \hspace{10mm} \Sigma_{\a \b} \bar{\psi}_{\gd} = 0 \, , \\[2mm]
		\bar{\Sigma}_{\ad \bd} \bar{\psi}_{\gd} = \ve_{\gd (\ad} \bar{\psi}_{\bd)} \, , \hspace{10mm} \bar{\Sigma}_{\ad \bd} \psi_{\g} = 0 \, .
	\end{align}
\end{subequations}

Another important example is symmetric and traceless tensors $\phi_{m_{1} \dots m_{s}}$, which are used to describe higher-spin currents. Using the $\sigma$-matrices we convert the spacetime indices to spinor indices (and vice-versa) according to the following rules:
\begin{subequations}
	\begin{align}
		\phi_{\a_{1} \dots \a_{s} \ad_{1} \dots \ad_{s}} &= (\s^{m_{1}})_{\a_{1} \ad_{1}} \dots (\s^{m_{s}})_{\a_{s} \ad_{s}} \phi_{m_{1} \dots m_{s}} \, , \\ 
		\phi_{m_{1} \dots m_{s}} &= \left(-\frac{1}{2}\right)^{s} (\tilde{\s}_{m_{1}})^{\ad_{1} \a_{1}} \dots (\tilde{\s}_{m_{s}})^{\ad_{s} \a_{s}} \phi_{\a_{1} \dots \a_{s} \ad_{1} \dots \ad_{s}} \, .
	\end{align}
\end{subequations}
By taking the trace in pairs of (un)dotted spinor indices, one may use  \eqref{Ch02App01-sigma-matrix-relations} to show that all such traces vanish due to $\phi$ with Lorentz indices being totally symmetric and traceless. Similarly, since $\phi_{m_{1} \dots m_{s}} = \phi_{(m_{1} \dots m_{s})}$, this implies that $\phi_{\a_{1} \dots \a_{s} \ad_{1} \dots \ad_{s}} = \phi_{(\a_{1} \dots \a_{s}) (\ad_{1} \dots \ad_{s})}$, i.e. it is (separately) symmetric in the undotted and dotted spinor indices respectively. Hence, totally symmetric and traceless tensors with Lorentz indices are in equivalence with totally symmetric and traceless spin-tensors in the $(s,s)$ representation (see also \cite{Buchbinder:1998qv} for a detailed discussion regarding spin-tensors in more general $(m,n)$ representations).

	\section{3D conventions and notation}\label{Appendix2B}

Many of the conventions for three-dimensions follow from the conventions used for four-dimensions, albeit with some minor modifications, see e.g. Appendix B of \cite{Kuzenko:2010rp}.

For the Minkowski metric we use the ``mostly plus'' convention: $\eta_{mn} = \text{diag}(-1,1,1)$. Spinor indices are then raised and lowered with the $\text{SL}(2,\mathbb{R})$ invariant anti-symmetric $\varepsilon$-tensor
\begin{subequations}
	\begin{align}
		\ve_{\a \b} = 
		\begingroup
		\setlength\arraycolsep{4pt}
		\begin{pmatrix}
			\, 0 & -1 \, \\
			\, 1 & 0 \,
		\end{pmatrix}
		\endgroup 
		\, , & \hspace{6mm}
		\ve^{\a \b} =
		\begingroup
		\setlength\arraycolsep{4pt}
		\begin{pmatrix}
			\, 0 & 1 \, \\
			\, -1 & 0 \,
		\end{pmatrix}
		\endgroup 
		\, , \hspace{6mm}
		\ve_{\a \g} \ve^{\g \b} = \d_{\a}{}^{\b} \, , \\[4mm]
		& \hspace{-8mm} \f_{\a} = \ve_{\a \b} \, \f^{\b} \, , \hspace{12mm} \f^{\a} = \ve^{\a \b} \, \f_{\b} \, .
	\end{align}
\end{subequations}
The Dirac gamma-matrices, $\gamma_{m}$, $m = 0,1,2$ are chosen to be real, and are expressed in terms of the Pauli matrices $\s_{i}$: 
\begin{subequations}
	\begin{align}
		\s_{1} &= 
		\begingroup
		\setlength\arraycolsep{4pt}
		\begin{pmatrix}
			\, 0 & 1 \, \\
			\, 1 & 0 \,
		\end{pmatrix}
		\endgroup 
		\, , &
		\s_{2} &= 
		\begingroup
		\setlength\arraycolsep{4pt}
		\begin{pmatrix}
			\, 0 & -\text{i} \, \\
			\, \text{i} & 0 \,
		\end{pmatrix}
		\endgroup 
		\, , &
		\s_{3} &= 
		\begingroup
		\setlength\arraycolsep{4pt}
		\begin{pmatrix}
			\, 1 & 0 \, \\
			\, 0 & -1 \,
		\end{pmatrix}
		\endgroup 
		\, ,
	\end{align}
\end{subequations}
so that $(\g^{m})_{\a}{}^{\b} = ( - \text{i} \s_{2}, \s_{1}, \s_{3} )$. We can raise and lower the indices on the $\g$-matrices as follows:
\begin{equation}
	(\g_{m})_{\a \b} = \ve_{\b \d} (\g_{m})_{\a}{}^{\d} \, , \hspace{10mm} (\g_{m})^{\a \b} = \ve^{\a \d} (\g_{m})_{\d}{}^{\b} \, .
\end{equation}
Note that the gamma-matrices with lowered indices may be obtained by starting with the 4D sigma-matrices \eqref{Ch02App01-sigma-matrices} and deleting the matrices with the index $m = 2$. The $\g$-matrices are traceless and symmetric
\begin{equation}
	(\g_{m})^{\a}{}_{\a} = 0 \, , \hspace{10mm} (\g_{m})_{\a \b} = (\g_{m})_{\b \a} \, ,
\end{equation} 
and also satisfy the Clifford algebra
\begin{equation}
	\{ \g_{m}, \g_{n} \} = 2 \eta_{mn} \, .
\end{equation}
For products of $\g$-matrices we have the following identities
\begin{subequations} \label{App2B-gamma matrix identities}
	\begin{align}
		(\g_{m})_{\a}{}^{\r} (\g_{n})_{\r}{}^{\b} &= \eta_{mn} \d_{\a}{}^{\b} + \e_{mnp} (\g^{p})_{\a}{}^{\b} \, , \\[2mm]
		\e_{m n p} (\g^{n})_{\a \b} (\g^{p})_{\g \d} &= \ve_{\g(\a} (\g_{m})_{\b)\d} + \ve_{\d(\a} (\g_{m})_{\b)\g} \, , \\[2mm]
		(\g_{m})_{\a}{}^{\r} (\g_{n})_{\r}{}^{\s} (\g_{p})_{\s}{}^{\b} &= \eta_{mn} (\g_{p})_{\a}{}^{\b} - \eta_{mp} (\g_{n})_{\a}{}^{\b} + \eta_{np} (\g_{m})_{\a}{}^{\b} + \e_{mnp} \d_{\a}{}^{\b} \, , \\[2mm]
		\text{tr}[ \g_{a} \g_{b} \g_{c} \g_{d} ] &= 2 \eta_{a b} \eta_{c d} - 2 \eta_{a c} \eta_{b d} + 2 \eta_{a d} \eta_{b c} \, ,
	\end{align}
\end{subequations}
where we have introduced the 3D Levi-Civita tensor $\e_{m n p}$, with the normalisation $\e^{012} = - \e_{012} = 1$. We also have the orthogonality and completeness relations for the $\g$-matrices
\begin{equation}
	(\g^{m})_{\a \b} (\g_{m})^{\r \s} = - \d_{\a}{}^{\r} \d_{\b}{}^{\s}  - \d_{\a}{}^{\s}  \d_{\b}{}^{\r} \, , \hspace{8mm} (\g_{m})_{\a \b} (\g_{n})^{\a \b} = -2 \eta_{mn} \, .
\end{equation}
The $\g$-matrices are used to swap from vector indices to spinor indices. For example, given some three-vector $x_{m}$, it may equivalently be expressed in terms of a symmetric second-rank spinor $x_{\a \b}$ as follows:
\begin{subequations}
	\begin{align}
		x_{\a \b} = (\g^{m})_{\a \b} x_{m}  \, , \hspace{5mm} x_{m} = - \frac{1}{2} (\g_{m})^{\a \b} x_{\a \b} \, , \\[2mm]
		\det [x_{\a \b}] = \frac{1}{2} x^{\a \b} x_{\a \b} = - x^{m} x_{m} = -x^{2} \, .
	\end{align}
\end{subequations}
The same conventions are also adopted for the spacetime partial derivatives $\partial_{m}$
\begin{subequations}
	\begin{align}
		\partial_{\a \b} = (\g^{m})_{\a \b} \partial_{m}  \, , \hspace{10mm} \partial_{m} = - \frac{1}{2} (\g_{m})^{\a \b} \partial_{\a \b} \, , \\[2mm]
		\partial_{m} x^{n} = \d_{m}^{n} \, , \hspace{10mm} \partial_{\a \b} x^{\r \s} = - \d_{\a}{}^{\r} \d_{\b}{}^{\s}  - \d_{\a}{}^{\s}  \d_{\b}{}^{\r} \, ,
	\end{align}
	\begin{equation}
		\x^{m} \partial_{m} = - \frac{1}{2} \x^{\a \b} \partial_{\a \b} \, .
	\end{equation}
\end{subequations}
In three dimensions, an antisymmetric tensor $X_{m n} = - X_{n m}$ is (Hodge) dual to a three-vector $X_{a}$ by the rule
\begin{align}
	X_{a} = \frac{1}{2} \e_{a m n} X^{m n} \, , \hspace{10mm} X_{m n} = - \e_{m n a} X^{a} \, . 
\end{align}
We can then, as in (2.170a), define a symmetric rank-2 spinor $X_{\a \b} = X_{\b \a}$ associated with the vector $X_{a}$ (and therefore the antisymmetric tensor $X_{m n}$) as follows:
\begin{align}
	X_{\a \b} := (\g^{a})_{\a \b} X_{a} = - \frac{1}{2} (\g^{a})_{\a \b} \e_{a m n} X^{m n} \, .
\end{align}
The algebraic objects $X_{m}$, $X_{m n}$ and $X_{\a \b}$ are in one-to-one correspondence with eachother. The inner products for these objects are related as follows:
\begin{align}
	- X^{m} P_{m} = \frac{1}{2} X^{m n} P_{m n} = \frac{1}{2} X^{\a \b} P_{\a \b} \, .
\end{align}
The Lorentz generators with two vector indices ($\Sigma_{m n} = - \Sigma_{n m}$), one vector index ($\Sigma_{m}$) and two spinor indices ($\Sigma_{\a \b} = \Sigma_{\b \a}$) are related to eachother by the rules
\begin{subequations}
	\begin{gather}
		\Sigma_{m n} = - \e_{ m n p } \Sigma^{p} \, , \hspace{10mm} \Sigma_{m} = \frac{1}{2} \e_{m a b} \Sigma^{a b} \, , \\[2mm]
		\Sigma_{\a \b} = (\g^{m})_{\a \b} \Sigma_{m} \, , \hspace{10mm} \Sigma_{m} = - \frac{1}{2} (\g_{m})^{\a \b} \Sigma_{\a \b} \, .
	\end{gather} 
\end{subequations}
The Lorentz generators act on a vector $V_{m}$ and a spinor $\psi_{\g}$ as follows:
\begin{align}
	\Sigma_{a b} V_{m} = 2 \eta_{m[a} V_{b]} \, , \hspace{10mm} \Sigma_{\a \b} \psi_{\g} = \ve_{\g (\a} \psi_{\b)} \, .
\end{align}

Similar to the 4D case, symmetric and traceless tensors $\phi_{m_{1} \dots m_{s}}$ are used to describe higher-spin currents. Using the $\gamma$-matrices we convert the spacetime indices to spinor indices (and vice-versa) according to the following rules:
\begin{subequations}
	\begin{align}
		\phi_{\a_{1} \dots \a_{2s}} &= (\g^{m_{1}})_{\a_{1} \a_{2}} \dots (\g^{m_{s}})_{\a_{2s-1} \a_{2s}} \phi_{m_{1} \dots m_{s}} \, , \\ 
		\phi_{m_{1} \dots m_{s}} &= \left(-\frac{1}{2}\right)^{s} (\g_{m_{1}})^{\a_{1} \a_{2}} \dots (\g_{m_{s}})^{\a_{2s-1} \a_{2s}} \phi_{\a_{1} \dots \a_{2s}} \, .
	\end{align}
\end{subequations}
By taking the trace in pairs of undotted spinor indices, one may use  \eqref{App2B-gamma matrix identities} to show that all such traces vanish due to $\phi$ with Lorentz indices being totally symmetric and traceless. Similarly, since $\phi_{m_{1} \dots m_{s}} = \phi_{(m_{1} \dots m_{s})}$, this implies that $\phi_{\a_{1} \dots \a_{2s}} = \phi_{(\a_{1} \dots \a_{2s})}$, i.e. it is symmetric in the undotted spinor indices. Hence, totally symmetric and traceless tensors with Lorentz indices are equivalent to totally symmetric and traceless spin-tensors. The story is similar for tensors of half-integer spin, which possess an overall odd number of undotted spinor indices. Upon conversion to Lorentz indices using the $\g$-matrices there is one spinor index left over and one must impose a $\gamma$-traceless constraint. The canonical example of such a half-integer spin current is the supersymmetry current, which we discuss in more detail in Chapter \ref{Chapter3}.

%
%


\end{subappendices}

\chapter{Correlation functions of conserved currents in 3D CFT} \label{Chapter3}
\graphicspath{{Images/3DCFT/}} 

In this chapter we develop a formalism to study the general structure of the three-point correlation functions of conserved currents in three-dimensional conformal field theory, assuming only the constraints imposed by conformal symmetry and conservation equations. Such a three-point function is denoted by
\begin{equation}
	\langle J^{}_{s_{1}}(x_{1}) \, J'_{s_{2}}(x_{2}) \, J''_{s_{3}}(x_{3}) \rangle \, .
\end{equation}
Here by $J^{}_{s}$ we denote a conserved current of integer or half-integer spin $s$. The formalism we present, based on the approach of Osborn and Petkou \cite{Osborn:1993cr}, is suitable for both integer and half-integer spin, and we reproduce all known results concerning the structure of three-point functions of integer-spin (bosonic) conserved currents. We also extend the results to three-point functions involving currents of an arbitrary half-integer spin and apply it to correlation functions involving conserved currents and scalar/spinor operators, thus covering essentially all possible three-point function in three-dimensional conformal field theory. The approach is exhaustive; first, we construct all possible structures for the three-point function for a given set of spins $s_1, s_2$ and $s_3$, consistent with its conformal properties. We then systematically identify the linearly independent structures and then, finally, impose the conservation equations and symmetries under permutations of spacetime points. As a result we obtain the three-point function in a form which can be explicitly presented for currents of arbitrary spins $s_1$, $s_2$, $s_3$ and is limited only by computer power. Due to these limitations we were able to carry out computations up to $s_{i} = 20$. It is found that in all cases with $s_{i} \leq 20$, including examples involving conserved {\it half-integer} spin currents, that the correlation function is fixed up to the following form:
\begin{equation}
	\langle J^{}_{s_{1}} J'_{s_{2}} J''_{s_{3}} \rangle = a_{1} \, \langle J^{}_{s_{1}} J'_{s_{2}} J''_{s_{3}} \rangle_{E_1} + a_{2} \, \langle J^{}_{s_{1}} J'_{s_{2}} J''_{s_{3}} \rangle_{E_2} + b \, \langle J^{}_{s_{1}} J'_{s_{2}} J''_{s_{3}} \rangle_{O} \, .
\end{equation}
where $\langle J^{}_{s_{1}} J'_{s_{2}} J''_{s_{3}} \rangle_{E_1}$ and $\langle J^{}_{s_{1}} J'_{s_{2}} J''_{s_{3}} \rangle_{E_2}$ are parity-even solutions (in the bosonic case corresponding to free bosonic and fermionic theories respectively), while $\langle J^{}_{s_{1}} J'_{s_{2}} J''_{s_{3}} \rangle_{O}$ is a parity-violating (or parity-odd) solution. Parity-odd solutions are of particular interest in three dimensions as they have been shown to correspond to Chern-Simons theories interacting with parity-violating matter \cite{Aharony:2011jz, Giombi:2011kc, Maldacena:2012sf, Jain:2012qi, GurAri:2012is, Aharony:2012nh, Giombi:2016zwa, Chowdhury:2017vel, Sezgin:2017jgm, Skvortsov:2018uru, Inbasekar:2019wdw}. Further, the existence of the parity-odd solution is dependent on a set of triangle inequalities:
\begin{align}
	s_{1} \leq s_{2} + s_{3} \, , && s_{2} \leq s_{1} + s_{3} \, , && s_{3} \leq s_{1} + s_{2} \, .
\end{align}
When the triangle inequalities are simultaneously satisfied there are two even solutions, and one odd solution. However, when any one of the above relations is not satisfied there are only two even solutions; the odd solution is incompatible with conservation equations. 

The results of this chapter are organised as follows. In Section \ref{Ch03-section3.1} we review the essentials of the group theoretic formalism used to construct correlation functions of 
primary operators in three dimensions. In Subsection \ref{Ch03-section3.3} we develop the formalism necessary to impose all constraints arising from conservation equations and point switch symmetries on three-point functions. 
In particular, we introduce an index-free, auxiliary spinor formalism which allows us to construct a generating function for the three-point functions, and we outline the important aspects of our computational approach. Section \ref{Ch03-section3.4} is then devoted to the analysis of three-point functions of conserved currents. In particular, we reproduce the known results previously proposed in~\cite{Maldacena:2011jn,Giombi:2011rz,Zhiboedov:2012bm}. We then use our formalism to analyse the structure of correlation functions involving fermionic currents. We present an explicit analysis for three-point correlation functions involving 
combinations of ``supersymmetry-like" spin-3/2 currents, the energy-momentum tensor and a conserved vector current. 
The results are then expanded to include higher-spin conserved currents. In Subsection \ref{Ch03-subsection3.4.3}, for completeness, we present a complete classification for the structure of three-point functions involving 
combinations of scalars, spinors and conserved higher-spin currents. The appendices \ref{Appendix3B}, \ref{Appendix3C} are devoted to mathematical conventions, various useful identities and extra results for three-point functions.

\section{Conformal building blocks}\label{Ch03-section3.1}

Consider 3D Minkowski space $\mathbb{M}^{3}$, parameterised by coordinates $ x^{m} $, where $m = 0, 1, 2$ are Lorentz indices. We consider the vector two-point functions \eqref{Ch02-Two-point building blocks 1} in the spinor representation, using the conventions outlined in Appendix \ref{Appendix2B}:
\begin{align}
	x_{12 \, \a \b} &= (\g^{m})_{\a \b} x_{12 \, m} \, , & x_{12}^{\a \b} &= (\g^{m})^{\a \b} x_{12 \, m} \, , & x_{12}^{2} &= - \frac{1}{2} x_{12}^{\a \b} x^{}_{12 \, \a \b} \, .
\end{align}
In this form the two-point functions possess the following useful properties:
\begin{align}  \label{Ch03-Two-point building blocks - properties 1} 
	x_{12 \, \a \b} = x_{12 \, \b \a} \, , \hspace{10mm} x_{12}^{\a \s} x^{}_{12 \, \s \b} = - x_{12}^{2} \d_{\b}^{\a} \, . 
\end{align}
Hence, we find
\begin{equation} \label{Ch03-Two-point building blocks 4}
	(x_{12}^{-1})^{\a \b} = - \frac{x_{12}^{\a \b}}{x_{12}^{2}} \, .
\end{equation}
We also introduce the normalised two-point functions, denoted by $\hat{x}_{12}$,
\begin{align} \label{Ch03-Two-point building blocks 3}
	\hat{x}_{12 \, \a \b} = \frac{x_{12 \, \a \b}}{( x_{12}^{2})^{1/2}} \, , \hspace{10mm} \hat{x}_{12}^{\a \s} \hat{x}^{}_{12 \, \s \b} = - \d_{\b}^{\a} \, . 
\end{align}
From here we can now construct an operator analogous to the conformal inversion tensor acting on the space of symmetric traceless spin-tensors of arbitrary rank. Given a two-point function $x$, we define the operator
\begin{equation} \label{Ch03-Higher-spin inversion operators a}
	\cI_{\a(k) \b(k)}(x) = \hat{x}_{(\a_{1} (\b_{1}} \dots \hat{x}_{ \a_{k}) \b_{k})}  \, ,
\end{equation}
along with its inverse
\begin{equation} \label{Ch03-Higher-spin inversion operators b}
	\cI^{\a(k) \b(k)}(x) = \hat{x}^{(\a_{1} (\b_{1}} \dots \hat{x}^{ \a_{k}) \b_{k})} \, .
\end{equation}
The spinor indices may be raised and lowered using the standard conventions as follows:
\begin{subequations}
	\begin{align}
		\cI_{\a(k)}{}^{\b(k)}(x) &= \ve^{\b_{1} \g_{1}} \dots \ve^{\b_{k} \g_{k}} \, \cI_{\a(k) \g(k)}(x) \, .
	\end{align}
\end{subequations}
Now due to the property
\begin{equation}
	\cI_{\a(k) \b(k)}(-x) = (-1)^{k} \cI_{\a(k) \b(k)}(x) \, ,
\end{equation}
the following identity holds for products of inversion tensors:
\begin{subequations} \label{Ch03-Higher-spin inversion operators - properties}
	\begin{align}
		\cI_{\a(k) \s(k)}(x_{12}) \, \cI^{\s(k) \b(k)}(x_{21}) &= \d_{(\a_{1}}^{(\b_{1}} \dots \d_{\a_{k})}^{\b_{k})} \, .
	\end{align}
\end{subequations}
The objects \eqref{Ch03-Higher-spin inversion operators a}, \eqref{Ch03-Higher-spin inversion operators b} prove to be essential in the construction of correlation functions of primary operators with arbitrary spin. Indeed, the vector representation of the inversion tensor may be recovered in terms of the spinor two-point functions as follows:
\begin{equation}
	I_{m n}(x) = - \frac{1}{2} \, \text{Tr}( \g_{m} \, \hat{x} \, \g_{n} \, \hat{x} ) \, .
\end{equation}
%
%

Given three distinct points in Minkowski space, $x_{i}$, with $i = 1,2,3$, we recall the conformally covariant three-point functions
\begin{align}
	X_{ij} &= \frac{x_{ik}}{x_{ik}^{2}} - \frac{x_{jk}}{x_{jk}^{2}} \, , & X_{ji} &= - X_{ij} \, ,  & X_{ij}^{2} &= \frac{x_{ij}^{2}}{x_{ik}^{2} x_{jk}^{2} } \, , 
\end{align}
where $(i,j,k)$ is a cyclic permutation of $(1,2,3)$. Converting to spinor notation, these objects may be represented as follows:
\begin{equation}
	X_{ij , \, \a \b} = (\g_{m})_{\a \b} X_{ij}^{m} \, , \hspace{10mm} X_{ij , \, \a \b} = - (x^{-1}_{ik})_{\a \s} x_{ij}^{\s \g} (x^{-1}_{kj})_{\g \b} \, .
\end{equation}
These objects satisfy properties similar to the two-point functions, as in \eqref{Ch03-Two-point building blocks - properties 1}. It is also convenient to define the normalised three-point functions, $\hat{X}_{ij}$, and the inverses, $(X_{ij}^{-1})$,
\begin{equation}
	\hat{X}_{ij , \, \a \b} = \frac{X_{ij , \, \a \b}}{( X_{ij}^{2})^{1/2}} \, , \hspace{10mm}	(X_{ij}^{-1})^{\a \b} = - \frac{X_{ij}^{\a \b}}{X_{ij}^{2}} \, .
\end{equation}  
Now given an arbitrary three-point building block, $X$, it is useful to construct the following higher-spin inversion operator:
\begin{equation}
	\cI_{\a(k) \b(k)}(X) = \hat{X}_{ (\a_{1} (\b_{1}} \dots \hat{X}_{\a_{k}) \b_{k})}  \, , \label{Ch03-Inversion tensor identities - three point functions a}
\end{equation}
along with its inverse
\begin{equation}
	\cI^{\a(k) \b(k)}(X) = \hat{X}^{(\a_{1} (\b_{1}} \dots \hat{X}^{ \a_{k}) \b_{k})} \, . \label{Ch03-Inversion tensor identities - three point functions b}
\end{equation}
These operators possess properties similar to the two-point higher-spin inversion operators \eqref{Ch03-Higher-spin inversion operators a}, \eqref{Ch03-Higher-spin inversion operators b}. There are also some useful algebraic identities relating the two- and three-point functions at various points, such as
\begin{subequations}  \label{Ch03-Inversion tensor identities - spinor case}
	\begin{align}
		\cI_{\a \s}(x_{13}) \, \cI^{\s \g}(x_{12}) \, \cI_{\g \b}(x_{23}) &= \cI_{\a \b}(X_{12}) \, , \\
		\cI^{\a \s}(x_{13}) \, \cI_{\s \g}(X_{12})  \, \cI^{\g \b}(x_{13}) &= \cI^{\a \b}(X^{I}_{23})  \, ,
	\end{align}
\end{subequations}
where $X^{I}_{\a \b} = \cI_{\a}{}^{\a'}(X) \, \cI_{\b}{}^{\b'}(X) \, X_{\a' \b'} = - X_{\a \b}$. These identities (and their cyclic permutations) are analogous to \eqref{Ch02-Inversion tensor identities - vector case 1}, \eqref{Ch02-Inversion tensor identities - vector case 2}, and also admit the following generalisations to higher-spins:
\begin{equation}
	\cI^{\a(k) \s(k)}(x_{13}) \, \cI_{\s(k) \g(k)}(X_{12}) \, \cI^{\g(k) \b(k)}(x_{13}) = \cI^{\a(k) \b(k)}(X^{I}_{23})  \, . \label{Ch03-Inversion tensor identities - higher spin case}
\end{equation}
In addition, similar to \eqref{Ch02-Inversion tensor identities - vector case 3}, there are also the following identities:
\begin{equation}
	\pa_{(1) \, \a \b} X_{12}^{ \g \d} = - \frac{2}{x_{13}^{2}} \, \cI_{(\a}{}^{\g}(x_{13}) \,  \cI_{\b)}{}^{\d}(x_{13}) \, , \hspace{5mm} \pa_{(2) \, \a \b} X_{12}^{ \g \d} = \frac{2}{x_{23}^{2}} \, \cI_{(\a}{}^{\g}(x_{23}) \,  \cI_{\b)}{}^{\d}(x_{23}) \, . \label{Ch03-Three-point building blocks - differential identities}
\end{equation}
These identities allow us to account for the fact that correlation functions of primary fields obey differential constraints which can arise due to conservation equations. Given a tensor field $\cT_{\cA}(X)$, there are the following differential identities which arise as a consequence of \eqref{Ch03-Three-point building blocks - differential identities}:
\begin{subequations}
	\begin{align}
		\pa_{(1) \, \a \b} \cT_{\cA}(X_{12}) &= \frac{1}{x_{13}^{2}} \, \cI_{\a}{}^{\g}(x_{13}) \,  \cI_{\b}{}^{\d}(x_{13}) \, \frac{ \pa}{ \pa X_{12}^{ \g \d}} \, \cT_{\cA}(X_{12}) \, ,  \label{Ch03-Three-point building blocks - differential identities 2} \\[2mm]
		\pa_{(2) \, \a \b} \cT_{\cA}(X_{12}) &= - \frac{1}{x_{23}^{2}} \, \cI_{\a}{}^{\g}(x_{23}) \,  \cI_{\b}{}^{\d}(x_{23}) \, \frac{ \pa}{ \pa X_{12}^{ \g \d}} \, \cT_{\cA}(X_{12}) \, . \label{Ch03-Three-point building blocks - differential identities 3}
	\end{align}
\end{subequations}
%

%

\subsection{Two-point correlation functions}\label{Ch03-section3.2}

Let $\phi_{\cA}$ be a primary field with dimension $\D$, where $\cA$ denotes a collection of Lorentz spinor indices. The two-point correlation function of $\phi_{\cA}$ is fixed by conformal symmetry to the form
\begin{equation} \label{Ch03-Two-point correlation function}
	\langle \phi_{\cA}(x_{1}) \, \phi^{\cB}(x_{2}) \rangle = c \, \frac{\cI_{\cA}{}^{\cB}(x_{12})}{(x_{12}^{2})^{\D}} \, , 
\end{equation} 
where $\cI$ is an appropriate representation of the inversion tensor and $c$ is a constant real parameter. The denominator of the two-point function is determined by the conformal dimension of $\phi$, which guarantees that the two-point function transforms with the appropriate weight under scale transformations. 

For two-point functions of conserved higher-spin currents, we use \eqref{Ch03-Higher-spin inversion operators a} and obtain the result
\begin{equation} 
	\langle J_{\a(2s)}(x_{1}) \, J^{\b(2s)}(x_{2}) \rangle = c \, \frac{\cI_{\a(2s)}{}^{\b(2s)}(x_{12})}{(x_{12}^{2})^{\D_{J}}} \, .
\end{equation} 
We then impose $(\g_{m})^{\a_{1} \a_{2}} \pa^{m}_{(1)} \langle J_{\a_{1} \a_{2} \a(2s-2)}(x_{1}) \, J^{\b(2s)}(x_{2}) \rangle = 0$ to obtain $\D_{J}= s+1$, which is consistent with the dimension of a conserved current of arbitrary spin-$s$. 

To simplify imposing the constraints due to conservation equations on two- and three-point functions of conserved higher-spin currents (which we recall are totally symmetric and traceless), we now introduce an essential technique which we will utilise throughout the rest of this thesis. The method is based on contracting the free tensor indices on the currents with commuting auxiliary spinors $u, v$ at $x_{1}$ and $x_{2}$ which satisfy
\begin{align}
	u^2 = \varepsilon_{\a \b} \, u^{\a} u^{\b}=0\,,  \hspace{10mm}
	v^{2} = \varepsilon_{\a \b} \, v^{\a} v^{\b}=0\, ,
\end{align}
such that
\begin{align}
	J_{s}(x_{1};u) = J_{\a_{1} \dots \a_{2s}}(x_{1}) \, u^{\a_{1}} \dots u^{\a_{2s}} \, ,
\end{align}
with a similar result for the current at $x_{2}$. The two-point function now has the following compact form:
\begin{equation} 
	\langle J_{s}(x_{1}; u) \, J_{s}(x_{2}; v)  \rangle = c \, \frac{( u \cdot \hat{x}_{12} \cdot v)^{2s} }{(x_{12}^{2})^{\D_{J}}} \, ,
\end{equation} 
where $u \cdot \hat{x}_{12} \cdot v = \hat{x}_{12 \, \a \b} u^{\a}  v^{\b}$. Hence, the two-point function is encoded in a polynomial of homogeneity $2s$ in the auxiliary spinors $u$ and $v$. The original ``tensor" two-point function is now obtained by acting on it with the partial derivative operators
\begin{align}
	\frac{\pa}{\pa \boldsymbol{u}^{\a(2s)}} = \frac{1}{(2s)!} \frac{\pa}{\pa u^{\a_{1}}} \dots \frac{\pa}{\pa u^{\a_{2s}}} \, , \hspace{10mm} \frac{\pa}{\pa \boldsymbol{v}^{\b(2s)}} = \frac{1}{(2s)!} \frac{\pa}{\pa v^{\b_{1}}} \dots \frac{\pa}{\pa v^{\b_{2s}}} \, . 
\end{align}
The tensor and polynomial forms of the two-point function are in one-to-one correspondence with eachother. Imposing conservation on $x_{1}$ is now tantamount to the condition
\begin{equation}
	D_{1} \langle J_{s}(x_{1}; u) \, J_{s}(x_{2}; v)  \rangle = 0 \, , \hspace{10mm} D_{1} = \frac{\pa}{\pa x_{(1) \a \b}} \frac{\pa}{\pa u^{\a}} \frac{\pa}{\pa u^{\b}} \, ,
\end{equation}
where the operator $D_{1}$ should be understood as a contraction of a partial derivative in $x_{1}$ with the $\gamma$-matrices. It is then a short calculation to demonstrate that the conservation condition is satisfied for $\D_{J} = s + 1$, as expected. We will comment more on this approach in Subsection \ref{Ch03-subsubsection3.3.3}.

\subsection{Three-point correlation functions}\label{Ch03-section3.3}

Now concerning three-point correlation functions, let $\phi$, $\psi$, $\pi$ be primary fields with scale dimensions $\D_{1}$, $\D_{2}$ and $\D_{3}$ respectively. The three-point function may be constructed using the general ansatz
\begin{align}
	\langle \phi_{\cA_{1}}(x_{1}) \, \psi_{\cA_{2}}(x_{2}) \, \pi_{\cA_{3}}(x_{3}) \rangle = \frac{ \cI^{(1)}{}_{\cA_{1}}{}^{\cA'_{1}}(x_{13}) \,  \cI^{(2)}{}_{\cA_{2}}{}^{\cA'_{2}}(x_{23}) }{(x_{13}^{2})^{\D_{1}} (x_{23}^{2})^{\D_{2}}}
	\; \cH_{\cA'_{1} \cA'_{2} \cA_{3}}(X_{12}) \, , \label{Ch03-Three-point function - general ansatz}
\end{align} 
where the tensor $\cH_{\cA_{1} \cA_{2} \cA_{3}}(X)$ encodes all information about the correlation function, and is related to the leading singular OPE coefficient \cite{Osborn:1993cr}. It is highly constrained by conformal symmetry as follows:
\begin{enumerate}
	\item[\textbf{(i)}] Under scale transformations of Minkowski space $x^{m} \mapsto x'^{m} = \l^{-2} x^{m}$, the three-point building blocks transform as $X^{m} \mapsto X'^{m} = \l^{2} X^{m}$. As a consequence, the correlation function transforms as 
	\begin{equation}
		\langle \phi_{\cA_{1}}(x_{1}') \, \psi_{\cA_{2}}(x_{2}') \, \pi_{\cA_{3}}(x_{3}') \rangle = (\l^{2})^{\D_{1} + \D_{2} + \D_{3}} \langle \phi_{\cA_{1}}(x_{1}) \, \psi_{\cA_{2}}(x_{2}) \,  \pi_{\cA_{3}}(x_{3}) \rangle \, ,
	\end{equation}
	which implies that $\cH$ obeys the scaling property
	\begin{equation}
		\cH_{\cA_{1} \cA_{2} \cA_{3}}(\l^{2} X) = (\l^{2})^{\D_{3} - \D_{2} - \D_{1}} \, \cH_{\cA_{1} \cA_{2} \cA_{3}}(X) \, , \hspace{5mm} \forall \l \in \mathbb{R} \, \backslash \, \{ 0 \} \, .
	\end{equation}
	This guarantees that the correlation function transforms correctly under scale transformations.
	
	\item[\textbf{(ii)}] If any of the fields $\phi$, $\psi$, $\pi$ obey differential equations, such as conservation laws in the case of conserved currents, then the tensor $\cH$ is also constrained by differential equations which may be derived with the aid of identities \eqref{Ch03-Three-point building blocks - differential identities 2}, \eqref{Ch03-Three-point building blocks - differential identities 3}.
	
	\item[\textbf{(iii)}] If any (or all) of the operators $\phi$, $\psi$, $\pi$ coincide, the correlation function possesses symmetries under permutations of spacetime points, e.g.
	\begin{equation}
		\langle \phi_{\cA_{1}}(x_{1}) \, \phi_{\cA_{2}}(x_{2}) \, \pi_{\cA_{3}}(x_{3}) \rangle = (-1)^{\e(\phi)} \langle \phi_{\cA_{2}}(x_{2}) \, \phi_{\cA_{1}}(x_{1}) \, \pi_{\cA_{3}}(x_{3}) \rangle \, ,
	\end{equation}
	where $\e(\phi)$ is the Grassmann parity of $\phi$. As a consequence, the tensor $\cH$ obeys constraints which will be referred to as ``point-switch identities".
	
\end{enumerate}
The constraints above fix the functional form of $\cH$ (and therefore the correlation function) up to finitely many independent parameters. Hence, using the general formula \eqref{Ch03-H ansatz}, the problem of computing three-point correlation functions is reduced to deriving the general structure of the tensor $\cH$ subject to the above constraints.

\subsubsection{Comments on differential constraints}\label{Ch03-subsubsection3.3.1}

An important aspect of the construction is that the fields in the correlation function are not treated equally. Depending on the way in which the general ansatz \eqref{Ch03-H ansatz} is constructed, it can be difficult to impose conservation equations on one of the three fields due to a lack of useful identities such as \eqref{Ch03-Three-point building blocks - differential identities 2}, \eqref{Ch03-Three-point building blocks - differential identities 3}. To illustrate this more clearly, consider the following example; suppose we want to determine the solution for the correlation function $\langle \phi_{\cA_{1}}(x_{1}) \, \psi_{\cA_{2}}(x_{2}) \, \pi_{\cA_{3}}(x_{3}) \rangle$, with the ansatz
\begin{equation} \label{Ch03-H ansatz}
	\langle \phi_{\cA_{1}}(x_{1}) \, \psi_{\cA_{2}}(x_{2}) \, \pi_{\cA_{3}}(x_{3}) \rangle = \frac{ \cI^{(1)}{}_{\cA_{1}}{}^{\cA'_{1}}(x_{13}) \,  \cI^{(2)}{}_{\cA_{2}}{}^{\cA'_{2}}(x_{23}) }{(x_{13}^{2})^{\D_{1}} (x_{23}^{2})^{\D_{2}}}
	\; \cH_{\cA'_{1} \cA'_{2} \cA_{3}}(X_{12}) \, . 
\end{equation} 
All information about this correlation function is encoded in the tensor $\cH$, however, this particular formulation of the three-point function prevents us from imposing conservation on the field $\pi$. To rectify this issue we reformulate the ansatz with $\pi$ at the front as follows:
\begin{equation} \label{Ch03-Htilde ansatz}
	\langle \pi_{\cA_{3}}(x_{3}) \, \psi_{\cA_{2}}(x_{2}) \, \phi_{\cA_{1}}(x_{1}) \rangle = \frac{ \cI^{(3)}{}_{\cA_{3}}{}^{\cA'_{3}}(x_{31}) \,  \cI^{(2)}{}_{\cA_{2}}{}^{\cA'_{2}}(x_{21}) }{(x_{31}^{2})^{\D_{3}} (x_{21}^{2})^{\D_{2}}}
	\; \tilde{\cH}_{\cA_{1} \cA'_{2} \cA'_{3} }(X_{23}) \, . 
\end{equation} 
In this case, all information about this correlation function is now encoded in the tensor $\tilde{\cH}$. Conservation on $\pi$ can now be imposed by treating $x_{3}$ as the first point with the aid of identities analogous to \eqref{Ch03-Three-point building blocks - differential identities}, \eqref{Ch03-Three-point building blocks - differential identities 2}, \eqref{Ch03-Three-point building blocks - differential identities 3} with $x_{1} \leftrightarrow x_{3}$. We now require an equation relating the tensors $\cH$ and $\tilde{\cH}$, which correspond to different representations of the same correlation function. Since the two ansatz above must be equal, we obtain
\begin{align} \label{Ch03-Htilde and H relation}
	\tilde{\cH}_{\cA_{1} \cA_{2}  \cA_{3} }(X_{23}) &= (x_{13}^{2})^{\D_{3} - \D_{1}} \bigg(\frac{x_{21}^{2}}{x_{23}^{2}} \bigg)^{\hspace{-1mm} \D_{2}} \, \cI^{(1)}{}_{\cA_{1}}{}^{\cA'_{1}}(x_{13}) \, \cI^{(2)}{}_{\cA_{2}}{}^{\cB_{2}}(x_{12}) \,  \cI^{(2)}{}_{\cB_{2}}{}^{\cA'_{2}}(x_{23}) \nonumber \\[-2mm]
	& \hspace{50mm} \times \cI^{(3)}{}_{\cA_{3}}{}^{\cA'_{3}}(x_{13}) \, \cH_{\cA'_{1} \cA'_{2} \cA'_{3}}(X_{12}) \, ,
\end{align}
where we have absorbed any signs due to Grassmann parity into the overall normalisation of $\tilde{\cH}$. At this point it is convenient to partition our solution into ``even" and ``odd" sectors as follows:
\begin{equation}
	\cH_{\cA_{1} \cA_{2}  \cA_{3} }(X) = \cH^{(+)}_{\cA_{1} \cA_{2}  \cA_{3} }(X) + \cH^{(-)}_{\cA_{1} \cA_{2}  \cA_{3} }(X) \, , 
\end{equation}
where $\cH^{(+)}$ contains all structures involving an even number of spinor metrics, $\ve_{\a \b}$, and $\cH^{(-)}$ contains structures involving an odd number of spinor metrics. With this choice of convention, as a consequence of \eqref{Ch03-Inversion tensor identities - spinor case}, \eqref{Ch03-Inversion tensor identities - higher spin case}, the following relation holds:
\begin{align} \label{Ch03-Hc and H relation}
	\cH^{(\pm)}_{\cA_{1} \cA_{2} \cA_{3}}(X^{I}_{23}) &= \pm \, (x_{13}^{2} X_{23}^{2})^{\D_{3} - \D_{2} - \D_{1}} \cI^{(1)}{}_{\cA_{1}}{}^{\cA'_{1}}(x_{13}) \, \cI^{(2)}{}_{\cA_{2}}{}^{\cA'_{2}}(x_{13}) \nonumber \\
	& \hspace{45mm} \times \cI^{(3)}{}_{\cA_{3}}{}^{\cA'_{3}}(x_{13}) \, \cH^{(\pm)}_{\cA'_{1} \cA'_{2} \cA'_{3}}(X_{12}) \, .
\end{align}
This equation essentially extends (2.14) in \cite{Osborn:1993cr} to spin-tensor representations, and allows us to construct an equation relating the different representations of the three-point function. After substituting \eqref{Ch03-Hc and H relation} directly into \eqref{Ch03-Htilde and H relation}, we apply identities such as \eqref{Ch03-Inversion tensor identities - spinor case} to obtain the following relation between $\cH$ and $\tilde{\cH}$:
\begin{equation} \label{Ch03-Htilde and Hc relation}
	\tilde{\cH}^{(\pm)}_{\cA_{1} \cA_{2} \cA_{3} }(X) = (X^{2})^{\D_{1} - \D_{3}} \, \cI^{(2)}{}_{\cA_{2}}{}^{\cA'_{2}}(X) \, \cH^{(\pm)}_{\cA_{1} \cA'_{2} \cA_{3}}(X^{I}) \, . 
\end{equation}
We see here that $\cI$ acts as an intertwining operator between the different representations of the three-point function. Once $\tilde{\cH}$ is obtained we can then impose conservation on $\pi$ as if it were located at the ``first point'', using identities analogous to \eqref{Ch03-Three-point building blocks - differential identities}. It is also important to note that the even and odd sectors of the correlation function are linearly independent, and therefore may be considered separately in the constraint analysis. Another result that follows from the properties \eqref{Ch03-Inversion tensor identities - three point functions a}, \eqref{Ch03-Inversion tensor identities - three point functions b} and \eqref{Ch03-Inversion tensor identities - higher spin case} of the inversion tensor is
\begin{align} \label{Ch03-H inversion}
	\cH^{(\pm)}_{\cA_{1} \cA_{2} \cA_{3}}(X^{I}) = \pm \, \cI^{(1)}{}_{\cA_{1}}{}^{\cA'_{1}}(X) \, \cI^{(2)}{}_{\cA_{2}}{}^{\cA'_{2}}(X) \, \cI^{(3)}{}_{\cA_{3}}{}^{\cA'_{3}}(X) \, \cH^{(\pm)}_{\cA'_{1} \cA'_{2} \cA'_{3}}(X) \, .
\end{align}
That is, ``even" structures are invariant under the action of $\cI$, while ``odd" structures are pseudo-invariant under the action of $\cI$.

If we now consider the correlation function of three conserved primaries $J^{}_{\a(I)}$, $J'_{\b(J)}$, $J''_{\g(K)}$, where $I=2s_{1}$, $J=2s_{2}$, $K=2s_{3}$, then the general ansatz is
\begin{align} \label{Ch03-Conserved correlator ansatz}
	\langle J^{}_{\a(I)}(x_{1}) \, J'_{\b(J)}(x_{2}) \, J''_{\g(K)}(x_{3}) \rangle = \frac{ \cI_{\a(I)}{}^{\a'(I)}(x_{13}) \,  \cI_{\b(J)}{}^{\b'(J)}(x_{23}) }{(x_{13}^{2})^{\D_{1}} (x_{23}^{2})^{\D_{2}}}
	\; \cH_{\a'(I) \b'(J) \g(K)}(X_{12}) \, ,
\end{align} 
where $\D_{i} = s_{i} + 1$. The constraints on $\cH$ are then as follows:
\begin{enumerate}
	\item[\textbf{(i)}] {\bf Homogeneity:}
	\begin{equation}
		\cH_{\a(I) \b(J) \g(K)}(\l^{2} X) = (\l^{2})^{\D_{3} - \D_{2} - \D_{1}} \, \cH_{\a(I) \b(J) \g(K)}(X) \, , \hspace{5mm} \forall \l \in \mathbb{R} \, \backslash \, \{ 0 \} \, .
	\end{equation}

	\item[\textbf{(ii)}] {\bf Differential constraints:} \\
	After application of the identities \eqref{Ch03-Three-point building blocks - differential identities 2}, \eqref{Ch03-Three-point building blocks - differential identities 3} we obtain the following constraints:
	\begin{subequations}
		\begin{align}
			\text{Conservation at $x_{1}$:} && \pa^{\a_{1} \a_{2}}_{X} \cH_{\a_{1} \a_{2} \a(I - 2) \b(J) \g(K)}(X) &= 0 \, , \\
			\text{Conservation at $x_{2}$:} && \pa^{\b_{1} \b_{2}}_{X} \cH_{\a(I) \b_{1} \b_{2} \b(J-2) \g(K)}(X) &= 0 \, , \\
			\text{Conservation at $x_{3}$:} && \pa^{\g_{1} \g_{2}}_{X} \tilde{\cH}_{\a(I) \b(J) \g_{1} \g_{2} \g(K-2)  }(X) &= 0 \, ,
		\end{align}
	\end{subequations}
	where 
	\begin{equation}
		\tilde{\cH}^{(\pm)}_{\a(I) \b(J) \g(K) }(X) = \pm \, (X^{2})^{\D_{1} - \D_{3}} \, \cI_{\b(J)}{}^{\b'(J)}(X) \, \cH^{(\pm)}_{\a(I) \b'(J) \g(K)}(-X) \, . 
	\end{equation}

	\item[\textbf{(iii)}] {\bf Point-switch symmetries:} \\
	If the fields $J$ and $J'$ coincide, then we obtain the following point-switch identity
	\begin{equation}
		\cH_{\a(I) \b(I) \g(K)}(X) = (-1)^{\e(J)} \cH_{\b(I) \a(I) \g(K)}(-X) \, ,
	\end{equation}
	where $\e(J)$ is the Grassmann parity of $J$. Likewise, if the fields $J$ and $J''$ coincide, then we obtain the constraint
	\begin{equation}
		\tilde{\cH}_{\a(I) \b(J) \g(I) }(X) = (-1)^{\e(J)} \cH_{\g(I) \b(J) \a(I)}(-X) \, .
	\end{equation}
\end{enumerate}
In practice, imposing the constraints above on correlation functions involving higher-spin currents quickly becomes unwieldy using the tensor formalism outlined above. In the next subsections we develop an index-free formalism to efficiently carry out the calculations.

\subsubsection{Generating function formalism}\label{Ch03-subsubsection3.3.3}

Suppose we must analyse the constraints on a general spin-tensor $\cH_{\cA_{1} \cA_{2} \cA_{3}}(X)$, where $\cA_{1} = \{ \a_{1}, ... , \a_{I} \}, \cA_{2} = \{ \b_{1}, ... , \b_{J} \}, \cA_{3} = \{ \g_{1}, ... , \g_{K} \}$ represent sets of totally symmetric spinor indices associated with the fields at points $x_{1}$, $x_{2}$ and $x_{3}$ respectively. We introduce sets of commuting auxiliary spinors for each point; $u$ at $x_{1}$, $v$ at $x_{2}$, and $w$ at $x_{3}$, where the spinors satisfy 
\begin{align}
	u^2 &= \varepsilon_{\a \b} \, u^{\a} u^{\b}=0\,,  &
	v^{2} &= \varepsilon_{\a \b} \, v^{\a} v^{\b}=0\,, & w^{2} &= \varepsilon_{\a \b} \, w^{\a} w^{\b}=0\,. 
	\label{Ch03-extra1}
\end{align}
Now if we define the objects
\begin{subequations}
	\begin{align}
		\boldsymbol{u}^{\cA_{1}} &\equiv \boldsymbol{u}^{\a(I)} = u^{\a_{1}} \dots u^{\a_{I}} \, , \\
		\boldsymbol{v}^{\cA_{2}} &\equiv \boldsymbol{v}^{\b(J)} = v^{\b_{1}} \dots v^{\b_{J}} \, , \\
		\boldsymbol{w}^{\cA_{3}} &\equiv \boldsymbol{w}^{\g(K)} = w^{\g_{1}} \dots w^{\g_{K}} \, ,
	\end{align}
\end{subequations}
then the generating polynomial for $\cH$ is constructed as follows:
\begin{equation} \label{Ch03-H - generating polynomial}
	\cH(X; u,v,w) = \,\cH_{ \cA_{1} \cA_{2} \cA_{3} }(X) \, \boldsymbol{u}^{\cA_{1}} \boldsymbol{v}^{\cA_{2}} \boldsymbol{w}^{\cA_{3}} \, . \\
\end{equation}
%
There is a one-to-one mapping between the space of symmetric traceless spin tensors and the polynomials constructed using the above method. Indeed, the tensor $\cH$ is extracted from the polynomial by acting on it with the following partial derivative operators:
\begin{subequations}
	\begin{align}
		\frac{\pa}{\pa \boldsymbol{u}^{\cA_{1}} } &\equiv \frac{\pa}{\pa \boldsymbol{u}^{\a(I)}} = \frac{1}{I!} \frac{\pa}{\pa u^{\a_{1}} } \dots \frac{\pa}{\pa u^{\a_{I}}}  \, , \\
		\frac{\pa}{\pa \boldsymbol{v}^{\cA_{2}} } &\equiv \frac{\pa}{\pa \boldsymbol{v}^{\b(J)}} = \frac{1}{J!} \frac{\pa}{\pa v^{\b_{1}} } \dots \frac{\pa}{\pa v^{\b_{J}}} \, , \\
		\frac{\pa}{\pa \boldsymbol{w}^{\cA_{3}} } &\equiv \frac{\pa}{\pa \boldsymbol{w}^{\g(K)}} = \frac{1}{K!} \frac{\pa}{\pa w^{\g_{1}} } \dots \frac{\pa}{\pa w^{\g_{K}}} \, . 
	\end{align}
\end{subequations}
The tensor $\cH$ is then extracted from the polynomial as follows:
\begin{equation}
	\cH_{\cA_{1} \cA_{2} \cA_{3}}(X) = \frac{\pa}{ \pa \boldsymbol{u}^{\cA_{1}} } \frac{\pa}{ \pa \boldsymbol{v}^{\cA_{2}}} \frac{\pa}{ \pa \boldsymbol{w}^{\cA_{3}} } \, \cH(X; u, v, w) \, .
\end{equation}
Auxiliary vectors/spinors are widely used 
in the construction of correlation functions throughout the literature (see e.g.~\cite{Giombi:2011rz, Costa:2011dw, Costa:2011mg, Stanev:2012nq, Zhiboedov:2012bm, Nizami:2013tpa, Elkhidir:2014woa}), however, usually the entire correlator is contracted with auxiliary variables and as a result one obtains a polynomial depending on all three spacetime points and the auxiliary spinors. In contrast, the approach outlined here contracts the auxiliary spinors with the tensor $\cH_{ \cA_{1} \cA_{2} \cA_{3} }(X)$, which depends on only the conformally covariant vector $X$. This is advantageous as it becomes straightforward to impose constraints on the correlation function (particularly conservation), since $\cH$ does not depend on any of the spacetime points explicitly.

The full three-point function may be translated into the auxiliary spinor formalism; recalling that $I = 2s_{1}$, $J = 2s_{2}$, $K = 2s_{3} $, first we define:
\begin{subequations}
	\begin{align}
		J^{}_{s_{1}}(x_{1}; u) & = J_{\a(I)}(x_{1}) \, \boldsymbol{u}^{\a(I)} \, , & J'_{s_{2}}(x_{2}; v) &= J_{\b(J)}(x_{2}) \, \boldsymbol{v}^{\a(J)} \, ,
	\end{align}
	\vspace{-10mm}
	\begin{align}
		J''_{s_{3}}(x_{3}; w) &= J_{\g(K)}(x_{3}) \, \boldsymbol{w}^{\g(K)} \, .
	\end{align}
\end{subequations}
The general ansatz for the three-point function is as follows:
\begin{align}
	\langle J^{}_{s_{1}}(x_{1}; u) \, J'_{s_{2}}(x_{2}; v) \, J''_{s_{3}}(x_{3}; w) \rangle = \frac{ \cI^{(I)}(x_{13}; u, \tilde{u}) \,  \cI^{(J)}(x_{23}; v, \tilde{v}) }{(x_{13}^{2})^{\D_{1}} (x_{23}^{2})^{\D_{2}}}
	\; \cH(X_{12}; \tilde{u},\tilde{v},w) \, ,
\end{align} 
where 
\begin{equation}
	\cI^{(s)}(x; u,\tilde{u}) \equiv \cI^{(s)}_{x}(u,\tilde{u}) = \boldsymbol{u}^{\a(s)} \cI_{\a(s)}{}^{\a'(s)}(x) \, \frac{\pa}{\pa \tilde{\boldsymbol{u}}^{\a'(s)}} \, ,
\end{equation}
is the inversion operator acting on polynomials degree $s$ in $\tilde{u}$, and $\D_{i} = s_{i} + 1$.
After converting the constraints summarised in the previous subsection into the auxiliary spinor formalism, we obtain:
\begin{enumerate}
	\item[\textbf{(i)}] {\bf Homogeneity:}
	\begin{equation}
		\cH(\l^{2} X ; u(I), v(J), w(K)) = (\l^{2})^{\D_{3} - \D_{2} - \D_{1}} \, \cH(X; u(I), v(J), w(K)) \, ,
	\end{equation}
	where we have used the notation $u(I)$, $v(J)$, $w(K)$ to keep track of the homogeneity of the auxiliary spinors $u$, $v$ and $w$.
	\item[\textbf{(ii)}] {\bf Differential constraints:}
	\begin{subequations} \label{Ch03-Conservation equations}
		\begin{align}
			\text{Conservation at $x_{1}$:} && \frac{\pa}{\pa X_{\a \b}} \frac{\pa}{\pa u^{\a}} \frac{\pa}{\pa u^{\b}} \, \cH(X; u(I), v(J), w(K)) &= 0 \, , \\
			\text{Conservation at $x_{2}$:} && \frac{\pa}{\pa X_{\a \b}} \frac{\pa}{\pa v^{\a}} \frac{\pa}{\pa v^{\b}} \, \cH(X; u(I), v(J), w(K)) &= 0 \, , \\
			\text{Conservation at $x_{3}$:} && \frac{\pa}{\pa X_{\a \b}} \frac{\pa}{\pa w^{\a}} \frac{\pa}{\pa w^{\b}} \, \tilde{\cH}(X; u(I), v(J), w(K)) &= 0 \, .
		\end{align}
	\end{subequations}
	In the auxiliary spinor formalism, $\tilde{\cH} = \tilde{\cH}^{(+)} + \tilde{\cH}^{(-)}$ is computed as follows:
	\begin{equation}
		\tilde{\cH}^{(\pm)}(X; u(I), v(J), w(K) ) = \pm \, (X^{2})^{\D_{1} - \D_{3}} \cI^{(J)}_{X}(v,\tilde{v}) \, \cH^{(\pm)}(-X; u(I), \tilde{v}(J), w(K)) \, , 
	\end{equation}
	\item[\textbf{(iii)}] {\bf Point switch symmetries:} \\
	If the fields $\phi$ and $\psi$ coincide (hence $I = J$), then we obtain the following point-switch constraint
	\begin{equation} \label{Ch03-Point switch A}
		\cH(X; u(I), v(I), w(K)) = (-1)^{\e(\phi)} \cH(-X; v(I), u(I), w(K)) \, ,
	\end{equation}
	where, again, $\e(\phi)$ is the Grassmann parity of $\phi$. Similarly, if the fields $\phi$ and $\pi$ coincide (hence $I = K$) then we obtain the constraint
	\begin{equation} \label{Ch03-Point switch B}
		\tilde{\cH}(X; u(I), v(J), w(I)) = (-1)^{\e(\phi)} \cH(-X; w(I), v(J), u(I)) \, .
	\end{equation}
\end{enumerate}

The approach outlined above proves to be useful as the polynomial \eqref{Ch03-H - generating polynomial} is now constructed out of scalar combinations of $X$, and the auxiliary spinors $u$, $v$ and $w$ with the appropriate homogeneity. At this point it is convenient to introduce the following ``primitive" structures:
\begin{subequations} \label{Ch03-Basis scalar structures}
	\begin{align}
		P_{1} &= \ve_{\a \b} v^{\a} w^{\b} \, , & P_{2} &= \ve_{\a \b} w^{\a} u^{\b} \, , & P_{3} &= \ve_{\a \b} u^{\a} v^{\b} \, , \\
		Q_{1} &= \hat{X}_{\a \b} v^{\a} w^{\b} \, , & Q_{2} &= \hat{X}_{\a \b} w^{\a} u^{\b} \, , & Q_{3} &= \hat{X}_{\a \b} u^{\a} v^{\b} \, , \\
		Z_{1} &= \hat{X}_{\a \b} u^{\a} u^{\b} \, , & Z_{2} &= \hat{X}_{\a \b} v^{\a} v^{\b} \, , & Z_{3} &= \hat{X}_{\a \b} w^{\a} w^{\b} \, .
	\end{align}
\end{subequations}
The most general ansatz for the polynomial $\cH$ is comprised of all possible combinations of the above structures which possess the correct homogeneity in $u$, $v$ and $w$. In general, it is a non-trivial technical problem to come up with an exhaustive list of possible solutions for the polynomial $\cH$ for a given set of spins. However, this problem is simplified by introducing a generating function for the polynomial $\cH(X; u, v, w)$:
\begin{align} \label{Ch03-Generating function}
	\cF(X; \G) &= X^{\d} P_{1}^{k_{1}} P_{2}^{k_{2}} P_{3}^{k_{3}} Q_{1}^{l_{1}} Q_{2}^{l_{2}} Q_{3}^{l_{3}} Z_{1}^{m_{1}} Z_{2}^{m_{2}} Z_{3}^{m_{3}} \, ,
\end{align}
where $\d = \D_{3} - \D_{2} - \D_{1}$, and the non-negative integers, $ \G = \{ k_{i}, l_{i}, m_{i}\}$, $i=1,2,3$, are solutions to the following linear system:
\begin{subequations} \label{Ch03-Diophantine equations}
	\begin{align}
		k_{2} + k_{3} + l_{2} + l_{3} + 2m_{1} &= I \, , \\
		k_{1} + k_{3} + l_{1} + l_{3} + 2m_{2} &= J \, , \\
		k_{1} + k_{2} + l_{1} + l_{2} + 2m_{3} &= K \, ,
	\end{align}
\end{subequations}
and $I = 2s_{1}$, $J = 2s_{2}$, $K = 2s_{3}$ specify the spin-structure of the correlation function. These equations are obtained by comparing the homogeneity of the auxiliary spinors $u$, $v$, $w$ in the generating function \eqref{Ch03-Generating function}, against the index structure of the tensor $\cH$. The solutions correspond to a linearly dependent basis of possible structures in which the polynomial $\cH$ can be decomposed. Using \textit{Mathematica}, it is straightforward to generate all possible solutions to \eqref{Ch03-Diophantine equations} for fixed (and in some cases arbitrary) values of the spins. 

Now let us assume there exists a finite number of solutions $\G_{i}$, $i = 1, ..., N$ to \eqref{Ch03-Diophantine equations} for a given choice of $I,J,K$. The set of solutions $\G = \{ \G_{i} \}$ may be partitioned into ``even" and ``odd" sets $\G^{+}$ and $\G^{-}$ respectively by counting the number of spinor metrics, $\ve_{\a \b}$, present in a particular solution. Since only the $P_{i}$ contain $\ve_{\a \b}$, we define
\begin{align}
	\G^{+} = \G|_{ \, k_{1} + k_{2} + k_{3} = \text{even}} \, , && \G^{-} = \G|_{ \, k_{1} + k_{2} + k_{3} = \text{odd}} \, .
\end{align}
Hence, the even solutions are those such that $k_{1} + k_{2} + k_{3} = \text{even}$ (i.e contains an even number of spinor metrics), while the odd solutions are those such that $k_{1} + k_{2} + k_{3} = \text{odd}$ (contains an odd number of spinor metrics).\footnote{This convention agrees with the known result that ``odd" solutions typically contain the Levi-Civita tensor, while the ``even" solutions do not.} Let $|\G^{+}| = N^{+}$ and $|\G^{-}| = N^{-}$, with $N = N^{+} + N^{-}$, then the most general ansatz for the polynomial $\cH$ in \eqref{Ch03-H - generating polynomial} is as follows:
\begin{subequations} \label{Ch03-H decomposition}
	\begin{equation}
		\cH(X; u, v, w) = \cH^{(+)}(X; u, v, w) + \cH^{(-)}(X; u, v, w) \, ,
	\end{equation}
	\vspace{-7mm}
	\begin{align}
		\cH^{(+)}(X; u, v, w) = \sum_{i=1}^{N^{+}} A_{i} \, \cF(X; \G^{+}_{i}) \, , && \cH^{(-)}(X; u, v, w) = \sum_{i=1}^{N^{-}} B_{i} \, \cF(X; \G^{-}_{i}) \, ,
	\end{align}
\end{subequations}
where $A_{i}$ and $B_{i}$ are a set of real constants. Since the even and odd sectors of the correlation function do not mix with eachother, they may be considered independently.

Using the above method it is straightforward to generate all possible structures for a given set of spins $\{s_{1}, s_{2}, s_{3} \}$, however, at this stage the solutions generated using this approach are linearly dependent. To form a linearly independent set of solutions we must systematically take into account the following non-linear relations between the primitive structures: 
\begin{subequations}
	\begin{align} \label{Ch03-Linear dependence 1}
		P_{1} Z_{1} + P_{2} Q_{3} + P_{3} Q_{2} &= 0 \, , \\
		P_{2} Z_{2} + P_{1} Q_{3} + P_{3} Q_{1} &= 0 \, , \\
		P_{3} Z_{3} + P_{1} Q_{2} + P_{2} Q_{1} &= 0 \, ,
	\end{align}
\end{subequations}
\vspace{-10mm}
\begin{subequations}
	\begin{align} \label{Ch03-Linear dependence 2}
		Q_{1} Z_{1} - Q_{2} Q_{3} - P_{2} P_{3} &= 0 \, , \\
		Q_{2} Z_{2} - Q_{1} Q_{3} - P_{1} P_{3} &= 0 \, , \\
		Q_{3} Z_{3} - Q_{1} Q_{2} - P_{1} P_{2} &= 0 \, ,
	\end{align}
\end{subequations}
\vspace{-10mm}
\begin{subequations}
	\begin{align} \label{Ch03-Linear dependence 3}
		Z_{2} Z_{3} + P_{1}^{2} - Q_{1}^{2} &= 0 \, , \\
		Z_{1} Z_{3} + P_{2}^{2} - Q_{2}^{2} &= 0 \, , \\
		Z_{1} Z_{2} + P_{3}^{2} - Q_{3}^{2} &= 0 \, ,
	\end{align}
\end{subequations}
\vspace{-10mm}
\begin{align} \label{Ch03-Linear dependence 4}
	P_{1} P_{2} P_{3} + P_{1} Q_{2} Q_{3} + P_{2} Q_{1} Q_{3} + P_{3} Q_{1} Q_{2} &= 0 \, .
\end{align}
This appears to be an exhaustive list of relations, and similar results have been obtained in other approaches which make use of auxiliary spinors \cite{Giombi:2011rz}. Applying the relations above to a set of linearly dependent polynomial structures is relatively straightforward to implement using Mathematica's built-in pattern matching capabilities. 

Now that we have taken care of linear-dependence, it now remains to impose conservation on all three points in addition to the various point-switch symmetries. Introducing the $P$, $Q$ and $Z$ objects proves to streamline this analysis significantly. First let us consider conservation; we define the following three differential operators:
\begin{align}
	D_{1} = \frac{\pa}{\pa X_{\a \b}} \frac{\pa}{\pa u^{\a}} \frac{\pa}{\pa u^{\b}} \, , && D_{2} = \frac{\pa}{\pa X_{\a \b}} \frac{\pa}{\pa v^{\a}} \frac{\pa}{\pa v^{\b}} \, , && D_{3} = \frac{\pa}{\pa X_{\a \b}} \frac{\pa}{\pa w^{\a}} \frac{\pa}{\pa w^{\b}} \, .
\end{align}
To impose conservation on $x_{1}$, (for either sector) we compute
\begin{align}
	D_{1} \cH(X; u,v,w) &= D_{1} \Bigg\{ \sum_{i=1}^{N} c_{i} \, \cF(X; \G_{i}) \Bigg\} \nonumber \\
	&= \sum_{i=1}^{N} c_{i} \, D_{1} \cF(X; \G_{i}) \, .
\end{align}
We then solve for the $c_{i}$ such that the result above vanishes. The explicit expression for $D_{1} \cF(X; \G)$ can be obtained computationally (which we will not present here as it is $\sim 200$ terms long), and hence, given a particular solution $\cF(X;\G_{i})$, we can compute $D_{1} \cF(X; \G_{i})$. The fact that $D_{1} \cF(X; \G)$ can also be expressed using the primitive structures \eqref{Ch03-Basis scalar structures} is due to the following reasoning: let $\mathbf{P}[ X(\d); u(I), v(J), w(K) ]$ represent the space of polynomials which are homogeneous degree $\d$ in $X$, $I$ in $u$, $J$ in $v$ and $K$ in $w$; any polynomial in this space can naturally be constructed in terms of the primitives \eqref{Ch03-Basis scalar structures}. The operator $D_{1}$ may then be interpreted as follows:
\begin{align}
	D_{1} : \mathbf{P}[ X(\d); u(I), v(J), w(K) ] \longmapsto \mathbf{P}[ X(\d-1); u(I-2), v(J), w(K) ] \, .
\end{align}
Hence, $D_{1}$ is a map from $\mathbf{P}[ X(\d); u(I), v(J), w(K) ]$ to $\mathbf{P}[ X(\d-1); u(I-2), v(J), w(K) ]$, that is, the space of polynomials homogeneous degree $\d-1$ in $X$, $I-2$ in $u$, $J$ in $v$ and $K$ in $w$. Any polynomial in this space can naturally be constructed using the same primitives defined in \eqref{Ch03-Basis scalar structures}. Analogous results also apply for $D_{2} \cF(X; \G)$.

To impose conservation on $x_{3}$ we must first obtain an explicit expression for $\tilde{\cH}$ in terms of $\cH$, that is, we must compute (e.g. for the even sector)
\begin{equation}
	\tilde{\cH}(X; u(I), v(J), w(K) ) = (X^{2})^{\D_{1} - \D_{3}} \cI^{(J)}_{X}(v,\tilde{v}) \, \cH(X; u(I), \tilde{v}(J), w(K)) \, .
\end{equation}
Recalling the fact that any solution for $\cH$ can be written in the form of the generating function $\cF(X; \G)$, we compute
\begin{align}
	\tilde{\cF}(X; \G) &= (X^{2})^{\D_{1} - \D_{3}} \cI^{(J)}_{X}(v,\tilde{v}) \, \cF(X; \G) \nonumber \\
	&= (X^{2})^{\D_{1} - \D_{3}} \cI^{(J)}_{X}(v,\tilde{v}) \, \big\{ X^{\d} P_{1}^{k_{1}} P_{2}^{k_{2}} P_{3}^{k_{3}} Q_{1}^{l_{1}} Q_{2}^{l_{2}} Q_{3}^{l_{3}} Z_{1}^{m_{1}} Z_{2}^{m_{2}} Z_{3}^{m_{3}} \big\} \nonumber \\
	&= X^{\D_{1} - \D_{2} - \D_{3}} P_{2}^{k_{2}} Q_{2}^{l_{2}} Z_{1}^{m_{1}}  Z_{3}^{m_{3}} \cI^{(J)}_{X}(v,\tilde{v}) \big\{ \tilde{P}_{1}^{k_{1}} \tilde{P}_{3}^{k_{3}} \tilde{Q}_{1}^{l_{1}} \tilde{Q}_{3}^{l_{3}} \tilde{Z}_{2}^{m_{2}} \big\} \, ,
\end{align}
where we have used $\sim$ to denote $\tilde{v}$ dependence. We now use the generalised Leibniz rule to expand the action of $\cI^{(J)}_{X}(v,\tilde{v}) := \cI^{(J)}_{X}$ on the monomials:
\begin{align}
	\cI^{(J)}_{X} \big\{ \tilde{P}_{1}^{k_{1}} \tilde{P}_{3}^{k_{3}} \tilde{Q}_{1}^{l_{1}} \tilde{Q}_{3}^{l_{3}} \tilde{Z}_{2}^{m_{2}} \big\} &= \frac{1}{J!} \, \cI_{X}^{J} \big\{ \tilde{P}_{1}^{k_{1}} \tilde{P}_{3}^{k_{3}} \tilde{Q}_{1}^{l_{1}} \tilde{Q}_{3}^{l_{3}} \tilde{Z}_{2}^{m_{2}} \big\} \nonumber \\
	&\hspace{-30mm}= \sum_{n_{1} + \dots + n_{5} = J} \frac{1}{J!} \, {J \choose n_{1}, \dots , n_{5} } ( \cI_{X}^{n_{1}} \tilde{P}_{1}^{k_{1}} ) ( \cI_{X}^{n_{2}} \tilde{P}_{3}^{k_{3}} ) ( \cI_{X}^{n_{3}} \tilde{Q}_{1}^{l_{1}} ) ( \cI_{X}^{n_{4}} \tilde{Q}_{3}^{l_{3}} ) ( \cI_{X}^{n_{5}} \tilde{Z}_{2}^{m_{2}} ) \nonumber \\
	&\hspace{-30mm}= \sum_{n_{1} + \dots + n_{5} = J} ( \cI_{X}^{(n_{1})} \tilde{P}_{1}^{k_{1}} ) ( \cI_{X}^{(n_{2})} \tilde{P}_{3}^{k_{3}} ) ( \cI_{X}^{(n_{3})} \tilde{Q}_{1}^{l_{1}} ) ( \cI_{X}^{(n_{4})} \tilde{Q}_{3}^{l_{3}} ) ( \cI_{X}^{(n_{5})} \tilde{Z}_{2}^{m_{2}} ) \nonumber \\
	&\hspace{-30mm}= ( \cI_{X}^{(k_{1})} \tilde{P}_{1}^{k_{1}} ) ( \cI_{X}^{(k_{3})} \tilde{P}_{3}^{k_{3}} ) ( \cI_{X}^{(l_{1})} \tilde{Q}_{1}^{l_{1}} ) ( \cI_{X}^{(l_{3})} \tilde{Q}_{3}^{l_{3}} ) ( \cI_{X}^{(2m_{2})} \tilde{Z}_{2}^{m_{2}} ) \, ,
\end{align}
where we have made use of the fact that $k_{1} + k_{3} + l_{1} + l_{3} + 2m_{2} = J$, while noting that all monomials are annihilated by sufficiently many subsequent actions of $\cI_{X}$. Now, by using the identities
\begin{subequations}
	\begin{gather}
		\cI_{X}(v,\tilde{v}) P_{1} = - Q_{1} \, ,  \hspace{10mm} \cI_{X}(v,\tilde{v}) P_{3} = Q_{3} \, , \\
		\cI_{X}(v,\tilde{v}) \, Q_{1} = - P_{1} \, , \hspace{10mm} \cI_{X}(v,\tilde{v}) \, Q_{3} = P_{3} \, , \\
		\cI_{X}^{(2)}(v,\tilde{v}) \, Z_{2} = Z_{2} \, , 
	\end{gather}
\end{subequations}
we finally obtain
\begin{align}
	\tilde{\cF}(X; \G) &= X^{\tilde{\d}} (-Q_{1})^{k_{1}} P_{2}^{k_{2}} Q_{3}^{k_{3}} (-P_{1})^{l_{1}} Q_{2}^{l_{2}} P_{3}^{l_{3}} Z_{1}^{m_{1}} Z_{2}^{m_{2}} Z_{3}^{m_{3}} \, , \nonumber \\
	&= (-1)^{ k_{1} + l_{1}} X^{\tilde{\d}} P_{1}^{l_{1}} P_{2}^{k_{2}} P_{3}^{l_{3}} Q_{1}^{k_{1}} Q_{2}^{l_{2}} Q_{3}^{k_{3}} Z_{1}^{m_{1}} Z_{2}^{m_{2}} Z_{3}^{m_{3}} \, ,
\end{align}
where $\tilde{\d} = \D_{1} - \D_{2} - \D_{3}$. Hence we arrive at the following result
\begin{equation}
	\tilde{\cF}(X; \G) = (-1)^{ k_{1} + l_{1}} \cF(X; \G)|_{\d \rightarrow \tilde{\d}, \, k_{1} \leftrightarrow l_{1}, \, k_{3} \leftrightarrow l_{3} } \, .
\end{equation}
Therefore the computation of $\tilde{\cH}$ is straightforward: we take each term in the ansatz for $\cH$ and make appropriate swaps of the primitive structures. This also simplifies imposing conservation at $x_{3}$, as we can now use the same generating function that we used for conservation at $x_{1}$ and $x_{2}$ as follows:
\begin{equation}
	D_{3} \tilde{\cF}(X; \G) = (-1)^{ k_{1} + l_{1}} D_{3} \cF(X; \G)|_{\d \rightarrow \tilde{\d}, \, k_{1} \leftrightarrow l_{1}, \, k_{3} \leftrightarrow l_{3} } \, .
\end{equation}
Now that we have exact expressions for $D_{i} \cF(X; \G)$, it remains to find out how point-switch symmetries act on the primitive structures. For permutation of spacetime points $x_{1}$ and $x_{2}$, we have $X \rightarrow - X$, $u \leftrightarrow v$. This results in the following replacement rules for the basis objects \eqref{Ch03-Basis scalar structures}:
\begin{subequations} \label{Ch03-Point switch A - basis}
	\begin{align} 
		P_{1} &\rightarrow - P_{2} \, , & P_{2} &\rightarrow -P_{1} \, , & P_{3} &\rightarrow -P_{3} \, , \\
		Q_{1} &\rightarrow - Q_{2} \, , & Q_{2} &\rightarrow - Q_{1} \, , & Q_{3} &\rightarrow - Q_{3} \, , \\
		Z_{1} &\rightarrow - Z_{2} \, , & Z_{2} &\rightarrow - Z_{1} \, , & Z_{3} &\rightarrow - Z_{3} \, .
	\end{align}
\end{subequations}
Likewise, for permutation of spacetime points $x_{1}$ and $x_{3}$ we have $X \rightarrow - X$, $u \leftrightarrow w$, resulting in the following replacements:
\begin{subequations} \label{Ch03-Point switch B - basis}
	\begin{align} 
		P_{1} &\rightarrow - P_{3} \, , & P_{2} &\rightarrow -P_{2} \, , & P_{3} &\rightarrow -P_{1} \, , \\
		Q_{1} &\rightarrow - Q_{3} \, , & Q_{2} &\rightarrow - Q_{2} \, , & Q_{3} &\rightarrow - Q_{1} \, , \\
		Z_{1} &\rightarrow - Z_{3} \, , & Z_{2} &\rightarrow - Z_{2} \, , & Z_{3} &\rightarrow - Z_{1} \, .
	\end{align}
\end{subequations}
We have now developed all the formalism necessary to analyse the structure of three-point correlation functions in 3D CFT. In the remaining sections of this chapter we will analyse the three-point functions of conserved higher-spin currents (for both integer and half-integer spin) using the following method:
\begin{enumerate}
	\item We construct all possible (linearly dependent) structures for $\cH(X; u,v,w)$ for a given set of spins, which is governed by the solutions to \eqref{Ch03-Diophantine equations}. The solutions are sorted into even and odd sectors and analysed separately.
	\item In each sector, we algorithmically apply the linear dependence relations \eqref{Ch03-Linear dependence 1}, \eqref{Ch03-Linear dependence 2}, \eqref{Ch03-Linear dependence 3}, \eqref{Ch03-Linear dependence 4} to the list of linearly dependent structures, which systematically reduces them to a linearly independent set. It works as follows: first we obtain the complete set of solutions for \eqref{Ch03-Diophantine equations}. For all structures in this list, we apply simplification rules such as e.g.
	\begin{subequations}
	\begin{align}
			P_{1}^{A} Z_{1}^{B} \rightarrow P_{1}^{A- \min(A,B)} Z_{1}^{B-\min(A,B)} ( - P_{2} Q_{3} - P_{3} Q_{2} )^{\min(A,B)} \, ,
	\end{align}
	\vspace{-10mm}
	\begin{align}
			Q_{1}^{A} Z_{1}^{B} \rightarrow  Q_{1}^{A-\min(A,B)} Z_{1}^{B-\min(A,B)} (  Q_{2} Q_{3} + P_{2} P_{3})^{\min(A,B)} \, ,
	\end{align}
	\vspace{-10mm}
	\begin{align}
			Z_{2}^{A} Z_{3}^{B} \rightarrow  Z_{2}^{A-\min(A,B)}Z_{3}^{B-\min(A,B)} ( -  P_{1}^{2} + Q_{1}^{2} )^{\min(A,B)} \, ,
	\end{align}
	\vspace{-10mm}
	\begin{align}
		P_{1}^{A} P_{2}^{B} P_{3}^{C} &\rightarrow P_{1}^{A-\min(A,B,C)} P_{2}^{B-\min(A,B,C)} P_{3}^{C-\min(A,B,C)} \nonumber \\
		& \hspace{15mm} \times ( - P_{1} Q_{2} Q_{3} -  P_{2} Q_{1} Q_{3} - P_{3} Q_{1} Q_{2} )^{\min(A,B,C)}  \, .
	\end{align}
	\end{subequations}
	This operation results in a list of equal size, except now each structure is expanded in terms of only the ``necessary" monomials (of course there is freedom of choice with regards to which monomials are eliminated). After obtaining this list we then delete all duplicate terms, which results in a smaller list. The above operations are repeated until a list with fixed minimum size is obtained. In general this process is sufficient to form a linearly independent ansatz for a given set of spins.
	
	\item Using the method outlined in Subsection \ref{Ch03-subsubsection3.3.3}, we impose the conservation equations \eqref{Ch03-Conservation equations} on each sector.
	\item Once the general form of the polynomial $\cH(X; u,v,w)$ (associated with the conserved three-point function $\langle J^{}_{s_{1}} J'_{s_{2}} J''_{s_{3}} \rangle$) is obtained for a given set of spins $\{s_{1},s_{2}, s_{3}\}$, we then impose any symmetries under permutation of spacetime points, that is, \eqref{Ch03-Point switch A} and \eqref{Ch03-Point switch B} (if applicable). In certain cases, imposing these constraints can eliminate the remaining structures.
\end{enumerate}
Due to computational limitations we carried out this explicit analysis up to $s_{i} = 20$, however, with more optimisation of the code and sufficient computational resources this approach could theoretically find the three-point function in an explicit form for arbitrarily high (fixed) spins. Since there are an enormous number of possible three-point functions with $s_{i} \leq 20$, we present the final results for $\cH(X; u,v,w)$ for some particularly interesting examples (the solutions become cumbersome to present beyond low spin cases). The main goal is to count the number of independent tensor structures after imposing all the constraints. 

The results in next section are organised as follows: in Section \ref{Ch03-section3.4} we analyse the correlation functions involving conserved low-spin currents. In particular, we analyse the three-point functions involving the energy-momentum tensor and conserved vector currents. Many of these results are known in the literature (see e.g. \cite{Osborn:1993cr,Giombi:2011rz,Zhiboedov:2012bm}) and they serve as a consistency test of our formalism. Then, in Subsection \ref{Ch03-subsubsection3.4.1.2}, we analyse the structure of three-point functions involving spin-3/2 currents, the energy-momentum tensor and conserved vector currents; these results are new and are naturally of interest in the context of superconformal field theories. Next, in Subsection \ref{Ch03-subsection3.4.2} we classify the general structure of three-point functions of conserved currents for arbitrary integer or half-integer spins, and we comment on their general features. Finally, in Subsection \ref{Ch03-subsection3.4.3}, we classify the structure of correlation functions involving combinations of higher-spin currents and scalar/spinor fields. We stress that the analysis is based only on symmetries and conservation equations and does not take into account any other features of local field theory. The results are completely analytic and we present explicit formula for $\cH(X; u,v,w)$ in all cases. For higher-spins, the results in the appendices \ref{Appendix3B} and \ref{Appendix3C} are presented as they appear in the Mathematica code.

\section{Three-point functions of conserved currents}\label{Ch03-section3.4}

We begin by analysing the structure of three-point functions involving conserved bosonic currents. As a test our approach we analyse the correlation functions involving low-spin currents such as the energy-momentum tensor and conserved vector current.

\subsection{Conserved low-spin currents}\label{Ch03-subsection3.4.1}
\subsubsection{Energy-momentum tensor and vector current correlators}\label{Ch03-subsubsection3.4.1.1}

The conserved currents which are fundamental in any conformal field theory are the conserved vector current, $V_{m}$, and the symmetric, traceless energy-momentum tensor, $T_{mn}$. The vector current has scale dimension $\Delta_{V} = 2$ and satisfies $\pa^{m} V_{m} = 0$, while the energy-momentum tensor has scale dimension $\Delta_{T} = 3$ and satisfies the conservation equation $\pa^{m} T_{mn} = 0$. Converting to spinor notation we have:
\begin{align}
	V_{\a_{1} \a_{2}}(x) = (\g^{m})_{\a_{1} \a_{2}} V_{m}(x) \, , && T_{\a_{1} \a_{2} \a_{3} \a_{4}}(x) = (\g^{m})_{(\a_{1} \a_{2}} (\g^{n})_{\a_{3} \a_{4})} T_{mn}(x) \, .
\end{align}
These objects possess fundamental information associated with internal and spacetime symmetries, hence, analysis of their three-point functions is of great importance. The general structure of correlation functions involving these fields have been widely studied throughout the literature of conformal field theory; here we present the solutions for them using our formalism. The possible three-point functions involving the conserved vector current and the energy-momentum tensor are:
\begin{align} \label{Ch03-Low-spin component correlators}
	\langle V_{\a(2)}(x_{1}) \, V_{\b(2)}(x_{2}) \, V_{\g(2)}(x_{3}) \rangle \, , &&  \langle V_{\a(2)}(x_{1}) \, V_{\b(2)}(x_{2}) \, T_{\g(4)}(x_{3}) \rangle \, , \\
	\langle T_{\a(4)}(x_{1}) \, T_{\b(4)}(x_{2}) \, V_{\g(2)}(x_{3}) \rangle \, , &&  \langle T_{\a(4)}(x_{1}) \, T_{\b(4)}(x_{2}) \, T_{\g(4)}(x_{3}) \rangle \, .
\end{align}
In all cases, we note that the triangle inequalities \eqref{Ch03-Triangle inequalities} are simultaneously satisfied, hence, we expect that each of these correlation functions should possess a parity-odd solution after imposing conservation on all three points. The analysis of these three-point functions is quite simple using our computational approach. Let us first consider $\langle V V V \rangle$; within the framework of our formalism we study the three-point function $\langle J^{}_{1} J'_{1} J''_{1} \rangle$.\\[5mm]
\textbf{Correlation function} $\langle J^{}_{1} J'_{1} J''_{1} \rangle$\textbf{:}\\[2mm]
The general ansatz for this correlation function, according to \eqref{Ch03-Conserved correlator ansatz} is
\begin{align}
	\langle J^{}_{\a(2)}(x_{1}) \, J'_{\b(2)}(x_{2}) \, J''_{\g(2)}(x_{3}) \rangle = \frac{ \cI_{\a(2)}{}^{\a'(2)}(x_{13}) \,  \cI_{\b(2)}{}^{\b'(2)}(x_{23}) }{(x_{13}^{2})^{2} (x_{23}^{2})^{2}}
	\; \cH_{\a'(2) \b'(2) \g(2)}(X_{12}) \, .
\end{align} 
Using the formalism outlined in Subsection \ref{Ch03-subsubsection3.3.3}, all information about this correlation function is encoded in the following polynomial:
\begin{align}
	\cH(X; u(2), v(2), w(2)) = \cH_{ \a(2) \b(2) \g(2) }(X) \, \boldsymbol{u}^{\a(2)}  \boldsymbol{v}^{\b(2)}  \boldsymbol{w}^{\g(2)} \, .
\end{align}
Using Mathematica we solve \eqref{Ch03-Diophantine equations} for the chosen spins and substitute each solution into the generating function \eqref{Ch03-Generating function}. This provides us with the following list of (linearly dependent) polynomial structures in the even and odd sectors respectively:
\begin{subequations}
	\begin{flalign}
		\textbf{Even:}& \hspace{5mm} \{Z_1 Z_2 Z_3,Q_3^2 Z_3,Q_2^2 Z_2,Q_1 Q_2 Q_3,Q_1^2 Z_1, \nonumber \\
		& \hspace{15mm} P_3^2 Z_3, P_2 P_3 Q_1,P_2^2 Z_2,P_1 P_3 Q_2,P_1 P_2 Q_3,P_1^2 Z_1 \}  \\
		\textbf{Odd:}& \hspace{5mm} \{P_3 Q_3 Z_3,P_3 Q_1 Q_2,P_2 Q_2 Z_2, \nonumber \\
		&\hspace{15mm} P_2 Q_1 Q_3,P_1 Q_2 Q_3,P_1 Q_1 Z_1,P_1 P_2 P_3 \}
	\end{flalign}
\end{subequations}
Next, we systematically apply the linear dependence relations \eqref{Ch03-Linear dependence 1} to these lists, reducing them to the following sets of linearly independent structures:
\begin{subequations}
	\begin{flalign}
		\textbf{Even:}& \hspace{5mm} \{P_2 P_3 Q_1,P_1 P_3 Q_2,P_1 P_2 Q_3,Q_1 Q_2 Q_3 \} \\
		\textbf{Odd:}& \hspace{5mm} \{ P_3 Q_1 Q_2,P_2 Q_1 Q_3,P_1 Q_2 Q_3 \}
	\end{flalign}
\end{subequations}
Note that application of the linear-dependence relations eliminates all terms involving $Z_{i}$ in this case. Next we construct an ansatz out of the linearly independent structures, see \eqref{Ch03-H decomposition}, where $\cH^{(\pm)}_{i}$ denotes a structure at position `$i$' in the even/odd list respectively. Then, after imposing conservation on all three points using the methods outlined in Subsection \ref{Ch03-subsubsection3.3.3}, we obtain a linear system of equations in the coefficients $A_{i}$ and $B_{i}$. We find the following final solutions for the even and odd sectors:
\begin{subequations}
	\begin{align}\label{Ch03-1-1-1}
		\textbf{Even:}& \hspace{5mm} \frac{A_1}{X^2} (P_2 P_3 Q_1+P_1 P_3 Q_2+Q_1 Q_2 Q_3 ) + \frac{A_2}{X^2} P_1 P_2 Q_3 \\
		\textbf{Odd:}& \hspace{5mm} \frac{B_1}{X^2} P_3 Q_1 Q_2 
	\end{align}
\end{subequations}
After imposing symmetries under permutation of spacetime points, e.g. $J=J'=J''$, the remaining structures must vanish unless the currents possess a flavour index associated with a non-Abelian symmetry group. In this case the vector currents take the form $V^{\bar{a}}_{\a \b}$, where $\bar{a}$ is a flavour index. Since the conformal building blocks do not carry any flavour indices, the three-point function of these currents, $\langle V^{\bar{a}}(x_{1}) V^{\bar{b}}(x_{2}) V^{\bar{c}}(x_{3}) \rangle$, factorises such that it is proportionate to an invariant tensor of the flavour group. Given generators $T^{\bar{a}}$ of the flavour group, we may define a completely anti-symmetric structure constant $f^{\bar{a} \bar{b} \bar{c}}$ by $[ T^{\bar{a}}, T^{\bar{b}} ] = \text{i} f^{\bar{a} \bar{b} \bar{c}} T^{\bar{c}}$. As a result of the anti-symmetry of the structure constant, the remaining structures in the three-point functions of vector currents with flavour indices are compatible with the point-switch symmetries, and therefore all three structures survive in general. 

The next example to consider is the mixed correlator $\langle V V T \rangle$. To study this case we may examine the correlation function $\langle J^{}_{1} J'_{1} J''_{2} \rangle$. \\[5mm]
\noindent
\textbf{Correlation function} $\langle J^{}_{1} J'_{1} J''_{2} \rangle$\textbf{:}\\[2mm]
Using the general formula, the ansatz for this three-point function:
\begin{align}
	\langle J^{}_{\a(2)}(x_{1}) \, J'_{\b(2)}(x_{2}) \, J''_{\g(4)}(x_{3}) \rangle = \frac{ \cI_{\a(2)}{}^{\a'(2)}(x_{13}) \,  \cI_{\b(2)}{}^{\b'(2)}(x_{23}) }{(x_{13}^{2})^{2} (x_{23}^{2})^{2}}
	\; \cH_{\a'(2) \b'(2) \g(4)}(X_{12}) \, .
\end{align} 
Using the formalism outlined in Subsection \ref{Ch03-subsubsection3.3.3}, all information about this correlation function is encoded in the following polynomial:
\begin{align}
	\cH(X; u(2), v(2), w(4)) = \cH_{ \a(2) \b(2) \g(4) }(X) \, \boldsymbol{u}^{\a(2)}  \boldsymbol{v}^{\b(2)}  \boldsymbol{w}^{\g(4)} \, .
\end{align}
After solving \eqref{Ch03-Diophantine equations}, we find the following linearly dependent polynomial structures in the even and odd sectors respectively:
\begin{subequations}
	\begin{align}
		\textbf{Even:}& \hspace{5mm} \{Z_1 Z_2 Z_3^2,Q_3^2 Z_3^2,Q_2^2 Z_2 Z_3,Q_1 Q_2 Q_3 Z_3,Q_1^2 Z_1 Z_3, \nonumber \\
		& \hspace{12.5mm} Q_1^2 Q_2^2, P_3^2 Z_3^2,P_1^2 Q_2^2,P_1^2 P_2^2, P_2 P_3 Q_1 Z_3,P_2^2 Z_2 Z_3, \nonumber \\
		& \hspace{15mm} P_2^2 Q_1^2,P_1 P_3 Q_2 Z_3,P_1 P_2 Q_3 Z_3,P_1 P_2 Q_1 Q_2,P_1^2 Z_1 Z_3 \} \\
		\textbf{Odd:}& \hspace{5mm} \{P_3 Q_3 Z_3^2,P_3 Q_1 Q_2 Z_3,P_2 Q_2 Z_2 Z_3,P_2 Q_1 Q_3 Z_3,P_2 Q_1^2 Q_2, \nonumber \\
		& \hspace{10mm} P_1 Q_2 Q_3 Z_3,P_1 Q_1 Z_1 Z_3,P_1 Q_1 Q_2^2,P_1 P_2 P_3
		Z_3,P_1 P_2^2 Q_1,P_1^2 P_2 Q_2 \}
	\end{align}
\end{subequations}
Next we systematically apply the linear dependence relations \eqref{Ch03-Linear dependence 1} to these lists, reducing them to the following linearly independent structures:
\begin{subequations}
	\begin{flalign}
		\textbf{Even:}& \hspace{5mm} \{ P_1^2 P_2^2,P_2^2 Q_1^2,P_1 P_2 Q_1 Q_2,P_1^2 Q_2^2,Q_1^2 Q_2^2 \} \\
		\textbf{Odd:}& \hspace{5mm} \{P_1 P_2^2 Q_1,P_1^2 P_2 Q_2,P_2 Q_1^2 Q_2,P_1 Q_1 Q_2^2 \}
	\end{flalign}
\end{subequations}
After constructing an appropriate ansatz for each sector, we then impose conservation on all three points using the methods outlined in Subsection \ref{Ch03-subsubsection3.3.3}. We obtain the following final solutions for the even and odd sectors:
\begin{subequations} \label{Ch03-1-1-2}
	\begin{align}
		\textbf{Even:}& \hspace{5mm} \frac{A_1}{X} \left( P_1^2 P_2^2 + P_1 P_2 Q_1 Q_2 \right) \nonumber \\
		& \hspace{20mm} +\frac{A_2}{X} \left(P_2^2 Q_1^2+3 P_1 P_2 Q_1 Q_2+P_1^2 Q_2^2-\sfrac{3}{5} Q_2^2
		Q_1^2\right) \\
		\textbf{Odd:}& \hspace{5mm} \frac{B_1}{X} \left(P_2 P_1^2 Q_2-P_1 Q_1 Q_2^2+P_2^2 P_1 Q_1-P_2 Q_1^2 Q_2\right)
	\end{align}
\end{subequations}
All structures survive after setting $J = J'$. Hence, this correlation function is fixed up to two independent even structures, and one odd structure.

Since the number of tensor structures rapidly increases with spin, we will skip the technical details for the other correlation functions and present only the linearly independent structures and the final results after imposing conservation. For $\langle T T V \rangle$ we may consider the correlation function $\langle J^{}_{2} J'_{2} J''_{1} \rangle$, for which we find the following linearly independent structures in the even and odd sectors:
\begin{subequations}
	\begin{align}
		\textbf{Even:}& \hspace{5mm} \{P_2 P_3^3 Q_1,P_1 P_3^3 Q_2,P_1 P_2 Q_3^3,Q_1 Q_2 Q_3^3, \nonumber \\
		&\hspace{25mm} P_3^2 Q_1 Q_2 Q_3,P_2 P_3 Q_1 Q_3^2,P_1 P_3 Q_2 Q_3^2 \} \\[2mm]
		\textbf{Odd:}& \hspace{5mm} \{ P_3^3 Q_1 Q_2,P_2 P_3^2 Q_1 Q_3,P_1 P_3^2 Q_2 Q_3, \nonumber \\
		&\hspace{25mm} P_3 Q_1 Q_2 Q_3^2,P_2 Q_1 Q_3^3,P_1 Q_2 Q_3^3 \}
	\end{align}
\end{subequations}
After imposing conservation on all three points we obtain the solutions:
\begin{subequations} \label{Ch03-2-2-1}
	\begin{align}
		\textbf{Even:}& \hspace{5mm} \frac{A_1}{X^4} \left(\sfrac{5}{3} P_1 P_2 Q_3^3+P_2 P_3 Q_1 Q_3^2+P_1 P_3 Q_2 Q_3^2+P_3^2 Q_1 Q_2 Q_3\right) \nonumber \\
		& \hspace{5mm} +\frac{A_2}{X^4} \Big( P_2 P_3^3
		Q_1+P_1 P_3^3 Q_2-6 P_2 P_3 Q_1 Q_3^2 \nonumber \\
		& \hspace{25mm} -6 P_1 P_3 Q_2 Q_3^2-\sfrac{23}{3} P_1 P_2 Q_3^3-\sfrac{7}{3} Q_1 Q_2 Q_3^3\Big) \\
		\textbf{Odd:}& \hspace{5mm} \frac{B_1}{X^4} \left(P_3^3 Q_1 Q_2-P_2 P_3^2 Q_1 Q_3-P_1 P_3^2 Q_2 Q_3-3 P_3 Q_1 Q_2 Q_3^2\right)
	\end{align}
\end{subequations}
In this case, all structures vanish after setting $J = J'$ and imposing the required symmetries under the exchange of $x_{1}$ and $x_{2}$. Hence, the correlation function $\langle T T V \rangle$ vanishes in any CFT. Finally, to study $\langle T T T \rangle$ we can analyse the correlation function $\langle J^{}_{2} J'_{2} J''_{2} \rangle$. For this three-point function the linearly independent structures are
\begin{subequations}
	\begin{align}
		\textbf{Even:}& \hspace{5mm} \{ P_1^2 P_2^2 Q_3^2,P_2^2 Q_1^2 Q_3^2, P_2^2 P_3^2 Q_1^2,P_1^2 P_3^2 Q_2^2,P_1 P_2 Q_1 Q_2
		Q_3^2, \nonumber \\
		&\hspace{15mm} P_1^2 Q_2^2 Q_3^2, P_2 P_3 Q_1^2 Q_2 Q_3,P_1 P_3 Q_1 Q_2^2 Q_3, P_3^2 Q_1^2 Q_2^2, Q_1^2 Q_2^2 Q_3^2 \} \\
		\textbf{Odd:}& \hspace{5mm} \{P_2 P_3^2 Q_1^2 Q_2,P_1 P_3^2 Q_1 Q_2^2,P_2^2 P_3 Q_1^2 Q_3,P_1^2 P_3 Q_2^2 Q_3, \nonumber \\
		&\hspace{15mm} P_3 Q_1^2 Q_2^2 Q_3, P_1 P_2^2 Q_1 Q_3^2,P_1^2 P_2 Q_2 Q_3^2,P_2 Q_1^2
		Q_2 Q_3^2,P_1 Q_1 Q_2^2 Q_3^2 \}
	\end{align}
\end{subequations}
After imposing conservation on all three points we obtain the solutions:
\begin{subequations} \label{Ch03-2-2-2}
	\begin{align}
		\textbf{Even:}& \hspace{5mm} \frac{A_1}{X^3} \Big(P_2^2 P_3^2 Q_1^2-\sfrac{30}{7} P_3^2 Q_2^2 Q_1^2+P_1^2 P_3^2 Q_2^2+\sfrac{3}{5} P_1^2 P_2^2 Q_3^2+Q_2^2
		Q_3^2 Q_1^2 \Big) \nonumber \\
		& \hspace{10mm} +\frac{A_2}{X^3} \Big(P_2^2 Q_3^2 Q_1^2+P_2 P_3 Q_2 Q_3 Q_1^2+P_1 P_2 Q_2 Q_3^2 Q_1 \nonumber \\
		& \hspace{30mm} +P_1 P_3 Q_2^2 Q_3 Q_1-\sfrac{9}{5} P_1^2 P_2^2 Q_3^2+P_1^2
		Q_2^2 Q_3^2\Big) \\
		\textbf{Odd:}& \hspace{5mm} \frac{B_1}{X^3} \left(P_2 P_3^2 Q_2 Q_1^2+P_3 Q_2^2 Q_3 Q_1^2+P_1 P_3^2 Q_2^2 Q_1\right)
	\end{align}
\end{subequations}
In all cases, it is clear that the general solutions are determined up to two independent even structures, and one odd structure. These results are consistent with \cite{Giombi:2011rz} in terms of the number of independent polynomial structures, however it is difficult to make a direct comparison.

\subsubsection{Spin-3/2 current correlators}\label{Ch03-subsubsection3.4.1.2}

In this section we will evaluate three-point functions involving conserved fermionic currents. 
To the best of our knowledge, these correlation functions have not been studied in much detail in the literature, particularly in three dimensions. The most important example of a fermionic conserved current is the supersymmetry current, $Q_{m,\a}$, which is prevalent in $\cN$-extended superconformal field theories. Such a field is primary with dimension $\D_{Q} = 5/2$, and satisfies the conservation equation $\pa^{m} Q_{m, \a } = 0$. In spinor notation, we have:
\begin{equation}
	Q_{\a(3)}(x) = (\g^{m})_{(\a_{1} \a_{2}} Q_{m, \a_{3})}(x) \, .
\end{equation}
Recall that in three-dimensional superconformal field theory, the supersymmetry current and the energy-momentum tensor are contained in the supercurrent multiplet, $\mathbf{J}_{\a(3)}(z)$, where $z^{\cA} = (x^{a}, \q^{\a})$ is a point in $\cN = 1$ superspace (the reader may skip to Subsection \ref{Ch05-subsection5.1.3} for more details about the supercurrent multiplet in three-dimensions, or see e.g. \cite{Buchbinder:2015qsa}). The supercurrent multiplet is a superconformal primary field of dimension $5/2$, and satisfies the conservation equation 
\begin{align}
	D^{\a} \mathbf{J}_{\a \b \g}(z) = 0 \, ,
\end{align}
where 
\begin{equation}
	D_{\a} = \frac{\pa}{\pa \q^{\a}} + \text{i} (\g^{m})_{\a \b} \q^{\b} \frac{\pa}{\pa x^{m}} \, ,
\end{equation}
is the standard spinor-covariant derivative \cite{Kuzenko:2010rp,Kuzenko:2011xg,Buchbinder:2015wia} in 3D $\cN=1$ superspace. The supersymmetry current, $Q_{\a(3)}$, and the energy-momentum tensor, $T_{\a(4)}$, are extracted from the $\cN=1$ supercurrent through ``bar-projection" as follows:
\begin{align}
	Q_{\a(3)}(x) = \mathbf{J}_{\a(3)}(z)|_{\q = 0} \, ,  && T_{\a(4)}(x) = D_{(\a_{1}} \mathbf{J}_{\a_{2} \a_{3} \a_{4} )}(z)|_{\q = 0} \, ,
\end{align}
Here, $\Phi(z)|_{\theta = 0}$, denotes setting the superspace variables $\theta^{\a}$ to zero. Likewise, the conserved vector current, $V_{\a(2)}$, is contained within the flavour current multiplet, $\mathbf{L}_{\a}(z)$. The flavour current is a superconformal primary field of dimension $3/2$ satisfying the superfield conservation equation $D^{\a} \mathbf{L}_{\a}(z) = 0$. The conserved vector current is extracted as follows:
\begin{align}
	V_{\a(2)}(x) = D_{(\a_{1}} \mathbf{L}_{\a_{2})}(z)|_{\q = 0} \, .
\end{align}
Since the supersymmetry current is a conserved current associated with supersymmetry transformations, it is interesting to study the correlation functions involving the supersymmetry current, the vector current, and the energy-momentum tensor. The two three-point functions involving $Q$, $V$ and $T$ which are of interest in $\cN=1$ superconformal field theories are:
\begin{align} \label{Ch03-Susy current correlators - 1}
	\langle Q_{\a(3)}(x_{1}) \, Q_{\b(3)}(x_{2}) \, V_{\g(2)}(x_{3}) \rangle \, , && \langle Q_{\a(3)}(x_{1}) \,  Q_{\b(3)}(x_{2}) \, T_{\g(4)}(x_{3}) \rangle \, .
\end{align}
These correlation functions are naturally contained in the following supersymmetric three-point functions
\begin{align}
	\langle \, \mathbf{J}_{\a(3)}(z_{1}) \, \mathbf{J}_{\b(3)}(z_{2}) \, \mathbf{L}_{\g}(z_{3}) \rangle \, , &&  \langle \, \mathbf{J}_{\a(3)}(z_{1}) \, \mathbf{J}_{\b(3)}(z_{2}) \, \mathbf{J}_{\g(3)}(z_{3}) \rangle \, .
\end{align}
In three-dimensions, $\langle \mathbf{J} \mathbf{J} \mathbf{L} \rangle$ vanishes, while $\langle \mathbf{J} \mathbf{J} \mathbf{J} \rangle$ is fixed up to a single tensor 
structure \cite{Nizami:2013tpa,Buchbinder:2015qsa, Buchbinder:2021gwu}. Therefore, since the component correlators \eqref{Ch03-Susy current correlators - 1} are obtained by 
bar-projecting the supersymmetric correlation functions, we find $\langle Q Q V \rangle = 0$, while $\langle Q Q T \rangle$ is fixed up to a single tensor structure. However, we must note that in this chapter we assume only conformal symmetry, not supersymmetry.
Therefore, the current $Q$ is henceforth denoted a ``supersymmetry-like" current; that is, it possess identical conformal properties to the supersymmetry current but is not necessarily equal to it.

In this subsection we present an explicit analysis of the general structure of the correlation functions involving $Q$, $V$ and $T$ that are compatible with the constraints of conformal symmetry and conservation equations. Let us first consider $\langle Q Q' V \rangle$, for which we may analyse the general structure of the correlation function $\langle J^{}_{3/2} J'_{3/2} J''_{1} \rangle$.

\newpage

\noindent
\textbf{Correlation function} $\langle J^{}_{3/2} J'_{3/2} J''_{1} \rangle$\textbf{:}\\[2mm]
Using the general formula, the ansatz for this three-point function:
\begin{align}
	\langle J^{}_{\a(3)}(x_{1}) \, J'_{\b(3)}(x_{2}) \, J''_{\g(2)}(x_{3}) \rangle = \frac{ \cI_{\a(3)}{}^{\a'(3)}(x_{13}) \,  \cI_{\b(3)}{}^{\b'(3)}(x_{23}) }{(x_{13}^{2})^{5/2} (x_{23}^{2})^{5/2}}
	\; \cH_{\a'(3) \b'(3) \g(2)}(X_{12}) \, .
\end{align} 
All physical information about this correlation function is encoded in the following polynomial:
\begin{align}
	\cH(X; u(3), v(3), w(2)) = \cH_{ \a(3) \b(3) \g(2) }(X) \, \boldsymbol{u}^{\a(3)}  \boldsymbol{v}^{\b(3)}  \boldsymbol{w}^{\g(2)} \, .
\end{align}
After solving \eqref{Ch03-Diophantine equations} to obtain the linearly dependent structures, we systematically applying the linear dependence relations \eqref{Ch03-Linear dependence 1} and find the following linearly independent structures in the parity-even and parity-odd sectors:
\begin{subequations}
	\begin{align}
		\textbf{Even:}& \hspace{5mm} \{P_3^2 Q_1 Q_2,P_2 P_3 Q_1 Q_3,P_1 P_3 Q_2 Q_3,P_1 P_2 Q_3^2,Q_1 Q_2 Q_3^2 \} \\
		\textbf{Odd:}& \hspace{5mm} \{P_2 P_3^2 Q_1,P_1 P_3^2 Q_2,P_3 Q_1 Q_2 Q_3,P_2 Q_1 Q_3^2,P_1 Q_2 Q_3^2 \}
	\end{align}
\end{subequations}
We then construct an appropriate ansatz for each sector and impose conservation on all three points to obtain the following final solutions for the even and odd sectors:
\begin{subequations} \label{Ch03-3/2-3/2-1}
	\begin{align}
		\textbf{Even:}& \hspace{5mm} \frac{A_1}{X^3} \Big(P_3^2 Q_1 Q_2+\sfrac{10}{9} P_1 P_2 Q_3^2-\sfrac{5}{9} Q_1 Q_2 Q_3^2 \Big) \nonumber \\
		& \hspace{10mm} +\frac{A_2}{X^3} \Big(\sfrac{17}{9} P_1 P_2 Q_3^2+P_2 P_3 Q_1 Q_3+P_1 P_3 Q_2 Q_3+\sfrac{5}{9}
		Q_1 Q_2 Q_3^2 \Big) \\
		\textbf{Odd:}& \hspace{5mm} \frac{B_1}{X^3} \big( P_2 P_3^2 Q_1+P_1 P_3^2 Q_2+3 P_3 Q_1 Q_2 Q_3 \big)
	\end{align}
\end{subequations}
Hence, we see that the correlation function $\langle J^{}_{3/2} J'_{3/2} J''_{1} \rangle$, and therefore $\langle Q Q' V \rangle$, is fixed up to two even structures and one odd structure. It may be show that all structures vanish for $J=J'$ as they do not-possess the correct symmetry under permutation of points $x_{1}$ and $x_{2}$, therefore we find that the correlation function $\langle Q Q V \rangle$ vanishes. 

Next we will analyse the general structure of $\langle Q Q T \rangle$, which is associated with the correlation function $\langle J^{}_{3/2} J'_{3/2} J''_{2} \rangle$ using the ansatz \eqref{Ch03-Conserved correlator ansatz}.\\

\noindent
\textbf{Correlation function} $\langle J^{}_{3/2} J'_{3/2} J''_{2} \rangle$\textbf{:}\\[2mm]
According to the general formula \eqref{Ch03-Conserved correlator ansatz}, the ansatz for this three-point function is:
\begin{align}
	\langle J^{}_{\a(3)}(x_{1}) \, J'_{\b(3)}(x_{2}) \, J''_{\g(4)}(x_{3}) \rangle = \frac{ \cI_{\a(3)}{}^{\a'(3)}(x_{13}) \,  \cI_{\b(3)}{}^{\b'(3)}(x_{23}) }{(x_{13}^{2})^{5/2} (x_{23}^{2})^{5/2}}
	\; \cH_{\a'(3) \b'(3) \g(4)}(X_{12}) \, .
\end{align} 
Using the formalism outlined in Subsection \ref{Ch03-subsubsection3.3.3}, we encode the structure of the three-point function into the following polynomial:
\begin{align}
	\cH(X; u(3), v(3), w(4)) = \cH_{ \a(3) \b(3) \g(4) }(X) \, \boldsymbol{u}^{\a(3)}  \boldsymbol{v}^{\b(3)}  \boldsymbol{w}^{\g(4)} \, .
\end{align}
We now solve \eqref{Ch03-Diophantine equations} and systematically apply the linear dependence relations \eqref{Ch03-Linear dependence 1} to the list of linearly dependent structures. As a result we obtain the following lists of linearly independent structures in the parity-even and parity-odd sectors:
\begin{subequations}
	\begin{align}
		\textbf{Even:}& \hspace{5mm} \{P_1^2 P_2^2 Q_3,P_2^2 Q_1^2 Q_3,P_1^2 Q_2^2 Q_3,Q_1^2 Q_2^2 Q_3, \nonumber \\
		&\hspace{25mm} P_1 P_2 Q_1 Q_2 Q_3, P_2 P_3 Q_1^2 Q_2,P_1 P_3 Q_1 Q_2^2 \} \\
		\textbf{Odd:}& \hspace{5mm} \{P_2^2 P_3 Q_1^2,P_1^2 P_3 Q_2^2,P_3 Q_1^2 Q_2^2,P_1 P_2^2 Q_1 Q_3, \nonumber \\
		& \hspace{25mm} P_1^2 P_2 Q_2 Q_3,P_2 Q_1^2 Q_2 Q_3,P_1 Q_1 Q_2^2 Q_3 \}
	\end{align}
\end{subequations}
After constructing an appropriate ansatz for each sector, we impose conservation on all three points and obtain the following final solutions in the parity-even and parity-odd sectors:
\begin{subequations} \label{Ch03-3/2-3/2-2}
	\begin{align}
		\textbf{Even:}& \hspace{5mm} \frac{A_1}{X^2} \Big(P_1^2 P_2^2 Q_3-\sfrac{15}{53} P_2^2 Q_1^2 Q_3+\sfrac{4}{53} P_1 P_2 Q_1 Q_2 Q_3-\sfrac{15}{53} P_1^2 Q_2^2 Q_3+\sfrac{5}{53} Q_1^2 Q_2^2
		Q_3 \Big) \nonumber \\
		& \hspace{10mm} +\frac{A_2}{X^2} \Big( P_2 P_3 Q_2 Q_1^2+\sfrac{14}{53} P_2^2 Q_3 Q_1^2+P_1 P_3 Q_2^2 Q_1 \nonumber \\
		&\hspace{30mm} +\sfrac{21}{53} P_1 P_2 Q_2 Q_3 Q_1 + \sfrac{14}{53} P_1^2
		Q_2^2 Q_3+\sfrac{13}{53} Q_2^2 Q_3 Q_1^2 \Big) \\
		\textbf{Odd:}& \hspace{5mm} \frac{B_1}{X^2} \big( P_1^2 P_3 Q_2^2 + P_2^2 P_3 Q_1^2-2 P_2 Q_2 Q_3 Q_1^2-2 P_1 Q_2^2 Q_3 Q_1-4 P_3 Q_2^2 Q_1^2\big)
	\end{align}
\end{subequations}
Hence, we note that the three-point function $\langle J^{}_{3/2} J'_{3/2} J''_{2} \rangle$, and therefore $\langle Q Q' T \rangle$, is fixed up to two independent ``even" structures and one ``odd" structure. In this case, both structures survive after imposing the symmetry under the exchange of $x_{1}$ and $x_{2}$, that is, when $J = J'$ (i.e. $Q = Q'$). Hence, the correlation function $\langle Q Q T \rangle$ is also fixed up to two even structures and a single odd structure.

\subsection{General structure of \texorpdfstring{$\langle J^{}_{s_{1}} J'_{s_{2}} J''_{s_{3}} \rangle$
	}{< J J' J'' >}}\label{Ch03-subsection3.4.2}

In this thesis we are particularly interested in the structure of three-point correlation functions involving conserved higher-spin currents. In three-dimensional conformal field theory, a conserved current of spin $s$ (integer or half-integer), is defined as a totally symmetric spin-tensor, $J_{\a_{1} \dots \a_{2s} }(x) = J_{(\a_{1} \dots \a_{2s}) }(x)$, satisfying a conservation equation of the form:
\begin{equation} \label{Ch03-3D Conserved current}
	(\gamma^{m})^{\a_{1} \a_{2}} \pa_{m} J_{\a_{1} \a_{2} \dots \a_{2s}} = 0 \, .
\end{equation}
Conserved currents are primary fields, as they possesses the following infinitesimal conformal transformation properties:
\begin{equation}
	\delta J_{\a_{1} \dots \a_{2s}}(x) = - \xi J_{\a_{1} \dots \a_{2s}}(x) - \Delta_{J} \, \s(x) J_{\a_{1} \dots \a_{2s}}(x) + 2s \, \omega_{( \a_{1} }{}^{\delta}(x) \, J_{\a_{2} \dots \a_{2s}) \delta}(x) \, , 
\end{equation}
where $\xi$ is a conformal Killing vector field, and $\s(x)$, $\omega_{\a \b}(x)$ are local parameters defined in terms of $\xi$, which are associated with local scale and combined Lorentz/special conformal transformations. The dimension $\Delta_{J}$ is uniquely fixed by the conservation condition \eqref{Ch03-3D Conserved current}, and it may be shown that this condition is superconformally covariant provided that $\Delta_{J} = s + 1$.

The three-point correlation functions of bosonic conserved currents have been extensively studied in 3D CFT \cite{Maldacena:2011jn,Giombi:2011rz,Zhiboedov:2012bm}. 
In particular, it has been shown that the general structure of the three-point correlation function $\langle J^{}_{s_{1}} J'_{s_{2}} J''_{s_{3}} \rangle$ (where all currents are bosonic) is fixed up to the following form:
\begin{equation}
	\langle J^{}_{s_{1}} J'_{s_{2}} J''_{s_{3}} \rangle = a_{1} \, \langle J^{}_{s_{1}} J'_{s_{2}} J''_{s_{3}} \rangle_{B} + a_{2} \, \langle J^{}_{s_{1}} J'_{s_{2}} J''_{s_{3}} \rangle_{F} + b \, \langle J^{}_{s_{1}} J'_{s_{2}} J''_{s_{3}} \rangle_{odd} \, .
\end{equation}
The solutions $\langle J^{}_{s_{1}} J'_{s_{2}} J''_{s_{3}} \rangle_{B}$, $\langle J^{}_{s_{1}} J'_{s_{2}} J''_{s_{3}} \rangle_{F}$ are generated by theories of a free-boson and free-fermion respectively, while the ``odd" structure, $\langle J^{}_{s_{1}} J'_{s_{2}} J''_{s_{3}} \rangle_{odd}$, is not generated by a free CFT; instead it is generated by a Chern-Simons theory interacting with parity-violating matter \cite{Aharony:2011jz, Giombi:2011kc, Maldacena:2012sf, GurAri:2012is, Aharony:2012nh, Jain:2012qi, Giombi:2016zwa, Chowdhury:2017vel, Sezgin:2017jgm, Skvortsov:2018uru, Inbasekar:2019wdw}. Furthermore, the existence of the odd solution depends on the following set of triangle inequalities:
\begin{align} \label{Ch03-Triangle inequalities}
	s_{1} &\leq s_{2} + s_{3} \, , & s_{2} &\leq s_{1} + s_{3} \, , & s_{3} &\leq s_{1} + s_{2} \, .
\end{align}
When the triangle inequalities are simultaneously satisfied, there are two even solutions and one odd solution, however, if any of the inequalities above are not satisfied then the odd solution is incompatible with current conservation.\footnote{Existence and uniqueness of the parity-odd solution (inside and outside triangle inequalities) has been proven in the ``light-like" limit in \cite{Giombi:2016zwa}. Similar arguments can be made to show there are only two forms for the parity-even solutions, these are sketched in \cite{Maldacena:2011jn}.} Furthermore, if any of the $J$, $J'$, $J''$ coincide (i.e. in cases where the currents are unique and have the same spin), then the resulting point-switch symmetries can kill off the remaining structures. 

By utilising the general formalism outlined in the previous sections we carried out an explicit analysis of three-point functions of conserved currents with $s_{i} \leq 20$. For currents of integer or half-integer spin, we found that the general structure of the three-point correlation function $\langle J^{}_{s_{1}} J'_{s_{2}} J''_{s_{3}} \rangle$ is fixed up to the following form:
\begin{equation}
	\langle J^{}_{s_{1}} J'_{s_{2}} J''_{s_{3}} \rangle = a_{1} \, \langle J^{}_{s_{1}} J'_{s_{2}} J''_{s_{3}} \rangle_{E_{1}} + a_{2} \, \langle J^{}_{s_{1}} J'_{s_{2}} J''_{s_{3}} \rangle_{E_{2}} + b \, \langle J^{}_{s_{1}} J'_{s_{2}} J''_{s_{3}} \rangle_{odd} \, .
\end{equation}
In this case the solutions $\langle J^{}_{s_{1}} J'_{s_{2}} J''_{s_{3}} \rangle_{E_{1}}$ and $\langle J^{}_{s_{1}} J'_{s_{2}} J''_{s_{3}} \rangle_{E_{2}}$ are parity-even, however, for currents of half-integer spin it is not clear whether these solutions necessarily correspond to theories of a free boson/fermion. The parity-odd solution is also present in three-point functions involving currents of half-integer spins, provided that the triangle inequalities are satisfied. Some further comments on the general structure of three-point functions are summarised below:
\begin{itemize}
	\item When the triangle inequalities are simultaneously satisfied, each polynomial structure in the solution for any three-point functions can be written as a product of at most 5 of the $P_{i}$, $Q_{i}$, with the $Z_{i}$ completely eliminated. 
	\item For the three-point functions $\langle J^{}_{s_{1}} J'_{s_{1}} J''_{s_{2}} \rangle$, for arbitrary (half-)integer $s_{1}$ and $s_{2}$: when the triangle inequalities are satisfied there are two even solutions and one odd solution, otherwise there are only two even solutions. After imposing $J=J'$ the solutions exist only when $s_{2}$ is an even integer. Note that for $s_{1} > s_{2}$ the triangle inequalities are always satisfied.
	\item For the three-point functions $\langle J_{s} \, J_{s} \, J_{s} \rangle$, with $s$ an integer, there are two even solutions and one odd solution, however they exist only for $s$ even. For $s$ odd the solutions survive only if the currents carry a flavour index associated with a non-Abelian symmetry group.
\end{itemize}
Another observation is that the triangle inequalities can be encoded in a discriminant, $\s$, which we define as follows:
\begin{align} \label{Ch03-Discriminant}
	\s(s_{1}, s_{2}, s_{3}) = q_{1} q_{2} q_{3} \, , \hspace{10mm} q_{i} = s_{i} - s_{j} - s_{k} - 1 \, ,
\end{align}
where $(i,j,k)$ is a cyclic permutation of $(1,2,3)$ (e.g. we have $q_{1} = s_{1} - s_{2} - s_{3} - 1$). For $\s(s_{1}, s_{2}, s_{3}) < 0$, there are two even solutions and one odd solution, while for $\s(s_{1}, s_{2}, s_{3}) \geq 0$ there are only two even solutions. 
The origin of this discriminant equation is actually quite simple within the framework of this formalism: recall that the correlation function can be encoded in a tensor $\cH$, which is a function of a single three-point covariant, $X$. There are three different (but equivalent) representations of a given correlation function, 
call them $\cH^{(i)}$, where the superscript $i$ denotes which point we set to act as the ``third point" in the ansatz \eqref{Ch03-H ansatz}. As shown in subsection \ref{Ch03-subsubsection3.3.1}, 
the representations are related by the intertwining operator $\cI$, with each $\cH^{(i)}$ being homogeneous degree $q_{i}$. After exhaustive analysis of the three-point functions with $s_{i} \leq 20$, 
a clear pattern emerges: the odd structure survives if and only if $\forall i$, $q_{i} < 0$. In other words, each $\cH^{(i)}$ must be a rational function of $X$ with homogeneity $q_{i} < 0$. The discriminant \eqref{Ch03-Discriminant} simply encodes information about whether the $\cH^{(i)}$ are simultaneously of negative homogeneity.

\subsection{Three-point functions involving scalars and spinors}\label{Ch03-subsection3.4.3}

In this section, for completeness, we analyse some of the important three-point correlation functions involving scalar and spinor fields. The results are interesting because the correlation functions can contain parity-odd solutions, with their existence depending on both triangle inequalities and the scale dimensions of the scalars/spinors. The correlation functions are analysed using the same methods as in the previous sections; the full classification of the results is presented below:
\begin{itemize}
	\item The three-point function $\langle \psi \, \psi' \, O \rangle$, where $\psi, \psi'$ are fundamental fermions and $O$ is a fundamental scalar, is fixed up to one even structure and one odd structure. All structures remain after imposing $\psi = \psi'$.
	\item The three-point function $\langle O \, O' J_{s} \rangle$, where $O,O'$ are fundamental scalars with dimension $\delta,\delta'$ respectively: for $\delta = \delta'$, there is a single even solution compatible with conservation which survives after imposing $O = O'$ only for even $s$. For $\delta \neq \delta'$ there are no solutions.
	\item The three-point function $\langle \psi \, \psi' J_{s} \rangle$, where $\psi,\psi$ are fundamental fermions with dimension $\delta,\delta'$ respectively: when $s=1$, the triangle inequalities are satisfied, and for $\delta = \delta'$ there are two even solutions and one odd solution. For $\delta \neq \delta'$ there is one even solution and one odd solution. In both cases, the three-point function vanishes after imposing $\psi = \psi'$. For $s >1$, the triangle inequalities are not satisfied, and for $\d = \d'$ there are two even solutions which survive after imposing $\psi = \psi'$ provided that $s$ is even. For $\d \neq \d'$ there are no solutions for general $s$.
	\item The three-point function $\langle \psi \, J^{}_{s} \, O \rangle$, for half-integer $s \geq 3/2$, where $\psi$ is a fundamental fermion with dimension $\delta$, and $O$ is a scalar with dimension $\delta'$: the triangle inequalities are not satisfied for any $s$, and in general there are no solutions after imposing conservation for arbitrary $\d, \d'$. However, there are two special cases; for $\delta = 3/2$ there is an even solution for $\delta' = 1$ and an odd solution for $\delta' = 2$.
	\item For three-point functions of the form $\langle J^{}_{s_{1}} J'^{}_{s_{2}} \, O \rangle$, where $O$ is a scalar field with dimension $\delta$, $s_{1}$ and $s_{2}$ must be simultaneously integer/half-integer for there to be a solution. For $s_{1} > s_{2}$, the triangle inequalities are not satisfied and there is no solution for general $\delta$, however, there is an even solution for $\delta = 1$, and an odd solution for $\delta = 2$. For $s_{1} = s_{2}$, the triangle inequalities are satisfied and there exists an even and odd solution for general $\delta$. The solutions also survive after imposing the symmetry $J=J'$.
	\item For three-point functions of the form $\langle \psi \, J^{}_{s_{1}} J'_{s_{2}} \rangle$, for half-integer $s_{1} \geq 3/2$ and integer $s_{2}$, where $\psi$ is a fundamental fermion with dimension $\delta$: for $\delta = 3/2$ there are two even solutions and one odd solution provided that the triangle inequalities are satisfied, otherwise, there are only two even solutions. In addition, for $\delta = 5/2$ there is one even solution and one odd solution when the triangle inequalities are satisfied, otherwise, there is a single odd solution. For general $\delta$, an even and odd solution exists if the triangle inequalities are satisfied, otherwise, there are no solutions. 
\end{itemize}
In Appendix \ref{Appendix3C} we present examples for some of the above cases. This concludes the analysis of three-point functions of conserved currents in three-dimensional conformal field theory.

\section{Summary of results}\label{Ch03-section3.5}

In this chapter we have systematically analysed the general structure of three-point functions of conserved currents of arbitrary integer and half-integer spins in three spacetime dimensions. To do this we generalised the construction of \cite{Osborn:1993cr} to a suitable spinor representation and augmented it with auxiliary spinor variables, which greatly simplifies the analysis for three-point functions of higher-spin currents. The main advantage of the approach is that the structure of the three-point function is reduced to determining the general form of a homogeneous polynomial which is built from simple monomial structures which are expressed in terms of the three-point conformally covariant building blocks. Since the conservation equations are formulated in terms of these covariants (as opposed to the spacetime points), the process of imposing the conservation equations is dramatically simplified. We demonstrated the effectiveness of the formalism by analysing in detail the three-point functions of the energy-momentum tensor and conserved vector currents, where we obtained the known results (in terms of the number of independent structures) \cite{Osborn:1993cr,Giombi:2011rz,Zhiboedov:2012bm,Giombi:2016zwa}. Next, we analysed the general structure of three-point functions involving supersymmetry currents. These results are new, and are naturally of interest in supersymmetric field theories. Finally, we proposed a general classification for the structure of three-point functions of conserved currents for arbitrary integer or half-integer spins by performing an explicit analysis for spins $s_{i} \leq 20$, which is a significant improvement over previous results in the literature \cite{Giombi:2011rz}. As a result of the analysis we provided strong evidence that three-point functions of conserved currents are fixed up to two parity even structures and one parity-violating structure, with the existence of the latter structure subject to a set of triangle inequalities in the spins.

Concerning the origin of the structures present in the three-point functions of higher-spin currents, let us recall that in three-dimensional conformal field theory the three-point function of the energy-momentum tensor is fixed up to two linearly independent parity-even structures and one parity-odd structure. The two parity-even structures are known to correspond to theories of free bosons and free fermions, which is consistent with the Maldacena-Zhiboedov theorem \cite{Maldacena:2011jn}. In this study it was shown that if a 3D CFT possesses higher-spin symmetry then the three-point functions of the energy-momentum tensor and higher-spin currents are equal to those of free field theories. The parity-odd structures, however, do not have an interpretation in terms of free field theories but are known to correspond to theories of a Chern-Simons gauge field interacting with parity-violating matter \cite{Aharony:2011jz, Giombi:2011kc, Maldacena:2012sf, Jain:2012qi, GurAri:2012is, Aharony:2012nh, Giombi:2016zwa, Chowdhury:2017vel, Sezgin:2017jgm, Skvortsov:2018uru, Inbasekar:2019wdw}. A similar story occurs for the three-point functions of higher-spin currents, which are also fixed up to two-independent parity-even structures and are known to correspond to free field theories. In terms of the number of independent structures appearing in the three-point functions of conserved currents, the analysis presented in this chapter is in agreement with the above results.

\begin{subappendices}
	\section{Examples: three-point functions of higher-spin currents} \label{Appendix3B}
\subsection*{Bosonic currents}
We now present explicit results for three-point functions involving bosonic higher-spin currents after conservation on all three points.

\noindent
\textbf{Correlation function} $\langle J^{}_{1} J'_{1} J''_{3} \rangle$\textbf{:} \hspace{3mm} $\sigma = 0$
\begin{flalign*}
	\hspace{5mm} \includegraphics[width=0.9\textwidth]{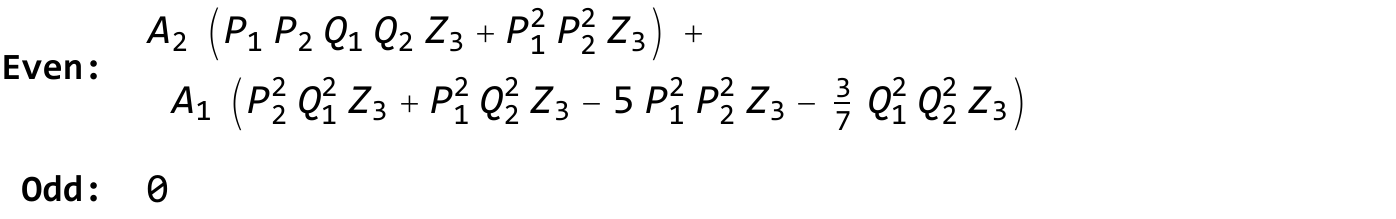} &&
\end{flalign*}
This is an instance in which one of the triangle inequalities is not satisfied, and we can see here that the odd solution vanishes. The even structures vanish after imposing $J = J'$. \\[4mm]
\textbf{Correlation function} $\langle J^{}_{1} J'_{1} J''_{4} \rangle$\textbf{:} \hspace{3mm} $\sigma > 0$
\begin{align}\label{Ch03-1-1-4}
	\includegraphics[width=0.9\textwidth, valign=c]{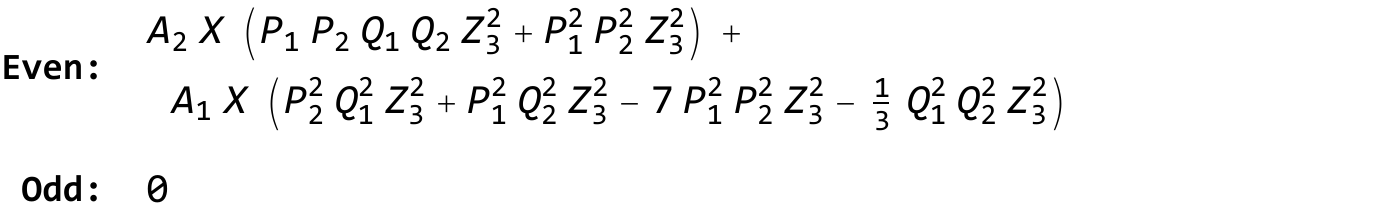}
\end{align} \\
\textbf{Correlation function} $\langle J^{}_{1} J'_{2} J''_{3} \rangle$\textbf{:} \hspace{3mm} $\sigma < 0$
\begin{align}\label{Ch03-1-2-3}
	\includegraphics[width=0.9\textwidth, valign=c]{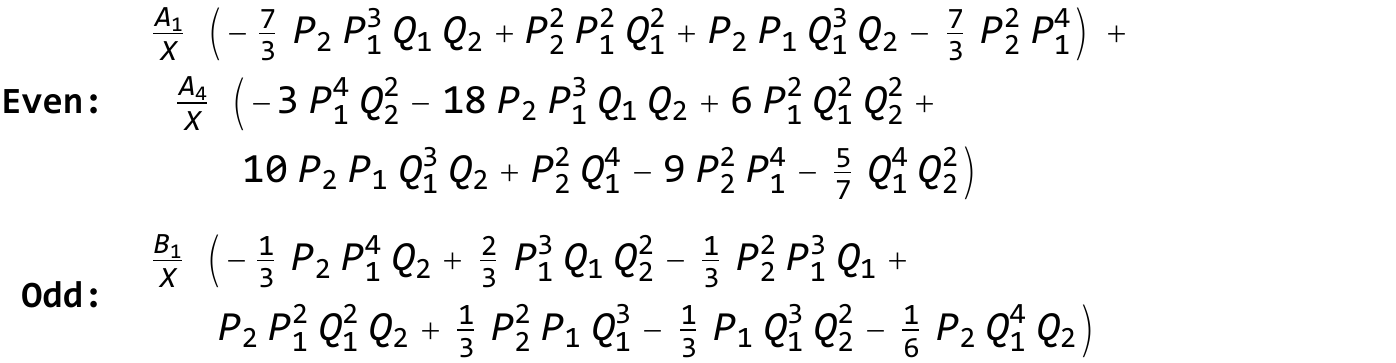}
\end{align} \\
\textbf{Correlation function} $\langle J^{}_{2} J'_{2} J''_{3} \rangle$\textbf{:} \hspace{3mm} $\sigma < 0$
\begin{align}\label{Ch03-2-2-3}
	\includegraphics[width=0.9\textwidth, valign=c]{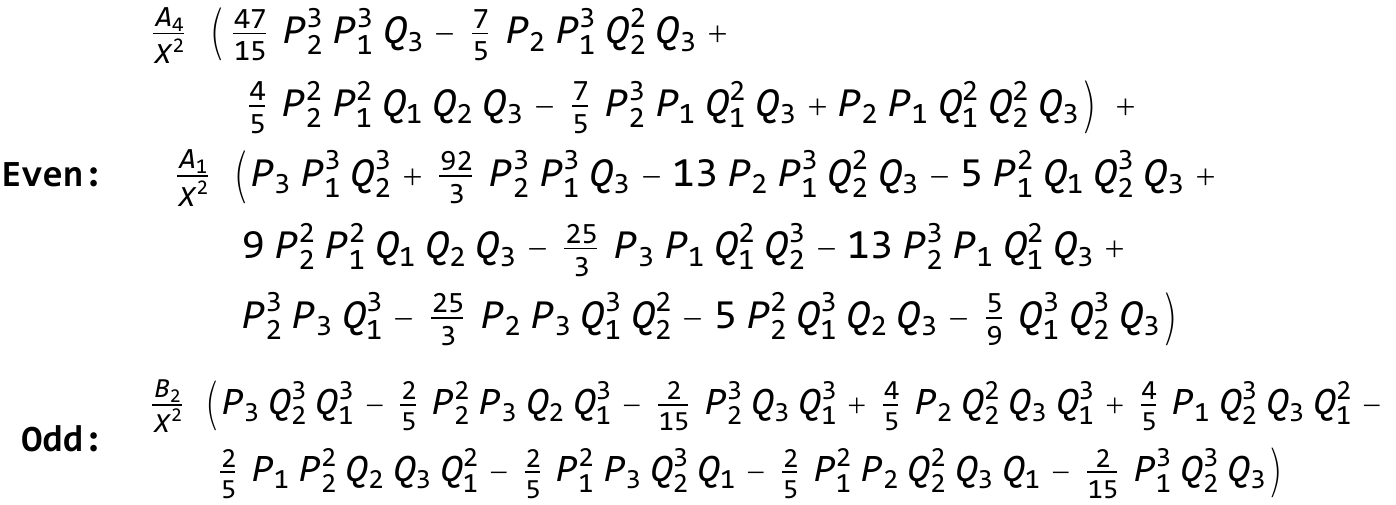}
\end{align} \\
\textbf{Correlation function} $\langle J^{}_{2} J'_{2} J''_{4} \rangle$\textbf{:} \hspace{3mm} $\sigma < 0$
\begin{align}\label{Ch03-2-2-4}
	\includegraphics[width=0.9\textwidth, valign=c]{4-4-8.pdf}
\end{align} \\
\textbf{Correlation function} $\langle J^{}_{2} J'_{2} J''_{5} \rangle$\textbf{:} \hspace{3mm} $\sigma = 0$
\begin{align}\label{Ch03-2-2-5}
	\includegraphics[width=0.9\textwidth, valign=c]{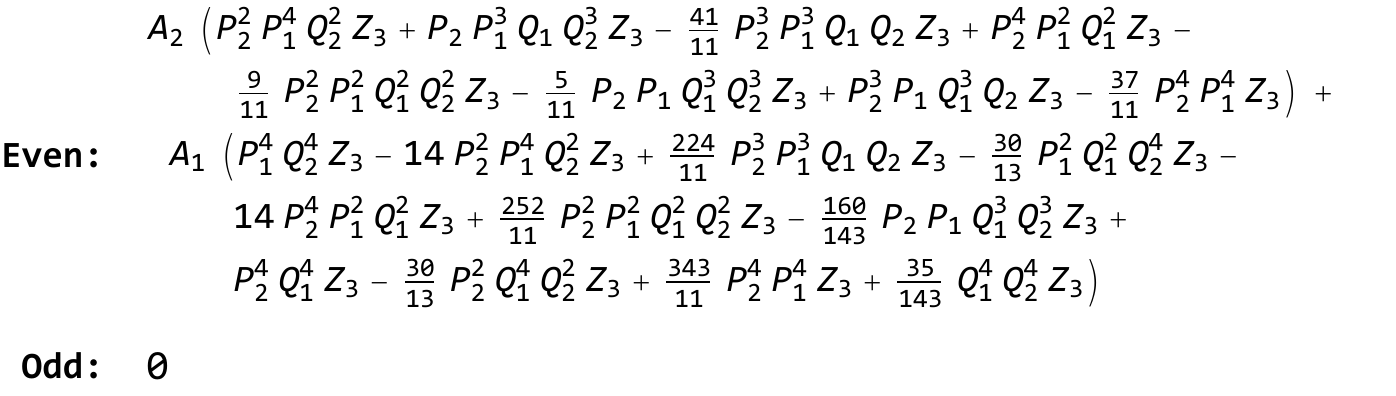}
\end{align} \\
\textbf{Correlation function} $\langle J^{}_{3} J'_{3} J''_{3} \rangle$\textbf{:} \hspace{3mm} $\sigma < 0$
\begin{align}\label{Ch03-3-3-3}
	\includegraphics[width=0.9\textwidth, valign=c]{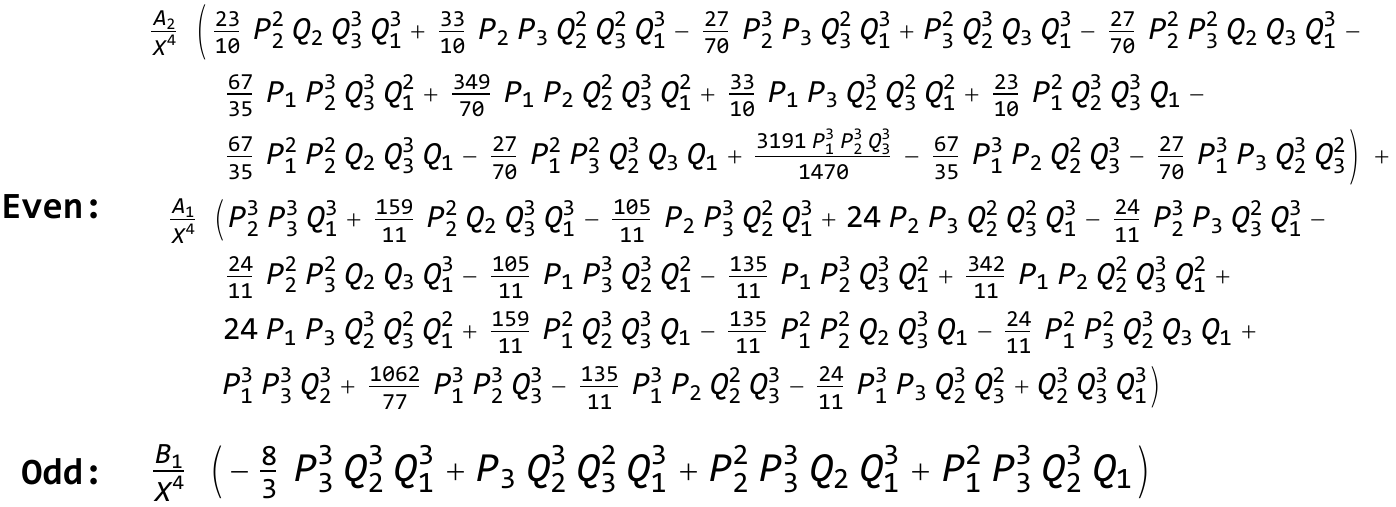}
\end{align} \\
\textbf{Correlation function} $\langle J^{}_{4} J'_{4} J''_{4} \rangle$\textbf{:} \hspace{3mm} $\sigma < 0$
\begin{align}\label{Ch03-4-4-4}
	\includegraphics[width=0.9\textwidth, valign=c]{8-8-8.pdf}
\end{align}

\newpage
\subsection*{Fermionic currents}

We now present some results for three-point correlation functions involving fermionic higher-spin currents.

\noindent
\textbf{Correlation function} $\langle J^{}_{3/2} J'_{3/2} J''_{3} \rangle$\textbf{:} \hspace{3mm} $\sigma < 0$
\begin{align}\label{Ch03-3/2-3/2-3}
	\includegraphics[width=0.9\textwidth, valign=c]{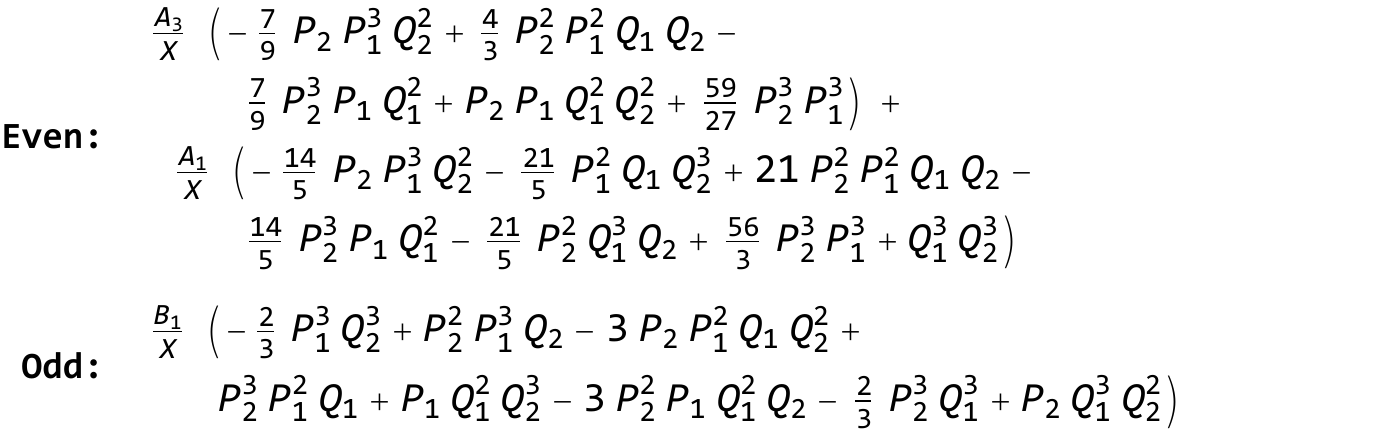}
\end{align} \\
Imposing the symmetry under permutation of spacetimes points $x_{1}$ and $x_{2}$, i.e. when $J = J'$, requires that the three-point function must vanish. \\[3mm]
\noindent\textbf{Correlation function} $\langle J^{}_{3/2} J'_{3/2} J''_{4} \rangle$\textbf{:} \hspace{3mm} $\sigma = 0$
\begin{align}\label{Ch03-3/2-3/2-4}
	\includegraphics[width=0.9\textwidth, valign=c]{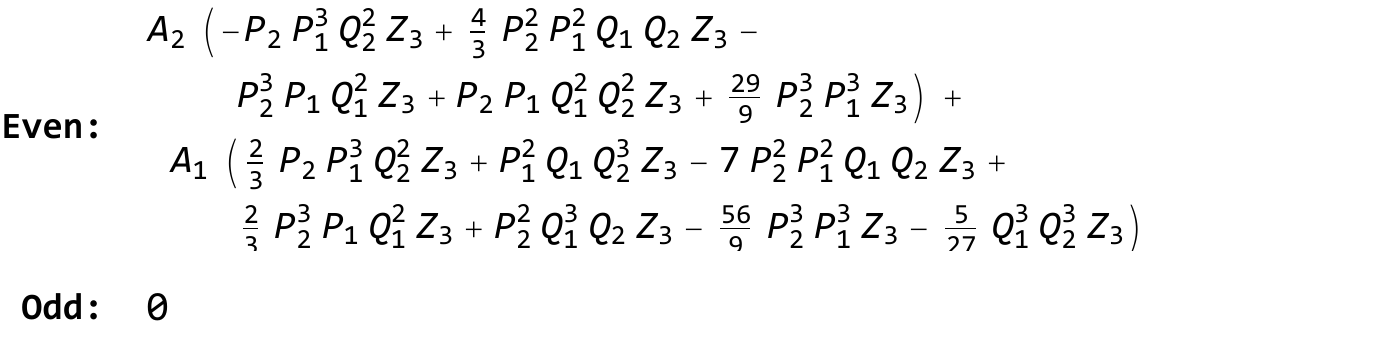}
\end{align} \\
This correlation function is compatible with the symmetry under permutation of spacetimes points $x_{1}$ and $x_{2}$. \\[3mm]
\textbf{Correlation function} $\langle J^{}_{5/2} J'_{3/2} J''_{1} \rangle$\textbf{:} \hspace{3mm} $\sigma < 0$
\begin{align}\label{Ch03-5/2-3/2-1}
	\includegraphics[width=0.9\textwidth, valign=c]{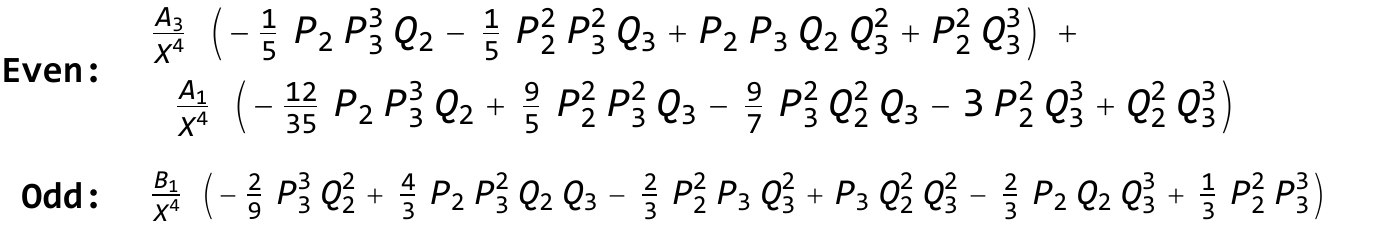}
\end{align} \\
\textbf{Correlation function} $\langle J^{}_{5/2} J'_{3/2} J''_{2} \rangle$\textbf{:} \hspace{3mm} $\sigma < 0$
\begin{align}\label{Ch03-5/2-3/2-2}
	\includegraphics[width=0.9\textwidth, valign=c]{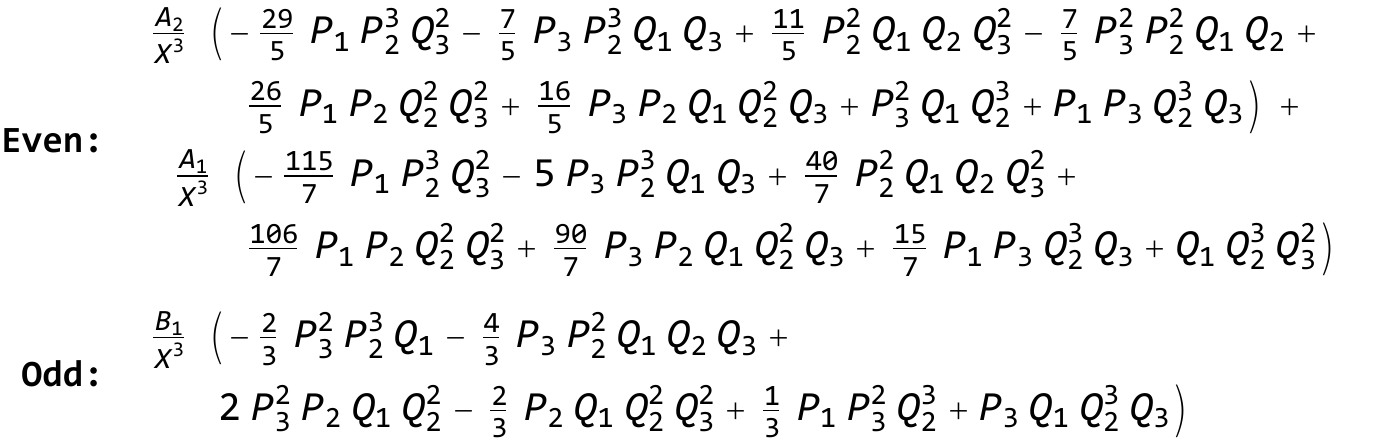}
\end{align} \\
\textbf{Correlation function} $\langle J^{}_{5/2} J'_{3/2} J''_{5} \rangle$\textbf{:} \hspace{3mm} $\sigma = 0$
\begin{align}\label{Ch03-5/2-3/2-5}
	\includegraphics[width=0.9\textwidth, valign=c]{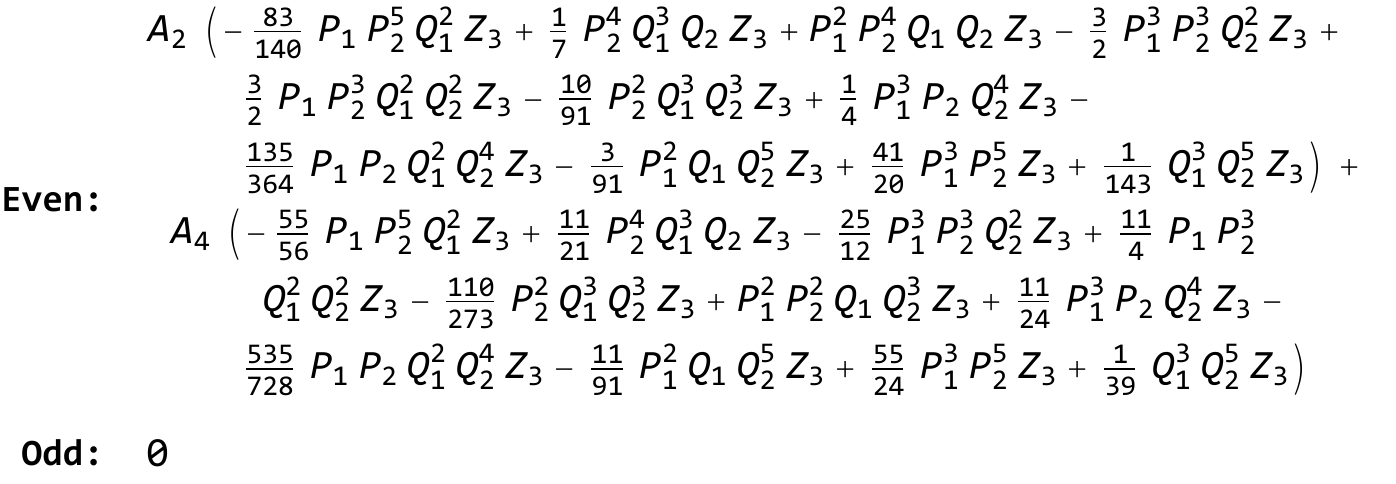}
\end{align} \\
\textbf{Correlation function} $\langle J^{}_{5/2} J'_{5/2} J''_{1} \rangle$\textbf{:} \hspace{3mm} $\sigma < 0$
\begin{align}\label{Ch03-5/2-5/2-1}
	\includegraphics[width=0.9\textwidth, valign=c]{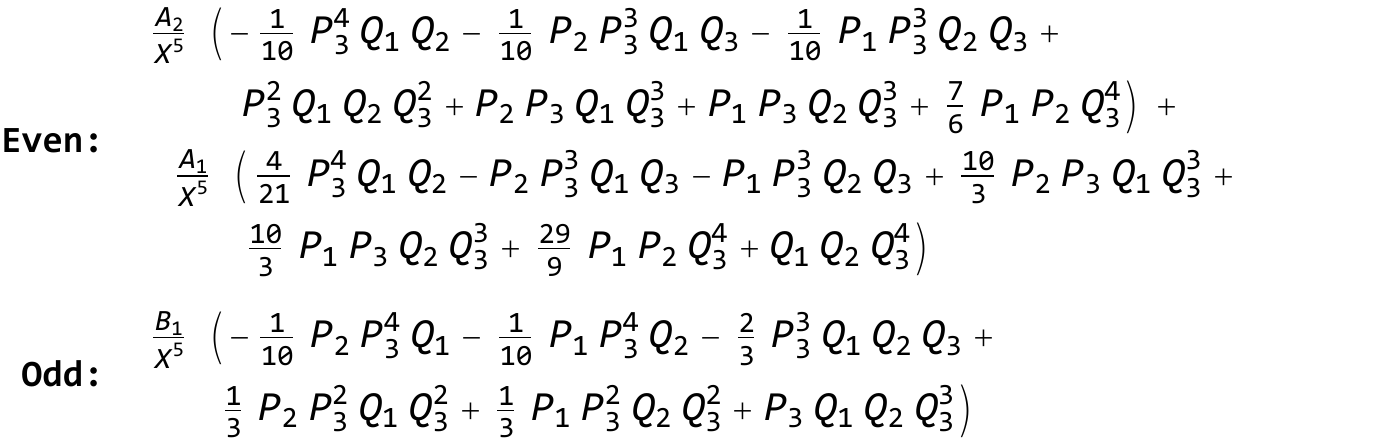}
\end{align} \\
\textbf{Correlation function} $\langle J^{}_{5/2} J'_{5/2} J''_{2} \rangle$\textbf{:} \hspace{3mm} $\sigma < 0$
\begin{align}\label{Ch03-5/2-5/2-2}
	\includegraphics[width=0.9\textwidth, valign=c]{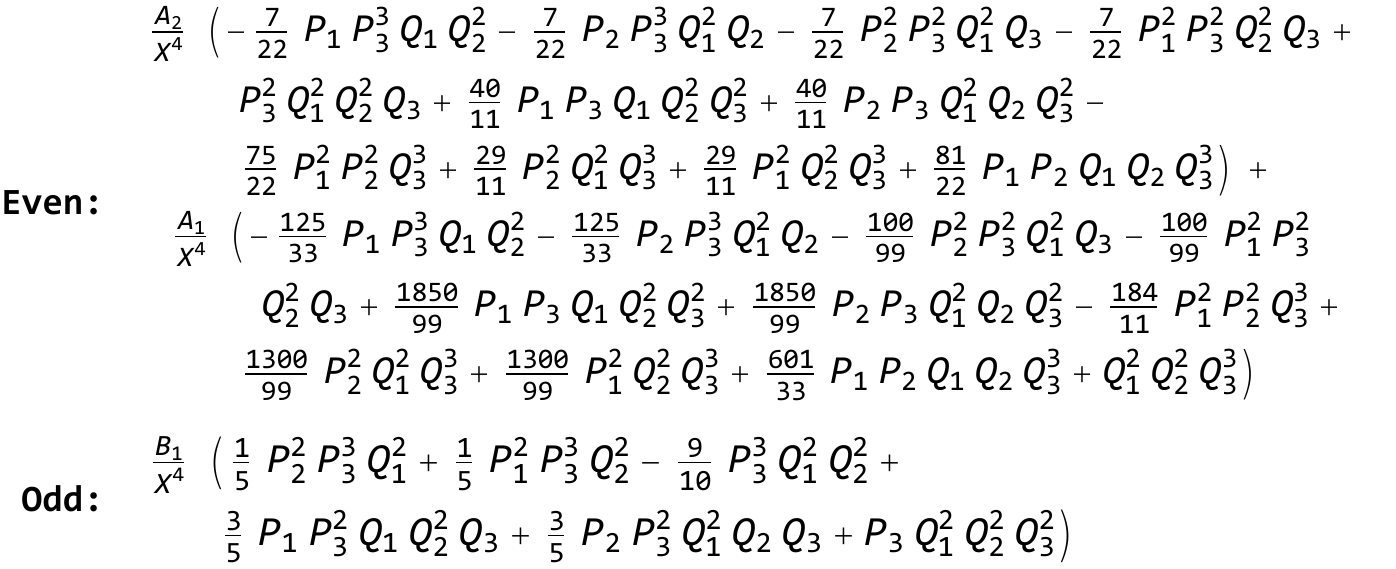}
\end{align} \\
\textbf{Correlation function} $\langle J^{}_{5/2} J'_{5/2} J''_{4} \rangle$\textbf{:} \hspace{3mm} $\sigma < 0$
\begin{align}\label{Ch03-5/2-5/2-4}
	\includegraphics[width=0.9\textwidth, valign=c]{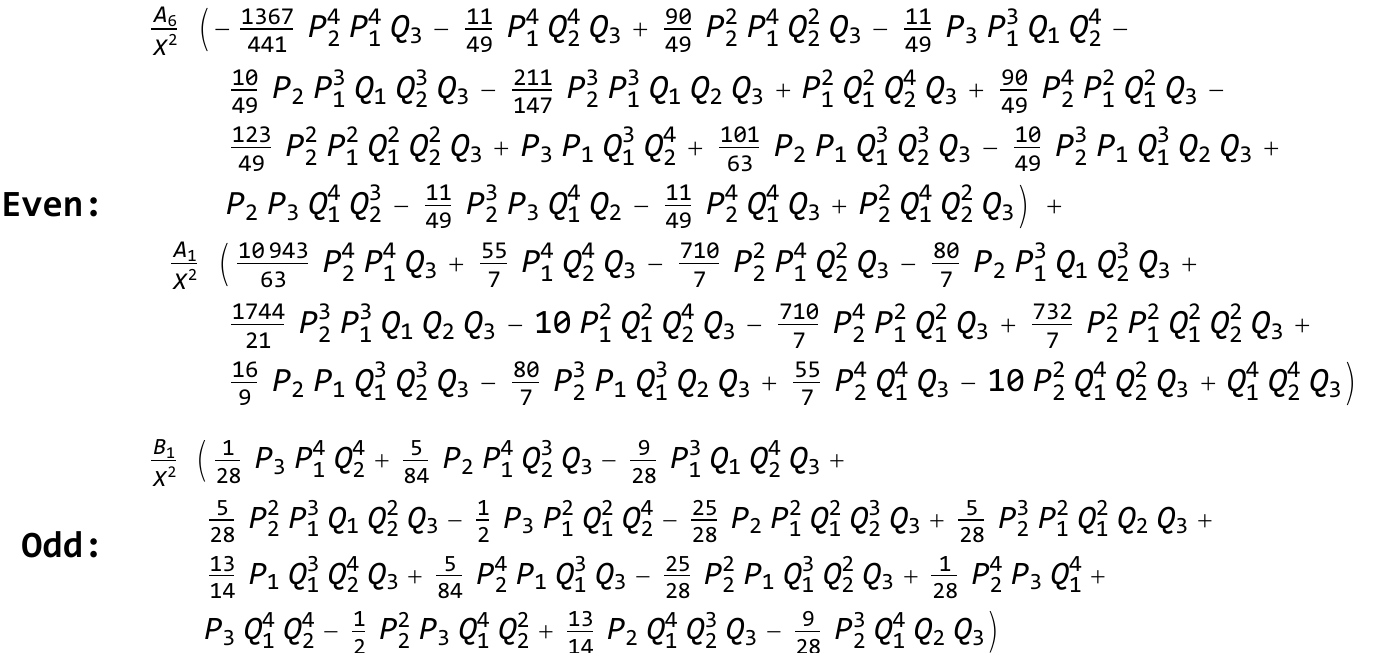}
\end{align} \\
The solutions quickly become cumbersome to present beyond these low-spin cases, however, our method effectively produces explicit results for any chosen spins within the confines of our computational limitations ($s_{i} \leq 20$).

\newpage
	\section{Examples: three-point functions involving scalars and spinors} \label{Appendix3C}
\noindent
\textbf{Correlation function} $\langle \psi \, \psi' \, O \rangle$\textbf{:}\\[2mm]
For $\Delta_{\psi} = \delta_{1}$, $\Delta_{\psi'} = \d'_{1}$, $\Delta_{O} = \d_{2}$, there is always one even and one odd solution
\begin{align}\label{Ch03-1/2-1/2-0}
	\includegraphics[width=0.9\textwidth, valign=c]{1-1-0.pdf}
\end{align} \\
\textbf{Correlation function} $\langle O \, O' J_{s} \rangle$\textbf{:}\\[2mm]
An even solution exists for $\Delta_{O} = \Delta_{O'} = \delta$:
\begin{align}\label{Ch03-0-0-2}
	\includegraphics[width=0.9\textwidth, valign=c]{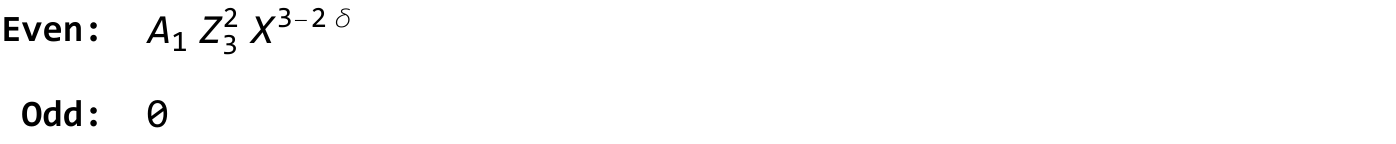}
\end{align}
In general, for $\langle O \, O' J_{s} \rangle$ there is always an even solution, however, it only survives the $O = O'$ point-switch symmetry for even $s$. \\[3mm]
\textbf{Correlation function} $\langle \psi \, \psi' J_{s} \rangle$\textbf{:}
\begin{align}\label{Ch03-1/2-1/2-2}
	\includegraphics[width=0.9\textwidth, valign=c]{1-1-4.pdf}
\end{align}
In this case the triangle inequalities are not satisfied and there are two even solutions. All structures survive after imposing $\psi = \psi'$. \\[3mm]
\textbf{Correlation function} $\langle J^{}_{1} J'_{1} \, O \rangle$\textbf{:}
\begin{align}\label{Ch03-1-1-0}
	\includegraphics[width=0.9\textwidth, valign=c]{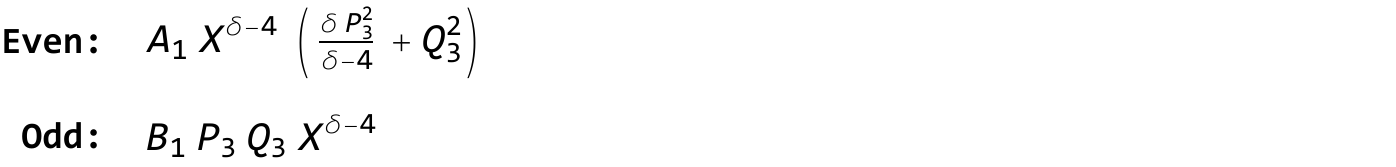}
\end{align}
In this case the triangle inequalities are satisfied and there is one even solution and one odd solution. The structures survive after imposing $J = J'$.

\newpage

\noindent
\textbf{Correlation function} $\langle J^{}_{2} J'_{1} \, O \rangle$\textbf{:}\\[2mm]
In this case there is a single even solution for $\D_{O} = 1$:
\begin{align}\label{Ch03-2-1-0-A}
	\includegraphics[width=0.9\textwidth, valign=c]{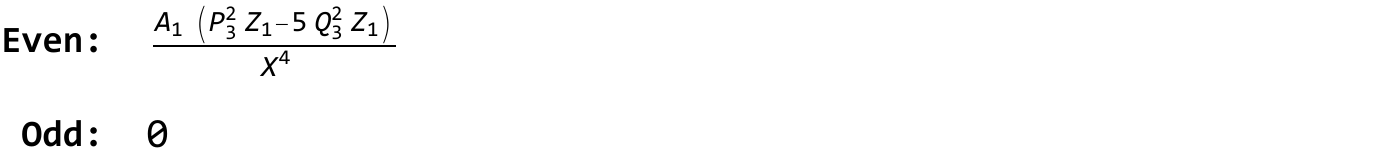}
\end{align}
There is also a single odd solution for $\D_{O} = 2$:
\begin{align}\label{Ch03-2-1-0-B}
	\includegraphics[width=0.9\textwidth, valign=c]{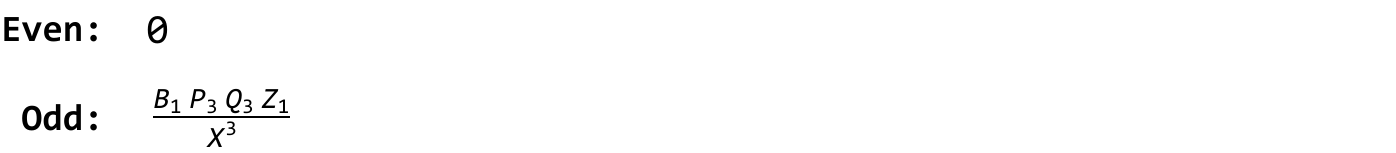}
\end{align}
However, there are no solutions for arbitrary $\D_{O}$. The same result was found in \cite{Giombi:2011rz}. \\[3mm]
\noindent
\textbf{Correlation function} $\langle J^{}_{3/2} J'_{3/2} \, O \rangle$\textbf{:}
\begin{align}\label{Ch03-3/2-3/2-0}
	\includegraphics[width=0.9\textwidth, valign=c]{3-3-0.pdf}
\end{align} \\
\textbf{Correlation function} $\langle J^{}_{2} J'_{2} \, O \rangle$\textbf{:}
\begin{align}\label{Ch03-2-2-0}
	\includegraphics[width=0.9\textwidth, valign=c]{4-4-0.pdf}
\end{align} \\
\textbf{Correlation function} $\langle \psi \, J_{3/2} \, O \rangle$\textbf{:}\\[2mm]
For $\Delta_{\psi} = 3/2$, there is an even solution for $\Delta_{O} = 1$:
\begin{align}\label{Ch03-1/2-3/2-0-A}
	\includegraphics[width=0.9\textwidth, valign=c]{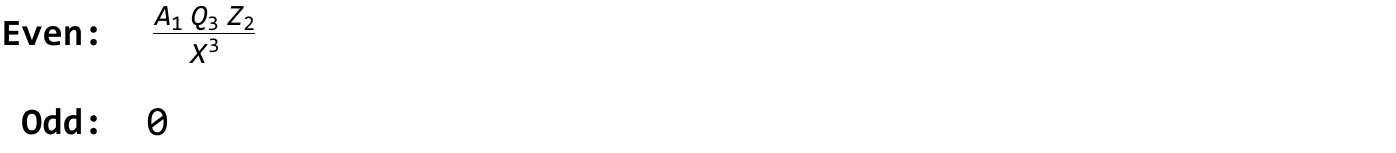}
\end{align}
There is also an odd solution for $\Delta_{O} = 2$:
\begin{align}\label{Ch03-1/2-3/2-0-B}
	\includegraphics[width=0.9\textwidth, valign=c]{1-3-0-B.pdf}
\end{align} \\
\textbf{Correlation function} $\langle \psi \, J^{}_{3/2} J'_{1} \rangle$\textbf{:}\\[2mm]
In this case there is an even and odd solution for general $\Delta_{\psi}$:
\begin{align}\label{Ch03-1/2-3/2-1-A}
	\includegraphics[width=0.9\textwidth, valign=c]{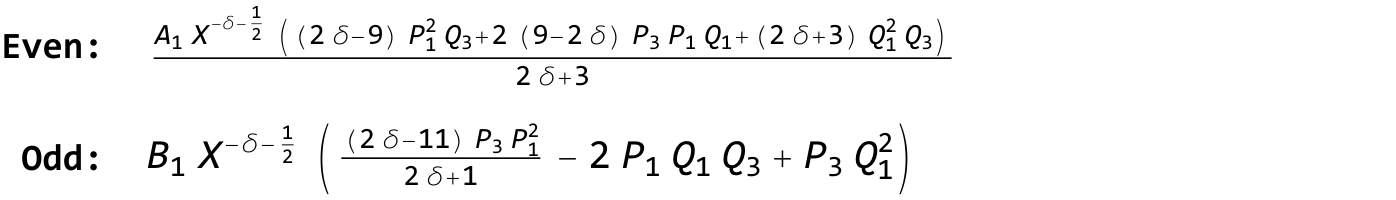}
\end{align} \\
There is also an additional even solution for $\Delta_{\psi} = 3/2$:
\begin{align}\label{Ch03-1/2-3/2-1-B}
	\includegraphics[width=0.9\textwidth, valign=c]{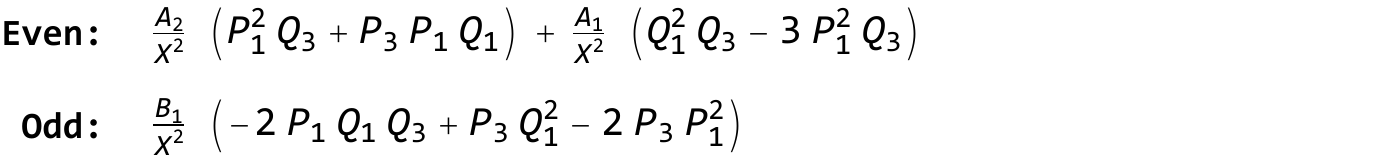}
\end{align} \\
\textbf{Correlation function} $\langle \psi \, J^{}_{3/2} J'_{2} \rangle$\textbf{:}\\[2mm]
In this case there is one even and one odd solution for general $\Delta_{\psi}$:
\begin{align}\label{Ch03-1/2-3/2-2-A}
	\includegraphics[width=0.9\textwidth, valign=c]{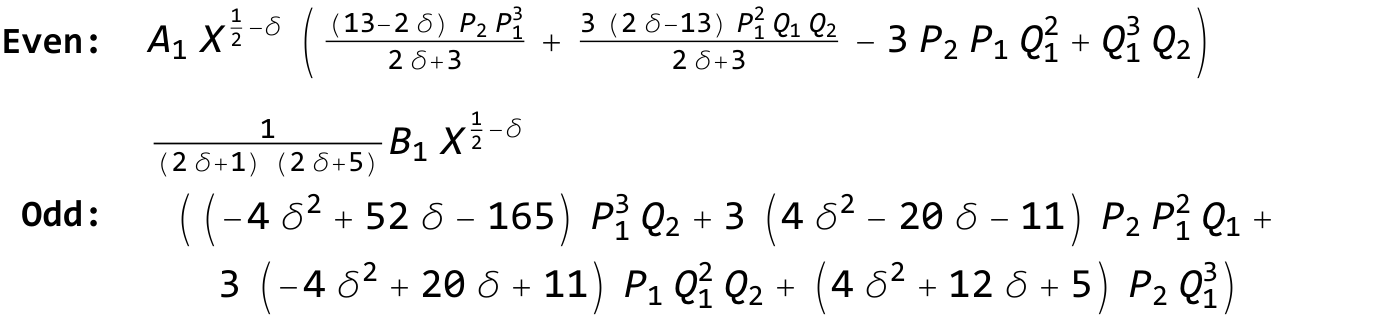}
\end{align}
However, there is an additional even solution for $\Delta_{\psi} = 3/2$:
\begin{align}\label{Ch03-1/2-3/2-2-B}
	\includegraphics[width=0.9\textwidth, valign=c]{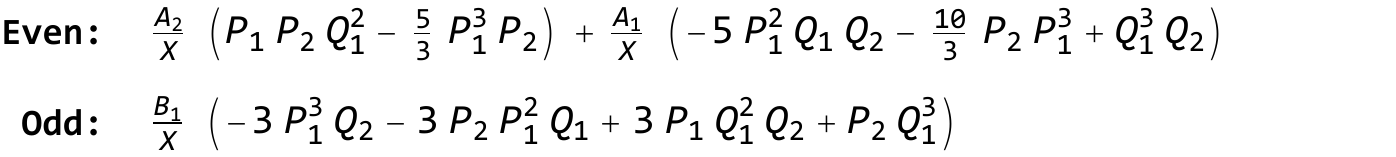}
\end{align}
%

%
\end{subappendices}

\chapter{Correlation functions of conserved currents in 4D CFT} \label{Chapter4}
\graphicspath{{Images/4DCFT/}} 

In this chapter our goal is to provide a complete classification of three-point functions of conserved currents $J_{\a(i) \ad(j)}$, with $i,j \geq 1$ in four-dimensional conformal field theory.
Such currents satisfy the conservation equation
\begin{equation}
	\pa^{\b \bd} J_{\b \a(i-1) \bd \ad(j-1)} = 0 \, ,
\end{equation}
and possess scale dimension $ \D_{J} = s + 2 $, where the spin $s$ is given by $s = \frac{1}{2}(i+j)$. 
To classify the possible three-point functions of currents $J_{\a(i) \ad(j)}$, we find it more convenient to parametrise them in terms of their spin, $s$, and an integer, $q=i-j$, as follows:
\begin{align} \label{Ch04-Current convention}
	J_{(s,q)} := J_{\a(s+\frac{q}{2}) \ad(s-\frac{q}{2})} \, .
\end{align}
With this convention $q$ is necessarily even/odd when $s$ is integer/half-integer valued. Note that the complex conjugate of $J_{(s,q)}$ is $J_{(s, -q)}$, hence, we introduce $\bar{J}_{(s,q)} := J_{(s, -q)}$ and view $q$ as being non-negative, taking values $q = 0, 1, ..., 2s - 2$. The case $q = 0$ corresponds to ``standard" bosonic conserved currents $J_{(s,0)} \equiv J_{\a(s) \ad(s)}$, with the vector current and energy-momentum tensor corresponding to the cases $s=1$ and $s=2$ respectively. Likewise, for $q=1$ we obtain pairs 
of (higher-spin) ``supersymmetry-like" currents $J_{(s,1)} \equiv J_{\a(s + \frac{1}{2}) \ad(s - \frac{1}{2})}$, $\bar{J}_{(s,1)} \equiv J_{(s,-1)} = J_{\a(s - \frac{1}{2}) \ad(s + \frac{1}{2})}$, where $s$ is necessarily half-integer valued. For example, by setting $s = \tfrac{3}{2}$ we obtain supersymmetry currents. In non-supersymmetric settings, the currents with $i=j$ (i.e. $q = 0$), were constructed explicitly in terms of free fields in  \cite{Craigie:1983fb}.

By using the notation \eqref{Ch04-Current convention} for the conserved currents there are essentially only two types of three-point functions to consider:
\begin{align} \label{Ch04-possible three point functions}
	\langle J^{}_{(s_{1}, q_{1})}(x_{1}) \, J'_{(s_{2}, q_{2})}(x_{2}) \, J''_{(s_{3}, q_{3})}(x_{3}) \rangle \, , \hspace{10mm} \langle J^{}_{(s_{1}, q_{1})}(x_{1}) \, 
	\bar{J}'_{(s_{2}, q_{2})}(x_{2}) \, J''_{(s_{3}, q_{3})}(x_{3}) \rangle \, .
\end{align}
Any other possible three-point functions are equivalent to these up to permutations of the points or complex conjugation. 
The main aim of this paper is to develop a general formalism to study the structure of the three-point correlation functions \eqref{Ch04-possible three point functions}, 
where we assume only the constraints imposed by conformal symmetry and conservation equations. In doing so we essentially provide a complete 
classification of all possible conserved three-point functions in 4D CFT. The three-point functions of currents with $q = 0, 1$ have been studied in 
e.g. \cite{Stanev:2012nq, Zhiboedov:2012bm, Elkhidir:2014woa, Buchbinder:2022cqp}. For bosonic conserved currents ($q_{i} = 0$), 
it is known that three-point functions of conserved currents with spins $s_{1}, s_{2}, s_{3}$ are fixed up to $2 \min(s_{1}, s_{2}, s_{3}) + 1$ solutions in general. We show that the same result also holds for three-point functions involving conserved currents with $q = 1$. The three-point functions of currents $J_{(s,q)}$ with $q \geq 2$, however, are relatively unexplored in the literature.

The formalism, which augments the approach of \cite{Osborn:1993cr} with auxiliary spinors, is suitable for constructing three-point functions of (conserved) primary operators in any Lorentz representation. Our approach is exhaustive in the sense that we construct all possible structures for the three-point function 
consistent with the conformal properties of the fields. We then systematically extract the linearly independent structures and impose the constraints arising from conservation equations, 
reality conditions, properties under inversion, and symmetries under permutations of spacetime points. The calculations are automated for arbitrary spins, and as a result we obtain the three-point function in an explicit form which can be presented up to our computational limit, $s_{i} = 10$. However, this limit is sufficient to propose a general classification of the results. 

Let us point out that although the formalism developed in this paper is conceptually similar to the one developed for 
three-dimensional CFT in~\cite{Buchbinder:2022mys}, there are two important differences. First, in three dimensions, the three-point functions of conserved 
currents can have at most three independent structures (two parity-even and one parity-odd), whereas in four dimensions the number of 
conserved structures (generically) grows linearly with the minimum spin. 
Second, for three-point functions in 3D CFT an important role is played by the triangle inequalities
\begin{align}
	s_{1} \leq s_{2} + s_{3} \, , && s_{2} \leq s_{1} + s_{3} \, , && s_{3} \leq s_{1} + s_{2} \, .
\end{align}
For three-point functions involving conserved currents which are within the triangle inequalities, there are two parity-even solutions and one parity-odd solution. 
However, if any of the triangle inequalities are not satisfied then the parity-odd solution is incompatible with conservation 
equations \cite{Giombi:2011rz,Maldacena:2011jn,Giombi:2016zwa,Jain:2021whr,Jain:2021gwa,Jain:2021vrv,Buchbinder:2022mys}. 
This statement has been proven in the light-cone limit in \cite{Maldacena:2011jn,Giombi:2016zwa} (see also \cite{Jain:2021whr} for results in 
momentum space). However, we found that in 4D CFT the triangle inequalities appear to have no significance for the number of independent conserved structures.

The content of this chapter is organised as follows. In Section \ref{Ch04-section2} we review the properties of the conformal building blocks used to construct correlation functions 
of primary operators in four dimensions. We then develop the formalism to construct three-point functions of primary operators of the form $J_{\a(i) \ad(j)}$, 
where we demonstrate how to impose all constraints arising from conservation equations, reality conditions and point switch symmetries. In particular, we utilise an index-free auxiliary spinor formalism 
to construct a generating function for the three-point functions, and we outline the pertinent aspects of our computational approach. In Section \ref{Ch04-section3}, we demonstrate 
our formalism by analysing the structure of three-point functions involving conserved vector currents, ``supersymmetry-like" currents and the energy-momentum tensor. 
We reproduce the known results previously found in~\cite{Osborn:1993cr,Stanev:2012nq,Zhiboedov:2012bm}. We then expand our discussion to include three-point functions of 
higher-spin currents belonging to any Lorentz representation, and provide a classification of the results.
For this the structure of the solutions is more easily identified by using the notation 
$J^{}_{(s,q)},\  \bar{J}_{(s,q)}$ for the currents as outlined above. In particular, we show that special attention is required for three-point functions of the form 
$\langle J^{}_{(s_{1},q)} \bar{J}'_{(s_{2},q)} J''_{(s_{3},0)} \rangle$ with $q \geq 2$. In this case the formula for the number of independent conserved structures is 
found to be quite non-trivial. In Appendix \ref{Appendix4B} we 
provide some examples of the three-point functions $\langle J^{}_{(s_{1},q)} \bar{J}'_{(s_{2},q)} J''_{(s_{3},0)} \rangle$ for which the number of independent 
conserved structures differs from $2 \min(s_{1}, s_{2}, s_{3}) + 1$.
Then, as a consistency check, in Appendix \ref{Appendix4C} we provide further examples of three-point functions involving scalars, spinors and conserved currents to compare against the results in \cite{Elkhidir:2014woa}.

%

\section{Conformal building blocks}\label{Ch04-section2}

In this section we will review the group theoretic formalism used to compute correlation functions of primary operators in four dimensional conformal field theories. For a more detailed introduction to the formalism as applied to correlation functions of bosonic primary fields see \cite{Osborn:1993cr}.

Consider 4D Minkowski space $\mathbb{M}^{4}$, parameterised by coordinates $ x^{m} $, where $m = 0, 1, 2, 3$ are Lorentz indices. For any two points, $x_{1}, x_{2}$, we construct the covariant two-point functions using the conventions outlined in Appendix \ref{Appendix2A}:
\begin{align}
	x_{12 \, \a \ad} &= (\s^{m})_{\a \ad} x_{12 \, m} \, , & x_{12}^{\ad \a} &= (\tilde{\s}^{m})^{\ad \a} x_{12 \, m} \, , & x_{12}^{2} &= - \frac{1}{2} x_{12}^{\ad \a} x^{}_{12 \, \a \ad} \, .
\end{align}
In this form the two-point functions possess the following useful properties:
\begin{align}  \label{Ch04-Two-point building blocks - properties 1} 
	x_{12}^{\ad \a} x^{}_{12 \, \b \ad} = - x_{12}^{2} \d_{\b}^{\a} \, , \hspace{10mm} x_{12}^{\ad \a} x_{12 \, \a \bd} = - x_{12}^{2} \d_{\bd}^{\ad} \, . 
\end{align}
Hence, we find
\begin{equation} \label{Ch04-Two-point building blocks 4}
	(x_{12}^{-1})^{\ad \a} = - \frac{x_{12}^{\ad \a}}{x_{12}^{2}} \, .
\end{equation}
We also introduce the normalised two-point functions, denoted by $\hat{x}_{12}$,
\begin{align} \label{Ch04-Two-point building blocks 3}
	\hat{x}_{12 \, \a \ad} = \frac{x_{12 \, \a \ad}}{( x_{12}^{2})^{1/2}} \, , \hspace{10mm} \hat{x}_{12}^{\ad \a} \hat{x}^{}_{12 \, \b \ad} = - \d_{\b}^{\a} \, . 
\end{align}
From here we can now construct an operator analogous to the conformal inversion tensor acting on the space of symmetric traceless tensors of arbitrary rank. Given a two-point function, $x$, we define the operator
\begin{equation} \label{Ch04-Higher-spin inversion operators a}
	\cI_{\a(k) \ad(k)}(x) = \hat{x}_{(\a_{1} (\ad_{1}} \dots \hat{x}_{ \a_{k}) \ad_{k})}  \, ,
\end{equation}
along with its inverse
\begin{equation} \label{Ch04-Higher-spin inversion operators b}
	\bar{\cI}^{\ad(k) \a(k)}(x) = \hat{x}^{(\ad_{1} (\a_{1}} \dots \hat{x}^{ \ad_{k}) \a_{k})} \, .
\end{equation}
The spinor indices may be raised and lowered using the standard conventions as follows:
\begin{subequations}
	\begin{align}
		\cI^{\a(k)}{}_{\ad(k)}(x) &= \ve^{\a_{1} \g_{1}} \dots \ve^{\a_{k} \g_{k}} \, \cI_{\g(k) \ad(k)}(x) \, , \\
		\bar{\cI}_{\ad(k)}{}^{\a(k)}(x) &= \ve_{\ad_{1} \gd_{1}} \dots \ve_{\ad_{k} \gd_{k}} \, \bar{\cI}^{\gd(k) \a(k)}(x) \, .
	\end{align}
\end{subequations}
Now due to the property
\begin{equation}
	\cI_{\a(k) \ad(k)}(-x) = (-1)^{k} \cI_{\a(k) \ad(k)}(x) \, ,
\end{equation}
we have the following useful relations:
\begin{subequations} \label{Ch04-Higher-spin inversion operators - properties}
	\begin{align}
		\cI_{\a(k) \ad(k)}(x_{12}) \, \bar{\cI}^{\ad(k) \b(k)}(x_{21}) &= \d_{(\a_{1}}^{(\b_{1}} \dots \d_{\a_{k})}^{\b_{k})} \, , \\
		\bar{\cI}^{\bd(k) \a(k)}(x_{12}) \, \cI_{\a(k) \ad(k)}(x_{21}) &= \d_{(\ad_{1}}^{(\bd_{1}} \dots \d_{\ad_{k})}^{\bd_{k})} \, .
	\end{align}
\end{subequations}
The objects \eqref{Ch04-Higher-spin inversion operators a}, \eqref{Ch04-Higher-spin inversion operators b} prove to be essential in the construction of correlation functions involving primary operators of arbitrary spins. Indeed, the vector representation of the inversion tensor may be recovered in terms of the spinor two-point functions as follows:
\begin{equation}
	I_{m n}(x) = - \frac{1}{2} \, \text{Tr}( \tilde{\s}_{m} \, \hat{x} \, \tilde{\s}_{n} \, \hat{x} ) \, .
\end{equation}
%
%

Now given three distinct points in Minkowski space, $x_{i}$, with $i = 1,2,3$, we define conformally covariant three-point functions in terms of the two-point functions as in \cite{Osborn:1993cr}
\begin{align}
	X_{ij} &= \frac{x_{ik}}{x_{ik}^{2}} - \frac{x_{jk}}{x_{jk}^{2}} \, , & X_{ji} &= - X_{ij} \, ,  & X_{ij}^{2} &= \frac{x_{ij}^{2}}{x_{ik}^{2} x_{jk}^{2} } \, , 
\end{align}
where $(i,j,k)$ is a cyclic permutation of $(1,2,3)$. For example, we have
\begin{equation}
	X_{12}^{m} = \frac{x_{13}^{m}}{x_{13}^{2}} - \frac{x_{23}^{m}}{x_{23}^{2}} \, , \hspace{10mm} X_{12}^{2} = \frac{x_{12}^{2}}{x_{13}^{2} x_{23}^{2} } \, .
\end{equation}
Let us recall that there are the following useful identities involving the two- and three-point functions and the conformal inversion tensor:
\begin{subequations}
	\begin{align} \label{Ch04-Inversion tensor identities - vector case 1}
		I_{m}{}^{a}(x_{13}) \, I_{a n}(x_{23}) &= I_{m}{}^{a}(x_{12}) \, I_{a n}(X_{31}) \, ,  & I_{m n}(x_{23}) \, X_{12}^{n} &= \frac{x_{12}^{2}}{x_{13}^{2}} \, X^{I}_{31 \, m} \, ,
	\end{align} \vspace{-6mm}
	\begin{align} \label{Ch04-Inversion tensor identities - vector case 2}
		I_{m}{}^{a}(x_{23}) \, I_{a n}(x_{13}) &= I_{m}{}^{a}(x_{21}) \, I_{a n}(X_{23}) \, ,  & I_{m n}(x_{13}) \, X_{12}^{n} &= \frac{x_{12}^{2}}{x_{23}^{2}} \, X^{I}_{23\, m} \, , 
	\end{align}
\end{subequations}
and the differential identities
\begin{align}
	\pa^{(1)}_{m} X_{12 \, n} = \frac{1}{x_{13}^{2}} I_{m n}(x_{13}) \, , \hspace{10mm} \pa^{(2)}_{m} X_{12 \, n} = - \frac{1}{x_{23}^{2}} I_{m n}(x_{23}) \, . \label{Ch04-Inversion tensor identities - vector case 3}
\end{align}
The three-point functions also may be represented in spinor notation as follows:
\begin{equation}
	X_{ij , \, \a \ad} = (\s_{m})_{\a \ad} X_{ij}^{m} \, , \hspace{10mm} X_{ij , \, \a \ad} = - (x^{-1}_{ik})_{\a \gd} x_{ij}^{\gd \g} (x^{-1}_{kj})_{\g \ad} \, .
\end{equation}
These objects satisfy properties similar to the two-point functions \eqref{Ch04-Two-point building blocks - properties 1}. Indeed, 
it is convenient to define the normalised three-point functions, $\hat{X}_{ij}$, and the inverses, $(X_{ij}^{-1})$
\begin{equation}
	\hat{X}_{ij , \, \a \ad} = \frac{X_{ij , \, \a \ad}}{( X_{ij}^{2})^{1/2}} \, , \hspace{10mm}	(X_{ij}^{-1})^{\ad \a} = - \frac{X_{ij}^{\ad \a}}{X_{ij}^{2}} \, .
\end{equation}  
Now given an arbitrary three-point building block $X$, it is also useful to construct the following higher-spin inversion operators:
\begin{equation}
	\cI_{\a(k) \ad(k)}(X) = \hat{X}_{ (\a_{1} (\ad_{1}} \dots \hat{X}_{\a_{k}) \ad_{k})}  \, ,
\end{equation}
along with its inverse
\begin{equation}
	\bar{\cI}^{\ad(k) \a(k)}(X) = \hat{X}^{(\ad_{1} (\a_{1}} \dots \hat{X}^{ \ad_{k}) \a_{k})} \, .
\end{equation}
These operators have properties similar to the two-point higher-spin inversion operators \eqref{Ch04-Higher-spin inversion operators a}, \eqref{Ch04-Higher-spin inversion operators b}. There are also some useful algebraic identities relating the two- and three-point functions at various points, such as
\begin{subequations} \label{Ch04-Inversion tensor identities - spinor case}
	\begin{align}
			\cI_{\a \gd}(x_{13}) \, \bar{\cI}^{\gd \g}(x_{12}) \, \cI_{\g \ad}(x_{23}) &= \cI_{\a \ad}(X_{12}) \, , \\
			\bar{\cI}^{\ad \g}(x_{13}) \, \cI_{\g \gd}(X_{12})  \, \bar{\cI}^{\gd \a}(x_{13}) &= \bar{\cI}^{\ad \a}(X^{I}_{23})  \, ,
	\end{align}
\end{subequations}
where $X^{I}_{\a \ad} = \cI_{\a}{}^{\ad'}(X) \, \bar{\cI}_{\ad}{}^{\a'}(X) \, X_{\a'\ad'} = - X_{\a \ad}$. These identities are analogous to \eqref{Ch04-Inversion tensor identities - vector case 1}, \eqref{Ch04-Inversion tensor identities - vector case 2}, and admit generalisations to higher-spins, for example
\begin{equation}
	\bar{\cI}^{\ad(k) \g(k)}(x_{13}) \, \cI_{\g(k) \gd(k)}(X_{12}) \, \bar{\cI}^{\gd(k) \a(k)}(x_{13}) = \bar{\cI}^{\ad(k) \a(k)}(X^{I}_{23})  \, . \label{Ch04-Inversion tensor identities - higher spin case}
\end{equation}
In addition, similar to \eqref{Ch04-Inversion tensor identities - vector case 3}, there are also the following useful identities:
\begin{equation}
	\pa^{(1)}_{\a \ad} X_{12}^{ \dot{\s} \s} = - \frac{2}{x_{13}^{2}} \, \cI_{\a}{}^{\dot{\s}}(x_{13}) \,  \bar{\cI}_{\ad}{}^{\s}(x_{13}) \, , \hspace{5mm} \pa^{(2)}_{\a \ad} X_{12}^{ \dot{\s} \s} = \frac{2}{x_{23}^{2}} \, \cI_{\a}{}^{\dot{\s}}(x_{23}) \,  \bar{\cI}_{\ad}{}^{\s}(x_{23}) \, . \label{Ch04-Three-point building blocks - differential identities}
\end{equation}
These identities allow us to account for the fact that correlation functions of primary fields can obey differential constraints which can arise due to conservation equations. Indeed, given a tensor field $\cT_{\cA}(X)$, there are the following differential identities which arise as a consequence of \eqref{Ch04-Three-point building blocks - differential identities}:
\begin{subequations}
	\begin{align}
		\pa_{(1) \, \a \ad} \cT_{\cA}(X_{12}) &= \frac{1}{x_{13}^{2}} \, \cI_{\a}{}^{\dot{\s}}(x_{13}) \,  \bar{\cI}_{\ad}{}^{\s}(x_{13}) \, \frac{ \pa}{ \pa X_{12}^{ \dot{\s} \s}} \, \cT_{\cA}(X_{12}) \, ,  \label{Ch04-Three-point building blocks - differential identities 2} \\[2mm]
		\pa_{(2) \, \a \ad} \cT_{\cA}(X_{12}) &= - \frac{1}{x_{23}^{2}} \, \cI_{\a}{}^{\dot{\s}}(x_{23}) \,  \bar{\cI}_{\ad}{}^{\s}(x_{23}) \, \frac{ \pa}{ \pa X_{12}^{ \dot{\s} \s}} \, \cT_{\cA}(X_{12}) \, . \label{Ch04-Three-point building blocks - differential identities 3}
	\end{align}
\end{subequations}
\subsection{Two-point correlation functions}\label{Ch04-section2.1}

Let $\F_{\cA}$ be a primary field with dimension $\D$, where $\cA$ denotes a collection of Lorentz spinor indices. The two-point correlation function of $\phi_{\cA}$ and its conjugate $\bar{\phi}^{\bar{\cA}}$ is fixed by conformal symmetry to the form
\begin{equation} \label{Ch04-two-point functions}
	\langle \phi_{\cA}(x_{1}) \, \bar{\phi}^{\bar{\cA}}(x_{2}) \rangle = c \, \frac{\cI_{\cA}{}^{\bar{\cA}}(x_{12})}{(x_{12}^{2})^{\D}} \, , 
\end{equation} 
where $\cI$ is an appropriate representation of the inversion tensor and $c$ is a constant complex parameter. The denominator of the two-point function is determined by the conformal dimension of $\phi_{\cA}$, which guarantees that the correlation function transforms with the appropriate weight under scale transformations.

For two-point functions of conserved higher-spin currents, we use \eqref{Ch04-two-point functions} and obtain the result
\begin{equation} 
	\langle J_{\a(s) \ad(s)}(x_{1}) \, J^{\b(s) \bd(s)}(x_{2}) \rangle = c \, \frac{\cI_{\a(s)}{}^{\bd(s)}(x_{12}) \, \bar{\cI}_{\ad(s)}{}^{\b(s)}(x_{12}) }{(x_{12}^{2})^{\D_{J}}} \, .
\end{equation} 
We may then impose $(\g_{m})^{\a_{1} \ad_{1}} \pa^{m}_{(1)} \langle J_{\a_{1} \a(s-1) \ad_{1} \ad(s-1)}(x_{1}) \, J^{\b(s) \bd(s)}(x_{2}) \rangle = 0$ to obtain $\D_{J}= s+2$, which is consistent with the dimension of a conserved current of arbitrary spin $s$. 

Similar to the 3D CFT case, we simplify imposing the constraints due to conservation equations on two- and three-point functions of conserved higher-spin currents by using auxiliary variables. In particular we introduce the commuting auxiliary spinors $u, \bar{u}$ and $v,\bar{v}$ at $x_{1}$ and $x_{2}$ which satisfy
\begin{subequations}
\begin{align}
	u^2 = \varepsilon_{\a \b} \, u^{\a} u^{\b}=0\,,  \hspace{10mm}
	v^{2} = \varepsilon_{\a \b} \, v^{\a} v^{\b}=0\, . \\
	\bar{u}^2 = \varepsilon_{\ad \bd} \, \bar{u}^{\ad} \bar{u}^{\bd}=0\,,  \hspace{10mm}
	\bar{v}^{2} = \varepsilon_{\ad \bd} \, \bar{v}^{\ad} \bar{v}^{\bd}=0\, .
\end{align}
\end{subequations}
We then contract the tensor indices on the currents with the auxiliary spinors so that
\begin{align}
	J_{s}(x_{1};u, \bar{u}) = J_{(\a_{1} \dots \a_{s})(\ad_{1} \dots \ad_{s}) }(x_{1}) \, u^{\a_{1}} \dots u^{\a_{s}} \bar{u}^{\ad_{1}} \dots \bar{u}^{\ad_{s}} \, ,
\end{align}
with a similar result for the current at $x_{2}$. The two-point function now has the following compact form:
\begin{equation} 
	\langle J_{s}(x_{1}; u) \, J_{s}(x_{2}; v)  \rangle = c \, \frac{( u \cdot \hat{x}_{12} \cdot \bar{v})^{s} ( v \cdot \hat{x}_{12} \cdot \bar{u})^{s} }{(x_{12}^{2})^{\D_{J}}} \, ,
\end{equation} 
where $u \cdot \hat{x}_{12} \cdot \bar{v} = \hat{x}_{12 \, \a \ad} u^{\a}  \bar{v}^{\ad}$. Hence, the two-point function is encoded in a polynomial of homogeneity $s$ in the auxiliary spinors $u,\bar{u}$ and $v,\bar{v}$. The original ``tensor" two-point function can be obtained by acting on it with suitable partial derivative operators, e.g.
\begin{align}
	\frac{\pa}{\pa \boldsymbol{U}^{\a(s) \ad(s)}} = \frac{1}{(s!)^{2}} \frac{\pa}{\pa u^{\a_{1}}} \dots \frac{\pa}{\pa u^{\a_{s}}} \frac{\pa}{\pa \bar{u}^{\ad_{1}}} \dots \frac{\pa}{\pa \bar{u}^{\ad_{s}}}\, ,
\end{align}
The tensor and polynomial forms of the two-point function are in one-to-one correspondence with eachother. Imposing conservation on $x_{1}$ is now tantamount to the condition
\begin{equation}
	D_{1} \langle J_{s}(x_{1}; u, \bar{u}) \, J_{s}(x_{2}; v, \bar{v})  \rangle = 0 \, , \hspace{10mm} D_{1} = \frac{\pa}{\pa x_{(1) \a \ad}} \frac{\pa}{\pa u^{\a}} \frac{\pa}{\pa \bar{u}^{\ad}} \, ,
\end{equation}
where the operator $D_{1}$ should be understood as a contraction of a partial derivative in $x_{1}$ with the $\sigma$-matrices. It can be shown that the conservation condition is satisfied for $\D_{J} = s + 2$, as expected of a conserved spin-$s$ current in four dimensions. 

We point out that conserved higher-spin currents in four dimensions were constructed in terms of free fields by Craigie, Dobrev and Todorov \cite{Craigie:1983fb}. The explicit computation of the two-point function of these currents was carried out recently by Kuzenko and Ponds \cite{Kuzenko:2022qeq}, where it was shown to be consistent with the CFT results.


\subsection{Three-point correlation functions}\label{Ch04-section2.2}

Now let $\phi$, $\psi$, $\pi$ be primary fields with scale dimensions $\D_{1}$, $\D_{2}$ and $\D_{3}$ respectively. The three-point function may be constructed using the general ansatz
\begin{align}
	\langle \phi_{\cA_{1}}(x_{1}) \, \psi_{\cA_{2}}(x_{2}) \, \pi_{\cA_{3}}(x_{3}) \rangle = \frac{ \cI^{(1)}{}_{\cA_{1}}{}^{\bar{\cA}_{1}}(x_{13}) \,  \cI^{(2)}{}_{\cA_{2}}{}^{\bar{\cA}_{2}}(x_{23}) }{(x_{13}^{2})^{\D_{1}} (x_{23}^{2})^{\D_{2}}}
	\; \cH_{\bar{\cA}_{1} \bar{\cA}_{2} \cA_{3}}(X_{12}) \, , \label{Ch04-Three-point function - general ansatz}
\end{align} 
where the tensor $\cH_{\bar{\cA}_{1} \bar{\cA}_{2} \cA_{3}}$ encodes all information about the correlation function, and is constrained by the conformal symmetry as follows:
\begin{enumerate}
	\item[\textbf{(i)}] Under scale transformations of Minkowski space $x^{m} \mapsto x'^{m} = \l^{-2} x^{m}$, the three-point building blocks transform as $X^{m} \mapsto X'^{m} = \l^{2} X^{m}$. As a consequence, the correlation function transforms as 
	\begin{equation}
		\langle \phi_{\cA_{1}}(x_{1}') \, \psi_{\cA_{2}}(x_{2}') \, \pi_{\cA_{3}}(x_{3}') \rangle = (\l^{2})^{\D_{1} + \D_{2} + \D_{3}} \langle \phi_{\cA_{1}}(x_{1}) \, \psi_{\cA_{2}}(x_{2}) \,  \pi_{\cA_{3}}(x_{3}) \rangle \, ,
	\end{equation}
	which implies that $\cH$ obeys the scaling property
	\begin{equation}
		\cH_{\bar{\cA}_{1} \bar{\cA}_{2} \cA_{3}}(\l^{2} X) = (\l^{2})^{\D_{3} - \D_{2} - \D_{1}} \, \cH_{\bar{\cA}_{1} \bar{\cA}_{2} \cA_{3}}(X) \, , \hspace{5mm} \forall \l \in \mathbb{R} \, \backslash \, \{ 0 \} \, .
	\end{equation}
	This guarantees that the correlation function transforms correctly under scale transformations.
	
	\item[\textbf{(ii)}] If any of the fields $\phi$, $\psi$, $\pi$ obey differential equations, such as conservation equations, then the tensor $\cH$ is also constrained by differential equations. Such constraints may be derived with the aid of identities \eqref{Ch04-Three-point building blocks - differential identities 2}, \eqref{Ch04-Three-point building blocks - differential identities 3}.
	
	\item[\textbf{(iii)}] If any (or all) of the operators $\phi$, $\psi$, $\pi$ coincide, the correlation function possesses symmetries under permutations of spacetime points, e.g.
	\begin{equation}
		\langle \phi_{\cA_{1}}(x_{1}) \, \phi_{\cA'_{1}}(x_{2}) \, \pi_{\cA_{3}}(x_{3}) \rangle = (-1)^{\e(\phi)} \langle \phi_{\cA'_{1}}(x_{2}) \, \phi_{\cA_{1}}(x_{1}) \, \pi_{\cA_{3}}(x_{3}) \rangle \, ,
	\end{equation}
	where $\e(\phi)$ is the Grassmann parity of $\phi$. As a consequence, the tensor $\cH$ obeys constraints which will be referred to as ``point-switch symmetries". A similar relation may also be derived for two fields which are related by complex conjugation.

\end{enumerate}

The constraints above fix the functional form of $\cH$ (and therefore the correlation function) up to finitely many independent parameters. Hence, using the general formula \eqref{Ch04-H ansatz}, the problem of computing three-point correlation functions is reduced to deriving the general structure of the tensor $\cH$ subject to the above constraints.	

\subsubsection{Comments on differential constraints}\label{Ch04-subsubsection2.2.1}

For three-point functions of conserved currents, we must impose conservation on all three space-time points. For $x_{1}$ and $x_{2}$, this process is simple due to the identities \eqref{Ch04-Three-point building blocks - differential identities 2}, \eqref{Ch04-Three-point building blocks - differential identities 3}, and the resulting conservation equations become equivalent to simple differential constraints on $\cH$. However, conservation on $x_{3}$ is more challenging due to a lack of useful identities analogous to \eqref{Ch04-Three-point building blocks - differential identities 2}, \eqref{Ch04-Three-point building blocks - differential identities 3} for $x_{3}$. To correctly impose conservation on $x_{3}$, consider the correlation function $\langle \phi_{\cA_{1}}(x_{1}) \, \psi_{\cA_{2}}(x_{2}) \, \pi_{\cA_{3}}(x_{3}) \rangle$, with the ansatz
\begin{equation} \label{Ch04-H ansatz}
	\langle \phi_{\cA_{1}}(x_{1}) \, \psi_{\cA_{2}}(x_{2}) \, \pi_{\cA_{3}}(x_{3}) \rangle = \frac{ \cI^{(1)}{}_{\cA_{1}}{}^{\bar{\cA}_{1}}(x_{13}) \,  \cI^{(2)}{}_{\cA_{2}}{}^{\bar{\cA}_{2}}(x_{23}) }{(x_{13}^{2})^{\D_{1}} (x_{23}^{2})^{\D_{2}}}
	\; \cH_{\bar{\cA}_{1} \bar{\cA}_{2} \cA_{3}}(X_{12}) \, . 
\end{equation} 
We now reformulate the ansatz with $\pi$ at the front
\begin{equation} \label{Ch04-Htilde ansatz}
	\langle \pi_{\cA_{3}}(x_{3}) \, \psi_{\cA_{2}}(x_{2}) \, \phi_{\cA_{1}}(x_{1}) \rangle = \frac{ \cI^{(3)}{}_{\cA_{3}}{}^{\bar{\cA}_{3}}(x_{31}) \,  \cI^{(2)}{}_{\cA_{2}}{}^{\bar{\cA}_{2}}(x_{21}) }{(x_{31}^{2})^{\D_{3}} (x_{21}^{2})^{\D_{2}}}
	\; \tilde{\cH}_{\cA_{1} \bar{\cA}_{2} \bar{\cA}_{3} }(X_{23}) \, . 
\end{equation} 
These two correlators are the same up to an overall sign due to Grassmann parity. Equating the two ansatz above yields the following relation:
\begin{align} \label{Ch04-Htilde and H relation}
	\begin{split}
		\tilde{\cH}_{\cA_{1} \bar{\cA}_{2} \bar{\cA}_{3} }(X_{23}) &= (x_{13}^{2})^{\D_{3} - \D_{1}} \bigg(\frac{x_{21}^{2}}{x_{23}^{2}} \bigg)^{\hspace{-1mm} \D_{2}} \, \cI^{(1)}{}_{\cA_{1}}{}^{\bar{\cA}_{1}}(x_{13}) \, \bar{\cI}^{(2)}{}_{\bar{\cA}_{2}}{}^{\cA'_{2}}(x_{12}) \,  \cI^{(2)}{}_{\cA'_{2}}{}^{\bar{\cA}'_{2}}(x_{23}) \\[-2mm]
		& \hspace{50mm} \times \bar{\cI}^{(3)}{}_{\bar{\cA}_{3}}{}^{\cA_{3}}(x_{13}) \, \cH_{\bar{\cA}_{1} \bar{\cA}'_{2} \cA_{3}}(X_{12}) \, . 
	\end{split}
\end{align}
Now suppose $\cH(X)$ (with indices suppressed) is composed of finitely many linearly independent tensor structures, $P_{i}(X)$, i.e $\cH(X) = \sum_{i} a_{i} P_{i}(X)$ where $a_{i}$ are constant complex parameters. We define $\bar{\cH}(X) = \sum_{i} \bar{a}_{i} \bar{P}_{i}(X)$, the conjugate of $\cH$, and also $\cH^{c}(X) = \sum_{i} a_{i} \bar{P}_{i}(X)$, which we denote as the complement of $\cH$. As a consequence of \eqref{Ch04-Inversion tensor identities - spinor case}, the following relation holds:
\begin{align} \label{Ch04-Hc and H relation}
	\begin{split}
		\cH^{c}_{\cA_{1} \cA_{2} \bar{\cA}_{3}}(X^{I}_{23}) &= (x_{13}^{2} X_{32}^{2})^{\D_{3} - \D_{2} - \D_{1}} \cI^{(1)}{}_{\cA_{1}}{}^{\bar{\cA}_{1}}(x_{13}) \, \cI^{(2)}{}_{\cA_{2}}{}^{\bar{\cA}_{2}}(x_{13}) \\
		& \hspace{45mm} \times \bar{\cI}^{(3)}{}_{\bar{\cA}_{3}}{}^{\cA_{3}}(x_{13}) \, \cH_{\bar{\cA}_{1} \bar{\cA}_{2} \cA_{3}}(X_{12}) \, .
	\end{split}
\end{align}
After inverting this equation and substituting it into \eqref{Ch04-Htilde and H relation}, we apply \eqref{Ch04-Inversion tensor identities - spinor case} to obtain an equation relating $\cH^{c}$ and $\tilde{\cH}$
\begin{equation} \label{Ch04-Htilde and Hc relation}
	\tilde{\cH}_{\cA_{1} \bar{\cA}_{2} \bar{\cA}_{3} }(X) = (X^{2})^{\D_{1} - \D_{3}} \, \bar{\cI}^{(2)}{}_{\bar{\cA}_{2}}{}^{\cA_{2}}(X) \, \cH^{c}_{\cA_{1} \cA_{2} \bar{\cA}_{3}}(X^{I}) \, . 
\end{equation}
Conservation on $x_{3}$ may now be imposed by using \eqref{Ch04-Three-point building blocks - differential identities 2}, with $ x_{1} \leftrightarrow x_{3}$. In principle, this procedure can be carried out for any configuration of the fields.

If we now consider the correlation function of three conserved primaries $J^{}_{\a(i_{1}) \ad(j_{1})}$, $J'_{\b(i_{2}) \bd(j_{2})}$, $J''_{\g(i_{3}) \gd(j_{3})}$, where $s_{1} = \tfrac{1}{2}(i_{1} + j_{1})$, $s_{2} = \tfrac{1}{2}( i_{2} + j_{2} )$, $s_{3} = \tfrac{1}{2}( i_{3} + j_{3} )$, then the general ansatz is
\begin{align} \label{Ch04-Conserved correlator ansatz}
	\langle \, J^{}_{\a(i_{1}) \ad(j_{1})}(x_{1}) \, J'_{\b(i_{2}) \bd(j_{2})}(x_{2}) \, J''_{\g(i_{3}) \gd(j_{3})}(x_{3}) \rangle &= \nonumber \\
	& \hspace{-55mm} \frac{1}{ (x_{13}^{2})^{\D_{1}} (x_{23}^{2})^{\D_{2}} } \, \cI_{\a(i_{1})}{}^{\ad'(i_{1})}(x_{13}) \, \bar{\cI}_{\ad(j_{1})}{}^{\a'(j_{1})}(x_{13}) \,  \cI_{\b(i_{2})}{}^{\bd'(i_{2})}(x_{23}) \, \bar{\cI}_{\bd(j_{2})}{}^{\b'(j_{2})}(x_{23}) \nonumber\\
	& \hspace{-25mm} \times \cH_{\a'(j_{1}) \ad'(i_{1}) \b'(j_{2}) \bd'(i_{2}) \g(i_{3}) \gd(j_{3})}(X_{12}) \, ,
\end{align} 
where $\D_{i} = s_{i} + 2$. The constraints on $\cH$ are then as follows:
\begin{enumerate}
	\item[\textbf{(i)}] {\bf Homogeneity:} \\
	Recall that $\cH$ is a homogeneous tensor field satisfying
	\begin{equation}
		\cH_{\a(j_{1}) \ad(i_{1}) \b(j_{2}) \bd(i_{2}) \g(i_{3}) \gd(j_{3})}(\l^{2} X) = (\l^{2})^{\D_{3} - \D_{2} - \D_{1}} \, \cH_{\a(j_{1}) \ad(i_{1}) \b(j_{2}) \bd(i_{2}) \g(i_{3}) \gd(j_{3})}(X) \, .
	\end{equation}
	It is often convenient to introduce $\hat{\cH}_{\a(j_{1}) \ad(i_{1}) \b(j_{2}) \bd(i_{2}) \g(i_{3}) \gd(j_{3})}(X)$, such that
	\begin{align}
		\cH_{\a(j_{1}) \ad(i_{1}) \b(j_{2}) \bd(i_{2}) \g(i_{3}) \gd(j_{3})}(X) &= X^{\D_{3} - \D_{2}- \D_{1}} \hat{\cH}_{\a(j_{1}) \ad(i_{1}) \b(j_{2}) \bd(i_{2}) \g(i_{3}) \gd(j_{3})}(X) \, ,
	\end{align}
	where $\hat{\cH}_{\a(j_{1}) \ad(i_{1}) \b(j_{2}) \bd(i_{2}) \g(i_{3}) \gd(j_{3})}(X)$ is homogeneous degree 0 in $X$, i.e.
	\begin{align}
		\hat{\cH}_{\a(j_{1}) \ad(i_{1}) \b(j_{2}) \bd(i_{2}) \g(i_{3}) \gd(j_{3})}( \l^{2} X) &= \hat{\cH}_{\a(j_{1}) \ad(i_{1}) \b(j_{2}) \bd(i_{2}) \g(i_{3}) \gd(j_{3})}(X) \, .
	\end{align}
	
	\item[\textbf{(ii)}] {\bf Differential constraints:} \\
	After application of the identities \eqref{Ch04-Three-point building blocks - differential identities 2}, \eqref{Ch04-Three-point building blocks - differential identities 3} we obtain the following constraints:
	\begin{subequations}
		\begin{align}
			\text{Conservation at $x_{1}$:} && \pa_{X}^{\a \ad} \cH_{\a \a(j_{1} -1) \ad \ad(i_{1} -1) \b(j_{2}) \bd(i_{2}) \g(i_{3}) \gd(j_{3})}(X) &= 0 \, , \\
			\text{Conservation at $x_{2}$:} && \pa_{X}^{\b \bd} \cH_{\a(j_{1}) \ad(i_{1}) \b \b(j_{2}-1) \bd \bd(i_{2}-1) \g(i_{3}) \gd(j_{3})}(X) &= 0 \, , \\
			\text{Conservation at $x_{3}$:} && \pa_{X}^{\g \gd} \tilde{\cH}_{\a(i_{1}) \ad(j_{1}) \b(j_{2}) \bd(i_{2}) \g \g(j_{3}-1) \gd \gd(i_{3}-1)}(X) &= 0 \, ,
		\end{align}
	\end{subequations}
	where
	\begin{align}
		\begin{split}
			\tilde{\cH}_{\a(i_{1}) \ad(j_{1}) \b(j_{2}) \bd(i_{2}) \g(j_{3}) \gd(i_{3})}(X) &= (X^{2})^{\D_{1} - \D_{3}} \, \cI_{\b(j_{2})}{}^{\bd'(j_{2})}(X) \, \bar{\cI}_{\bd(i_{2})}{}^{\b'(i_{2})}(X) \\
			& \hspace{15mm}\times \cH^{c}_{\a(i_{1}) \ad(j_{1}) \b'(i_{2}) \bd'(j_{2}) \g(j_{3}) \gd(i_{3})}(X^{I}) \, . 
		\end{split}
	\end{align}

	\item[\textbf{(iii)}] {\bf Point-switch symmetries:} \\
	If the fields $J$ and $J'$ coincide, then we obtain the following point-switch identity
	\begin{equation}
		\cH_{\a(i_{1}) \ad(j_{1}) \b(i_{1}) \bd(j_{1}) \g(i_{3}) \gd(j_{3})}(X) = (-1)^{\e(J)} \cH_{\a(i_{1}) \ad(j_{1}) \b(i_{1}) \bd(j_{1}) \g(i_{3}) \gd(j_{3})}(-X) \, ,
	\end{equation}
	where $\e(J)$ is the Grassmann parity of $J$. Likewise, if the fields $J$ and $J''$ coincide, then we obtain the constraint
	\begin{equation}
		\tilde{\cH}_{\a(i_{1}) \ad(j_{1}) \b(j_{2}) \bd(i_{2}) \g(j_{1}) \gd(i_{1})}(X) = (-1)^{\e(J)} \cH_{\g(i_{1}) \gd(j_{1}) \b(j_{2}) \bd(i_{2}) \a(j_{1}) \ad(i_{1}) }(-X)  \, .
	\end{equation}

	\item[\textbf{(iv)}] {\bf Reality condition:} \\
	If the fields in the correlation function belong to the $(s,s)$ representation, then the three-point function must satisfy the reality condition
	\begin{equation}
		\cH_{\a(i_{1}) \ad(i_{1}) \b(i_{2}) \bd(i_{2}) \g(i_{3}) \gd(i_{3})}(X) = \bar{\cH}_{\a(i_{1}) \ad(i_{1}) \b(i_{2}) \bd(i_{2}) \g(i_{3}) \gd(i_{3})}(X)  \, .
	\end{equation}
	Similarly, if the fields at $J$, $J'$ at $x_{1}$ and $x_{2}$ respectively possess the same spin and are conjugate to each other, i.e. $J' = \bar{J}$, we must impose a combined reality/point-switch condition using the following constraint
	\begin{equation}
		\cH_{\a(i_{1}) \ad(j_{1}) \b(j_{1}) \bd(i_{1}) \g(i_{3}) \gd(j_{3})}(X) = \bar{\cH}_{\b(i_{1}) \bd(j_{1}) \a(j_{1}) \ad(i_{1})  \g(i_{3}) \gd(j_{3})}(-X) \, ,
	\end{equation}

\end{enumerate}
Working with the tensor formalism is quite messy and complicated in general, hence, to simplify the analysis we will utilise auxiliary spinors to carry out the computations.

\subsubsection{Generating function formalism}\label{Ch04-subsubsection2.2.2}

Analogous to the approach of \cite{Buchbinder:2022mys} we utilise auxiliary spinors to streamline the calculations. Consider a general spin-tensor $\cH_{\cA_{1} \cA_{2} \cA_{3}}(X)$, where $\cA_{1} = \{ \a(i_{1}), \ad(j_{1}) \}, \cA_{2} = \{ \b(i_{2}), \bd(j_{2}) \}, \cA_{3} = \{ \g(i_{3}), \gd(j_{3}) \}$ represent sets of totally symmetric spinor indices associated with the fields at points $x_{1}$, $x_{2}$ and $x_{3}$ respectively. We introduce sets of commuting auxiliary spinors for each point; $ U = \{ u, \bar{u} \}$ at $x_{1}$, $ V = \{ v, \bar{v} \}$ at $x_{2}$, and $W = \{ w, \bar{w} \}$ at $x_{3}$, where the spinors satisfy 
\begin{align}
	u^2 &= \varepsilon_{\a \b} \, u^{\a} u^{\b}=0\,, & \bar{u}^2& = \varepsilon_{\ad \bd} \, \bar{u}^{\ad} \bar{u}^{\bd}=0\,,  &
	v^{2} &= \bar{v}^{2} = 0\,, & w^{2} &= \bar{w}^{2} = 0\,. 
	\label{Ch04-extra1}
\end{align}
Now if we define the objects
\begin{subequations}
	\begin{align}
		\boldsymbol{U}^{\cA_{1}} &\equiv \boldsymbol{U}^{\a(i_{1}) \ad(j_{1})} = u^{\a_{1}} \dots u^{\a_{i_{1}}} \bar{u}^{\ad_{1}} \dots \bar{u}^{\ad_{j_{1}}} \, , \\
		\boldsymbol{V}^{\cA_{2}} &\equiv \boldsymbol{V}^{\b(i_{2}) \bd(j_{2})} = v^{\b_{1}} \dots v^{\b_{i_{2}}} \bar{v}^{\bd_{1}} \dots \bar{v}^{\bd_{j_{2}}} \, , \\
		\boldsymbol{W}^{\cA_{3}} &\equiv \boldsymbol{W}^{\g(i_{3}) \gd(j_{3})} = w^{\g_{1}} \dots w^{\g_{i_{3}}} \bar{w}^{\gd_{1}} \dots \bar{w}^{\gd_{j_{3}}} \, ,
	\end{align}
\end{subequations}
then the generating polynomial for $\cH$ is constructed as follows:
\begin{equation} \label{Ch04-H - generating polynomial}
	\cH(X; U, V, W) = \cH_{ \cA_{1} \cA_{2} \cA_{3} }(X) \, \boldsymbol{U}^{\cA_{1}} \boldsymbol{V}^{\cA_{2}} \boldsymbol{W}^{\cA_{3}} \, . \\
\end{equation}
The tensor $\cH$ can then be extracted from the polynomial by acting on it with the following partial derivative operators:
\begin{subequations}
	\begin{align}
		\frac{\pa}{\pa \boldsymbol{U}^{\cA_{1}} } &\equiv \frac{\pa}{\pa \boldsymbol{U}^{\a(i_{1}) \ad(j_{1})} } = \frac{1}{i_{1}!j_{1}!} \frac{\pa}{\pa u^{\a_{1}} } \dots \frac{\pa}{\pa u^{\a_{i_{1}}}} \frac{\pa}{\pa \bar{u}^{\ad_{1}}} \dots \frac{\pa }{\pa \bar{u}^{\ad_{j_{1}}}} \, , \\
		\frac{\pa}{\pa \boldsymbol{V}^{\cA_{2}} } &\equiv \frac{\pa}{\pa \boldsymbol{V}^{\b(i_{2}) \bd(j_{2})} } = \frac{1}{i_{2}!j_{2}!} \frac{\pa}{\pa v^{\b_{1}} } \dots \frac{\pa}{\pa v^{\b_{i_{2}}}} \frac{\pa}{\pa \bar{v}^{\bd_{1}}} \dots \frac{\pa }{\pa \bar{v}^{\bd_{j_{2}}}} \, , \\
		\frac{\pa}{\pa \boldsymbol{W}^{\cA_{3}} } &\equiv \frac{\pa}{\pa \boldsymbol{W}^{\g(i_{3}) \gd(j_{3})} } = \frac{1}{i_{3}!j_{3}!} \frac{\pa}{\pa w^{\g_{1}} } \dots \frac{\pa}{\pa w^{\g_{i_{3}}}} \frac{\pa}{\pa \bar{w}^{\gd_{1}}} \dots \frac{\pa }{\pa \bar{w}^{\gd_{j_{3}}}} \, . 
	\end{align}
\end{subequations}
The tensor $\cH$ is then extracted from the polynomial as follows:
\begin{equation}
	\cH_{\cA_{1} \cA_{2} \cA_{3}}(X) = \frac{\pa}{ \pa \boldsymbol{U}^{\cA_{1}} } \frac{\pa}{ \pa \boldsymbol{V}^{\cA_{2}}} \frac{\pa}{ \pa \boldsymbol{W}^{\cA_{3}} } \, \cH(X; U, V, W) \, .
\end{equation}
The polynomial $\cH$, \eqref{Ch04-H - generating polynomial}, is now constructed out of scalar combinations of $X$, and the auxiliary spinors $U$, $V$ and $W$ with the appropriate homogeneity. Such a polynomial can be constructed out of the following monomials:
\begin{subequations} \label{Ch04-Basis scalar structures}
	\begin{align} 
		P_{1} &= \ve_{\a \b} v^{\a} w^{\b} \, , & P_{2} &= \ve_{\a \b} w^{\a} u^{\b} \, , & P_{3} &= \ve_{\a \b} u^{\a} v^{\b} \, , \\
		\bar{P}_{1} &= \ve_{\ad \bd} \bar{v}^{\ad} \bar{w}^{\bd} \, , & \bar{P}_{2} &= \ve_{\ad \bd} \bar{w}^{\ad} \bar{u}^{\bd} \, , & \bar{P}_{3} &= \ve_{\ad \bd} \bar{u}^{\ad} \bar{v}^{\bd} \, , \\[1mm]
		Q_{1} &= \hat{X}_{\a \ad} \, v^{\a} \bar{w}^{\ad} \, ,  &  Q_{2} &= \hat{X}_{\a \ad} \, w^{\a} \bar{u}^{\ad} \, ,  &  Q_{3} &= \hat{X}_{\a \ad} \, u^{\a} \bar{v}^{\ad} \, , \\
		\bar{Q}_{1} &= \hat{X}_{\a \ad} \, w^{\a} \bar{v}^{\ad} \, ,  &  \bar{Q}_{2} &= \hat{X}_{\a \ad} \, u^{\a} \bar{w}^{\ad} \, ,  &  \bar{Q}_{3} &= \hat{X}_{\a \ad} \, v^{\a} \bar{u}^{\ad} \, , \\[1mm]
		Z_{1} &= \hat{X}_{\a \ad} \, u^{\a} \bar{u}^{\ad}  \, , & Z_{2} &= \hat{X}_{\a \ad} \, v^{\a} \bar{v}^{\ad} \, , & Z_{3} &= \hat{X}_{\a \ad} \, w^{\a} \bar{w}^{\ad} \, .
	\end{align}
\end{subequations}
To construct linearly independent structures for a given three-point function, one must also take into account the following linear dependence relations between the monomials:
\begin{subequations} \label{Ch04-Linear dependence 1}
	\begin{align} 
		Z_{2} Z_{3} + P_{1} \bar{P}_{1} - Q_{1} \bar{Q}_{1} &= 0 \, , \\
		Z_{1} Z_{3} + P_{2} \bar{P}_{2} - Q_{2} \bar{Q}_{2} &= 0 \, , \\
		Z_{1} Z_{2} + P_{3} \bar{P}_{3} - Q_{3} \bar{Q}_{3} &= 0 \, ,
	\end{align}
\end{subequations}
\vspace{-10mm}
\begin{subequations} \label{Ch04-Linear dependence 2}
	\begin{align} 
		Z_{1} P_{1} + P_{2} \bar{Q}_{3} + P_{3} Q_{2} &= 0 \, , & Z_{1} \bar{P}_{1} + \bar{P}_{2} Q_{3} + \bar{P}_{3} \bar{Q}_{2} &= 0  \, , \\
		Z_{2} P_{2} + P_{3} \bar{Q}_{1} + P_{1} Q_{3} &= 0 \, , & Z_{2} \bar{P}_{2} + \bar{P}_{3} Q_{1} + \bar{P}_{1} \bar{Q}_{3} &= 0 \, , \\
		Z_{3} P_{3} + P_{1} \bar{Q}_{2} + P_{2} Q_{1} &= 0 \, , & Z_{3} \bar{P}_{3} + \bar{P}_{1} Q_{2} + \bar{P}_{2} \bar{Q}_{1} &= 0 \, .
	\end{align}
\end{subequations}
\vspace{-10mm}
\begin{subequations} \label{Ch04-Linear dependence 3}
	\begin{align} 
		Z_{1} Q_{1} + \bar{P}_{2} P_{3} - \bar{Q}_{2} \bar{Q}_{3} &= 0 \, , & Z_{1} \bar{Q}_{1} + P_{2} \bar{P}_{3} - Q_{2} Q_{3} &= 0  \, , \\
		Z_{2} Q_{2} + \bar{P}_{3} P_{1} - \bar{Q}_{3} \bar{Q}_{1} &= 0 \, , & Z_{2} \bar{Q}_{2} + P_{3} \bar{P}_{1} - Q_{3} Q_{1} &= 0 \, , \\
		Z_{3} Q_{3} + \bar{P}_{1} P_{2} - \bar{Q}_{1} \bar{Q}_{2} &= 0 \, , & Z_{3} \bar{Q}_{3} + P_{1} \bar{P}_{2} - Q_{1} Q_{2} &= 0 \, .
	\end{align}
\end{subequations}
These allow elimination of the combinations $Z_{i} Z_{j}$, $Z_{i} P_{i}$, $Z_{i} \bar{P}_{i}$, $Z_{i} Q_{i}$, $Z_{i} \bar{Q}_{i}$. There are also the following relations involving triple products:
\begin{subequations} \label{Ch04-Linear dependence 4}
	\begin{align} 
		P_{1} \bar{P}_{2} \bar{P}_{3} + \bar{P}_{1} Q_{2} \bar{Q}_{3} + \bar{P}_{2} \bar{Q}_{3} \bar{Q}_{1} + \bar{P}_{3} Q_{1} Q_{2} &= 0 \, , \\
		P_{2} \bar{P}_{3} \bar{P}_{1} + \bar{P}_{2} Q_{3} \bar{Q}_{1} + \bar{P}_{3} \bar{Q}_{1} \bar{Q}_{2} + \bar{P}_{1} Q_{2} Q_{3} &= 0 \, , \\
		P_{3} \bar{P}_{1} \bar{P}_{2} + \bar{P}_{3} Q_{1} \bar{Q}_{2} + \bar{P}_{1} \bar{Q}_{2} \bar{Q}_{3} + \bar{P}_{2} Q_{3} Q_{1} &= 0 \, ,
	\end{align}
\end{subequations}
\vspace{-10mm}
\begin{subequations} \label{Ch04-Linear dependence 5}
	\begin{align} 
		\bar{P}_{1} P_{2} P_{3} + P_{1} \bar{Q}_{2} Q_{3} + P_{2} Q_{3} Q_{1} + P_{3} \bar{Q}_{1} \bar{Q}_{2} &= 0 \, , \\
		\bar{P}_{2} P_{3} P_{1} + P_{2} \bar{Q}_{3} Q_{1} + P_{3} Q_{1} Q_{2} + P_{1} \bar{Q}_{2} \bar{Q}_{3} &= 0 \, , \\
		\bar{P}_{3} P_{1} P_{2} + P_{3} \bar{Q}_{1} Q_{2} + P_{1} Q_{2} Q_{3} + P_{2} \bar{Q}_{3} \bar{Q}_{1} &= 0 \, ,
	\end{align}
\end{subequations}
\vspace{-10mm}
\begin{subequations} \label{Ch04-Linear dependence 6}
	\begin{align} 
		\bar{P}_{1} P_{2} \bar{Q}_{3} - P_{1} \bar{P}_{2} Q_{3} + \bar{Q}_{1} \bar{Q}_{2} \bar{Q}_{3} - Q_{1} Q_{2} Q_{3} &= 0 \, , \\
		\bar{P}_{2} P_{3} \bar{Q}_{1} - P_{2} \bar{P}_{3} Q_{1} + \bar{Q}_{1} \bar{Q}_{2} \bar{Q}_{3} - Q_{1} Q_{2} Q_{3} &= 0 \, , \\
		\bar{P}_{3} P_{1} \bar{Q}_{2} - P_{3} \bar{P}_{1} Q_{2} + \bar{Q}_{1} \bar{Q}_{2} \bar{Q}_{3} - Q_{1} Q_{2} Q_{3} &= 0 \, ,
	\end{align}
\end{subequations}
which allow for elimination of the products $P_{i} \bar{P}_{j} \bar{P}_{k}$, $\bar{P}_{i} P_{j} P_{k}$, $\bar{P}_{i} P_{j} \bar{Q}_{k}$. These relations (which appear to be exhaustive) are sufficient to reduce any set of structures in a given three-point function to a linearly independent set.

The task now is to construct a complete list of possible (linearly independent) solutions for the polynomial $\cH$ for a given set of spins. This process is simplified by introducing a generating function, $\cF(X; U, V, W \, | \, \G)$, defined as follows:
\begin{align} \label{Ch04-Generating function}
	\cF(X; U,V,W \, | \, \G) &= P_{1}^{k_{1}} P_{2}^{k_{2}} P_{3}^{k_{3}} \, 	\bar{P}_{1}^{\bar{k}_{1}} \bar{P}_{2}^{\bar{k}_{2}} \bar{P}_{3}^{\bar{k}_{3}} Q_{1}^{l_{1}} Q_{2}^{l_{2}} Q_{3}^{l_{3}} \, \bar{Q}_{1}^{\bar{l}_{1}} \bar{Q}_{2}^{\bar{l}_{2}} \bar{Q}_{3}^{\bar{l}_{3}} Z_{1}^{r_{1}} Z_{2}^{r_{2}} Z_{3}^{r_{3}} \, ,
\end{align}
where the non-negative integers, $ \G =  \bigcup_{i \in \{1,2,3\} }  \{ k_{i}, \bar{k}_{i}, l_{i}, \bar{l}_{i}, r_{i}\}$, are solutions to the following linear system:
\begin{subequations} \label{Ch04-Diophantine equations}
	\begin{align}
		k_{2} + k_{3} + r_{1} + l_{3} + \bar{l}_{2} &= i_{1} \, , &  \bar{k}_{2} + \bar{k}_{3} + r_{1} + \bar{l}_{3} + l_{2} &= j_{1} \, , \\
		k_{1} + k_{3} + r_{2} + l_{1} + \bar{l}_{3} &= i_{2} \, , &  \bar{k}_{1} + \bar{k}_{3} + r_{2} + \bar{l}_{1} + l_{3} &= j_{2} \, , \\
		k_{1} + k_{2} + r_{3} + l_{2} + \bar{l}_{1} &= i_{3} \, , &  \bar{k}_{1} + \bar{k}_{2} + r_{3} + \bar{l}_{2} + l_{1} &= j_{3} \, .
	\end{align}
\end{subequations}
Here $i_{1}, i_{2}, i_{3}$, $j_{1}, j_{2}, j_{3}$ are fixed integers corresponding to the spin representations of the fields in the three-point function. From here it is convenient to define
\begin{equation}
	\D s = \frac{1}{2}( i_{1} + i_{2} + i_{3} - j_{1} - j_{2} - j_{3} ) \, .
\end{equation}
Using the system of equations \eqref{Ch04-Diophantine equations}, we obtain
\begin{align}
	\D s =  k_{1} + k_{2} + k_{3} - \bar{k}_{1} - \bar{k}_{2} - \bar{k}_{3}  \, ,
\end{align}
in addition to
\begin{subequations}
	\begin{align}
		&k_{1} + k_{2} + k_{3} \leq \min( i_{1} + i_{2}, i_{1} + i_{3}, i_{2} + i_{3} ) \, , \\
		&\bar{k}_{1} + \bar{k}_{2} + \bar{k}_{3} \leq \min( j_{1} + j_{2}, j_{1} + j_{3}, j_{2} + j_{3} ) \, .
	\end{align}
\end{subequations}
Hence, the conditions for a given three-point function to be non-vanishing are
\begin{align} \label{Ch04-e1}
	- \min( j_{1} + j_{2}, j_{1} + j_{3}, j_{2} + j_{3} ) \leq \D s \leq \min( i_{1} + i_{2}, i_{1} + i_{3}, i_{2} + i_{3} ) \, .
\end{align}
Indeed, this is the same condition found in \cite{Elkhidir:2014woa}. Now given a finite number of solutions $\G_{I}$, $I = 1, ..., N$ to \eqref{Ch04-Diophantine equations} for a particular choice of $i_{1}, i_{2}, i_{3}, j_{1}, j_{2}, j_{3}$, the most general ansatz for the polynomial $\cH$ in \eqref{Ch04-H - generating polynomial} is as follows:
\begin{equation}
	\cH(X; U, V, W) = X^{\D_{3} - \D_{2} - \D_{1}} \sum_{I=1}^{N} a_{I} \cF(X; U, V, W \, | \, \G_{I}) \, ,
\end{equation}
where $a_{I}$ are a set of complex constants. Hence, constructing the most general ansatz for the generating polynomial $\cH$ is now equivalent to finding all non-negative integer solutions $\G_{I}$ of \eqref{Ch04-Diophantine equations}. Once this ansatz has been obtained, the linearly independent structures can be found by systematically applying the linear dependence relations \eqref{Ch04-Linear dependence 1}--\eqref{Ch04-Linear dependence 6}.

Let us now recast the constraints on the three-point function into the auxiliary spinor formalism. Recalling that $s_{1} = \tfrac{1}{2}(i_{1} + j_{1})$, $s_{2} = \tfrac{1}{2}(i_{2} + j_{2})$, $s_{3} = \tfrac{1}{2}( i_{3} + j_{3})$, first we define:
\begin{subequations}
	\begin{align}
		J^{}_{s_{1}}(x_{1}; U) & = J^{}_{\a(i_{1}) \ad(j_{1})}(x_{1}) \, \boldsymbol{U}^{\a(i_{1}) \ad(j_{1})} \, , \\
		J'_{s_{2}}(x_{2}; V) &= J'_{\a(i_{2}) \ad(j_{2})}(x_{2}) \, \boldsymbol{V}^{\a(i_{2}) \ad(j_{2})} \, , \\
		J''_{s_{3}}(x_{3}; W) &= J''_{\g(i_{3}) \gd(j_{3})}(x_{3}) \, \boldsymbol{W}^{\g(i_{3}) \gd(j_{3})} \, ,
	\end{align}
\end{subequations}
where, to simplify notation, we denote $J^{}_{(s, q)} \equiv J^{}_{s}$. The general ansatz can be converted easily into the auxiliary spinor formalism, and is of the form
\begin{align}
	\begin{split}
		\langle J^{}_{s_{1}}(x_{1}; U) \, J'_{s_{2}}(x_{2}; V) \, J''_{s_{3}}(x_{3}; W) \rangle &= \frac{ \cI^{(i_{1}, j_{1})}(x_{13}; U, \tilde{U}) \,  \cI^{(i_{2}, j_{2})}(x_{23}; V, \tilde{V}) }{(x_{13}^{2})^{\D_{1}} (x_{23}^{2})^{\D_{2}}} \\
		& \hspace{5mm} \times \cH(X_{12}; \tilde{U},\tilde{V}, W) \, ,
	\end{split}
\end{align} 
where $\D_{i} = s_{i} + 2$. The generating polynomial, $\cH(X; U, V, W)$, is defined as
\begin{align}
	\cH(X; U,V,W) = \cH_{\a(i_{1}) \ad(j_{1}) \b(i_{2}) \bd(j_{2}) \g(i_{3}) \gd(j_{3})}(X) \, \boldsymbol{U}^{\a(i_{1}) \ad(j_{1})} \boldsymbol{V}^{\b(i_{2}) \bd(j_{2})} \boldsymbol{W}^{\g(i_{3}) \gd(j_{3})} \, ,
\end{align}
where
\begin{align}
	\cI^{(i,j)}(x; U, \tilde{U}) &\equiv \cI^{(i,j)}_{x}(U,\tilde{U}) = \boldsymbol{U}^{\a(i) \ad(j)} \cI_{\a(i)}{}^{\ad'(i)}(x) \, \bar{\cI}_{\ad(j)}{}^{\a'(j)}(x) \, \frac{\pa}{\pa \tilde{\boldsymbol{U}}^{\a'(j) \ad'(i)}} \, ,
\end{align}
is the inversion operator acting on polynomials degree $(i,j)$ in $(\tilde{\bar{u}}, \tilde{u})$. It should also be noted that $\tilde{\boldsymbol{U}}$ has index structure conjugate to $\boldsymbol{U}$. Sometimes we will omit the indices $(i,j)$ to streamline the notation. After converting the constraints summarised in the previous subsection into the auxiliary spinor formalism, we obtain:
\begin{enumerate}
	\item[\textbf{(i)}] {\bf Homogeneity:} \\
	Recall that $\cH$ is a homogeneous polynomial satisfying the following scaling property:
	\begin{align}
		\begin{split}
			\cH(\l^{2} X; U(i_{1},j_{1}), V(i_{2},j_{2}), W(i_{3},j_{3})) &= \\ 
			& \hspace{-35mm} (\l^{2})^{\D_{3} - \D_{2} - \D_{1}} \, \cH(X; U(i_{1},j_{1}), V(i_{2},j_{2}), W(i_{3},j_{3})) \, ,
		\end{split}
	\end{align}
	where we have used the notation $U(i_{1},j_{1})$, $V(i_{2},j_{2})$, $W(i_{3},j_{3})$ to keep track of homogeneity in the auxiliary spinors $(u, \bar{u})$, $(v,\bar{v})$ and $(w,\bar{w})$. For compactness we will suppress the homogeneities of the auxiliary spinors in the results.

	\item[\textbf{(ii)}] {\bf Differential constraints:} \\
	First, define the following three differential operators:
	\begin{align} \label{Ch04-Derivative operators}
		D_{1} = \pa_{X}^{\a \ad} \frac{\pa}{\pa u^{\a}} \frac{\pa}{\pa \bar{u}^{\ad}} \, , && D_{2} =\pa_{X}^{\a \ad} \frac{\pa}{\pa v^{\a}} \frac{\pa}{\pa \bar{v}^{\ad}} \, , && D_{3} = \pa_{X}^{\a \ad} \frac{\pa}{\pa w^{\a}} \frac{\pa}{\pa \bar{w}^{\ad}} \, .
	\end{align}
	Conservation on all three points may be imposed using the following constraints:
	\begin{subequations} \label{Ch04-Ch04.1-Conservation equations}
		\begin{align}
			\text{Conservation at $x_{1}$:} && D_{1} \, \cH(X; U,V,W) &= 0 \, , \\[1mm]
			\text{Conservation at $x_{2}$:} && D_{2} \, \cH(X; U,V,W) &= 0 \, , \\[1mm]
			\text{Conservation at $x_{3}$:} && D_{3} \, \tilde{\cH}(X; U,V,W) &= 0 \, ,
		\end{align}
	\end{subequations}
	where, in the auxiliary spinor formalism, $\tilde{\cH}$ is computed as follows:
	\begin{equation}
		\tilde{\cH}(X; U,V,W) = (X^{2})^{\D_{1} - \D_{3}} \cI_{X}(V,\tilde{V}) \, \cH^{c}(-X; U, \tilde{V}, W) \, .
	\end{equation}
	Using the properties of the inversion tensor, it can be shown, using an identical approach to 3D CFT case, that this transformation is equivalent to the following replacement rules for the building blocks:
	\begin{subequations}
		\begin{align} 
			P_{1} &\rightarrow Q_{1} \, , & P_{2} &\rightarrow - \bar{P}_{2} \, , & P_{3} &\rightarrow - \bar{Q}_{3} \, \\
			\bar{P}_{1} &\rightarrow \bar{Q}_{1} \, , & \bar{P}_{2} &\rightarrow - P_{2} \, , & \bar{P}_{3} &\rightarrow - Q_{3} \, \\
			Q_{1} &\rightarrow - P_{1} \, , & Q_{2} &\rightarrow \bar{Q}_{2} \, , & Q_{3} &\rightarrow \bar{P}_{3} \, \\
			\bar{Q}_{1} &\rightarrow - \bar{P}_{1} \, , & \bar{Q}_{2} &\rightarrow Q_{2} \, , & \bar{Q}_{3} &\rightarrow P_{3} \, \\
			Z_{1} &\rightarrow Z_{1} \, , & Z_{2} &\rightarrow - Z_{2} \, , & Z_{3} &\rightarrow Z_{3} \, .
		\end{align}
	\end{subequations}
	For imposing conservation equations on three-point functions, one must act on the generating function \eqref{Ch04-Generating function} with the operators \eqref{Ch04-Derivative operators}. For this, the following identities for the derivatives of the monomials $Q_{i}, Z_{i}$ are useful:
	\begin{subequations}
		\begin{align}
			\pa_{X}^{\a \ad} Q_{1} &= - \frac{1}{X} \big( 2 v^{\a} \bar{w}^{\ad} + \hat{X}^{\ad \a} Q_{1} \big) \, ,\\
			\pa_{X}^{\a \ad} Q_{2} &= - \frac{1}{X} \big( 2 w^{\a} \bar{u}^{\ad} + \hat{X}^{\ad \a} Q_{2} \big) \, ,\\
			\pa_{X}^{\a \ad} Q_{3} &= - \frac{1}{X} \big( 2 u^{\a} \bar{v}^{\ad} + \hat{X}^{\ad \a} Q_{3} \big) \, ,
		\end{align}
	\end{subequations}
	\vspace{-5mm}
	\begin{subequations}
		\begin{align}
			\pa_{X}^{\a \ad} Z_{1} &= - \frac{1}{X} \big( 2 u^{\a} \bar{u}^{\ad} + \hat{X}^{\ad \a} Z_{1} \big) \, ,\\
			\pa_{X}^{\a \ad} Z_{2} &= - \frac{1}{X} \big( 2 v^{\a} \bar{v}^{\ad} + \hat{X}^{\ad \a} Z_{2} \big) \, ,\\
			\pa_{X}^{\a \ad} Z_{3} &= - \frac{1}{X} \big( 2 w^{\a} \bar{w}^{\ad} + \hat{X}^{\ad \a} Z_{3} \big) \, .
		\end{align}
	\end{subequations}
	Analogous identities for derivatives of $\bar{Q}_{i}$ may be obtained by complex conjugation.
	
	
	
	\item[\textbf{(iii)}] {\bf Point switch symmetries:} \\
	If the fields $J$ and $J'$ coincide (hence $i_{1} = i_{2}$, $j_{1} = j_{2}$), then we obtain the following point-switch constraint
	\begin{equation} \label{Ch04-Point switch A}
		\cH(X; U,V,W) = (-1)^{\e(J)} \cH(-X; V,U,W) \, ,
	\end{equation}
	where $\e(J)$ is the Grassmann parity of $J$. Similarly, if the fields $J$ and $J''$ coincide (hence $i_{1} = i_{3}$, $j_{1} = j_{3}$) then we obtain the constraint
	\begin{equation} \label{Ch04-Point switch B}
		\tilde{\cH}(X; U, V, W) = (-1)^{\e(J)} \cH(-X; W, V, U) \, .
	\end{equation}

	\item[\textbf{(iv)}] {\bf Reality condition:} \\
	If the fields in the correlation function belong to the $(s,s)$ representation, then the three-point function must satisfy the reality condition
	\begin{equation} \label{Ch04-Reality condition}
		\bar{\cH}(X; U, V, W) = \cH(X; U, V, W) \, .
	\end{equation}
	Similarly, if the fields at $J$, $J'$ at $x_{1}$ and $x_{2}$ respectively possess the same spin and are conjugate to each other, i.e. $J' = \bar{J}$, we must impose a combined reality/point-switch condition using the following constraint
	\begin{equation} \label{Ch04-Reality/switch condition}
		\cH(X; U, V, W) = \bar{\cH}(-X; V, U, W) \, ,
	\end{equation}
	%

\end{enumerate}
%


\subsubsection{Parity-even and parity-odd structures}\label{Ch04-subsubsection2.2.3}


Recall from Chapter \ref{Chapter2} that whenever parity is a symmetry of a given CFT, so too is invariance under inversions. 
An inversion transformation $\cI$ maps fields in the $(i, j)$ representation onto fields in the complex conjugate representation, $(j, i)$.\footnote{For a detailed discussion of parity transformations in 4D CFT, see \cite{Elkhidir:2014woa}.} In particular, if the fields in a given three-point function belong to the $(s, s)$ representation then it is possible to construct linear combinations of structures for the three-point function which are eigenfunctions of the inversion operator. We denote these as parity-even and parity-odd solutions respectively. Indeed, given a tensor $\cH_{\cA_{1} \cA_{2} \cA_{3}}(X) = X^{\D_{3} - \D_{2}- \D_{1}} \hat{\cH}_{\cA_{1} \cA_{2} \cA_{3}}(X)$, the following inversion formula holds:
\begin{align}
	\hat{\cH}^{c}_{\bar{\cA}_{1} \bar{\cA}_{2} \bar{\cA}_{3}}(X^{I}) = \bar{\cI}^{(1)}{}_{\bar{\cA}_{1}}{}^{\cA_{1}}(X) \, \bar{\cI}^{(2)}{}_{\bar{\cA}_{2}}{}^{\cA_{2}}(X) \, \bar{\cI}^{(3)}{}_{\bar{\cA}_{3}}{}^{\cA_{3}}(X) \, \hat{\cH}_{\cA_{1} \cA_{2} \cA_{3}}(X) \, .
\end{align}
Hence, we notice that under $\cI$, $\cH$ transforms into the complex conjugate representation. An analogous formula can be derived using the auxiliary spinor formalism. Given a polynomial $\cH(X; U, V, W) = X^{\D_{3} - \D_{2}- \D_{1}} \hat{\cH}(X; U,V,W)$, the following holds:
\begin{align} \label{Ch04-H inversion formula}
	\hat{\cH}^{c}(X^{I}; U, V, W) = \cI_{X}(U,\tilde{U}) \, \cI_{X}(V,\tilde{V}) \, \cI_{X}(W,\tilde{W}) \, \hat{\cH}(X; \tilde{U}, \tilde{V}, \tilde{W}) \, .
\end{align}
This formula is easily understood by noting that the monomials \eqref{Ch04-Basis scalar structures} have simple transformation properties under $\cI$:
\begin{align} \label{Ch04-Inversion of basis structures}
	P_{i} &\xrightarrow{\cI} - \bar{P}_{i} \, , & Q_{i} &\xrightarrow{\cI} \bar{Q}^{I}_{i} \, , & Z_{i} &\xrightarrow{\cI} Z^{I}_{i} \, ,
\end{align}
with analogous rules applying for the building blocks $\bar{P}_{i}, \bar{Q}_{i}$. Since, for primary fields in the $(s,s)$ representation the three-point maps onto itself under inversion, it is possible to classify the parity-even and parity-odd structures in $\cH$ using \eqref{Ch04-H inversion formula}. By letting $\hat{\cH}(X) = \hat{\cH}^{(+)}(X) + \hat{\cH}^{(-)}(X)$, we have  
\begin{align} \label{Ch04-H inversion classification}
	\hat{\cH}^{(\pm)}(X^{I}; U, V, W) = \pm \, \cI_{X}(U,\tilde{U}) \, \cI_{X}(V,\tilde{V}) \, \cI_{X}(W,\tilde{W}) \, \hat{\cH}^{(\pm)}(X; \tilde{U}, \tilde{V}, \tilde{W}) \, .
\end{align}
Structures satisfying the above property are defined as parity-even/odd for overall sign $+/-$. This is essentially the same approach used to classify parity-even and parity-odd
three-point functions in 3D CFT, which proves to be equivalent to the classification based on the absence/presence of the Levi-Civita pseudo-tensor. 
However, it is crucial to note that in three dimensions any combination of the basic monomial structures comprising $\cH$ are naturally eigenfunctions of the inversion operator. 
The same is not necessarily true for three-point functions in four dimensions due to \eqref{Ch04-Inversion of basis structures}, as the monomials \eqref{Ch04-Basis scalar structures} now map onto their complex conjugates. Hence, we are required to take non-trivial linear combinations of the basic structures
and use the linear dependence relations~\eqref{Ch04-Linear dependence 1}--\eqref{Ch04-Linear dependence 6} to form eigenfunctions of the inversion operator. Our classification of parity-even/odd solutions obtained this way is in complete agreement with the results found in~\cite{Stanev:2012nq}.


\section{Three-point functions of conserved currents}\label{Ch04-section3}

In the next subsections we analyse the structure of three-point functions involving conserved currents in 4D CFT. 
We classify, using computational methods, all possible three-point functions involving the conserved currents $J_{(s,q)}$, $\bar{J}_{(s,q)}$ for $s_{i} \leq 10$. 
In particular, we determine the general structure and the number of independent solutions present in the three-point functions \eqref{Ch04-possible three point functions}. 
As pointed out in the introduction, the number of independent conserved structures generically grows linearly with the minimum spin and the solution for the 
function $\cH(X; U, V, W)$ quickly becomes too long and complicated even for relatively low spins. 
Thus, although our method allows us to find $\cH(X; U, V, W)$ in a very explicit form for arbitrary spins (limited only by computer power), 
we find it practical to present the solutions when 
there is a small number of structures. Such examples involving low spins are discussed in Subsection~\ref{Ch04-subsection3.1}.
In Subsection~\ref{Ch04-subsection3.2} we state the classification for arbitrary spins. Some additional examples are presented in the appendices \ref{Appendix4A}, \ref{Appendix4B}, \ref{Appendix4C}.

\subsection{Conserved low-spin currents}\label{Ch04-subsection3.1}

We begin our analysis by considering correlation functions involving conserved low-spin currents such as the energy-momentum tensor, vector current, and ``supersymmetry-like" currents in 4D CFT. Many of these results are known throughout the literature (see e.g. \cite{Osborn:1993cr,Erdmenger:1996yc,Stanev:2012nq,Zhiboedov:2012bm}), but we derive them again here to demonstrate our approach.

\subsubsection{Energy-momentum tensor and vector current correlators}\label{Ch04-subsubsection3.1.1}

The fundamental bosonic conserved currents in any conformal field theory are the conserved vector current, $V_{m}$, and the symmetric, traceless energy-momentum tensor, $T_{mn}$. The vector current has scale dimension $\Delta_{V} = 3$ and satisfies $\pa^{m} V_{m} = 0$, while the energy-momentum tensor has scale dimension $\Delta_{T} = 4$ and satisfies the conservation equation $\pa^{m} T_{mn} = 0$. Converting to spinor notation using the conventions outlined in Appendix \ref{Appendix2A}, we have:
\begin{align}
	V_{\a \ad}(x) = (\s^{m})_{\a \ad} V_{m}(x) \, , && T_{(\a_{1} \a_{2}) (\ad_{1} \ad_{2})}(x) = (\s^{m})_{(\a_{1} (\ad_{1}} (\g^{n})_{\a_{2}) \ad_{2})} T_{mn}(x) \, .
\end{align}
These objects possess fundamental information associated with internal and spacetime symmetries, hence, their three-point functions are of great importance. The possible three-point functions involving the conserved vector current and the energy-momentum tensor are:
\begin{subequations}
	\begin{align} \label{Ch04-Low-spin component correlators}
		\langle V_{\a \ad}(x_{1}) \, V_{\b \bd}(x_{2}) \, V_{\g \gd}(x_{3}) \rangle \, , &&  \langle V_{\a \ad}(x_{1}) \, V_{\b \bd}(x_{2}) \, T_{\g(2) \gd(2)}(x_{3}) \rangle \, , \\
		\langle T_{\a(2) \ad(2)}(x_{1}) \, T_{\b(2) \bd(2)}(x_{2}) \, V_{\g \gd}(x_{3}) \rangle \, , &&  \langle T_{\a(2) \ad(2)}(x_{1}) \, T_{\b(2) \bd(2)}(x_{2}) \, T_{\g(2) \gd(2)}(x_{3}) \rangle \, .
	\end{align}
\end{subequations}
Let us first consider $\langle V V V \rangle$. By using the notation for the currents $J^{}_{(s,q)}$, $\bar{J}^{}_{(s,q)}$, this corresponds to the general three-point function $\langle J^{}_{(1,0)} J'_{(1,0)} J''_{(1,0)} \rangle$.\\

\newpage

\noindent\textbf{Correlation function} $\langle J^{}_{(1,0)} J'_{(1,0)} J''_{(1,0)} \rangle$\textbf{:}\\[2mm]
The general ansatz for this correlation function, according to \eqref{Ch04-Conserved correlator ansatz} is
\begin{align}
	\langle J^{}_{\a \ad}(x_{1}) \, J'_{\b \bd}(x_{2}) \, J''_{\g \gd}(x_{3}) \rangle = \frac{ \cI_{\a}{}^{\ad'}(x_{13}) \, \bar{\cI}_{\ad}{}^{\a'}(x_{13}) \,  \cI_{\b}{}^{\bd'}(x_{23}) \, \bar{\cI}_{\bd}{}^{\b'}(x_{23}) }{(x_{13}^{2})^{3} (x_{23}^{2})^{3}}
	\; \cH_{\a' \ad' \b' \bd' \g \gd}(X_{12}) \, .
\end{align} 
Using the formalism outlined in Subsection \ref{Ch04-subsubsection2.2.2}, all information about this correlation function is encoded in the following polynomial:
\begin{align}
	\cH(X; U, V, W) = \cH_{ \a \ad \b \bd \g \gd }(X) \, \boldsymbol{U}^{\a \ad}  \boldsymbol{V}^{\b \bd}  \boldsymbol{W}^{\g \gd} \, .
\end{align}
Using Mathematica we solve \eqref{Ch04-Diophantine equations} for the chosen spin representations of the currents and substitute each solution into the generating function \eqref{Ch04-Generating function}. This provides us with the following list of (linearly dependent) polynomial structures:
%
\begin{align}
	\begin{split}
		&\big\{ Q_1 Q_2 Q_3,Z_1 Z_2 Z_3,P_3 Q_2 \bar{P}_1,P_1 Z_1 \bar{P}_1,P_1 Q_3 \bar{P}_2,P_2 Z_2 \bar{P}_2,P_2 Q_1 \bar{P}_3, P_3 Z_3 \bar{P}_3, \\ 
		& \hspace{20mm}  Q_1 Z_1 \bar{Q}_1,P_3 \bar{P}_2 \bar{Q}_1, Q_2 Z_2 \bar{Q}_2,P_1 \bar{P}_3 \bar{Q}_2,Q_3 Z_3 \bar{Q}_3,P_2 \bar{P}_1
		\bar{Q}_3,\bar{Q}_1 \bar{Q}_2 \bar{Q}_3 \big\}\,.
	\end{split}
\end{align}
Next, we systematically apply the linear dependence relations \eqref{Ch04-Linear dependence 1} to these lists, reducing them to the following sets of linearly independent structures:
%
\begin{align}
	\big\{ Q_1 Q_2 Q_3,P_3 Q_2 \bar{P}_1,P_1 Q_3 \bar{P}_2,P_2 Q_1 \bar{P}_3,\bar{Q}_1 \bar{Q}_2 \bar{Q}_3 \big\}\,.
\end{align}
Note that application of the linear-dependence relations eliminates all terms involving $Z_{i}$ in this case.  Since this correlation function is composed of fields in the $(s,s)$ representation, the solutions for the three-point function may be split up into parity-even and parity-odd contributions. To do this we construct linear combinations for the polynomial $\hat{\cH}(X; U, V, W)$ which are even/odd under inversion in accordance with \eqref{Ch04-H inversion classification}:
%
%
\begin{align}
	\begin{split}
		&A_1  (\bar{Q}_1 \bar{Q}_2 \bar{Q}_3+Q_1
		Q_2 Q_3 ) + A_2  (P_3 Q_2 \bar{P}_1-\bar{Q}_1 \bar{Q}_2 \bar{Q}_3 ) +A_3  (P_1 Q_3 \bar{P}_2-\bar{Q}_1 \bar{Q}_2
		\bar{Q}_3 ) \\
		& \hspace{20mm} +A_4  (P_2 Q_1 \bar{P}_3-\bar{Q}_1 \bar{Q}_2 \bar{Q}_3 )+ B_1  (Q_1 Q_2 Q_3-\bar{Q}_1 \bar{Q}_2 \bar{Q}_3 ) \, .
	\end{split}
\end{align}
We note here (and in all other examples) that the parity-even contributions possess the complex coefficients $A_{i}$, while the parity-odd solutions possess the complex coefficients $B_{i}$. It can be explicitly checked that these structures possess the appropriate transformation properties. Next, since the correlation function is overall real, we must impose the reality condition \eqref{Ch04-Reality condition}. As a result, we find that the parity-even coefficients $A_{i}$ are purely real, i.e., $A_{i} = a_{i}$, while the parity-odd coefficients $B_{i}$ are purely imaginary, i.e., $B_{i} = \text{i} b_{i}$.

We must now impose the conservation of the currents. Following the procedure outlined in Subsection \ref{Ch04-subsubsection2.2.2} we obtain a linear system in the coefficients $a_{i}$, $b_{i}$ which can be easily solved computationally. We find the following solution for $\cH(X;U,V,W)$ consistent with conservation on all three points:
%
%
\begin{align}
	\begin{split}
		& \frac{a_1}{X^3} \big(Q_1 Q_2 Q_3 + 2 P_1 Q_3 \bar{P}_2-\bar{Q}_1 \bar{Q}_2 \bar{Q}_3 \big) \\[1mm]
		& \hspace{10mm} +\frac{a_2}{X^3} \big( P_3 Q_2
		\bar{P}_1-3 P_1 Q_3 \bar{P}_2+P_2 Q_1 \bar{P}_3+\bar{Q}_1 \bar{Q}_2 \bar{Q}_3 \big) \\[1mm]
		& \hspace{25mm} + \frac{\text{i} b_1}{X^3} \big( Q_1 Q_2 Q_3- \bar{Q}_1 \bar{Q}_2
		\bar{Q}_3 \big) \, .
	\end{split}
\end{align}
The only remaining constraints to impose are symmetries under permutations of spacetime points, which apply when the currents in the three-point function are identical, i.e. when $J=J'=J''$. After imposing \eqref{Ch04-Point switch A}, \eqref{Ch04-Point switch B}, only the structure corresponding to the coefficient $b_{1}$ survives. 
However, the $a_{1}$, $a_{2}$ structures can exist if the currents are non-abelian. 
This is consistent with the results of \cite{Osborn:1993cr, Erdmenger:1996yc, Stanev:2012nq}.\footnote{The coefficient $b_1$ is related to the chiral anomaly of the CFT under consideration when it is coupled to a background vector field \cite{Erdmenger:1996yc}. This anomaly exists in chiral theories which are not invariant under parity and, thus, admit a parity-odd contribution.}

The next example to consider is the mixed correlator $\langle V V T \rangle$. To study this case we may examine the correlation function 
$\langle J^{}_{(1,0)} J'_{(1,0)} J''_{(2,0)} \rangle$. \\[5mm]
\noindent
\textbf{Correlation function} $\langle J^{}_{(1,0)} J'_{(1,0)} J''_{(2,0)} \rangle$\textbf{:}\\[2mm]
Using the general formula, the ansatz for this three-point function is:
\begin{align}
	\begin{split}
		\langle J^{}_{\a \ad}(x_{1}) \, J'_{\b \bd}(x_{2}) \, J''_{\g(2) \gd(2)}(x_{3}) \rangle &= \frac{ \cI_{\a}{}^{\ad'}(x_{13}) \, \bar{\cI}_{\ad}{}^{\a'}(x_{13}) \,  \cI_{\b}{}^{\bd'}(x_{23}) \, \bar{\cI}_{\bd}{}^{\b'}(x_{23})}{(x_{13}^{2})^{3} (x_{23}^{2})^{3}} \\
		& \hspace{20mm} \times \, \cH_{\a' \ad' \b' \bd' \g(2) \gd(2)}(X_{12}) \, .
	\end{split}
\end{align} 
All information about this correlation function is encoded in the following polynomial:
\begin{align}
	\cH(X; U, V, W) = \cH_{ \a \ad \b \bd \g(2) \gd(2) }(X) \, \boldsymbol{U}^{\a \ad}  \boldsymbol{V}^{\b \bd}  \boldsymbol{W}^{\g(2) \gd(2)} \, .
\end{align}
After solving \eqref{Ch04-Diophantine equations}, we find the following linearly dependent polynomial structures:
%
\begin{align}
	\begin{split}
		&\big\{ P_1 P_2 \bar{P}_1 \bar{P}_2,P_2 Q_1 Q_2 \bar{P}_1,P_3 Q_2 Z_3 \bar{P}_1,P_3 Z_3^2 \bar{P}_3,  Q_3 Z_3^2 \bar{Q}_3, Z_1 Z_2 Z_3^2,\\
		& \hspace{10mm} P_1 Q_3 Z_3 \bar{P}_2, P_2 Z_2 Z_3 \bar{P}_2, Q_1 Q_2 Q_3 Z_3, P_2 Q_1 Z_3 \bar{P}_3,Q_1 Z_1 Z_3 \bar{Q}_1, \\
		& \hspace{20mm} P_2 Q_1 \bar{P}_2 \bar{Q}_1, P_3 Z_3 \bar{P}_2 \bar{Q}_1,Q_2 Z_2 Z_3 \bar{Q}_2,P_1 Q_2 \bar{P}_1 \bar{Q}_2,P_1 Z_3 \bar{P}_3
		\bar{Q}_2, \\
		& \hspace{30mm} Q_1 Q_2 \bar{Q}_1 \bar{Q}_2, P_1 \bar{P}_2 \bar{Q}_1 \bar{Q}_2,P_1 Z_1 Z_3 \bar{P}_1,P_2 Z_3 \bar{P}_1 \bar{Q}_3,Z_3
		\bar{Q}_1 \bar{Q}_2 \bar{Q}_3 \big\}\,.
	\end{split}
\end{align}
We now systematically apply the linear dependence relations \eqref{Ch04-Linear dependence 1}--\eqref{Ch04-Linear dependence 6} to obtain the linearly independent structures
%
\begin{align}
	\big\{ P_2 Q_1 Q_2 \bar{P}_1,P_1 P_2 \bar{P}_1 \bar{P}_2,P_2 Q_1 \bar{P}_2 \bar{Q}_1,P_1 Q_2 \bar{P}_1 \bar{Q}_2,Q_1 Q_2
	\bar{Q}_1 \bar{Q}_2,P_1 \bar{P}_2 \bar{Q}_1 \bar{Q}_2 \big\}\,.
\end{align}
Next, we construct the following parity-even and parity-odd linear combinations which comprise the polynomial $\hat{\cH}(X; U, V, W)$:
%
\begin{align}
	\begin{split}
		& A_1  (P_2 Q_1 Q_2 \bar{P}_1+P_1 \bar{P}_2 \bar{Q}_1
		\bar{Q}_2 )+A_2 P_1 P_2 \bar{P}_1 \bar{P}_2+A_3 P_2 Q_1 \bar{P}_2 \bar{Q}_1 \\
		& \hspace{5mm} +A_4 P_1 Q_2 \bar{P}_1 \bar{Q}_2+A_5 Q_1 Q_2 \bar{Q}_1 \bar{Q}_2+B_1 (P_2 Q_1 Q_2 \bar{P}_1-P_1 \bar{P}_2
		\bar{Q}_1 \bar{Q}_2 ) \, .
	\end{split}
\end{align}
%
We now impose conservation on all three points to obtain the final solution for $\cH(X;U,V,W)$
\begin{align} \label{Ch04-1-1-2}
	\begin{split}
		&\frac{a_1}{X^2} \Big( P_2 Q_1 Q_2 \bar{P}_1+P_1 Q_2 \bar{P}_1
		\bar{Q}_2+P_2 Q_1 \bar{P}_2 \bar{Q}_1+P_1 \bar{P}_2 \bar{Q}_1 \bar{Q}_2-\sfrac{2}{3} Q_1 Q_2 \bar{Q}_1
		\bar{Q}_2 \Big) \\[1mm]
		& \hspace{5mm} +\frac{a_2}{X^2} \Big(-\sfrac{1}{2} P_2 Q_1 \bar{P}_2 \bar{Q}_1-\sfrac{1}{2} P_1 Q_2 \bar{P}_1 \bar{Q}_2+P_1 P_2 \bar{P}_1
		\bar{P}_2+\sfrac{1}{3} Q_1 Q_2 \bar{Q}_1 \bar{Q}_2 \Big) \\[1mm]
		& \hspace{15mm} +\frac{\text{i} b_1}{X^2} \Big( P_2 Q_1 Q_2 \bar{P}_1- P_1 \bar{P}_2 \bar{Q}_1 \bar{Q}_2 \Big) \, .
	\end{split}
\end{align}
In this case, only the parity-even structures (proportional to $a_{1}$ and $a_{2}$) survive after setting $J = J'$. Hence, this correlation function is fixed up to two independent 
parity-even structures with real coefficients.

The number of polynomial structures increases rapidly for increasing $s_{i}$, and for the three-point functions $\langle T T V \rangle$, $\langle T T T \rangle$ we will present only the linearly independent structures and the final results after imposing parity, reality, and conservation on all three points. For $\langle T T V \rangle$ we may consider the correlation function $\langle J^{}_{(2,0)} J'_{(2,0)} J''_{(1,0)} \rangle$, which is constructed from the following list of linearly independent structures:
\begin{align}
	\begin{split}
		& \big\{ P_3 Q_1 Q_2 Q_3 \bar{P}_3,P_3^2 Q_2 \bar{P}_1 \bar{P}_3,P_2 P_3 Q_1 \bar{P}_3^2,Q_1 Q_2 Q_3^2 \bar{Q}_3, P_1 Q_3^2 \bar{P}_2 \bar{Q}_3, \\
		& \hspace{20mm} P_3 Q_2 Q_3 \bar{P}_1 \bar{Q}_3, P_2 Q_1 Q_3 \bar{P}_3 \bar{Q}_3,P_3 \bar{P}_3 \bar{Q}_1 \bar{Q}_2
		\bar{Q}_3,Q_3 \bar{Q}_1 \bar{Q}_2 \bar{Q}_3^2 \big\}\,.
	\end{split}
\end{align}
We now construct linearly independent parity-even and parity-odd solutions consistent with \eqref{Ch04-H inversion classification}. Then, after imposing all the constraints due to reality and conservation, we obtain the final solution for $\cH(X; U,V,W)$:
\begin{align}
	\begin{split}
		&\frac{a_1}{X^5} \Big(3 P_1 Q_3^2 \bar{P}_2 \bar{Q}_3+P_3 Q_1 Q_2 Q_3 \bar{P}_3+2 P_3 Q_2 Q_3 \bar{P}_1 \bar{Q}_3 \\
		& \hspace{10mm} +2 P_2 Q_1 Q_3
		\bar{P}_3 \bar{Q}_3 +P_3 \bar{P}_3 \bar{Q}_1 \bar{Q}_2 \bar{Q}_3+\sfrac{7}{2} Q_1 Q_2 Q_3^2 \bar{Q}_3-\sfrac{7}{2} Q_3 \bar{Q}_1
		\bar{Q}_2 \bar{Q}_3^2 \Big) \\[1mm]
		&+\frac{a_2}{X^5} \Big(P_3^2 Q_2 \bar{P}_1 \bar{P}_3+P_2 P_3 Q_1 \bar{P}_3^2-6 P_3 Q_2 Q_3
		\bar{P}_1 \bar{Q}_3-2 P_3 \bar{P}_3 \bar{Q}_1 \bar{Q}_2 \bar{Q}_3 \\
		& \hspace{10mm}-7 P_1 Q_3^2 \bar{P}_2 \bar{Q}_3-6 P_2 Q_1 Q_3 \bar{P}_3
		\bar{Q}_3+\sfrac{17}{2} Q_3 \bar{Q}_1 \bar{Q}_2 \bar{Q}_3^2-\sfrac{21}{2} Q_1 Q_2 Q_3^2 \bar{Q}_3 \Big) \\[1mm]
		& \hspace{5mm} +\frac{\text{i} b_1}{X^5} \Big( 
		P_3 Q_1 Q_2 Q_3 \bar{P}_3- P_3 \bar{P}_3 \bar{Q}_1 \bar{Q}_2 \bar{Q}_3-\sfrac{3}{2} Q_1 Q_2 Q_3^2 \bar{Q}_3+\sfrac{3}{2} 
		Q_3 \bar{Q}_1 \bar{Q}_2 \bar{Q}_3^2 \Big) \, .
	\end{split}
\end{align}
After setting $J = J'$ and imposing the required symmetries under the exchange of $x_{1}$ and $x_{2}$ we find that $b_{1} = 0$, while $a_{1}, a_{2}$ remain unconstrained. Hence, the correlation function $\langle T T V \rangle$ is fixed up to two parity-even structures with real coefficients. 

The final fundamental three-point function to study is $\langle T T T \rangle$, and for this we analyse the correlation function $\langle J^{}_{(2,0)} J'_{(2,0)} J''_{(2,0)} \rangle$. In this case there are 15 linearly independent structures to consider:
\begin{align}
	\begin{split}
		&\big\{Q_1^2 Q_2^2 Q_3^2,P_3 Q_1 Q_2^2 Q_3 \bar{P}_1,P_3^2 Q_2^2 \bar{P}_1^2,P_1 Q_1 Q_2 Q_3^2 \bar{P}_2,P_1^2 Q_3^2
		\bar{P}_2^2,P_2 Q_1^2 Q_2 Q_3 \bar{P}_3, \\
		& \hspace{10mm} P_3 Q_1 Q_2 \bar{P}_3 \bar{Q}_1 \bar{Q}_2,P_2 Q_1 Q_3
		\bar{P}_2 \bar{Q}_1 \bar{Q}_3,P_1 Q_2 Q_3 \bar{P}_1 \bar{Q}_2 \bar{Q}_3,Q_1 Q_2 Q_3 \bar{Q}_1 \bar{Q}_2 \bar{Q}_3, \\
		& \hspace{15mm} P_2^2 Q_1^2 \bar{P}_3^2,P_3 Q_2 \bar{P}_1 \bar{Q}_1 \bar{Q}_2 \bar{Q}_3,P_1 Q_3 \bar{P}_2 \bar{Q}_1 \bar{Q}_2 \bar{Q}_3,P_2 Q_1 \bar{P}_3 \bar{Q}_1 \bar{Q}_2
		\bar{Q}_3,\bar{Q}_1^2 \bar{Q}_2^2 \bar{Q}_3^2 \big\}\,.
	\end{split}
\end{align}
From these structures we construct linear combinations that are even/odd under parity, analogous to the previous examples. Then, after imposing reality and conservation on all three points we obtain the following solution for $\cH(X; U,V,W)$:
%
\begin{align}
	\begin{split}
		&\frac{a_1}{X^4}
		\Big(Q_1^2 Q_2^2 Q_3^2+2 P_1^2 Q_3^2 \bar{P}_2^2 -2 Q_1 Q_2  Q_3 \bar{Q}_1 \bar{Q}_2 \bar{Q}_3 \\
		& \hspace{10mm}+2 P_1 Q_1 Q_2  Q_3^2 \bar{P}_2 -2
		P_1 Q_3 \bar{P}_2 \bar{Q}_1 \bar{Q}_2 \bar{Q}_3 +\bar{Q}_1^2 \bar{Q}_2^2 \bar{Q}_3^2 \Big) \\[1mm]
		&+\frac{a_2}{X^4} \Big(P_2 Q_1^2 Q_2 Q_3 \bar{P}_3 +P_3 Q_1 Q_2^2 Q_3 \bar{P}_1 -\sfrac{17}{3} P_1 Q_1 Q_2 Q_3^2 \bar{P}_2 +2 P_2 Q_1 Q_3
		\bar{P}_2 \bar{Q}_1 \bar{Q}_3  \\
		& \hspace{10mm} +3 Q_1 Q_2 Q_3 \bar{Q}_1 \bar{Q}_2 \bar{Q}_3+P_2 Q_1 \bar{P}_3 \bar{Q}_1 \bar{Q}_2 \bar{Q}_3
		-\sfrac{20}{3} P_1^2 Q_3^2 \bar{P}_2^2-3 \bar{Q}_1^2 \bar{Q}_2^2 \bar{Q}_3^2 \\
		& \hspace{15mm} +2 P_1 Q_2 Q_3 \bar{P}_1 \bar{Q}_2 \bar{Q}_3+P_3
		Q_2 \bar{P}_1 \bar{Q}_1 \bar{Q}_2 \bar{Q}_3+\sfrac{23}{3} P_1 Q_3 \bar{P}_2 \bar{Q}_1 \bar{Q}_2
		\bar{Q}_3 \Big) \\[1mm]
		&+\frac{a_3}{X^4} \Big( P_3^2 Q_2^2 \bar{P}_1^2 + \sfrac{19}{3}
		P_1^2 Q_3^2 \bar{P}_2^2  +\sfrac{16}{3} P_1 Q_1 Q_2 Q_3^2 \bar{P}_2  -2 Q_1 Q_2 Q_3 \bar{Q}_1 \bar{Q}_2 \bar{Q}_3  \\
		& \hspace{8mm} + P_2^2 Q_1^2 \bar{P}_3^2 -2 P_2 Q_1 Q_3 \bar{P}_2 \bar{Q}_1 \bar{Q}_3  -2 P_1 Q_2 Q_3 \bar{P}_1 \bar{Q}_2 \bar{Q}_3  -\sfrac{22}{3} P_1 Q_3 \bar{P}_2 \bar{Q}_1 \bar{Q}_2
		\bar{Q}_3  \\
		& \hspace{13mm} -3 P_3 Q_1 Q_2 \bar{P}_3
		\bar{Q}_1 \bar{Q}_2-2 P_3 Q_2 \bar{P}_1 \bar{Q}_1 \bar{Q}_2 \bar{Q}_3-2 P_2 Q_1 \bar{P}_3 \bar{Q}_1 \bar{Q}_2 \bar{Q}_3 +3 \bar{Q}_1^2 \bar{Q}_2^2 \bar{Q}_3^2 \Big) \\[1mm]
		&+\frac{\text{i} b_1}{X^4} \Big( Q_1^2 Q_2^2 Q_3^2+2  P_1 Q_1 Q_2 Q_3^2 \bar{P}_2 + \bar{Q}_1^2 \bar{Q}_2^2 \bar{Q}_3^2 \\
		& \hspace{25mm} -2  Q_1 Q_2 Q_3 \bar{Q}_1 \bar{Q}_2 \bar{Q}_3 -2  P_1 Q_3 \bar{P}_2 \bar{Q}_1 \bar{Q}_2 \bar{Q}_3 \Big) \\[1mm]
		&+\frac{\text{i} b_2}{X^4} \Big( P_2 Q_1^2 Q_2 Q_3 \bar{P}_3 + P_3 Q_1 Q_2^2 Q_3 \bar{P}_1 - 3 P_1 Q_1 Q_2 Q_3^2 \bar{P}_2 + Q_1 Q_2 Q_3 \bar{Q}_1 \bar{Q}_2 \bar{Q}_3 \\
		& \hspace{15mm} - P_2 Q_1 \bar{P}_3 \bar{Q}_1 \bar{Q}_2 \bar{Q}_3 - P_3 Q_2 \bar{P}_1 \bar{Q}_1 \bar{Q}_2 \bar{Q}_3+3  P_1 Q_3 \bar{P}_2 \bar{Q}_1 \bar{Q}_2 \bar{Q}_3 - \bar{Q}_1^2 \bar{Q}_2^2
		\bar{Q}_3^2 \Big) \, .
	\end{split}
\end{align}
In this case only three of the structures (corresponding to the real coefficients $a_{1}, a_{2}, a_{3}$) survive the point-switch symmetries upon exchange of $x_{1}$, $x_{2}$ and $x_{3}$. Hence, $\langle T T T \rangle$ is fixed up to three parity-even structures with real coefficients.

In all cases we note that the number of independent structures (prior to imposing exchange symmetries) is $2 \min(s_{1}, s_{2}, s_{3}) + 1$ in general, 
where $\min(s_{1}, s_{2}, s_{3}) + 1$ are parity-even and $\min(s_{1}, s_{2}, s_{3})$ are parity-odd.
These results are in agreement with \cite{Osborn:1993cr,Stanev:2012nq, Stanev:2013eha, Stanev:2013qra, Zhiboedov:2012bm} in terms of the number of independent 
structures, however, our construction of the three-point function is quite different.

\subsubsection{Spin-3/2 current correlators}\label{subsubsection3.1.2}

In this section we will evaluate three-point functions involving conserved fermionic currents. The most important examples of fermionic conserved currents in 4D CFT are the supersymmetry currents, $Q_{m,\a}$, $\bar{Q}_{m,\ad}$, which appear in $\cN$-extended superconformal field theories \cite{Buchbinder:1998qv}. Such fields are primary with dimension $\D_{Q} = \D_{\bar{Q}} = 7/2$, and satisfy the conservation equations $\pa^{m} Q_{m, \a } = 0$, $\pa^{m} \bar{Q}_{m, \ad } = 0$. In spinor notation, we have:
\begin{equation}
	Q_{\a \ad, \b}(x) = (\s^{m})_{\a \ad} Q_{m,\b}(x) \, , \hspace{10mm} \bar{Q}_{\a \ad, \bd}(x) = (\s^{m})_{\a \ad} \bar{Q}_{m,\bd}(x) \, .
\end{equation}
%
The correlation functions involving supersymmetry currents, vector currents, and the energy-momentum tensor are of fundamental importance. 
The four possible three-point functions involving $Q$, $V$ and $T$ which are of interest in $\cN=1$ superconformal field theories are:
\begin{subequations} \label{Ch04-Susy current correlators - 1}
	\begin{align}
		\langle Q_{\a(2) \ad}(x_{1}) \, Q_{\b(2) \bd}(x_{2}) \, V_{\g \gd}(x_{3}) \rangle \, , && \langle Q_{\a(2) \ad}(x_{1}) \,  Q_{\b(2) \bd}(x_{2}) \, T_{\g(2) \gd(2)}(x_{3}) \rangle \, , \\
		\langle Q_{\a(2) \ad}(x_{1}) \, \bar{Q}_{\b \bd(2)}(x_{2}) \, V_{\g \gd}(x_{3}) \rangle \, , && \langle Q_{\a(2) \ad}(x_{1}) \,  \bar{Q}_{\b(2) \bd}(x_{2}) \, T_{\g(2) \gd(2)}(x_{3}) \rangle \, .
	\end{align}
\end{subequations}
%

These three-point functions were analysed in \cite{Buchbinder:2022cqp} using a similar approach, but we present them again here for completeness and to demonstrate our general formalism. Note that in the subsequent analysis we assume only conformal symmetry, not supersymmetry. 

We now present an explicit analysis of the general structure of correlation functions involving $Q$, $\bar{Q}$, $V$ and $T$ that are compatible with the constraints of conformal symmetry and conservation equations. Using our conventions for the currents, we recall that $Q \equiv J_{(3/2,1)}$, $\bar{Q} \equiv \bar{J}_{(3/2,1)}$. Let us first consider $\langle Q Q V \rangle$, for which we may analyse the general structure of the correlation function $\langle J^{}_{(3/2,1)} J'_{(3/2,1)} J''_{(1,0)} \rangle$. \\[5mm]
\noindent
\textbf{Correlation function} $\langle J^{}_{(3/2,1)} J'_{(3/2,1)} J''_{(1,0)} \rangle$\textbf{:}\\[2mm]
Using the general formula, the ansatz for this three-point function:
\begin{align}
	\begin{split}
		\langle J^{}_{\a(2) \ad}(x_{1}) \, J'_{\b(2) \bd}(x_{2}) \, J''_{\g \gd}(x_{3}) \rangle &= \frac{ \cI_{\a(2)}{}^{\ad'(2)}(x_{13}) \, \bar{\cI}_{\ad}{}^{\a'}(x_{13} ) \,  \cI_{\b(2)}{}^{\bd'(2)}(x_{23}) \,  \bar{\cI}_{\bd}{}^{\b'}(x_{23}) }{(x_{13}^{2})^{7/2} (x_{23}^{2})^{7/2}} \\ 
		& \hspace{5mm} \times \cH_{\a' \ad'(2) \b' \bd'(2) \g \gd}(X_{12}) \, .
	\end{split}
\end{align} 
Using the formalism outlined in Subsection \ref{Ch04-subsubsection2.2.2}, all information about this correlation function is encoded in the following polynomial:
\begin{align}
	\cH(X; U, V, W) = \cH_{ \a \ad(2) \b \bd(2) \g \gd }(X) \, \boldsymbol{U}^{\a \ad(2)}  \boldsymbol{V}^{\b \bd(2)}  \boldsymbol{W}^{\g \gd} \, .
\end{align}
After solving \eqref{Ch04-Diophantine equations}, we find the following linearly dependent polynomial structures in the even and odd sectors respectively:
%
\begin{align}
	\begin{split}
		&\big\{Q_2 Z_1 Z_2 \bar{P}_1,Q_2 Q_3 Z_2 \bar{P}_2,Q_1 Q_2 Q_3 \bar{P}_3,Z_1 Z_2 Z_3 \bar{P}_3,P_3 Q_2 \bar{P}_1 \bar{P}_3, \\
		& \hspace{5mm} P_1 Z_1 \bar{P}_1 \bar{P}_3,P_1 Q_3 \bar{P}_2 \bar{P}_3, P_2 Z_2 \bar{P}_2 \bar{P}_3,P_2 Q_1 \bar{P}_3^2,P_3 Z_3 \bar{P}_3^2, P_1 \bar{P}_3^2 \bar{Q}_2, \\
		& \hspace{10mm} Q_1 Z_1 \bar{P}_3 \bar{Q}_1,P_3 \bar{P}_2 \bar{P}_3 \bar{Q}_1,Q_2 Z_2 \bar{P}_3 \bar{Q}_2, Z_1 Z_2 \bar{P}_2 \bar{Q}_1, Q_2 Q_3 \bar{P}_1 \bar{Q}_3,\\
		& \hspace{15mm} Q_3 Z_3 \bar{P}_3 \bar{Q}_3,P_2 \bar{P}_1 \bar{P}_3 \bar{Q}_3,Z_1 \bar{P}_1
		\bar{Q}_1 \bar{Q}_3,Q_3 \bar{P}_2 \bar{Q}_1 \bar{Q}_3,\bar{P}_3 \bar{Q}_1 \bar{Q}_2 \bar{Q}_3 \big\}\,.
	\end{split}
\end{align}
Next we systematically apply the linear dependence relations \eqref{Ch04-Linear dependence 1}--\eqref{Ch04-Linear dependence 6} and obtain the following linearly independent structures:
%
\begin{align}
	\big\{ Q_1 Q_2 Q_3 \bar{P}_3,P_3 Q_2 \bar{P}_1 \bar{P}_3,P_2 Q_1 \bar{P}_3^2,Q_2 Q_3 \bar{P}_1 \bar{Q}_3,Q_3 \bar{P}_2 \bar{Q}_1
	\bar{Q}_3,\bar{P}_3 \bar{Q}_1 \bar{Q}_2 \bar{Q}_3 \big\}\,.
\end{align}
We now impose conservation on all three points and find that the solution for $\cH(X; U,V,W)$ is unique up to a complex coefficient, $A_{1} = a_{1} + \text{i} \tilde{a}_{1}$:
%
\begin{align} \label{Ch04-QQV}
	\begin{split}
		&\frac{A_1}{X^4} \Big(Q_1 Q_2 Q_3 \bar{P}_3 + \sfrac{5}{9} P_2 Q_1 \bar{P}_3^2+\sfrac{5}{9} P_3 Q_2 \bar{P}_1 \bar{P}_3 \\
		& \hspace{15mm} -\sfrac{1}{9} \bar{P}_3 \bar{Q}_1 \bar{Q}_2 \bar{Q}_3-\sfrac{2}{9} Q_2 Q_3 \bar{P}_1 \bar{Q}_3-\sfrac{2}{9} Q_3 \bar{P}_2 \bar{Q}_1
		\bar{Q}_3 \Big) \, .
	\end{split}
\end{align}
However, this three-point function is not compatible with the point-switch symmetry associated with setting $J=J'$. Therefore we conclude that the three-point function $\langle Q Q V \rangle$ must vanish in general. 

\vspace{2mm}

\noindent
\textbf{Correlation function} $\langle J^{}_{(3/2,1)} \bar{J}'_{(3/2,1)} J''_{(1,0)} \rangle$\textbf{:}\\[2mm]
Using the general formula we obtain the following ansatz:
\begin{align}
	\begin{split}
		\langle J^{}_{\a(2) \ad}(x_{1}) \, J'_{\b \bd(2)}(x_{2}) \, J''_{\g \gd}(x_{3}) \rangle &= \frac{ \cI_{\a(2)}{}^{\ad'(2)}(x_{13}) \, \bar{\cI}_{\ad}{}^{\a'}(x_{13}) \,  \cI_{\b}{}^{\bd'}(x_{23}) \,  \bar{\cI}_{\bd(2)}{}^{\b'(2)}(x_{23}) }{(x_{13}^{2})^{7/2} (x_{23}^{2})^{7/2}} \\
		& \hspace{15mm} \times \cH_{\a' \ad'(2) \b'(2) \bd' \g \gd}(X_{12}) \, .
	\end{split}
\end{align} 
The tensor three-point function is encoded in the following polynomial:
\begin{align}
	\cH(X; U, V, W) = \cH_{ \a \ad(2) \b(2) \bd  \g \gd }(X) \, \boldsymbol{U}^{\a \ad(2)}  \boldsymbol{V}^{\b(2) \bd}  \boldsymbol{W}^{\g \gd} \, .
\end{align}
After solving \eqref{Ch04-Diophantine equations}, we find the following linearly dependent polynomial structures:
%
\begin{align}
	\begin{split}
		&\big\{ Q_1 Q_2 Z_1 Z_2,P_3 Q_2 Z_2 \bar{P}_2,P_1 Z_1 Z_2 \bar{P}_2,P_3 Q_1 Q_2 \bar{P}_3,P_1 Q_1 Z_1 \bar{P}_3, \\
		& \hspace{10mm} P_1 P_3 \bar{P}_2
		\bar{P}_3, Q_1 Q_2 Q_3 \bar{Q}_3,Z_1 Z_2 Z_3 \bar{Q}_3,P_3 Q_2 \bar{P}_1 \bar{Q}_3,P_1 Z_1 \bar{P}_1 \bar{Q}_3, \\
		& \hspace{17mm} P_1 Q_3
		\bar{P}_2 \bar{Q}_3, P_2 Z_2 \bar{P}_2 \bar{Q}_3, P_2 Q_1 \bar{P}_3 \bar{Q}_3,P_3 Z_3 \bar{P}_3 \bar{Q}_3,Q_1 Z_1 \bar{Q}_1
		\bar{Q}_3, \\
		& \hspace{22mm} P_3 \bar{P}_2 \bar{Q}_1 \bar{Q}_3,Q_2 Z_2 \bar{Q}_2 \bar{Q}_3, P_1 \bar{P}_3 \bar{Q}_2 \bar{Q}_3,Q_3 Z_3
		\bar{Q}_3^2,P_2 \bar{P}_1 \bar{Q}_3^2,\bar{Q}_1 \bar{Q}_2 \bar{Q}_3^2 \big\}\,.
	\end{split}
\end{align}
Next we systematically apply the linear dependence relations \eqref{Ch04-Linear dependence 1} to this list, which results in the following linearly independent structures:
%
\begin{align}
	\big\{P_3 Q_1 Q_2 \bar{P}_3,Q_1 Q_2 Q_3 \bar{Q}_3,P_3 Q_2 \bar{P}_1 \bar{Q}_3,P_1 Q_3 \bar{P}_2 \bar{Q}_3,P_2 Q_1 \bar{P}_3
	\bar{Q}_3,\bar{Q}_1 \bar{Q}_2 \bar{Q}_3^2\big\}\,.
\end{align}
We now construct the ansatz for this three-point function using the linearly independent structures above. After imposing conservation on all three points the final solution is
%
\begin{align} \label{Ch04-QQbV}
	\begin{split}
		& \frac{A_1}{X^4}
		\Big(P_3 Q_1 Q_2 \bar{P}_3 + \sfrac{3}{2} P_1 Q_3 \bar{P}_2 \bar{Q}_3-\sfrac{3}{4} \bar{Q}_1 \bar{Q}_2\bar{Q}_3^2\Big) \\[1mm]
		& \hspace{5mm} +\frac{A_2}{X^4} \Big(P_1 Q_3 \bar{P}_2 \bar{Q}_3-\bar{Q}_1 \bar{Q}_2 \bar{Q}_3^2+Q_1 Q_2 Q_3 \bar{Q}_3\Big) \\[1mm]
		& \hspace{10mm} +\frac{A_3}{X^4} \Big(P_3 Q_2 \bar{P}_1 \bar{Q}_3-\sfrac{1}{2} P_1 Q_3 \bar{P}_2 \bar{Q}_3+P_2 Q_1 \bar{P}_3
		\bar{Q}_3+\sfrac{3}{4} \bar{Q}_1 \bar{Q}_2 \bar{Q}_3^2\Big)\,.
	\end{split}
\end{align}
Therefore we see that the correlation function $\langle J^{}_{(3/2,1)} \bar{J}'_{(3/2,1)} J''_{(1,0)} \rangle$ and, hence, $\langle Q \bar{Q} V \rangle$, is fixed up to three independent complex coefficients. After imposing the combined point-switch/reality condition on $Q$ and $\bar{Q}$, we find that the complex coefficients $A_{i}$ must be purely imaginary, i.e., $A_{i} = \text{i} \tilde{a}_{i}$. Hence, the correlation function $\langle Q \bar{Q} V \rangle$ is fixed up to three independent real parameters.

Next we determine the general structure of $\langle Q Q T \rangle$ and $\langle Q \bar{Q} T \rangle$, which are associated with the correlation functions $\langle J^{}_{(3/2,1)} J'_{(3/2,1)} J''_{(2,0)} \rangle$, $\langle J^{}_{(3/2,1)} \bar{J}'_{(3/2,1)} J''_{(2,0)} \rangle$ respectively using our general formalism. Since the number of structures grows rapidly with spin, we will simply present the final results after conservation. For $\langle J^{}_{(3/2,1)} J'_{(3/2,1)} J''_{(2,0)} \rangle$ we obtain a single independent structure (up to a complex coefficient):
%
\begin{align} \label{Ch04-QQT}
	\begin{split}
		&\frac{A_1}{X^3} \Big(Q_1 Q_3 Q_2^2 \bar{P}_1+\sfrac{7}{4} P_3 Q_2^2 \bar{P}_1^2+\sfrac{1}{2} P_1 Q_3 Q_2 \bar{P}_1
		\bar{P}_2 \\
		& \hspace{10mm} -\sfrac{5}{4} Q_1 Q_3 Q_2 \bar{P}_2 \bar{Q}_1 -5 Q_1 Q_2 \bar{P}_3 \bar{Q}_1 \bar{Q}_2-\sfrac{7}{2} Q_2 \bar{P}_1
		\bar{Q}_1 \bar{Q}_2 \bar{Q}_3 \\
		& \hspace{20mm} +\sfrac{1}{2} P_1 Q_3 \bar{P}_2^2 \bar{Q}_1+\sfrac{7}{4} P_2 Q_1 \bar{P}_2 \bar{P}_3
		\bar{Q}_1-\sfrac{5}{4} \bar{P}_2 \bar{Q}_1^2 \bar{Q}_2 \bar{Q}_3 \Big)\,.
	\end{split}
\end{align}
This solution is manifestly compatible with the point-switch symmetry resulting from setting $J = J'$, hence, $\langle Q Q T \rangle$ is unique up to a complex parameter. On the other hand, for $\langle J^{}_{(3/2,1)} \bar{J}'_{(3/2,1)} J''_{(2,0)} \rangle$ we obtain four independent conserved structures proportional to complex coefficients
%
\begin{align} \label{Ch04-QQbT}
	\begin{split}
		& \frac{A_1}{X^3} \Big( Q_1^2 Q_3 Q_2^2 + \sfrac{6}{7} P_1 Q_2 \bar{P}_1
		\bar{Q}_2 \bar{Q}_3 +\sfrac{6}{7} P_2 Q_1 \bar{P}_2 \bar{Q}_1 \bar{Q}_3 \\
		& \hspace{30mm} + \sfrac{6}{7} P_1 \bar{P}_2 \bar{Q}_1 \bar{Q}_2 \bar{Q}_3-\sfrac{10}{7} Q_1 Q_2 \bar{Q}_1 \bar{Q}_2 \bar{Q}_3\Big) \\[1mm]
		& +\frac{A_2}{X^3} \Big(P_2 Q_2 Q_1^2 \bar{P}_3+P_3 Q_2^2 Q_1 \bar{P}_1-P_2 Q_1 \bar{P}_2
		\bar{Q}_1 \bar{Q}_3 \\
		& \hspace{15mm} -P_1 Q_2 \bar{P}_1 \bar{Q}_2 \bar{Q}_3-P_1 \bar{P}_2 \bar{Q}_1 \bar{Q}_2 \bar{Q}_3+Q_2 Q_1 \bar{Q}_1 \bar{Q}_2 \bar{Q}_3 \Big) \\[1mm]
		&+\frac{A_3}{X^3} \Big(P_1 Q_1 Q_2 Q_3 \bar{P}_2-\sfrac{3}{7} P_2 Q_1 \bar{P}_2 \bar{Q}_1
		\bar{Q}_3-\sfrac{13}{14} P_1 \bar{P}_2 \bar{Q}_1 \bar{Q}_2 \bar{Q}_3 \\
		& \hspace{25mm} -\sfrac{3}{7} P_1 Q_2 \bar{P}_1 \bar{Q}_2
		\bar{Q}_3+\sfrac{3}{14} Q_1 Q_2 \bar{Q}_1 \bar{Q}_2 \bar{Q}_3 \Big) + \frac{A_4}{X^3} P_1^2 Q_3 \bar{P}_2^2 \,.
	\end{split}
\end{align}
After imposing the combined point-switch/reality condition, we find that the complex coefficients $A_{i}$ must be purely real. Hence, the three-point function $\langle Q \bar{Q} T \rangle$ is fixed up to four independent real parameters. The results \eqref{Ch04-QQV}, \eqref{Ch04-QQbV}, \eqref{Ch04-QQT}, \eqref{Ch04-QQbT} are in agreement with those found in \cite{Buchbinder:2022cqp}.



\subsection{General structure of three-point functions}\label{Ch04-subsection3.2}


In four dimensions, three-point correlation functions of bosonic higher-spin conserved currents have been analysed in the following 
publications~\cite{Stanev:2012nq, Zhiboedov:2012bm} (see \cite{Osborn:1998qu,Kuzenko:1999pi, Buchbinder:2022kmj} for supersymmetric results). For three-point functions involving 
bosonic currents $J_{(s,0)} = J_{\a(s) \ad(s)}$, the general structure of the three-point function $\langle J^{}_{(s_{1},0)} J'_{(s_{2},0)} J''_{(s_{3},0)} \rangle$ was found to be fixed up to the 
following form \cite{Maldacena:2011jn, Giombi:2011rz, Zhiboedov:2012bm}:
\begin{equation}
	\langle  J^{}_{(s_{1},0)} J'_{(s_{2},0)} J''_{(s_{3},0)}  \rangle = \sum_{I=1}^{2 \min(s_{1}, s_{2}, s_{3}) + 1} a_{I} \, \langle  J^{}_{(s_{1},0)} J'_{(s_{2},0)} J''_{(s_{3},0)}  \rangle_{I} \, ,
\end{equation}
where $a_{I}$ are real coefficients and $\langle J^{}_{(s_{1},0)} J'_{(s_{2},0)} J''_{(s_{3},0)} \rangle_{I}$ are linearly independent conserved structures.\footnote{Note that if the reality condition is not imposed, the three-point function is fixed up to $2 \min(s_{i}) + 1$ structures with complex coefficients.} 
Among the $2 \min(s_{1}, s_{2}, s_{3}) + 1$ structures, $\min(s_{1}, s_{2}, s_{3}) + 1$ are parity-even while $\min(s_{1}, s_{2}, s_{3}) $ are parity odd. For correlation functions involving identical fields we must also impose point-switch symmetries. The following classification holds:
\begin{itemize}
	\item For three-point functions $\langle J^{}_{(s,0)} J'_{(s,0)} J''_{(s,0)} \rangle$ there are $2 s + 1$ conserved structures, $s+1$ being parity even and $s$ being parity odd.  
	When the fields coincide, i.e. $J = J' $ the number of structures is reduced to the $s+1$ parity-even structures in the case when the spin $s$ is even,
	or to the $s$ parity-odd structures in the case when $s$ is odd. 
	
	\item For three-point functions $\langle J^{}_{(s_{1},0)} J'_{(s_{1},0)} J''_{(s_{2},0)} \rangle$, there are $2 \min(s_{1}, s_{2}) + 1$ conserved structures, 
	$\min(s_{1}, s_{2}) +1$ being parity even and $\min(s_{1}, s_{2}) $ being parity odd.  
	For $J = J'$, the number of structures is reduced to the $\min(s_{1}, s_{2})+1$ parity-even structures in the case when the spin $s_2$ is even, or to the $\min(s_{1}, s_{2})$ parity-odd structures in the case when $s_2$ is odd.
\end{itemize}
Note that the above classification is consistent with the results of \cite{Stanev:2012nq}, and we have explicitly reproduced them up to $s_{i} = 10$ in our computational approach. 

Now let us discuss three-point functions involving currents with $q = 1$, which define ``supersymmetry-like" fermionic higher-spin currents. 
The possible correlation functions that we can construct from these are 
$\langle J^{}_{(s_{1}, 1)} \, J'_{(s_{2}, 1)} \, J''_{(s_{3}, 0)}\rangle$ and $\langle J^{}_{(s_{1}, 1)} \, \bar{J}'_{(s_{2}, 1)} \, J''_{(s_{3}, 0)} \rangle$. 
Note that for $s_{1} = s_{2} = 3/2$ and $s_{3} = 1,2$ we obtain the familiar three-point functions \eqref{Ch04-Susy current correlators - 1}. 
Based on our computational analysis we found that the three-point function $\langle J^{}_{(s_{1}, 1)} \, J'_{(s_{2}, 1)} \, J''_{(s_{3}, 0)}\rangle$ is fixed up 
to a unique structure after conservation in general. On the other hand, we found that three-point functions of the form $\langle J^{}_{(s_{1}, 1)} \, \bar{J}'_{(s_{2}, 1)} \, J''_{(s_{3}, 0)} \rangle$ 
are fixed up to $2 \min(s_{1}, s_{2}, s_{3}) + 1$ independent conserved structures. It's important to note that for these three-point functions there is no notion of parity-even/odd 
structures. 

The remainder of this section is dedicated to classifying the number of independent structures in the general three-point functions
\begin{align} \label{Ch04-Possible conserved three-point functions}
	\langle J^{}_{(s_{1}, q_{1})} \, J'_{(s_{2}, q_{2})} \, J''_{(s_{3}, q_{3})} \rangle \, , && \langle J^{}_{(s_{1}, q_{1})} \, \bar{J}'_{(s_{2}, q_{2})} \, J''_{(s_{3}, q_{3})} \rangle  \, ,
\end{align}
for arbitrary $(s_i, q_i)$. We investigated the general structure of these three-point functions up to $s_{i} = 10$. Provided that the inequalities \eqref{Ch04-e1} are satisfied, we conjecture that the following classification holds in general:
\begin{itemize}
	\item For three-point functions $\langle J^{}_{(s_{1},q_{1})} J'_{(s_{2},q_{2})} J''_{(s_{3}, q_{3})} \rangle$, $\langle J^{}_{(s_{1}, q_{1})} \bar{J}'_{(s_{2}, q_{2})} J''_{(s_{3}, q_{3})} \rangle$ with $q_{1} \neq q_{2} \neq q_{3}$, there is a unique solution in general. Similarly, the three-point function is also unique for the cases: i) $q_{1} = 0$, $q_{2} \neq q_{3}$, and ii) $q_{1} = q_{2} = 0$ with $q_{3} \neq 0$.
	
	\item For three-point functions $\langle J^{}_{(s_{1},q)} J'_{(s_{2},q)} J''_{(s_{3},0)} \rangle$ there is a unique solution up to a complex coefficient. However, for the case where $s_{1} = s_{2}$ (fermionic or bosonic) and 
	$J = J'$, the structure survives the resulting point-switch symmetry only when $s_{3}$ is an 
	even integer.
	
	\item For three-point functions $\langle J^{}_{(s_{1},q)} \bar{J}'_{(s_{2},q)} J''_{(s_{3},0)} \rangle$ we obtain quite a non-trivial result which we will now explain.
	The number of structures, $N(s_{1}, s_{2}, s_{3};q)$, obeys the following formula:
	\begin{equation} \label{Ch04-Number of structures}
		N(s_{1}, s_{2}, s_{3}; q) = 2 \min(s_{1}, s_{2}, s_{3}) + 1 - \text{max} \big( \tfrac{q}{2} - | s_{3} - \min(s_{1}, s_{2}) | , 0 \big) \, ,
	\end{equation}
	where $s_{1}, s_{2}$ are simultaneously integer/half-integer, for integer $s_{3}$.
	This formula can be obtained with the following method. 
	Let us fix $s_{1}$, $s_{2}$ and let $q \geq 2$, where we recall that $q$ is necessarily even/odd when $s$ is integer/half-integer valued, and also, since $J_{(s,q)} := J_{\a(s+\frac{q}{2}) \ad(s-\frac{q}{2})}$ it follows that the maximal allowed value of $q$ is  $2 \min(s_{1}, s_{2}) - 2$. By varying $s_{3}$ and computing the resulting conserved three-point function, 
	one can notice that if $s_3$ lies within the interval
	%
	\begin{equation} \label{Ch04-weird interval}
		\min(s_{1}, s_{2}) - \frac{q}{2} < s_{3} < \min(s_{1}, s_{2}) + \frac{q}{2} \, ,
	\end{equation}
	then the number of structures is decreased from $2 \min(s_{1}, s_{2}, s_{3}) + 1$ by 
	\begin{equation}
		\d N(s_{1}, s_{2}, s_{3}; q)= \frac{q}{2} - | s_{3} - \min(s_{1}, s_{2}) |  \,. 
		\label{e2}
	\end{equation}
	For $s_{3}$ outside the interval \eqref{Ch04-weird interval} there is always $2 \min(s_{1}, s_{2}, s_{3}) + 1$ structures in general.
	It should also be noted that \eqref{Ch04-Number of structures} is also valid for $q = 0,1$ (by virtue of the $\max()$ function). In these cases the additional 
	term does not contribute and we obtain $N(s_{1}, s_{2}, s_{3}; 0) = N(s_{1}, s_{2}, s_{3}; 1) = 2 \min(s_{1}, s_{2}, s_{3}) + 1$. Explicit solutions for particular cases are presented in Appendix \ref{Appendix4B}.
	
	Below we tabulate the number of structures in the conserved three-point functions 
	$\langle J^{}_{(s_{1},q)} \bar{J}'_{(s_{2},q)} J''_{(s_{3},0)} \rangle$ for some fixed $s_{1}, s_{2}$ while varying $q$ and $s_{3}$. The coloured values are contained in the interval \eqref{Ch04-weird interval} defined by $s_{1}$, $s_{2}$ and $q$. We have used colour to identify the pattern in the number of structures. Analogous tables can be constructed for any choice of $s_{1}$, $s_{2}$ and it is easy to see that the results are consistent with the general formula \eqref{Ch04-Number of structures}, which appears to hold for all such correlators within the bounds of our computational limitations $(s_{i} \leq 10)$.
	\begin{center}
		\captionof{table}{No. of structures in $\langle J^{}_{(s_{1},q)} \bar{J}'_{(s_{2},q)} J''_{(s_{3},0)} \rangle$ for $s_{1} = 5$, $s_{2} = 6$.\label{Tab:T1}}
		\vspace{5mm}
		{\def\arraystretch{1.3}
			{\setlength{\tabcolsep}{3mm}
				\begin{tabular}{c*{10}{c}}
					\toprule
					& \multicolumn{9}{c}{$s_{3}$} \\
					\cmidrule(lr){2-10}
					\hspace{3mm} $q$ \hspace{3mm} & 1 & 2 & 3 & 4 & 5 & 6 & 7 & 8 & 9 \\
					\cmidrule(lr){1-1}
					\cmidrule(lr){2-10}
					0 & 3 & 5 & 7 & 9 & 11 & 11 & 11 & 11  & 11 \\[1mm]
					2 & 3 & 5 & 7 & 9 & \textcolor{red}{10} & 11 & 11 & 11  & 11 \\[1mm]
					4 & 3 & 5 & 7 & \textcolor{red}{8} & \textcolor{blue}{9} & \textcolor{red}{10} & 11 & 11 & 11\\[1mm]
					6 & 3 & 5 & \textcolor{red}{6} & \textcolor{blue}{7} & \textcolor{teal}{8} & \textcolor{blue}{9} & \textcolor{red}{10} & 11 & 11\\[1mm]
					8 & 3& \textcolor{red}{4} & \textcolor{blue}{5} & \textcolor{teal}{6} & \textcolor{violet}{7} & \textcolor{teal}{8} & \textcolor{blue}{9} & \textcolor{red}{10} & 11\\[1mm]
					\bottomrule
		\end{tabular}}}
		\caption*{ \textcolor{red}{$\bullet$} $\d N = 1$ \hspace{5mm} \textcolor{blue}{$\bullet$} $\d N = 2$ \hspace{5mm} \textcolor{teal}{$\bullet$} $\d N = 3$ \hspace{5mm} \textcolor{violet}{$\bullet$} $\d N = 4$}
	\end{center}

	\begin{center}
		\captionof{table}{No. of structures in $\langle J^{}_{(s_{1},q)} \bar{J}'_{(s_{2},q)} J''_{(s_{3},0)} \rangle$ for $s_{1} = 9/2$, $s_{2} = 11/2.$\label{Tab:T2}}
		\vspace{5mm}
		{\def\arraystretch{1.3}
			{\setlength{\tabcolsep}{3mm}
				\begin{tabular}{ c*{9}{c}}
					\toprule
					& \multicolumn{8}{c}{$s_{3}$} \\
					\cmidrule(lr){2-9}
					\hspace{3mm} $q$ \hspace{3mm} & 1 & 2 & 3 & 4 & 5 & 6 & 7 & 8 \\
					\cmidrule(lr){1-1}
					\cmidrule(lr){2-9}
					1 & 3 & 5 & 7 & 9 & 10 & 10 & 10 & 10 \\[1mm]
					3 & 3 & 5 & 7 & \textcolor{red}{8} & \textcolor{red}{9} & 10 & 10 & 10 \\[1mm]
					5 & 3 & 5 & \textcolor{red}{6} & \textcolor{blue}{7} & \textcolor{blue}{8} & \textcolor{red}{9} & 10 & 10 \\[1mm]
					7 & 3 & \textcolor{red}{4} & \textcolor{blue}{5} & \textcolor{teal}{6} & \textcolor{teal}{7} & \textcolor{blue}{8} & \textcolor{red}{9} & 10 \\[1mm]
					\bottomrule
		\end{tabular}}}
		\caption*{ \textcolor{red}{$\bullet$} $\d N = 1$ \hspace{5mm} \textcolor{blue}{$\bullet$} $\d N = 2$ \hspace{5mm} \textcolor{teal}{$\bullet$} $\d N = 3$}
	\end{center}

	\item For three-point functions $\langle J^{}_{(s_{1},q)} \bar{J}'_{(s_{1},q)} J''_{(s_{2},0)} \rangle$ the number of structures adheres to the formula \eqref{Ch04-Number of structures}. However, for $J = J'$, we must impose the combined point-switch/reality condition. After imposing this constraint we find that the free complex parameters must be purely real/imaginary for $s_{2}$ even/odd. 
	
\end{itemize}
The above classification appears to be complete, and we have not found any other permutations of fields/spins which give rise to new results.

\section{Summary of results}

In this chapter we undertook the analysis of three-point functions of conserved currents in four dimensional conformal field theory, where it is shown that the results are quite different compared to the 3D CFT case. In particular we generalised the formalism of Osborn and Petkou \cite{Osborn:1993cr} to primary fields in an arbitrary Lorentz representation in four dimensions. For conserved vector-like currents in the $(s,s)$ representation, we demonstrated that the three-point functions of conserved currents are fixed up to $2 \min(s_{i}) + 1$ structures, as shown in \cite{Stanev:2012nq,Zhiboedov:2012bm} some time ago. We then extended these results by proposing a classification for the general structure of three-point function for currents of currents in an arbitrary Lorentz representation up to a suitably high computational bound $(s_{i} \leq 10)$. These results are qualitatively new, and in particular we showed that for some three-point functions the number of independent structures deviates from $2 \min(s_{i}) + 1$.

Concerning the origin of the structures present in the three-point functions of higher-spin currents, let us recall that in four-dimensional conformal field theory the three-point function of the energy-momentum tensor is fixed up to three linearly independent parity-even structures and two parity-odd structures. The parity-even structures are known to correspond to theories of free bosons, free fermions, and a free abelian vector field respectively \cite{Osborn:1993cr}, which may be understood in terms of the four-dimensional extension of the Maldacena-Zhiboedov theorem, developed by Alba and Diab \cite{Alba:2015upa}. The parity-odd structures, however, do not seem to have an interpretation in terms of free field theories, and it remains an open problem to understand their origin. For higher-spin currents in $(s,s)$ representations we have shown that additional structures can arise in their three-point functions (increasing linearly with the minimum spin). It was conjectured by Zhiboedov \cite{Zhiboedov:2012bm} (see also Alba and Diab \cite{Alba:2015upa}) that the additional structures could correspond to theories of $(j,0)$ (anti-)self-dual tensors \cite{Siegel:1988gd}. Conserved currents were constructed for theories involving such fields in \cite{Gelfond:2006be}. For conserved currents belonging to more general $(m,n)$ representations, it is unclear why there is a reduction in the number of independent conserved structures in their three-point functions. It should also be noted that the analysis undertaken by Alba and Diab \cite{Alba:2015upa} does not apply to such currents.

\begin{subappendices}
	\section{Examples: three-point functions of bosonic higher-spin currents}\label{Appendix4A}
In this appendix we provide some examples of three-point functions of bosonic (vector-like) higher-spin currents, $\langle J^{}_{s_{1}} J'_{s_{2}} J''_{s_{3}} \rangle \equiv \langle J^{}_{(s_{1},0)} J'_{(s_{2},0)} J''_{(s_{3},0)} \rangle$, which are in general fixed up to $\min(s_{1}, s_{2}, s_{3}) + 1$ parity-even structures $(a_{i})$ and $\min(s_{1}, s_{2}, s_{3})$ parity-odd structures $(b_{i})$. In many cases the results are too lengthy to typeset in LaTeX, and so we present the outputs as they appear in Mathematica. \\[4mm]
\noindent
\textbf{Correlation function} $\langle J^{}_{1} J'_{1} J''_{3} \rangle$\textbf{:} \hspace{3mm}
\begin{flalign}
	\hspace{5mm} \includegraphics[width=0.7\textwidth]{1-1-3.pdf} &&
\end{flalign}
\textbf{Correlation function} $\langle J^{}_{1} J'_{1} J''_{4} \rangle$\textbf{:} \hspace{3mm}
\begin{flalign}
	\hspace{5mm} \includegraphics[width=0.7\textwidth]{1-1-4.pdf} &&
\end{flalign}
\newpage

A general feature of three-point functions is that the solutions reach a fixed length once the triangle inequalities are saturated. For example, if we consider the solution for $\langle J^{}_{1} J'_{1} J''_{2} \rangle$ given by \eqref{Ch04-1-1-2} and compare it to the two examples above, it is clear that the general structure of $\langle J^{}_{1} J'_{1} J''_{s} \rangle$ is similar to $\langle J^{}_{1} J'_{1} J''_{2} \rangle$ except multiplied by a factor of $Z_{3}^{s-2}$ (with appropriately modified coefficients). With this in mind it is possible to guess the general solution for $\langle J^{}_{1} J'_{1} J''_{s} \rangle$ for arbitrary $s$
\begin{align}
	& a_{1}
	X^{s-4} Z_3^{s-2} \Big( P_2 Q_1 Q_2 \bar{P}_1+P_1 \bar{P}_2 \bar{Q}_1 \bar{Q}_2 + \sfrac{2}{s} P_1 Q_2 \bar{P}_1 \bar{Q}_2 \\
	& \hspace{40mm} +\sfrac{2}{s} P_2 Q_1 \bar{P}_2
		\bar{Q}_1-\sfrac{4}{s (s+1)} Q_1 Q_2 \bar{Q}_1
		\bar{Q}_2\Big) \nonumber \\[2mm]
	& + a_2 X^{s-4} Z_3^{s-2} \Big(P_1 P_2 \bar{P}_1 \bar{P}_2-\sfrac{1}{s}P_2 Q_1 \bar{P}_2 \bar{Q}_1-\sfrac{1}{s} P_1 Q_2 \bar{P}_1 \bar{Q}_2+\sfrac{2}{s (s+1)} Q_1 Q_2 \bar{Q}_1 \bar{Q}_2\Big) \nonumber \\[2mm]
	& +\text{i} b_1 X^{s-4} Z_3^{s-2} \left(P_2 Q_1 Q_2 \bar{P}_1- P_1 \bar{P}_2
	\bar{Q}_1 \bar{Q}_2\right) \nonumber
\end{align}
The solution satisfies conservation on all three-points, as expected. In principle one can follow a similar procedure to find three-point functions of conserved currents for any two fixed spins and an arbitrary third spin. However, one must keep in mind that the number of solutions grows linearly with the minimum spin and, hence, they are quite difficult to construct in general. \\[4mm]
\textbf{Correlation function} $\langle J^{}_{1} J'_{2} J''_{3} \rangle$\textbf{:} \hspace{3mm}
\begin{flalign}
	\hspace{5mm} \includegraphics[width=0.75\textwidth]{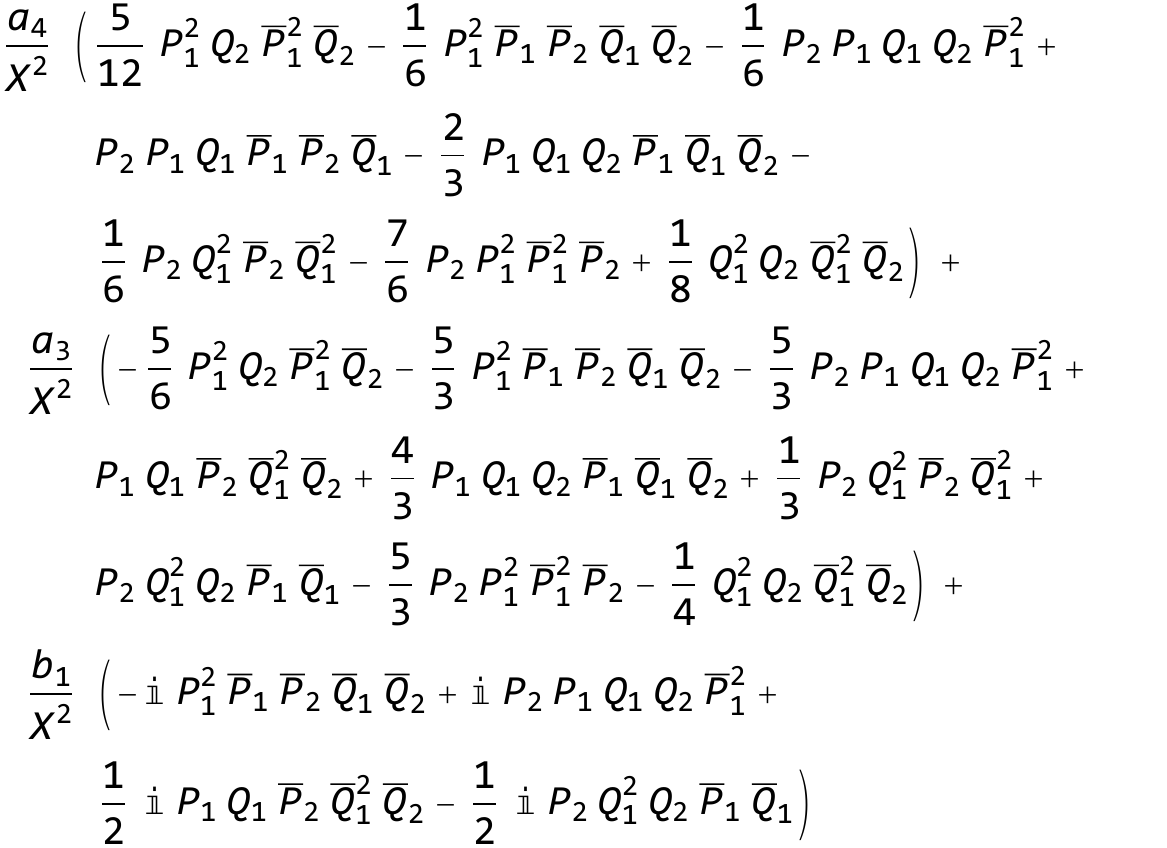} &&
\end{flalign}

\newpage

\noindent\textbf{Correlation function} $\langle J^{}_{2} J'_{2} J''_{3} \rangle$\textbf{:} \hspace{3mm}
\begin{flalign}
	\hspace{5mm} \includegraphics[width=0.95\textwidth]{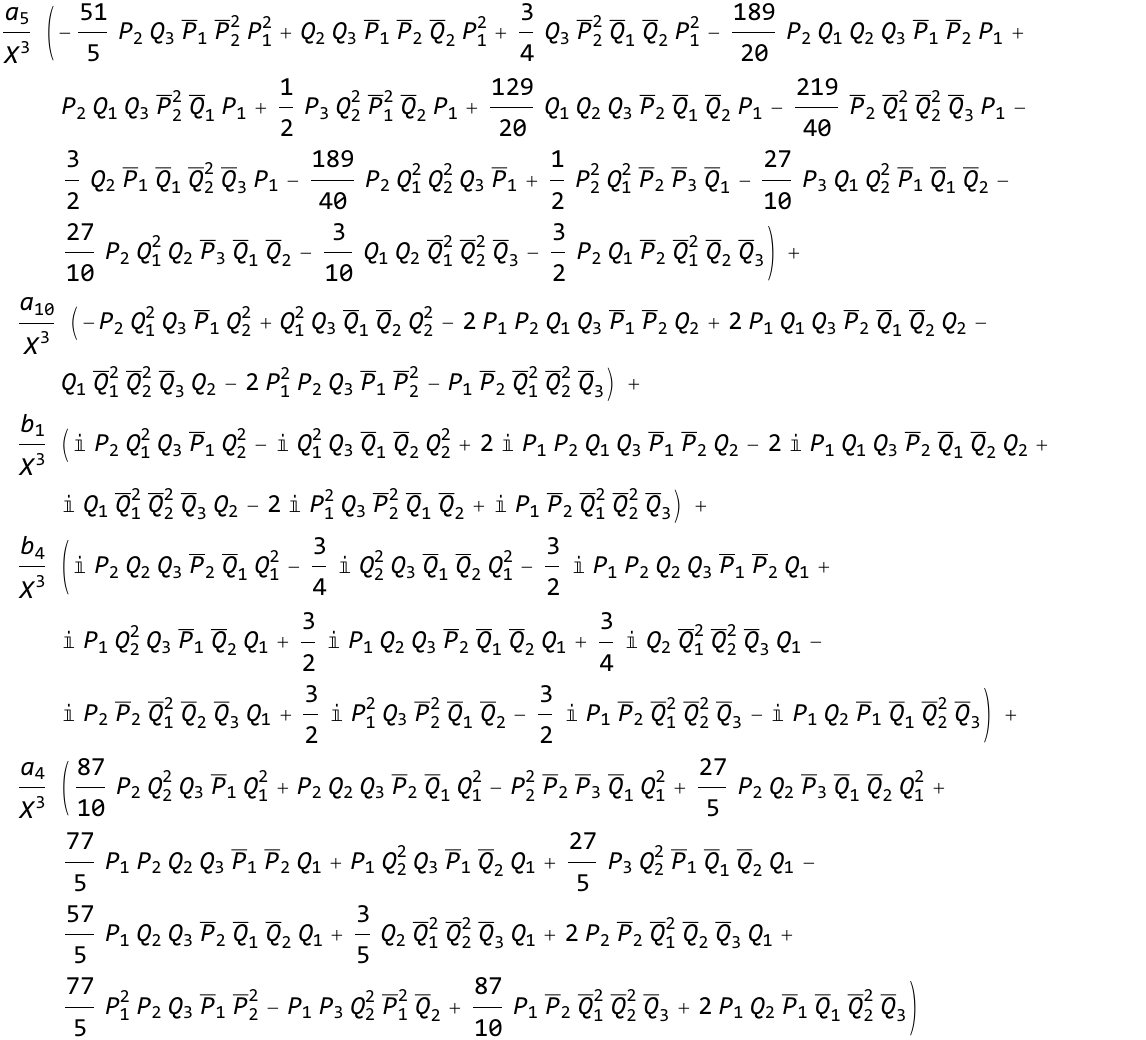} &&
\end{flalign}\\
\vfill

\newpage

\noindent\textbf{Correlation function} $\langle J^{}_{2} J'_{2} J''_{4} \rangle$\textbf{:} \hspace{3mm}
\begin{align}
	\includegraphics[width=0.9\textwidth, valign=c]{2-2-4.pdf}
\end{align} \\
\vfill

\newpage

\noindent\textbf{Correlation function} $\langle J^{}_{2} J'_{3} J''_{4} \rangle$\textbf{:} \hspace{3mm}
\begin{align}
	\includegraphics[width=0.9\textwidth, valign=c]{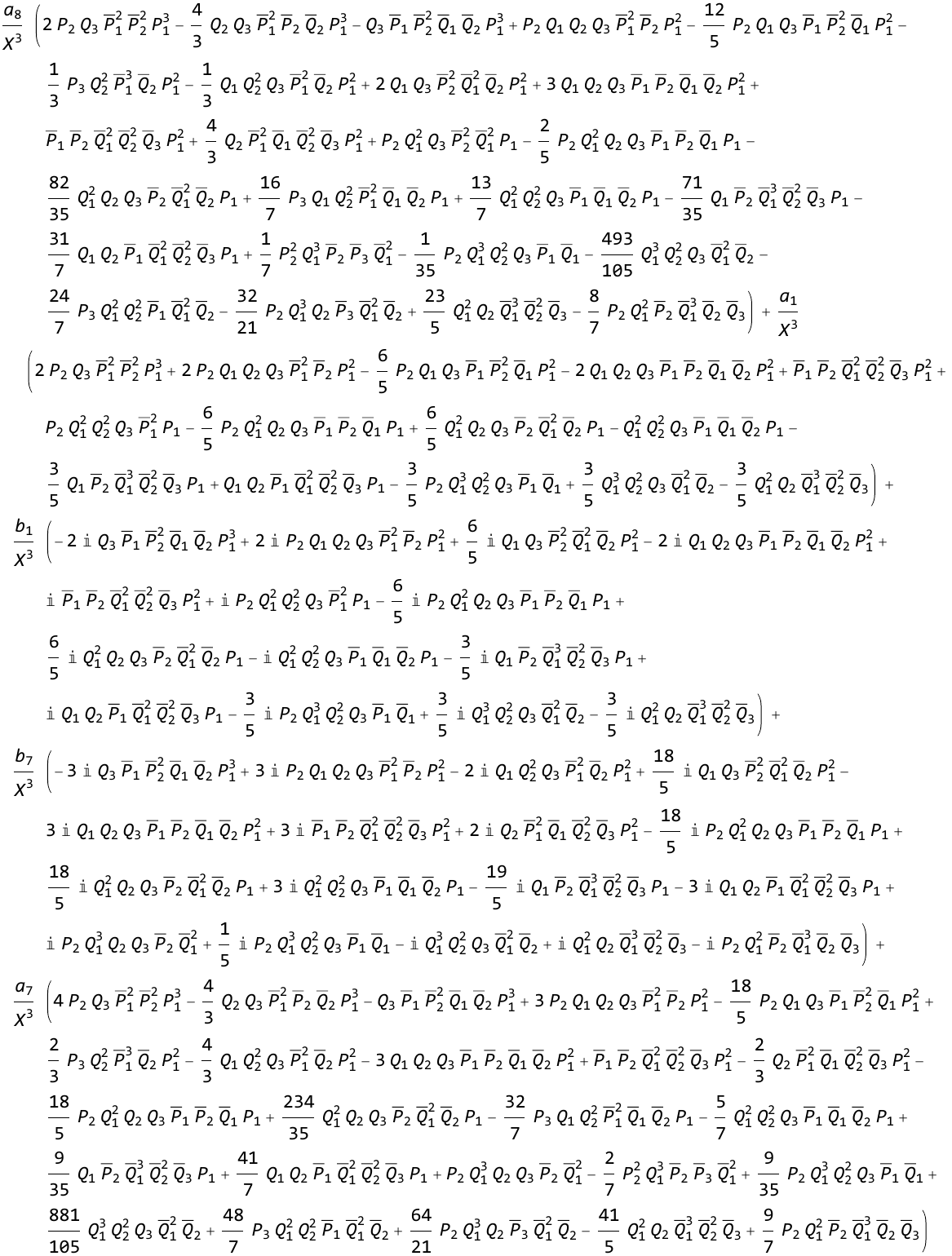}
\end{align} \\
\vfill
\newpage

\noindent\textbf{Correlation function} $\langle J^{}_{3} J'_{3} J''_{3} \rangle$\textbf{:} \hspace{3mm}
\begin{align}
	\includegraphics[width=0.9\textwidth, valign=c]{3-3-3.pdf}
\end{align} \\
\vfill

\newpage


	\section{Examples: three-point functions $\langle J^{}_{(s_{1},q)} \bar{J}'_{(s_{2},q)} J''_{(s_{3},0)} \rangle$} \label{Appendix4B}
In this appendix we provide some examples of three-point functions $\langle J^{}_{(s_{1},q)} \bar{J}'_{(s_{2},q)} J''_{(s_{3},0)} \rangle$. We compute two of the examples presented in Tables \ref{Tab:T1} \& \ref{Tab:T2} to illustrate the peculiar change in the number of independent conserved structures for particular values of $q$. Due to the large size of the solutions for increasing $s_{i}$, we present only some simple cases.

First let us consider the three-point function $\langle J^{}_{(s_{1},q)} \bar{J}'_{(s_{2},q)} J''_{(s_{3},0)} \rangle$ with $s_{1} = 5$, $s_{2} = 6$, $q = 8$ and $s_{3} = 2$. Using our formalism, all information about this correlation function is encoded in the following polynomial:
\begin{align}
	\cH(X; U, V, W) = \cH_{\a(1) \ad(9) \b(10) \bd(2) \g(2) \gd(2) }(X)  \, \boldsymbol{U}^{\a(1) \ad(9)}  \, \boldsymbol{V}^{\b(10) \bd(2)} \, \boldsymbol{W}^{\g(2) \gd(2)} \, .
\end{align}
There are 13 possible linearly independent structures that can be constructed in this case:
\begin{align}
	\begin{split}
		& \big\{ P_1 P_3 Q_1^2 Q_2 \bar{P}_3^2 \bar{Q}_3^6,P_1 Q_1^2 Q_2 Q_3 \bar{P}_3 \bar{Q}_3^7, P_3 Q_1 Q_2 \bar{P}_1 \bar{Q}_1 \bar{Q}_3^8, \\
		& \hspace{5mm} P_3 Q_1^2 Q_2 \bar{P}_3 \bar{Q}_1 \bar{Q}_3^7, P_1 Q_1 Q_2 Q_3 \bar{P}_1 \bar{Q}_3^8, P_1 P_3 Q_1 Q_2 \bar{P}_1 \bar{P}_3 \bar{Q}_3^7, \\
		& \hspace{10mm} P_1 Q_1 Q_3 \bar{P}_2 \bar{Q}_1 \bar{Q}_3^8,P_2 Q_1^2
		\bar{P}_3 \bar{Q}_1 \bar{Q}_3^8,P_1 \bar{P}_1 \bar{Q}_1 \bar{Q}_2 \bar{Q}_3^9, Q_1 \bar{Q}_1^2 \bar{Q}_2 \bar{Q}_3^9, \\
		& \hspace{15mm} P_1
		P_3 Q_2 \bar{P}_1^2 \bar{Q}_3^8, Q_1^2 Q_2 Q_3 \bar{Q}_1
		\bar{Q}_3^8, P_1^2 Q_3 \bar{P}_1 \bar{P}_2 \bar{Q}_3^8 \big\} \, .
	\end{split}
\end{align}
We now impose conservation on all three points. The following solution is obtained:
\begin{align}
	\begin{split}
		& \frac{A_1}{X^{11}} \Big(\frac{22}{189} Q_1 \bar{Q}_1^2 \bar{Q}_2 \bar{Q}_3^9-\frac{11}{63} P_1 \bar{P}_1 \bar{Q}_1 \bar{Q}_2
		\bar{Q}_3^9+\frac{11}{56} P_1^2 Q_3 \bar{P}_1 \bar{P}_2 \bar{Q}_3^8 \\
		& \hspace{15mm} -\frac{121}{378} Q_1^2 Q_2 Q_3 \bar{Q}_1
		\bar{Q}_3^8-\frac{143}{756} P_1 Q_1 Q_3 \bar{P}_2 \bar{Q}_1 \bar{Q}_3^8-\frac{11}{14} P_1 Q_1^2 Q_2 Q_3
		\bar{P}_3 \bar{Q}_3^7 \\
		& \hspace{25mm}+\frac{1}{2} P_3 Q_1^2 Q_2 \bar{P}_3 \bar{Q}_1 \bar{Q}_3^7+P_1 P_3 Q_1^2 Q_2 \bar{P}_3^2
		\bar{Q}_3^6 \Big) \\[2mm]
		& + \frac{A_2}{X^{11}} \Big(\frac{31}{54} Q_1 \bar{Q}_1^2 \bar{Q}_2 \bar{Q}_3^9-\frac{11}{18} P_1 \bar{P}_1
		\bar{Q}_1 \bar{Q}_2 \bar{Q}_3^9+\frac{11}{16} P_1^2 Q_3 \bar{P}_1 \bar{P}_2 \bar{Q}_3^8 \\
		& \hspace{15mm} -\frac{10}{27} Q_1^2 Q_2 Q_3 \bar{Q}_1 \bar{Q}_3^8-\frac{143}{216} P_1 Q_1 Q_3 \bar{P}_2 \bar{Q}_1 \bar{Q}_3^8-\frac{1}{4} P_2 Q_1^2 \bar{P}_3 \bar{Q}_1 \bar{Q}_3^8 \\
		& \hspace{25mm} -\frac{1}{4} P_1 Q_1^2 Q_2 Q_3 \bar{P}_3 \bar{Q}_3^7+P_1 P_3 Q_1 Q_2 \bar{P}_1
		\bar{P}_3 \bar{Q}_3^7-\frac{3}{4} P_3 Q_1^2 Q_2 \bar{P}_3 \bar{Q}_1 \bar{Q}_3^7 \Big) \\[2mm]
		& + \frac{A_3}{X^{11}}  \Big( \frac{2}{3} Q_1
		\bar{Q}_1^2 \bar{Q}_2 \bar{Q}_3^9-P_1 \bar{P}_1 \bar{Q}_1 \bar{Q}_2 \bar{Q}_3^9+P_1 Q_1 Q_2 Q_3 \bar{P}_1
		\bar{Q}_3^8 \\
		& \hspace{25mm} + P_1^2 Q_3 \bar{P}_1 \bar{P}_2 \bar{Q}_3^8-\frac{2}{3} Q_1^2 Q_2 Q_3 \bar{Q}_1
		\bar{Q}_3^8-\frac{2}{3} P_1 Q_1 Q_3 \bar{P}_2 \bar{Q}_1 \bar{Q}_3^8\Big) \\[2mm]
		& + \frac{A_4}{X^{11}} \Big( \frac{43}{27} Q_1
		\bar{Q}_1^2 \bar{Q}_2 \bar{Q}_3^9-\frac{8}{9} P_1 \bar{P}_1 \bar{Q}_1 \bar{Q}_2 \bar{Q}_3^9+P_1 P_3 Q_2
		\bar{P}_1^2 \bar{Q}_3^8 \\
		& \hspace{15mm} +P_1^2 Q_3 \bar{P}_1 \bar{P}_2 \bar{Q}_3^8-\frac{97}{54} Q_1^2 Q_2 Q_3 \bar{Q}_1
		\bar{Q}_3^8-2 P_3 Q_1 Q_2 \bar{P}_1 \bar{Q}_1 \bar{Q}_3^8 \\
		& \hspace{25mm} -\frac{44}{27} P_1 Q_1 Q_3 \bar{P}_2 \bar{Q}_1
		\bar{Q}_3^8-\frac{1}{2} P_2 Q_1^2 \bar{P}_3 \bar{Q}_1 \bar{Q}_3^8 \Big) \, ,
	\end{split}
\end{align}
where $A_{i}$ are complex coefficients. Hence we see that this three-point function is fixed up to four independent conserved structures. Recall that for $q = 0$, the three-point function $\langle J^{}_{(s_{1},q)} \bar{J}'_{(s_{2},q)} J''_{(s_{3},0)} \rangle$ reduces to a three-point function of vector-like currents. Hence, we should expect $2 \min(s_{1}, s_{2}, s_{3})+ 1 = 5$ independent structures. Similar results can be obtained for other values of $q$ and $s_{3}$, which are contained in Table \ref{Tab:T1}.

Next, let us consider the three-point function $\langle J^{}_{(s_{1},q)} \bar{J}'_{(s_{2},q)} J''_{(s_{3},0)} \rangle$ with $s_{1} = 9/2$, $s_{2} = 11/2$, $q = 7$ and $s_{3} = 2$. All information about this correlation function is encoded in the following polynomial:
\begin{align}
	\cH(X; U, V, W) = \cH_{\a(1) \ad(8) \b(9) \bd(2) \g(2) \gd(2) }(X)  \, \boldsymbol{U}^{\a(1) \ad(8)}  \, \boldsymbol{V}^{\b(9) \bd(2)} \, \boldsymbol{W}^{\g(2) \gd(2)} \, .
\end{align}
In this case there are also 13 possible linearly independent structures:
\begin{align}
	\begin{split}
		& \big\{ P_1 P_3 Q_1^2 Q_2 \bar{P}_3^2 \bar{Q}_3^5,P_1 Q_1^2 Q_2 Q_3 \bar{P}_3 \bar{Q}_3^6, P_1
		P_3 Q_2 \bar{P}_1^2 \bar{Q}_3^7, \\
		& \hspace{8mm} P_3 Q_1^2 Q_2 \bar{P}_3 \bar{Q}_1 \bar{Q}_3^6,P_1 Q_1 Q_2 Q_3 \bar{P}_1 \bar{Q}_3^7, P_1 Q_1 Q_3 \bar{P}_2 \bar{Q}_1 \bar{Q}_3^7,   \\
		& \hspace{15mm} Q_1^2 Q_2 Q_3 \bar{Q}_1 \bar{Q}_3^7,P_3 Q_1 Q_2 \bar{P}_1 \bar{Q}_1 \bar{Q}_3^7, P_1 P_3 Q_1 Q_2 \bar{P}_1 \bar{P}_3 \bar{Q}_3^6, \\
		& \hspace{20mm} P_2 Q_1^2 \bar{P}_3 \bar{Q}_1 \bar{Q}_3^7,P_1 \bar{P}_1 \bar{Q}_1 \bar{Q}_2 \bar{Q}_3^8, P_1^2 Q_3 \bar{P}_1 \bar{P}_2 \bar{Q}_3^7, Q_1 \bar{Q}_1^2 \bar{Q}_2 \bar{Q}_3^8 \big\} \, .
	\end{split}
\end{align}
We now impose conservation on all three points, and the following solution is obtained:
\begin{align}
	\begin{split}
		& \frac{A_{1}}{X^{10}} \Big( \frac{5}{36} Q_1 \bar{Q}_1^2 \bar{Q}_2 \bar{Q}_3^8-\frac{5}{24} P_1 \bar{P}_1 \bar{Q}_1 \bar{Q}_2
		\bar{Q}_3^8+\frac{5}{21} P_1^2 Q_3 \bar{P}_1 \bar{P}_2 \bar{Q}_3^7 \\
		& \hspace{15mm}-\frac{25}{72} Q_1^2 Q_2 Q_3 \bar{Q}_1
		\bar{Q}_3^7-\frac{115}{504} P_1 Q_1 Q_3 \bar{P}_2 \bar{Q}_1 \bar{Q}_3^7-\frac{5}{6} P_1 Q_1^2 Q_2 Q_3 \bar{P}_3
		\bar{Q}_3^6 \\
		& \hspace{25mm} + \frac{1}{2} P_3 Q_1^2 Q_2 \bar{P}_3 \bar{Q}_1 \bar{Q}_3^6+P_1 P_3 Q_1^2 Q_2 \bar{P}_3^2
		\bar{Q}_3^5 \Big) \\[2mm]
		& +\frac{A_{2}}{X^{10}} \Big(\frac{7}{12} Q_1 \bar{Q}_1^2 \bar{Q}_2 \bar{Q}_3^8-\frac{5}{8} P_1 \bar{P}_1
		\bar{Q}_1 \bar{Q}_2 \bar{Q}_3^8+\frac{5}{7} P_1^2 Q_3 \bar{P}_1 \bar{P}_2 \bar{Q}_3^7 \\
		& \hspace{15mm} -\frac{3}{8} Q_1^2 Q_2 Q_3 \bar{Q}_1 \bar{Q}_3^7-\frac{115}{168} P_1 Q_1 Q_3 \bar{P}_2 \bar{Q}_1 \bar{Q}_3^7-\frac{1}{4} P_2 Q_1^2
		\bar{P}_3 \bar{Q}_1 \bar{Q}_3^7 \\
		& \hspace{25mm} -\frac{1}{4} P_1 Q_1^2 Q_2 Q_3 \bar{P}_3 \bar{Q}_3^6+P_1 P_3 Q_1 Q_2 \bar{P}_1
		\bar{P}_3 \bar{Q}_3^6-\frac{3}{4} P_3 Q_1^2 Q_2 \bar{P}_3 \bar{Q}_1 \bar{Q}_3^6\Big) \\[2mm]
		& +\frac{A_{3}}{X^{10}} \Big(\frac{2}{3} Q_1
		\bar{Q}_1^2 \bar{Q}_2 \bar{Q}_3^8-P_1 \bar{P}_1 \bar{Q}_1 \bar{Q}_2 \bar{Q}_3^8+P_1 Q_1 Q_2 Q_3 \bar{P}_1
		\bar{Q}_3^7 \\
		& \hspace{15mm} +P_1^2 Q_3 \bar{P}_1 \bar{P}_2 \bar{Q}_3^7-\frac{2}{3} Q_1^2 Q_2 Q_3 \bar{Q}_1
		\bar{Q}_3^7-\frac{2}{3} P_1 Q_1 Q_3 \bar{P}_2 \bar{Q}_1 \bar{Q}_3^7\Big) \\[2mm]
		& +\frac{A_{4}}{X^{10}} \Big(\frac{19}{12} Q_1
		\bar{Q}_1^2 \bar{Q}_2 \bar{Q}_3^8-\frac{7}{8} P_1 \bar{P}_1 \bar{Q}_1 \bar{Q}_2 \bar{Q}_3^8+P_1 P_3 Q_2
		\bar{P}_1^2 \bar{Q}_3^7 \\
		& \hspace{15mm} +P_1^2 Q_3 \bar{P}_1 \bar{P}_2 \bar{Q}_3^7-\frac{43}{24} Q_1^2 Q_2 Q_3 \bar{Q}_1
		\bar{Q}_3^7-2 P_3 Q_1 Q_2 \bar{P}_1 \bar{Q}_1 \bar{Q}_3^7 \\
		& \hspace{25mm} -\frac{13}{8} P_1 Q_1 Q_3 \bar{P}_2 \bar{Q}_1 \bar{Q}_3^7-\frac{1}{2} P_2 Q_1^2 \bar{P}_3 \bar{Q}_1 \bar{Q}_3^7\Big) \, ,
	\end{split}
\end{align}
where $A_{i}$ are complex coefficients. Hence we see that this three-point function is fixed up to four independent conserved structures. Recall that for the $q = 1$ case we expect $2 \min(s_{1}, s_{2}, s_{3})+ 1 = 5$ independent structures. Similar results are obtained for other values of $q$ and $s_{3}$, and with further testing we obtain Table \ref{Tab:T2}.


	\section{Examples: three-point functions involving scalars and spinors} \label{Appendix4C}
In this appendix we provide some examples of three-point functions involving scalars, spinors and a conserved tensor operator. The results here serve as a consistency check against those presented in \cite{Osborn:1993cr,Elkhidir:2014woa}.

\noindent
\textbf{Correlation function} $\langle O \, O' J_{(s,0)} \rangle$\textbf{:}\\[2mm]
Let $O$, $O'$ be scalar operators of dimension $\D_{1}$ and $\D_{2}$ respectively. We consider the three-point function $\langle O \, O' J_{(s,0)} \rangle$. According to the formula \eqref{Ch04-e1}, a three-point function can be constructed only if $J$ is in the $(s,s)$ representation. Using the general formula, the ansatz for this three-point function is:
\begin{align}
	\langle O(x_{1}) \, O'(x_{2}) \, J_{\g(s) \gd(s)}(x_{3}) \rangle &= \frac{ 1 }{(x_{13}^{2})^{\D_{1}} (x_{23}^{2})^{\D_{2}}} \, \cH_{\g(s) \gd(s)}(X_{12}) \, .
\end{align} 
All information about this correlation function is encoded in the following polynomial:
\begin{align}
	\cH(X; W) = \cH_{ \g(s) \gd(s) }(X)  \, \boldsymbol{W}^{\g(s) \gd(s)} \, .
\end{align}
We recall that $\cH$ satisfies the homogeneity property $\cH(X) = X^{s+2 - \D_{1} - \D_{2}} \hat{\cH}(X)$, where $\hat{\cH}(X)$ is homogeneous degree $0$. The only possible structure for $\hat{\cH}(X)$ is:
\begin{align}
	\hat{\cH}(X; W) = A \, Z_{3}^{s} \, .
\end{align}
where $A$ is a complex coefficient. After imposing conservation on $x_{3}$ using the methods outlined in subsection \ref{Ch04-subsubsection2.2.2}, we find 
\begin{align}
	D_{3} \tilde{\cH}(X; W) = A \left(\D_1-\D _2\right) (-1)^{s+1} s (s+1) Z_3^{s-1} X^{\D _1-\D_2-s-3} \, .
\end{align}
Hence, we find that this three-point function is compatible with conservation on $x_{3}$ only for $\D_{1} = \D_{2}$. When the scalars $O$, $O'$ coincide, then the solution satisfies the point-switch symmetry associated with exchanging $x_{1}$ and $x_{2}$ only for even $s$. This result is in agreement with \cite{Elkhidir:2014woa}. \\[5mm]
\noindent
\textbf{Correlation function} $\langle \psi \, \bar{\psi}' J_{(s,q)} \rangle$\textbf{:}\\[2mm]
Let $\psi$, $\bar{\psi}'$ be spinor operators of dimension $\D_{1}$ and $\D_{2}$ respectively. We now consider the three-point function $\langle \psi \, \bar{\psi}' J_{(s,q)} \rangle$. According to the formula \eqref{Ch04-e1}, a three-point function can be constructed only if $J$ belongs to the representations $(s,s)$, $(s-1,s+1)$ or $(s+1,s-1)$ (the latter two corresponding to $q = 2$). First consider the $(s,s)$ representation. Using the general formula, the ansatz for this three-point function is:
\begin{align}
	\langle \psi_{\a}(x_{1}) \, \bar{\psi}_{\bd}'(x_{2}) \, J_{\g(s) \gd(s)}(x_{3}) \rangle &= \frac{ \cI_{\a}{}^{\ad'}(x_{13}) \, \bar{\cI}_{\bd}{}^{\b'}(x_{23}) }{(x_{13}^{2})^{\D_{1}} (x_{23}^{2})^{\D_{2}}} \, \cH_{\ad' \b' \g(s) \gd(s)}(X_{12}) \, .
\end{align} 
All information about this correlation function is encoded in the following polynomial:
\begin{align}
	\cH(X; U, V, W) = \cH_{ \ad \b \g(s) \gd(s) }(X)  \, \boldsymbol{U}^{\ad}  \boldsymbol{V}^{\b } \boldsymbol{W}^{\g(s) \gd(s)} \, .
\end{align}
We recall that $\cH$ satisfies the homogeneity property $\cH(X) = X^{s+2 - \D_{1} - \D_{2}} \hat{\cH}(X)$, where $\hat{\cH}(X)$ is homogeneous degree $0$. In this case there are two possible linearly independent structures for $\hat{\cH}(X)$:
\begin{align}
	\hat{\cH}(X; U, V, W) = A_{1} P_{2} \bar{P}_{1} Z_{3}^{s-1}  + A_{2} \, Q_{1} \bar{Q}_{2} Z_{3}^{s-1} \, ,
\end{align}
where $A_{1}$ and $A_{2}$ are complex coefficients. After imposing conservation on $x_{3}$ using the methods outlined in Subsection \ref{Ch04-subsubsection2.2.2}, we find 
\begin{align}
	\begin{split}
		D_{3} \tilde{\cH}(X; U,V,W) &= \left(\D _1-\D_2\right) (-1)^{s+1}  \big\{
		(A_1 + (s^2+s-1 ) A_2 ) \, Q_2 \bar{P}_1 \\
		& \hspace{20mm} + ( (s^2+s-1) A_1 + A_2 ) \bar{P}_2 \bar{Q}_1 \big\} Z_3^{s-2} X^{\D _1-\D_2-s-3} \, .
	\end{split}
\end{align}
Hence, we find that this three-point function is automatically compatible with conservation on $x_{3}$ for $\D_{1} = \D_{2}$. For $\D_{1} \neq \D_{2}$ it is simple to see that conservation is satisfied only for $s=1$, which results in $A_{1} = - A_{2}$ and, hence, the solution is unique. However, for $s>1$ there is no solution in general. In the case where $\psi = \psi'$, we also have to impose the combined point-switch/reality condition, which results in the coefficients $A_{i}$ being purely real/imaginary for $s$ even/odd. This result is consistent with \cite{Elkhidir:2014woa}.

Now let us consider the $(s+1,s-1)$ representation, with $s > 1$. Note that the analysis for $(s-1, s+1)$ is essentially identical and will be omitted. Using the general formula, the ansatz for this three-point function is:
\begin{align}
	\langle \psi_{\a}(x_{1}) \, \bar{\psi}_{\bd}'(x_{2}) \, J_{\g(s+1) \gd(s-1)}(x_{3}) \rangle &= \frac{ \cI_{\a}{}^{\ad'}(x_{13}) \, \bar{\cI}_{\bd}{}^{\b'}(x_{23}) }{(x_{13}^{2})^{\D_{1}} (x_{23}^{2})^{\D_{2}}} \, \cH_{\ad' \b' \g(s+1) \gd(s-1)}(X_{12}) \, .
\end{align} 
All information about this correlation function is encoded in the following polynomial:
\begin{align}
	\cH(X; U, V, W) = \cH_{ \ad \b \g(s+1) \gd(s-1) }(X)  \, \boldsymbol{U}^{\ad}  \boldsymbol{V}^{\b } \boldsymbol{W}^{\g(s+1) \gd(s-1)} \, .
\end{align}
We recall that $\cH$ satisfies the homogeneity property $\cH(X) = X^{s+2 - \D_{1} - \D_{2}} \hat{\cH}(X)$, where $\hat{\cH}(X)$ is homogeneous degree $0$. In this case there is only one possible structure for $\hat{\cH}(X)$:
\begin{align}
	\hat{\cH}(X; U, V, W) = A \, P_{2} \bar{Q}_{1} Z_{3}^{s-1}  \, ,
\end{align}
where $A$ is a complex coefficient. After imposing conservation on $x_{3}$ using the methods outlined in Subsection \ref{Ch04-subsubsection2.2.2}, we find 
\begin{align}
	D_{3} \tilde{\cH}(X;  U, V, W) &= A \left(\D_1-\D _2-1 \right) (-1)^{s+1} ( s^2 + s - 2 ) \, \bar{P}_1 \bar{P}_2 Z_3^{s-2}
	X^{\D _1-\D _2-s-3} \, .
\end{align}
Hence, we find that this three-point function is automatically compatible with conservation on $x_{3}$ for $\D_{2} = \D_{1} - 1$. For $\D_{2} \neq \D_{1} - 1$ there is no solution in general (recall that $s>1$). This result is also consistent with \cite{Elkhidir:2014woa}.


\end{subappendices}

\chapter{Correlation functions of conserved supercurrents in 3D $\cN = 1$ SCFT}\label{Chapter5}
\graphicspath{{Images/3DSCFT/}}

In Chapter \ref{Chapter3} we demonstrated that in three-dimensional conformal field theories the general structure of the three-point correlation functions of conserved currents, $\langle J^{}_{s_{1}} J'_{s_{2}} J''_{s_{3}} \rangle$, where $J^{}_{s}$ denotes a conserved current of arbitrary spin-$s$, is fixed up to the following form~\cite{Giombi:2011rz, Maldacena:2011jn}:
\begin{equation}
	\langle J^{}_{s_{1}} J'_{s_{2}} J''_{s_{3}} \rangle = a_{1} \langle J^{}_{s_{1}} J'_{s_{2}} J''_{s_{3}} \rangle_{E_{1}} + a_{2} \langle J^{}_{s_{1}} J'_{s_{2}} J''_{s_{3}} \rangle_{E_{2}} + b \langle J^{}_{s_{1}} J'_{s_{2}} J''_{s_{3}} \rangle_{O} \, ,
\end{equation}
where $\langle J^{}_{s_{1}} J'_{s_{2}} J''_{s_{3}} \rangle_{E_{1}}$, $\langle J^{}_{s_{1}} J'_{s_{2}} J''_{s_{3}} \rangle_{E_{2}}$ are parity-even solutions corresponding to free field theories, and $\langle J^{}_{s_{1}} J'_{s_{2}} J''_{s_{3}} \rangle_{O}$ is a parity-violating, or parity-odd solution which is known to not be generated by a free CFT. The existence of the parity-odd solution is subject to the following triangle inequalities on the spins:
\begin{align}
	s_{1} \leq s_{2} + s_{3} \, , && s_{2} \leq s_{1} + s_{3} \, , && s_{3} \leq s_{1} + s_{2} \, .
\end{align}
The parity-odd solutions are of particular interest in three dimensions as they have been shown to arise in theories of a Chern-Simons gauge field interacting with parity-violating matter \cite{Aharony:2011jz, Giombi:2011kc, Maldacena:2012sf, Jain:2012qi, GurAri:2012is, Aharony:2012nh, Giombi:2016zwa, Chowdhury:2017vel, Sezgin:2017jgm, Skvortsov:2018uru, Inbasekar:2019wdw}.  The next natural question to ask is: what are the implications of supersymmetry on the general structure of three-point correlation functions?

The intent of the present chapter is to answer this question by providing a complete classification of the three-point functions of conserved supercurrents in 3D $\cN=1$ superconformal field theory. To achieve this we develop a general formalism to construct the three-point function
\begin{equation} \label{3D N=1 three-point function}
	\langle \mathbf{J}^{}_{s_{1}}(z_{1}) \, \mathbf{J}'_{s_{2}}(z_{2}) \, \mathbf{J}''_{s_{3}}(z_{3}) \rangle \, ,
\end{equation}
where $z_{1}, z_{2}, z_{3}$ are points in 3D $\cN=1$ Minkowski superspace, and the superfield $\mathbf{J}_{s}(z)$ is a conserved higher-spin supercurrent of superspin-$s$ (integer or half-integer). These currents are primary superfields transforming in an irreducible representation of the 3D $\cN=1$ superconformal algebra, $\mathfrak{so}(3,2|1) \cong \mathfrak{osp}(1|2;\mathbb{R})$. They are described by totally symmetric spin-tensors of rank $2s$, $\mathbf{J}_{\a_{1} ... \a_{2s}}(z) = \mathbf{J}_{(\a_{1} ... \a_{2s})}(z)$, where $s >0$ is an arbitrary (half-)integer, and satisfy the following superfield conservation equation:
\begin{equation} \label{Conserved supercurrent}
	D^{\a_{1}} \mathbf{J}_{\a_{1} \a_{2} ... \a_{2s}}(z) = 0\, .
\end{equation}
Here, $D^{\a}$ denotes the covariant spinor derivative in 3D $\cN=1$ Minkowski superspace. As a result of the superfield conservation equation \eqref{Conserved supercurrent}, conserved supercurrents have scale dimension $\D_{\mathbf{J}} = s + 1$ (saturating the unitary bound), and at the component level contain independent conserved currents of spin-$s$ and $s+\tfrac{1}{2}$ respectively. The most important examples of conserved supercurrents in superconformal field theory are the supercurrent and flavour current multiplets, corresponding to the cases $s=\tfrac{3}{2}$ and $s = \tfrac{1}{2}$ respectively (for a review of the properties of supercurrent and flavour current multiplets in three-dimensional theories, see \cite{Buchbinder:2015qsa,Korovin:2016tsq} and the references there-in). The supercurrent multiplet contains the energy-momentum tensor and the supersymmetry current.\footnote{In $\cN$-extended superconformal theories, the supercurrent multiplet also contains the $R$-symmetry currents.} Likewise, the flavour current multiplet contains a conserved vector current. The three-point correlation functions of these currents contain important physical information about a given superconformal field theory and are highly constrained by superconformal symmetry. 

The general structure of three-point correlation functions of conserved (higher-spin) currents in 3D $\cN=1$ superconformal field theory was proposed in~\cite{Nizami:2013tpa} to be fixed up to the 
following form:
%
\begin{equation}
	\langle \mathbf{J}^{}_{s_{1}} \mathbf{J}'_{s_{2}} \mathbf{J}''_{s_{3}} \rangle = a \, \langle \mathbf{J}^{}_{s_{1}} \mathbf{J}'_{s_{2}} \mathbf{J}''_{s_{3}} \rangle_{E} + b \, \langle \mathbf{J}^{}_{s_{1}} \mathbf{J}'_{s_{2}} \mathbf{J}''_{s_{3}} \rangle_{O} \, ,
	\label{zh1}	
\end{equation}
where $\langle \mathbf{J}^{}_{s_{1}} \mathbf{J}'_{s_{2}} \mathbf{J}''_{s_{3}} \rangle_{E}$ is a parity-even solution, and $\langle \mathbf{J}^{}_{s_{1}} \mathbf{J}'_{s_{2}} \mathbf{J}''_{s_{3}} \rangle_{O}$ 
is a parity-odd solution. However, the analysis of \cite{Nizami:2013tpa}, which only studied cases with $s_{i} \leq 2$, was insufficient to determine the conditions under which the parity-odd solution appears. Indeed, in \cite{Buchbinder:2021gwu} it was pointed out that there is a tension between supersymmetry and existence of parity-odd structures in the three-point functions of conserved currents, which means that the coefficient $b$ in~\eqref{zh1} 
vanishes in many correlators. Hence, the aim of this chapter is to provide a complete classification for when the parity-odd structures are allowed and when they are not. The main results of this chapter are as follows: for the three-point functions of conserved currents which are overall Grassmann-even (bosonic) in superspace, the existence of the parity-odd solution is subject to the superspin triangle inequalities
\begin{align}
	s_{1} \leq s_{2} + s_{3} \, , && s_{2} \leq s_{1} + s_{3} \, , && s_{3} \leq s_{1} + s_{2} \, .
	\label{zh2}	
\end{align}
When the triangle inequalities are simultaneously satisfied there is one even solution and one odd solution, however, if any of the above inequalities
are not satisfied then the odd solution is incompatible with the superfield conservation equations. For the three-point functions which are overall Grassmann-odd (fermionic) in superspace, we construct an explicit solution for the parity-even structure and we prove that the parity-odd structure vanishes for arbitrary superspins. Our classification is in perfect agreement with the results previously derived
in~\cite{Buchbinder:2015qsa, Buchbinder:2021gwu} for the three-point functions of the energy-momentum tensor and conserved vector currents. They belong to the supermultiplets 
of superspins $s=\tfrac{3}{2}$ and $s = \tfrac{1}{2}$ respectively and, hence, their three-point functions are Grassmann-odd in superspace. 
Based on the general classification for arbitrary superspins presented in this chapter, these three-point functions do not possess parity-odd
contributions, which is in agreement with the earlier results \cite{Nizami:2013tpa,Buchbinder:2015qsa}. Our classification is also in agreement with the result presented in \cite{Buchbinder:2021qlb} for the three-point function 
of a conserved supercurrent of superspin-2; this three-point function is Grassmann-even in superspace and contains a parity-odd contribution.

This chapter is organised as follows. First, in Section \ref{Ch05-section5.1} we outline some aspects of superconformal field theory in three-dimensions. In particular, we introduce the superconformal Killing vectors fields of three-dimensional $\cN=1$ Minkowski superspace and the transformation properties of superconformal primary fields. Then, in Section \ref{Ch05-section5.2} we review the essentials of the group theoretic formalism used to construct correlation functions of 
primary superfields in 3D $\cN=1$ SCFT. The preliminaries presented in sections \ref{Ch05-section5.1}, \ref{Ch05-section5.2} make use of the coset construction of 3D $\cN=1$ Minkowski space presented in \cite{Buchbinder:2015qsa} (see also \cite{Park:1999cw}). In Subsection \ref{Ch05-subsection5.2.2} we outline a method to impose all constraints arising from superfield conservation equations and point-switch symmetries on three-point functions of conserved higher-spin supercurrents. 
In particular, we introduce an index-free, auxiliary spinor formalism which allows us to construct a generating function for the three-point functions and we outline the important aspects of our computational approach. As a test of our approach, we present an explicit analysis for three-point correlation functions involving 
combinations of supercurrent and flavour current multiplets, reproducing the known results \cite{Buchbinder:2015qsa, Buchbinder:2021gwu}. 
Next, in Subsection \ref{Ch05-subsection5.3.2}, we expand the analysis to three-point functions of conserved higher-spin supercurrents, where we present a complete classification of the general structure. Many examples are provided in the appendices \ref{Appendix5B}, \ref{Appendix5C}. For completeness, in Subsection \ref{Ch05-subsection5.3.3} we present a classification for three-point correlation functions involving 
combinations of scalar superfields and conserved higher-spin supercurrents. Finally, in Section \ref{Ch05-section5.4}, we present the analytic construction of Grassmann-odd three-point functions of conserved supercurrents based on a method of irreducible decomposition. In particular, we prove that these three-point functions do not contain a parity violating contribution for arbitrary superspins. The construction of the parity-even sector is shown to reduce to solving a system of linear homogeneous equations with a tri-diagonal matrix of co-rank one, which we solve explicitly for arbitrary superspins.

\vfill

\section{Aspects of superconformal symmetry in three-dimensions}\label{Ch05-section5.1}

This section is dedicated to introducing fundamental building blocks required to undertake the analysis of three-point functions of conserved supercurrents in 3D $\cN=1$ superconformal field theory. It is not intended to be a complete discussion, however, most of the technical details have been laid out entirely in the provided references. 

\subsection{Superconformal Killing vectors}\label{Ch05-subsection5.1.1}

Let us begin by reviewing superconformal transformations and the transformation properties of primary superfields in 3D $\cN = 1$ Minkowski superspace. This section closely follows the group theoretic constructions of 3D $\cN=1$ superspace which have been detailed by Buchbinder, Kuzenko, Park, Tartaglino-Mazzucchelli \cite{Kuzenko:2006mv,Kuzenko:2010rp,Kuzenko:2010bd,Kuzenko:2011rd,Kuzenko:2011xg,Kuzenko:2012tb,Kuzenko:2016qwo,Buchbinder:2015qsa,Buchbinder:2015wia} (see also the work of Park \cite{Park:1999cw,Park:1999pd}, who initiated the study of correlation functions in superconformal field theory). Here we will review the pertinent aspects of the formalism required to construct superconformally covariant three-point functions. 

Consider 3D, $\cN=1$ Minkowski superspace $\mathbb{M}^{3 | 2}$, parameterised by coordinates $z^{A} = (x^{a} , \q^{\a})$, where $a = 0,1,2$, $\a = 1,2$ are Lorentz and spinor indices respectively. We consider infinitesimal superconformal transformations, $z^{A} \rightarrow z^{A} + \d z^{A}$, in superspace
\begin{equation}
	\d z^{A} = \x z^{A}  \hspace{3mm} \Longleftrightarrow \hspace{3mm} \d x^{a} = \x^{a}(z) + \text{i} (\g^{a})_{\a \b} \, \x^{\a}(z) \, \q^{\b} \, , 
	\hspace{8mm} \d \q^{\a} = \x^{\a}(z) , \, 
	\label{new1}	
\end{equation}
which are associated with the real first-order differential operator
\begin{equation}
	\x = \x^{A}(z) \, D_{A} = \x^{a}(z) \, \partial_{a} + \x^{\a}(z) D_{\a} \, , \hspace{5mm} D_{A} = (\pa_{a}, D_{\a}) \, , \label{Superconformal Killing vector field}
\end{equation}
where $D_{\a}$ is the standard covariant spinor derivative
\begin{equation} \label{Ch05.1-covariant spinor derivative}
	D_{\a} = \frac{\pa}{\pa \q^{\a}} +  \text{i} (\g^{m})_{\a \b} \q^{\b} \frac{\pa}{\pa x^{m}} \, , \hspace{10mm} \{ D_{\a}, D_{\b} \} = 2 \text{i} (\gamma^{m})_{\a \b} \pa_{m} \, .
\end{equation}
By requiring the operator $\xi$ to satisfy the ``master equation" $[\x , D_{\a} ] \propto D_{\b}$, as a consequence we obtain
\begin{equation}
	D_{\a} \xi^{b} = 2 \text{i} (\g^{b})_{\a \b} \xi^{\b} \hspace{5mm} \implies \hspace{5mm} \x^{\a} = \frac{\text{i}}{6} D_{\b} \x^{\a \b} \, .
\end{equation}
An implication of this condition is that the vector component, $\xi^{\a \b}(z)$, is an ordinary conformal Killing vector (which parametrically depends on $\q$)
\begin{equation}
	\partial_{a} \x_{b} + \partial_{b} \x_{a} = \frac{2}{3} \eta_{a b} \partial_{c} \x^{c} \, .
	\label{new2}	
\end{equation}
For this reason the solutions to this equation are called the superconformal Killing vector fields of Minkowski superspace \cite{Buchbinder:1998qv,Kuzenko:2010rp}. They span a Lie algebra isomorphic to the superconformal algebra $\mathfrak{osp}(1 | 2 ; \mathbb{R})$. The components of the operator $\x$ were calculated explicitly in \cite{Park:1999cw}, and are found to be \cite{Buchbinder:2015qsa}
\begin{subequations}
	\begin{align}
		\begin{split}
			\x^{\a \b} &= a^{\a \b} - \omega^{\a}{}_{\g} x^{\g \b} - x^{\a \g} \omega_{\g}{}^{\b} + \s x^{\a \b} + 4 \text{i} \e^{(\a} \q^{\b)} \\
			& \hspace{20mm} + x^{\a \g} x^{\b \d} b_{\g \d} + \text{i} b_{\d}^{(\a } x^{\b) \d} \q^{2} - 4 \text{i} \eta_{\g} x^{\g(\a} \q^{\b)} \, , \label{Ch05.1-Superconformal killing vector - component 1}
		\end{split}
	\end{align}
	\vspace{-5mm}
	\begin{align}	
		\x^{\a} &= \e^{\a} - \omega^{\a}{}_{\b} \q^{\b} + \frac{1}{2} \s \q^{\a} + b_{\b \g} x^{\b \a} \q^{\g} + \eta_{\b} ( \text{i} \q^{\b} \q^{\a} - x^{\b \a} ) \, , \label{Ch05.1Superconformal killing vector - component 2}
	\end{align}
	\begin{equation}
		a_{\a \b} = a_{\b \a} \, , \hspace{10mm} \omega_{\a \b} = \omega_{\b \a} \, , \hspace{5mm} \omega^{\a}{}_{\a} = 0 \, , \hspace{10mm} b_{\a \b} = b_{\b \a} \, .
	\end{equation}
\end{subequations}
The bosonic parameters correspond to infinitesimal translations ($a_{\a \b}$), 
Lorentz transformations ($\omega_{\a \b}$), scale transformations ($\s$) and special conformal transformations ($b_{\a \b}$) respectively, while the
fermionic parameters correspond to $Q$-supersymmetry ($\e^{\a}$) and $S$-supersymmetry ($\eta^{\a}$) transformations \cite{Buchbinder:2015qsa}. Furthermore, the identity $D_{[\a} \x_{\b]} \propto \ve_{\a \b} $ implies that 
\begin{subequations}
	\begin{gather}
		[ \x , D_{\a} ] = - ( D_{\a} \x^{\b}) D_{\b} = \hat{\omega}_{\a}{}^{\b}(z) D_{\b} - \frac{1}{2} \s(z) D_{\a} \, , \\
	\hat{\omega}_{\a \b}(z) := - D_{(\a} \x_{\b)} = - \frac{1}{4} \pa^{\g}{}_{(\a} \xi_{\b) \g} \, , \\
		\s(z) := D_{\a} \x^{\a} = \frac{1}{3} \pa_{a} \xi^{a} \, . 
		\label{Ch05.1-new4}
	\end{gather}
\end{subequations}
The local parameters $\hat{\omega}^{\a \b}(z)$, $\s(z)$ are interpreted as being associated with combined special-conformal/Lorentz and scale transformations respectively, and appear in the transformation formulae for primary tensor superfields. For later use we also introduce the $z$-dependent $S$-supersymmetry parameter
\begin{equation}
	\eta_{\a}(z) = -\frac{\text{i}}{2} D_{\a} \s(z) \,.
	\label{new5}
\end{equation}
Explicit calculations of the local parameters give \cite{Park:1999cw,Buchbinder:2015qsa}
\begin{subequations}
	\begin{align}
		\hat{\omega}^{\a \b}(z) &= \omega^{\a \b} - x^{\g (\a} b^{\b)}_{\g} + 2 \text{i} \eta^{(\a} \q^{\b)} - \frac{\text{i}}{2} b^{\a \b} \q^{2} \, , \label{Ch05.1-Local parameter 1} \\ 
		\s(z) &= \s + b_{\a \b} x^{\a \b} + 2 \text{i} \q^{\a} \eta_{\a} \, , \label{Ch05.1-Local parameter 3} \\
		\eta_{\a}(z) &= \eta_{\a} - b_{\a \b} \q^{\b} \, . \label{Ch05.1-Local parameter 4}
	\end{align}
\end{subequations}
The local parameters $\hat{\omega}^{\a \b}(z)$, $\s(z)$ are analogous to the local parameters $\hat{\omega}_{mn}(x)$ and $\sigma(x)$ which we found introduced in the non-supersymmetric case in Chapter \ref{Chapter2}.

Once the general form of the superconformal Killing vector is obtained one may then read off the associated generators according to the rule
\begin{align}
	\d z^{M} = - \text{i} \omega^{A} G_{A} z^{M} \, .
\end{align}
In particular, the $Q$-supersymmetry transformations $z^{A} \rightarrow z'^{A}$ act on the superspace coordinates $(x^{a},\q^{\a})$ according to the rule
\begin{align}
	x'^{\a \b} = x^{\a \b} + \text{i} ( \e^{\a} \q^{\b} + \e^{\b} \q^{\a} ) \, , \hspace{10mm} \q'^{\a} = \q^{\a} + \e^{\a} \, ,
\end{align}
from which one obtains the supercharges
\begin{align}
	Q_{\a} = \text{i} \frac{\pa}{\pa \q^{\a}} + (\g^{m})_{\a \b} \q^{\b} \frac{\pa}{\pa x^{m}} \, .
\end{align}
These operators commute with the covariant spinor-derivatives, $\{ Q_{\a}, D_{\b} \} = 0$, and satisfy the (anti-)commutation relation
\begin{align}
	\{ Q_{\a}, Q_{\b} \} = 2 \text{i} (\g^{m})_{\a \b} \pa_{m} \, .
\end{align}
The results for the other superconformal generators are well-known and we omit them here as they are not explicitly relevant to the construction of superconformally covariant three-point functions. See e.g. \cite{Park:1999cw,Nizami:2013tpa} for the explicit form of the generators and some discussion concerning finite superconformal transformations.

For the construction of correlation functions an important superconformal transformation analogous to the inversion in ordinary CFT is the so called ``superinversion" transformation, which is given by the following rule:
\begin{align}
	\boldsymbol{x}'_{\a \b} = -(\boldsymbol{x}^{-1})_{\a \b} \, , \hspace{10mm} \theta'_{\a} = -(\boldsymbol{x}^{-1})_{\a \b} \theta^{\beta} \, .
\end{align}
This transformation is an extension of the conformal inversion to 3D $\cN=1$ superspace \cite{Buchbinder:1998qv,Park:1999cw,Buchbinder:2015qsa}. Similar to the results presented in \cite{Buchbinder:1998qv}, the superinversion only scales the infinitesimal supersymmetric interval (3D extension of the Volkov-Akulov one-form)
\begin{align}
	e^{\a \b} = \text{d} x^{\a \b} + \text{i} \text{d} \q^{\a} \q^{\b} - \text{i} \q^{\a} \text{d} \q^{\b} \, ,
\end{align}
hence, it is a superconformal transformation. Furthermore, it is well-known that the full superconformal group can be generated by considering combinations of supertranslations, Lorentz transformations, dilatations and superinversion \cite{Park:1999pd,Buchbinder:2015qsa}. In particular, special superconformal transformations may be realised by a superinversion, followed by a $Q$-supersymmetry transformation, followed by another superinversion.



\subsection{Primary superfields}\label{Ch05-subsection5.1.2}

Now consider a tensor superfield $\F_{\cA}(z)$ transforming in an irreducible representation of the Lorentz group with respect to the index $\cA$. Such a superfield is called primary with dimension $\D$ if it possesses the following superconformal transformation properties \cite{Buchbinder:2015qsa}
\begin{equation}
	\d \F_{\cA} = - \x \F_{\cA} - \D \s(z) \F_{\cA} + \hat{\omega}^{\a \b}(z) (\Sigma_{\a \b})_{\cA}{}^{\cB} \F_{\cB} \,,
	\label{new6}
\end{equation}
where $\x$ is the superconformal Killing vector, $\s(z)$, $\hat{\omega}^{\a \b}(z)$ are $z$-dependent parameters associated with $\x$ which represent combined local scale, special conformal, Lorentz and $S$-supersymmetry transformations respectively. The matrix $\Sigma_{\a \b}$ is a Lorentz generator.

\subsection{Conserved supercurrents}\label{Ch05-subsection5.1.3}


In this chapter our main goal is to analyse the structure of three-point correlation functions involving conserved supercurrents of an arbitrary integer or half-integer superspin, known as higher-spin supercurrents. The study of conserved currents has been carried out by many authors, starting with Howe, Stelle and Townsend \cite{Howe:1981qj}, who introduced the general structure of conserved supercurrents in 4D $\cN=1$ theories. 

In 3D $\cN=1$ theories, a conserved higher-spin supercurrent of superspin-$s$ (integer or half-integer), is defined as a totally symmetric spin-tensor of rank-$2s$, $\mathbf{J}_{\a_{1} \dots \a_{2s} }(z) = \mathbf{J}_{(\a_{1} \dots \a_{2s}) }(z) = \mathbf{J}_{\a(2s) }(z)$, satisfying a conservation equation of the form:
\begin{equation} \label{Ch05.1-Conserved current}
	D^{\a_{1}} \mathbf{J}_{\a_{1} \a_{2} \dots \a_{2s}}(z) = 0 \, ,
\end{equation}
where $D^{\a}$ is the covariant spinor derivative \eqref{Ch05.1-covariant spinor derivative}. Conserved currents are primary superfields as they possesses the following infinitesimal superconformal transformation properties \cite{Buchbinder:1998qv,Park:1999cw,Buchbinder:2015qsa}:
\begin{equation}
	\delta \mathbf{J}_{\a_{1} \dots \a_{2s}}(z) = - \xi \mathbf{J}_{\a_{1} \dots \a_{2s}}(z) - \Delta_{\mathbf{J}} \, \s(z) \, \mathbf{J}_{\a_{1} \dots \a_{2s}}(z) + 2s \, \hat{\omega}_{( \a_{1} }{}^{\delta}(z) \, \mathbf{J}_{\a_{2} \dots \a_{2s}) \delta}(z) \, .
\end{equation}
The dimension $\Delta_{\mathbf{J}}$ is constrained by the conservation condition \eqref{Ch05.1-Conserved current} to $\D_{\mathbf{J}} = s+1$. Higher-spin supercurrent superfields possess the following component structure:
\begin{equation}
	\mathbf{J}_{\a(2s)}(z) = J^{(0)}_{\a(2s)}(x) + J^{(1)}_{\a(2s+1)}(x) \, \q^{\a_{2s+1}} + \tilde{J}^{(1)}_{(\a_{1} ... \a_{2s-1}}(x) \, \q^{}_{\a_{2s})}  + J^{(2)}_{\a(2s)}(x) \, \q^{2} \, .
\end{equation}
After imposing \eqref{Ch05.1-Conserved current}, a short calculation gives $\tilde{J}^{(1)} = 0$, while $J^{(2)}$ is a function of $J^{(0)}_{\a(2s)}$.
On the other hand, the components $J^{(0)}$, $J^{(1)}$ satisfy the following conservation equations:
\begin{equation}
	\pa^{\a_{1} \a_{2}} J^{(0)}_{\a_{1} \a_{2} \a(2s-2)}(x) = 0 \, , \hspace{10mm} \pa^{\a_{1} \a_{2}} J^{(1)}_{\a_{1} \a_{2} \a(2s -1)}(x) = 0 \, .
\end{equation}
Hence, at the component level, a higher-spin supercurrent of superspin-$s$ contains conserved conformal currents of spin-$s$ and spin-$(s+\tfrac{1}{2})$ respectively. Higher-spin supercurrents in 3D $\cN=1$ theories have been explicitly constructed in \cite{Nizami:2013tpa}.

Just as the energy-momentum tensor and flavour currents are fundamental conserved currents in any conformal field theory, so too are their supersymmetric analogues, the supercurrent and flavour current multiplets. The 3D $\cN=1$ conformal supercurrent, corresponding to \eqref{Ch05.1-Conserved current} with $s = 3/2$, is a primary dimension $5/2$ totally symmetric spin-tensor $\mathbf{J}_{\a \b \g}$, which contains the three-dimensional 
energy-momentum tensor along with the supersymmetry current \cite{Buchbinder:1998qv,Komargodski:2010rb,Korovin:2016tsq}. It obeys the conservation equation
\begin{equation}
	D^{\a} \mathbf{J}_{\a \b \g} = 0 \, ,
\end{equation}
and has the following superconformal transformation law:
\begin{equation}
	\d \mathbf{J}_{\a \b \g} = - \x \mathbf{J}_{\a \b \g} - \frac{5}{2} \s(z) \mathbf{J}_{\a \b \g} + 3 \hat{\omega}(z)^{\d}_{(\a} \mathbf{J}_{\b \g) \d} \, .
\end{equation}
The $\cN=1$ supercurrent may be derived from, for example, supergravity prepotential approaches \cite{Buchbinder:1998qv} or the superfield Noether procedure \cite{Magro:2001aj,Kuzenko:2010ni} (which is somewhat similar to the analysis presented in Chapter \ref{Chapter2}). The supercurrent contains the energy-momentum tensor and the supersymmetry current in addition to the R-symmetry current (in $\cN$-extended theories). These component currents can be extracted by the process of bar projection \cite{Buchbinder:1998qv} as follows:
\begin{align}
	T_{\a_{1} \a_{2} \a_{3} \a_{4}}(x) = D_{(\a_{1}} \mathbf{J}_{\a_{2} \a_{3} \a_{4})}(z) |_{\q = 0} \, , \hspace{10mm} Q_{\a_{1} \a_{2} \a_{3} }(x) = \mathbf{J}_{\a_{1} \a_{2} \a_{3}}(z)|_{\q=0} \, .
\end{align}
The component currents satisfy the conservation equations
\begin{align}
	(\g_{m})^{\a_{1} \a_{2}} \pa^{m} T_{\a_{1} \a_{2} \a_{3} \a_{4}}(x) = 0 \, , \hspace{10mm} (\g_{m})^{\a_{1} \a_{2}} \pa^{m}  Q_{\a_{1} \a_{2} \a_{3} }(x) = 0 \, ,
\end{align}
as expected from conserved currents in conformal field theories.

Another essential current in 3D $\cN=1$ superconformal field theory is the flavour current multiplet, corresponding to $s = 1/2$ in \eqref{Ch05.1-Conserved current}. The flavour current superfield, which is typically associated with internal symmetries, is a primary dimension $3/2$ spinor superfield 
$\mathbf{L}_{\a}$ obeying the conservation equation\footnote{The structure and the conservation equations for flavour current multiplets follow from the structure of unconstrained prepotentials for vector multiplets~\cite{Siegel:1979fr, Gates:1983nr, Hitchin:1986ea, Zupnik:1988en, Zupnik:1988wa, Zupnik:1999iy, Zupnik:2009zn}.}
\begin{equation}
	D^{\a} \mathbf{L}_{\a} = 0 \, . \label{Ch05.1-N=1 flavour current conservation equation}
\end{equation}
It transforms covariantly under the superconformal group as
\begin{equation}
	\d \mathbf{L}_{\a} = -\x \mathbf{L}_{\a} - \frac{3}{2} \s(z) \mathbf{L}_{\a} + \hat{\omega}(z)_{\a}{}^{\b} \mathbf{L}_{\b} \, . 
\end{equation}
In this case there is only one conserved component current
\begin{align}
	V_{\a_{1} \a_{2}}(x) = D_{(\a_{1}} \mathbf{L}_{\a_{2})}(z) |_{\q = 0} \, , \hspace{10mm} (\g_{m})^{\a_{1} \a_{2}} \pa^{m} V_{\a_{1} \a_{2}} = 0 \, .
\end{align}
We can also consider the case when there are several flavour current multiplets corresponding to a simple flavour group. Such currents are of the form $\mathbf{L}_{\a} = \mathbf{L}_{\a}^{\bar{a}} \,  \Sigma^{\bar{a}}$, where $\bar{a}$ is a flavour index and $\Sigma^{\bar{a}}$ are the generators of a simple flavour group.


\section{Superconformal building blocks}\label{Ch05-section5.2}


Our aim now is to provide an overview of the construction of two- and three-point functions of primary superfields, analogous to the approach presented in chapters \ref{Chapter3} and \ref{Chapter4} for conformal field theory in three and four spacetime dimensions. We have omitted discussion of generating superfunctionals, free field constructions of conserved supercurrents, and superconformal Ward identities, however, the symmetry arguments presented in Chapter \ref{Chapter2} for the conformal bootstrap also generalise to correlation functions of superconformal primary fields in superspace. The group theoretic formalism which we utilise to construct three-point functions in 3D $\cN=1$ superspace which are covariant with respect to superconformal transformations has been worked out entirely in \cite{Buchbinder:2015qsa} (see also \cite{Park:1999cw,Nizami:2013tpa}). Analogous to ordinary CFT, the three-point function is constructed such that it is invariant under both (super)translations, Lorentz transformations, scale transformations, and superinversion, which naturally extends to invariance under the entire superconformal group. See \cite{Park:1997bq, Osborn:1998qu, Park:1999pd, Park:1999cw} for the initial works on the construction of correlation functions in three- and four-dimensional superconformal field theory. In this section we utilise the construction presented in \cite{Buchbinder:2015qsa,Buchbinder:2015wia,Kuzenko:2016cmf}.

Analogous to the non-supersymmetric construction presented in Chapter \ref{Chapter3}, given two points in superspace $z_{1}$ and $z_{2}$ we can define the two-point functions
\begin{equation}
	\boldsymbol{x}_{12}^{\alpha \beta} = (x_{1} - x_{2})^{\alpha \beta} + 2 \text{i} \theta^{(\alpha}_{1} \theta^{\beta)}_{2} - \text{i} \theta^{\a}_{12} \theta^{\b}_{12} \, ,  \hspace{10mm} \theta^{\alpha}_{12} = \theta_{1}^{\alpha} - \theta_{2}^{\alpha} \, . \label{Ch05.1-Two-point building blocks 1}
\end{equation}
The two-point functions transform under the superconformal group as follows
\begin{subequations}
	\begin{align}
		\tilde{\d} \boldsymbol{x}_{12}^{\a \b} &= - \bigg( \hat{\omega}^{\a}{}_{\g}(z_{1}) - \frac{1}{2} \d^{\a}{}_{\g} \, \s(z_{1}) \bigg) \boldsymbol{x}_{12}^{\g \b} - \boldsymbol{x}_{12}^{\a \g} \bigg( \hat{\omega}_{\g}{}^{\b}(z_{2}) - \frac{1}{2} \d_{\g}{}^{\b} \s(z_{2}) \bigg) \, , \label{Ch05.1-Two-point building blocks 1 - transformation law 1} \\[2mm]
		\tilde{\d} \q_{12 }^{\a} &= - \bigg( \hat{\omega}^{\a}{}_{\b}(z_{1}) - \frac{1}{2} \d^{\a}{}_{\b} \, \s(z_{1}) \bigg) \q_{12}^{\b} - \boldsymbol{x}_{12}^{\a \b} \, \eta_{\b}(z_{2}) \,. \label{Ch05.1-Two-point building blocks 1 - transformation law 2}
	\end{align}
\end{subequations}
Here the total variation $\tilde{\d}$ is defined by its action on an $n$-point function $\F(z_{1},...,z_{n})$ as
\begin{equation}
	\tilde{\d} \F(z_{1},...,z_{n}) = \sum_{i=1}^{n} \x_{z_{i}} \F(z_{1},...,z_{n}) \, . \label{Ch05.1-Total variation} 
\end{equation}
We note that only \eqref{Ch05.1-Two-point building blocks 1 - transformation law 1} transforms covariantly under superconformal transformations, as \eqref{Ch05.1-Two-point building blocks 1 - transformation law 2} 
contains an inhomogeneous piece in its transformation law. Therefore the two-point functions $\theta_{ij}$ are not used explicitly in the construction of two- or three-point correlation functions. 

Now due to the useful property $\boldsymbol{x}_{21}^{\a \b} = - \boldsymbol{x}_{12}^{\b \a}$, the two-point function \eqref{Ch05.1-Two-point building blocks 1} can be split into symmetric and antisymmetric parts as follows:
\begin{equation}
	\boldsymbol{x}_{12}^{\a \b} = x_{12}^{\a \b} + \frac{\text{i}}{2} \ve^{\alpha \beta} \theta^{2}_{12} \, , \hspace{10mm} \q_{12}^{2} = \q_{12}^{\a} \q^{}_{12 \, \a} \, . \label{Ch05.1-Two-point building blocks 1 - properties 1}
\end{equation}
The symmetric component
\begin{equation}
	x_{12}^{\a \b} = (x_{1} - x_{2})^{\alpha \beta} + 2 \text{i} \theta^{(\alpha}_{1} \theta^{\beta)}_{2} \, , \label{Ch05.1-Two-point building blocks 1 - properties 2}
\end{equation}
is recognised as the bosonic part of the standard two-point superspace interval. The two-point functions also possess the property
\begin{align}  \label{Ch05.1-Two-point building blocks - properties 1} 
	\boldsymbol{x}_{12}^{\a \s} \boldsymbol{x}^{}_{21 \, \s \b} = \boldsymbol{x}_{12}^{2} \d_{\b}^{\a} \, , \hspace{10mm} \boldsymbol{x}_{12}^{2} = - \frac{1}{2} \boldsymbol{x}_{12}^{\a \b}  \boldsymbol{x}^{}_{12  \, \a \b} \, .
\end{align}
Hence, we find
\begin{equation} \label{Ch05.1-Two-point building blocks 4}
	(\boldsymbol{x}_{12}^{-1})^{\a \b} = - \frac{\boldsymbol{x}_{12}^{ \b \a}}{\boldsymbol{x}_{12}^{2}} \, .
\end{equation}
It is now useful to introduce the normalised two-point functions, denoted by $\hat{\boldsymbol{x}}_{12}$,
\begin{align} \label{Ch05.1-Two-point building blocks 3}
	\hat{\boldsymbol{x}}_{12 \, \a \b} = \frac{\boldsymbol{x}_{12 \, \a \b}}{( \boldsymbol{x}_{12}^{2})^{1/2}} \, , \hspace{10mm} \hat{\boldsymbol{x}}_{12}^{\a \s} \hat{\boldsymbol{x}}^{}_{21 \, \s \b} = \d_{\b}^{\a} \, . 
\end{align}
Under superconformal transformations, $\boldsymbol{x}_{12}^{2}$ transforms with local scale parameters, while \eqref{Ch05.1-Two-point building blocks 3} transforms with local Lorentz parameters
\begin{subequations}
	\begin{align}
		\tilde{\d} \boldsymbol{x}_{12}^{2} &= ( \s(z_{1}) + \s(z_{2}) ) \, \boldsymbol{x}_{12}^{2} \, , \label{Ch05.1-Two-point building blocks 2 - transformation law 1} \\
		\tilde{\d} \hat{\boldsymbol{x}}_{12}^{\a \b} &= - \hat{\omega}^{\a}{}_{\g}(z_{1}) \, \hat{\boldsymbol{x}}_{12}^{\g \b} - \hat{\boldsymbol{x}}_{12}^{\a \g} \, \hat{\omega}_{\g}{}^{\b}(z_{2}) \, . \label{Ch05.1-Two-point building blocks 3 - transformation law 1}
	\end{align}
\end{subequations}
Due to these properties, the object $\hat{\boldsymbol{x}}_{12}$ may be regarded as a parallel transport operator from the point $z_{1}$ to $z_{2}$, analogous to the inversion tensor in the non-supersymmetric case. There are also the following differential identities for the action of covariant spinor derivatives on the two-point functions:
\begin{equation} 
	D_{(1) \g} \boldsymbol{x}_{12}^{\a \b} = - 2 \text{i} \q^{\b}_{12} \d_{\g}^{\a} \, , \hspace{10mm} D_{(1) \a} \boldsymbol{x}_{12}^{\a \b} = - 4 \text{i} \q^{\b}_{12} \, , \label{Ch05.1-Two-point building blocks 1 - differential identities}
\end{equation}
where $D_{(i) \a}$ acts on the superspace point $z_{i}$. From here we can now construct an operator analogous to the conformal inversion tensor acting on the space of symmetric traceless spin-tensors of arbitrary rank. Given a two-point function $\boldsymbol{x}$, we define the operator
\begin{equation} \label{Ch05.1-Higher-spin inversion operators a}
	\cI_{\a(k) \b(k)}(\boldsymbol{x}) = \hat{\boldsymbol{x}}_{(\a_{1} (\b_{1}} \dots \hat{\boldsymbol{x}}_{ \a_{k}) \b_{k})}  \, ,
\end{equation}
along with its inverse
\begin{equation} \label{Ch05.1-Higher-spin inversion operators b}
	\cI^{\a(k) \b(k)}(\boldsymbol{x}) = \hat{\boldsymbol{x}}^{(\a_{1} (\b_{1}} \dots \hat{\boldsymbol{x}}^{ \a_{k}) \b_{k})} \, .
\end{equation}
The spinor indices may be raised and lowered using the standard conventions as follows:
\begin{align}
	\cI_{\a(k)}{}^{\b(k)}(\boldsymbol{x}) &= \ve^{\b_{1} \g_{1}} \dots \ve^{\b_{k} \g_{k}} \, \cI_{\a(k) \g(k)}(\boldsymbol{x}) \, .
\end{align}
Now due to the property
\begin{equation}
	\cI_{\a(k) \b(k)}(-\boldsymbol{x}) = (-1)^{k} \cI_{\a(k) \b(k)}(\boldsymbol{x}) \, ,
\end{equation}
the following identity holds for products of inversion tensors:
\begin{align} \label{Ch05.1-Higher-spin inversion operators - properties}
	\cI_{\a(k) \s(k)}(\boldsymbol{x}_{12}) \, \cI^{\s(k) \b(k)}(\boldsymbol{x}_{21}) &= \d_{(\a_{1}}^{(\b_{1}} \dots \d_{\a_{k})}^{\b_{k})} \, .
\end{align}
The objects \eqref{Ch05.1-Higher-spin inversion operators a}, \eqref{Ch05.1-Higher-spin inversion operators b} prove to be essential in the construction of correlation functions of primary operators with arbitrary spin. In particular \eqref{Ch05.1-Higher-spin inversion operators a} satisfies the following property under superinversion transformations $\boldsymbol{i}:(\boldsymbol{x}, \theta) \rightarrow (\boldsymbol{x}',\theta')$ 
\begin{align}
		\cI_{\a}{}^{\a'}(\boldsymbol{x}_{1}) \, \cI_{\b}{}^{\b'}(\boldsymbol{x}_{2}) \, \cI_{\a' \b'}(\boldsymbol{x}_{12}) = \cI_{\a \b}(\boldsymbol{x}'_{12}) \, .
\end{align}
This is similar to \eqref{Ch02-Inversion tensor transformation} in the non-supersymmetric case. 

Note that the vector representation of the inversion tensor may be recovered in terms of the spinor two-point functions as follows:
\begin{equation}
	I_{m n}(\boldsymbol{x}_{12}) = - \frac{1}{2} \, \text{Tr}( \g_{m} \, \hat{\boldsymbol{x}}_{12} \, \g_{n} \, \hat{\boldsymbol{x}}_{12} ) \, .
\end{equation}
It satisfies the fundamental properties
\begin{align}
	I_{mn}(\boldsymbol{x}_{12}) = I_{nm}(\boldsymbol{x}_{21}) \, , \hspace{10mm}  	I_{mp}(\boldsymbol{x}_{12}) 	I^{pn}(\boldsymbol{x}_{21}) = \d_{m}{}^{n} \, ,
\end{align}
and in the purely bosonic case it reduces to the ordinary inversion tensor
\begin{equation}
	I_{m n}(x_{12}) = I_{m n}(\boldsymbol{x}_{12})|_{\q = 0} = \eta_{mn} - 2 \hat{x}_{12 m} \hat{x}_{12 n} \, .
\end{equation}
%


Essential to the analysis of three-point correlation functions are three-point covariants/building blocks. Indeed, given three superspace points, $z_{1}, z_{2}, z_{3}$, one can define the objects, $\cZ_{k} = ( \boldsymbol{X}_{ij} , \Q_{ij} )$ as follows:
\begin{align} \label{Ch05.1-Three-point building blocks 1}
		\boldsymbol{X}_{ij \, \a \b} &= -(\boldsymbol{x}_{ik}^{-1})_{\a \g}  \boldsymbol{x}_{ij}^{\g \d} (\boldsymbol{x}_{kj}^{-1})_{\d \b} \, , \hspace{5mm} \Q_{ij \, \a} = (\boldsymbol{x}_{ik}^{-1})_{\a \b} \q_{ki}^{\b} - (\boldsymbol{x}_{jk}^{-1})_{\a \b} \q_{kj}^{\b} \, ,
\end{align}
where the labels $(i,j,k)$ are a cyclic permutation of $(1,2,3)$. These objects possess the fundamental property $\boldsymbol{X}_{ij \, \a \b} = - \boldsymbol{X}_{ji \, \b \a}$. As a consequence, the three-point building blocks~\eqref{Ch05.1-Three-point building blocks 1} possess properties similar to those of the two-point building blocks
\begin{align} 
	\boldsymbol{X}_{ij}^{\a \s} \boldsymbol{X}^{}_{ji \, \s \b} = \boldsymbol{X}_{ij}^{2} \d_{\b}^{\a} \, , \hspace{5mm} \boldsymbol{X}_{ij}^{2} = - \frac{1}{2} \boldsymbol{X}_{ij}^{\a \b}  \boldsymbol{X}^{}_{ij \, \a \b} \, .
\end{align}
Hence, we find
\begin{equation}
	(\boldsymbol{X}_{ij}^{-1})^{\a \b} = - \frac{\boldsymbol{X}_{ij}^{ \b \a}}{\boldsymbol{X}_{ij}^{2}} \, .
\end{equation}
It is also useful to note that one may decompose $\boldsymbol{X}_{ij}$ into symmetric and anti-symmetric parts similar to \eqref{Ch05.1-Two-point building blocks 1 - properties 1} as follows:
\begin{equation}
	\boldsymbol{X}_{ij  \, \a \b} = X_{ij  \, \a \b} - \frac{\text{i}}{2} \ve_{\a \b} \Q_{ij}^{2} \, , \hspace{10mm} X_{ij  \, \a \b} = X_{ij  \, \b \a} \, , \label{Ch05.1-Three-point building blocks 1a - properties 3}
\end{equation}
where the symmetric spin-tensor, $X_{ij  \, \a \b}$, can be equivalently represented by the three-vector $X_{ij \, m} = - \frac{1}{2} (\g_{m})^{\a \b} X_{ij  \, \a \b}$. Since the building blocks possess the same properties up to cyclic permutations of the points, we will only examine the properties of $\boldsymbol{X}_{12}$ and $\Q_{12}$, as these objects appear most frequently in our analysis of correlation functions. One can compute
\begin{equation}
	\boldsymbol{X}_{12}^{2} = - \frac{1}{2} \boldsymbol{X}_{12}^{\a \b}  \boldsymbol{X}_{12 \, \a \b}^{} = \frac{\boldsymbol{x}_{12}^{2}}{\boldsymbol{x}_{13}^{2} \boldsymbol{x}_{23}^{2}} \, , \hspace{10mm}  \Q_{12}^{2} = \Q^{\a}_{12} \Q^{}_{12 \, \a} \, . \label{Ch05.1-Three-point building blocks 2}
\end{equation}
The building block $\boldsymbol{X}_{12}$ also possesses the following superconformal transformation properties:
\begin{subequations}
	\begin{align}
		\tilde{\d} \boldsymbol{X}_{12 \, \a \b} &= \hat{\omega}_{\a}{}^{\g}(z_{3}) \boldsymbol{X}_{12 \, \g \b} + \boldsymbol{X}_{12 \, \a \g} \, \hat{\omega}^{\g}{}_{\b}(z_{3}) - \s(z_{3}) \boldsymbol{X}_{12 \, \a \b} \, , \label{Ch05.1-Three-point building blocks 1a - transformation law 1} \\[2mm]
		\tilde{\d} \Q_{12 \, \a} &= \Big( \hat{\omega}_{\a}{}^{\b}(z_{3}) - \frac{1}{2} \, \d_{\a}{}^{\b} \s(z_{3}) \Big) \Q_{12 \, \b}  \, , \label{Ch05.1-Three-point building blocks 1a - transformation law 2}
	\end{align}
\end{subequations}
and, therefore
\begin{equation}
	\tilde{\d} \boldsymbol{X}_{12}^{2} = - 2 \s(z_{3}) \boldsymbol{X}_{12}^{2} \, , \hspace{10mm} \tilde{\d} \Q_{12}^{2} = - \s(z_{3}) \, \Q_{12}^{2} \, , \label{Ch05.1-Three-point building blocks 2 - transformation law 1}
\end{equation}
i.e. $(\boldsymbol{X}_{12}$, $\Q_{12})$ is superconformally covariant at $z_{3}$. As a consequence, one can identify the three-point superconformal invariant
\begin{equation}
	\boldsymbol{J} = \frac{\Q_{12}^{2}}{\sqrt{\boldsymbol{X}_{12}^{2}}} \hspace{5mm} \Longrightarrow \hspace{5mm} \tilde{\d} \boldsymbol{J} = 0 \, ,
\end{equation}
which proves to be invariant under permutations of the superspace points, i.e.
\begin{equation}
	\boldsymbol{J} = \frac{\Q_{12}^{2}}{\sqrt{\boldsymbol{X}_{12}^{2}}} = \frac{\Q_{31}^{2}}{\sqrt{\boldsymbol{X}_{31}^{2}}} = \frac{\Q_{23}^{2}}{\sqrt{\boldsymbol{X}_{23}^{2}}}  \, . \label{Ch05.1-Superconformal invariants}
\end{equation}
This is an important result initially found by Park \cite{Park:1997bq,Park:1999cw}. Recall that in three-dimensional conformal field theory it is impossible to construct a conformal invariant from three points; such an invariant is only possible in supersymmetric theories in superspace (similar results are also found in four dimensional superconformal field theories \cite{Osborn:1998qu}).

Analogous to the two-point functions, it is also useful to introduce the normalised three-point building blocks, denoted by $\hat{\boldsymbol{X}}_{ij}$, $\hat{\Q}_{ij}$, 
\begin{align} \label{Ch05.1-Normalised three-point building blocks}
	\hat{\boldsymbol{X}}_{ij \, \a \b} = \frac{\boldsymbol{X}_{ij \, \a \b}}{( \boldsymbol{X}_{ij}^{2})^{1/2}} \, , \hspace{10mm} \hat{\Q}_{ij}^{\a} = \frac{ \Q_{ij}^{\a} }{(\boldsymbol{X}_{ij}^{2})^{1/4}} \, ,
\end{align}
such that
\begin{align}
	\hat{\boldsymbol{X}}_{ij}^{\a \s} \hat{\boldsymbol{X}}^{}_{ji \, \s \b} = \d_{\b}^{\a} \, , \hspace{10mm} \boldsymbol{J} = \hat{\Q}_{ij}^{2} \, .
\end{align}
Compared with the standard three-point building blocks, \eqref{Ch05.1-Three-point building blocks 1}, the objects \eqref{Ch05.1-Normalised three-point building blocks} transform only with local Lorentz parameters. Now given an arbitrary three-point building block, $\boldsymbol{X}$, let us construct the following higher-spin inversion operator:
\begin{equation}
	\cI_{\a(k) \b(k)}(\boldsymbol{X}) = \hat{\boldsymbol{X}}_{ (\a_{1} (\b_{1}} \dots \hat{\boldsymbol{X}}_{\a_{k}) \b_{k})}  \, , \label{Ch05.1-Inversion tensor identities - three point functions a}
\end{equation}
along with its inverse
\begin{equation}
	\cI^{\a(k) \b(k)}(\boldsymbol{X}) = \hat{\boldsymbol{X}}^{(\a_{1} (\b_{1}} \dots \hat{\boldsymbol{X}}^{ \a_{k}) \b_{k})} \, . \label{Ch05.1-Inversion tensor identities - three point functions b}
\end{equation}
These operators possess similar properties to the higher-spin two-point inversion operators \eqref{Ch05.1-Higher-spin inversion operators a}, \eqref{Ch05.1-Higher-spin inversion operators b}, and are essential to the analysis of three-point correlation functions involving higher-spin primary superfields. In particular, one can prove the following useful identities involving $\boldsymbol{X}_{ij}$ and $\Q_{ij}$ at different superspace points:
\begin{subequations}
	\begin{align}
		\cI_{\a}{}^{\s}(\boldsymbol{x}_{13}) \, \cI_{\b}{}^{\g}(\boldsymbol{x}_{13}) \, \cI_{\s \g}(\boldsymbol{X}_{12}) &= \cI_{\a \b}(\boldsymbol{X}^{I}_{23})  \, , \label{Ch05.1-Three-point building blocks 1a - properties 1}\\[2mm]
		\cI_{\a}{}^{\g}(\boldsymbol{x}_{13}) \, \hat{\Q}_{12 \, \g} &= \hat{\Q}^{I}_{23 \, \a} \, , \label{Ch05.1-Three-point building blocks 1a - properties 2}
	\end{align}
\end{subequations}
where we have defined
\begin{equation}
	\hat{\boldsymbol{X}}{}^{I}_{ij \, \a \b} = - \hat{\boldsymbol{X}}_{ij \, \a \b} \, , \hspace{10mm} \hat{\Q}^{I}_{ij \a} = \cI_{\a \b}(\boldsymbol{X}_{ij}) \, \hat{\Q}^{\b}_{ij}\, .
	\label{Ch05.1-zh3}
\end{equation}
Using the inversion operators above, the identity \eqref{Ch05.1-Three-point building blocks 1a - properties 1} (and cyclic permutations) admits the following generalisation
\begin{equation}
	\cI_{\a(k)}{}^{\s(k)}(\boldsymbol{x}_{13}) \, \cI_{\b(k)}{}^{\g(k)}(\boldsymbol{x}_{13}) \, \cI_{\s(k) \g(k)}(\boldsymbol{X}_{12}) = \cI_{\a(k) \b(k)}(\boldsymbol{X}^{I}_{23})  \, . \label{Ch05.1-Inversion tensor identities - higher spin case}
\end{equation}
Due to the transformation properties \eqref{Ch05.1-Three-point building blocks 1a - transformation law 1}, \eqref{Ch05.1-Three-point building blocks 1a - transformation law 2} it is often useful to make the identifications $(\boldsymbol{X}_{1}, \Q_{1}) := (\boldsymbol{X}_{23}, \Q_{23})$, $(\boldsymbol{X}_{2}, \Q_{2}) := (\boldsymbol{X}_{31}, \Q_{31})$, $(\boldsymbol{X}_{3}, \Q_{3}) := (\boldsymbol{X}_{12}, \Q_{12})$, in which case we have e.g. $\boldsymbol{X}_{21} = - \boldsymbol{X}_{3}^{\text{T}}$; we will switch between these notations when convenient. Let us now introduce the following analogues of the covariant spinor derivative and supercharge operators involving the three-point objects:
\begin{equation}
	\cD_{(i) \a} = \frac{\partial}{\partial \Q^{\a}_{i}} + \text{i} (\g^{m})_{\a \b} \Q^{\b}_{i} \frac{\partial}{\partial X^{m}_{i}} \, , \hspace{5mm} \cQ_{(i) \a} = \text{i} \frac{\partial}{\partial \Q^{\a}_{i}} + (\g^{m})_{\a \b} \Q^{\b}_{i} \frac{\partial}{\partial X^{m}_{i}} \, , \label{Ch05.1-Supercharge and spinor derivative analogues}
\end{equation}
which obey the standard commutation relations
\begin{equation}
	\big\{ \cD_{(i) \a} , \cD_{(i) \b} \big\} = \big\{ \cQ_{(i) \a} , \cQ_{(i) \b} \big\} = 2 \text{i} \, (\g^{m})_{\a \b} \frac{\partial}{\partial X^{m}_{i}} \, .
\end{equation}
Some useful identities involving~\eqref{Ch05.1-Supercharge and spinor derivative analogues} are, e.g.
\begin{equation}
	\cD_{(3) \g} \boldsymbol{X}_{3 \, \a \b} = - 2 \text{i} \ve_{\g \b} \Q_{3 \, \a} \, , \hspace{5mm} \cQ_{(3) \g} \boldsymbol{X}_{3 \, \a \b} = - 2 \ve_{\g \a} \Q_{3 \, \b} \, . \label{Ch05.1-Three-point building blocks 1a - differential identities 1}
\end{equation}
We must also account for the fact that correlation functions of primary superfields obey differential constraints as a result of superfield conservation equations. Using \eqref{Ch05.1-Two-point building blocks 1 - differential identities} we obtain the following identities
\begin{subequations}
	\begin{align}
		D_{(1) \g} \boldsymbol{X}_{3 \, \a \b} &= 2 \text{i} (\boldsymbol{x}^{-1}_{13})_{\a \g} \Q_{3 \, \b} \, , \hspace{5mm} D_{(1) \a} \Q_{3 \, \b} = - (\boldsymbol{x}_{13}^{-1})_{\b \a} \, , \label{Ch05.1-Three-point building blocks 1c - differential identities 1}\\[2mm]
		D_{(2) \g} \boldsymbol{X}_{3 \, \a \b} &= 2 \text{i} (\boldsymbol{x}^{-1}_{23})_{\b \g} \Q_{3 \, \b} \, , \hspace{5mm} D_{(2) \a} \Q_{3 \, \b} = (\boldsymbol{x}_{23}^{-1})_{\b \a} \, . \label{Ch05.1-Three-point building blocks 1c - differential identities 2}
	\end{align}
\end{subequations}
Now given a function $f(\boldsymbol{X}_{3} , \Q_{3})$, there are the following differential identities which arise as a consequence of \eqref{Ch05.1-Three-point building blocks 1a - differential identities 1}, \eqref{Ch05.1-Three-point building blocks 1c - differential identities 1} and \eqref{Ch05.1-Three-point building blocks 1c - differential identities 2}:
\begin{subequations}
	\begin{align}
		D_{(1) \g} f(\boldsymbol{X}_{3} , \Q_{3}) &= (\boldsymbol{x}_{13}^{-1})_{\a \g} \cD_{(3)}^{\a} f(\boldsymbol{X}_{3} , \Q_{3}) \, ,  \label{Ch05.1-Three-point building blocks 1c - differential identities 3} \\[2mm]
		D_{(2) \g} f(\boldsymbol{X}_{3} , \Q_{3}) &= \text{i} (\boldsymbol{x}_{23}^{-1})_{\a \g} \cQ_{(3)}^{\a} f(\boldsymbol{X}_{3} , \Q_{3}) \, .  \label{Ch05.1-Three-point building blocks 1c - differential identities 4}
	\end{align}
\end{subequations}
These will prove to be essential for imposing differential constraints on three-point correlation functions of primary superfields. In particular it allows for conservation equations on each point to be realised as differential constraints on the general form of $\cH(\boldsymbol{X}, \Q)$, which depends only on the three-point superconformal covariants.

\subsection{Two-point correlation functions}\label{Ch05-subsection5.2.1}

Let $\F_{\cA}$ be a primary superfield with dimension $\D$, where $\cA$ denotes a collection of Lorentz spinor indices. The two-point correlation function of $\F_{\cA}$ is fixed by superconformal symmetry to the form
\begin{equation} \label{Ch05.1-Two-point correlation function}
	\langle \F_{\cA}(z_{1}) \, \F^{\cB}(z_{2}) \rangle = c \, \frac{\cI_{\cA}{}^{\cB}(\boldsymbol{x}_{12})}{(\boldsymbol{x}_{12}^{2})^{\D}} \, , 
\end{equation} 
where $\cI$ is an appropriate representation of the inversion tensor and $c$ is a constant real parameter. The denominator of the two-point function is determined by the conformal dimension of $\F_{\cA}$, which guarantees that the correlation function transforms with the appropriate weight under scale transformations. The superconformal covariance of of the two-point function may be seen by considering its infinitesimal transformation properties. For conserved supercurrents, it may be shown using the differential identities \eqref{Ch05.1-Two-point building blocks 1 - differential identities} that the dimension of the two-point function is fixed to $\D = s + 1$. This process is simplified considerably by introducing auxiliary variables, however, we omit the discussion here as it proceeds almost identically to the 3D CFT case.

\subsection{Three-point correlation functions}\label{Ch05-subsection5.2.2}

In this subsection we will review the various properties of three-point correlation functions in 3D $\cN=1$ superconformal field theory. First we present the superfield ansatz presented in \cite{Buchbinder:2015qsa} (which was first introduced by Park in \cite{Park:1999cw}). We then develop an index free auxiliary spinor formalism to simplify the overall form of three-point function for arbitrary spins.

For three-point correlation functions, let $\F$, $\J$ and $\P$ be primary superfields with scale dimensions $\D_{1}$, $\D_{2}$ and $\D_{3}$ respectively. The three-point function may be 
constructed using the general superconformally covariant ansatz \cite{Buchbinder:2015qsa}
\begin{align}
	\langle \F_{\cA_{1}}(z_{1}) \, \J_{\cA_{2}}(z_{2}) \, \P_{\cA_{3}}(z_{3}) \rangle = \frac{ \cI^{(1)}{}_{\cA_{1}}{}^{\cA'_{1}}(\boldsymbol{x}_{13}) \,  \cI^{(2)}{}_{\cA_{2}}{}^{\cA'_{2}}(\boldsymbol{x}_{23}) }{(\boldsymbol{x}_{13}^{2})^{\D_{1}} (\boldsymbol{x}_{23}^{2})^{\D_{2}}}
	\; \cH_{\cA'_{1} \cA'_{2} \cA_{3}}(\boldsymbol{X}_{12}, \Q_{12}) \, . \label{Ch05.1-Three-point function - general ansatz}
\end{align} 
The proof of superconformal covariance is analogous to approach presented in Subsection \ref{Ch02-subsection2.3.3} (see also \cite{Park:1999cw}). It may also be shown by considering the total variation of \eqref{Ch05.1-Three-point function - general ansatz} and using the infinitesimal transformation properties of the two- and three-point building blocks. The tensor $\cH_{\cA_{1} \cA_{2} \cA_{3}}(\boldsymbol{X},\Q)$, which is now a function of two superconformally covariant building blocks, encodes all information about the correlation function, and is related to the leading singular OPE coefficient \cite{Osborn:1993cr}. It is highly constrained by superconformal symmetry as follows:
\begin{enumerate}
	\item[\textbf{(i)}] Under scale transformations of $\mathbb{M}^{3|2}$, $z = ( x, \q ) \mapsto z' = ( \l^{-2} x, \l^{-1} \q )$, hence, the three-point covariants transform as $( \boldsymbol{X}, \Q) \mapsto ( \boldsymbol{X}', \Q') = ( \l^{2} \boldsymbol{X}, \l \Q )$. As a consequence, the correlation function transforms as 
	\begin{equation}
		\langle \F_{\cA_{1}}(z_{1}') \, \J_{\cA_{2}}(z_{2}') \, \P_{\cA_{3}}(z_{3}') \rangle = (\l^{2})^{\D_{1} + \D_{2} + \D_{3}} \langle \F_{\cA_{1}}(z_{1}) \, \J_{\cA_{2}}(z_{2}) \,  \P_{\cA_{3}}(z_{3}) \rangle \, ,
	\end{equation}
	which implies that $\cH$ obeys the scaling property
	\begin{equation}
		\cH_{\cA_{1} \cA_{2} \cA_{3}}( \l^{2} \boldsymbol{X}, \l \Q) = (\l^{2})^{\D_{3} - \D_{2} - \D_{1}} \, \cH_{\cA_{1} \cA_{2} \cA_{3}}(\boldsymbol{X}, \Q) \, , \hspace{5mm} \forall \l \in \mathbb{R} \, \backslash \, \{ 0 \} \, .
	\end{equation}
	This guarantees that the correlation function transforms correctly under scale transformations.
	
	\item[\textbf{(ii)}] If any of the fields $\F$, $\J$, $\P$ obey differential equations, such as conservation laws in the case of conserved currents, then the tensor $\cH$ is also constrained by differential equations which may be derived with the aid of identities \eqref{Ch05.1-Three-point building blocks 1c - differential identities 3}, \eqref{Ch05.1-Three-point building blocks 1c - differential identities 4}.
	
	\item[\textbf{(iii)}] If any (or all) of the operators $\F$, $\J$, $\P$ coincide, the correlation function possesses symmetries under permutations of spacetime points, e.g.
	\begin{equation}
		\langle \F_{\cA_{1}}(z_{1}) \, \F_{\cA_{2}}(z_{2}) \, \P_{\cA_{3}}(z_{3}) \rangle = (-1)^{\e(\F)} \langle \F_{\cA_{2}}(z_{2}) \, \F_{\cA_{1}}(z_{1}) \, \P_{\cA_{3}}(z_{3}) \rangle \, ,
	\end{equation}
	where $\e(\F)$ is the Grassmann parity of $\F$. As a consequence, the tensor $\cH$ obeys constraints which will be referred to as ``point-switch identities".
	
\end{enumerate}
The constraints above fix the functional form of $\cH$ (and therefore the correlation function) up to finitely many independent parameters. Hence, using the general formula \eqref{Ch05.1-H ansatz}, the problem of computing three-point correlation functions is reduced to deriving the general structure of the tensor $\cH$ subject to the above constraints.

\subsubsection{Comments on differential constraints}\label{Ch05-subsection5.2.2.1}

We must also consider imposing differential constraints on the third superspace point for the general ansatz \eqref{Ch05.1-H ansatz}. We follow a similar approach to the 3D CFT case; consider the correlation function $\langle \F_{\cA_{1}}(z_{1}) \, \J_{\cA_{2}}(z_{2}) \, \P_{\cA_{3}}(z_{3}) \rangle$, with the ansatz
\begin{equation} \label{Ch05.1-H ansatz}
	\langle \F_{\cA_{1}}(z_{1}) \, \J_{\cA_{2}}(z_{2}) \, \P_{\cA_{3}}(z_{3}) \rangle = \frac{ \cI^{(1)}{}_{\cA_{1}}{}^{\cA'_{1}}(\boldsymbol{x}_{13}) \,  \cI^{(2)}{}_{\cA_{2}}{}^{\cA'_{2}}(\boldsymbol{x}_{23}) }{(\boldsymbol{x}_{13}^{2})^{\D_{1}} (\boldsymbol{x}_{23}^{2})^{\D_{2}}}
	\; \cH_{\cA'_{1} \cA'_{2} \cA_{3}}(\boldsymbol{X}_{12}, \Q_{12}) \, . 
\end{equation} 
We now reformulate the ansatz with $\P$ at the front as follows:
\begin{equation} \label{Ch05.1-Htilde ansatz}
	\langle \P_{\cA_{3}}(z_{3}) \, \J_{\cA_{2}}(z_{2}) \, \F_{\cA_{1}}(z_{1}) \rangle = \frac{ \cI^{(3)}{}_{\cA_{3}}{}^{\cA'_{3}}(\boldsymbol{x}_{31}) \,  \cI^{(2)}{}_{\cA_{2}}{}^{\cA'_{2}}(\boldsymbol{x}_{21}) }{(\boldsymbol{x}_{31}^{2})^{\D_{3}} (\boldsymbol{x}_{21}^{2})^{\D_{2}}}
	\; \tilde{\cH}_{\cA_{1} \cA'_{2} \cA'_{3} }(\boldsymbol{X}_{23}, \Q_{23}) \, . 
\end{equation} 
Conservation on $\P$ can now be imposed by treating $z_{3}$ as the first point with the aid of identities analogous to \eqref{Ch05.1-Three-point building blocks 1c - differential identities 3}, \eqref{Ch05.1-Three-point building blocks 1c - differential identities 4}. After relating the two ansatz above, we obtain the following equation relating the two different representations of the correlation function:
\begin{align} \label{Ch05.1-Htilde and H relation}
	\tilde{\cH}_{\cA_{1} \cA_{2}  \cA_{3} }(\boldsymbol{X}_{23}, \Q_{23}) &= (\boldsymbol{x}_{13}^{2})^{\D_{3} - \D_{1}} \bigg(\frac{\boldsymbol{x}_{21}^{2}}{\boldsymbol{x}_{23}^{2}} \bigg)^{\hspace{-1mm} \D_{2}} \, \cI^{(1)}{}_{\cA_{1}}{}^{\cA'_{1}}(\boldsymbol{x}_{13}) \, \cI^{(2)}{}_{\cA_{2}}{}^{\cB_{2}}(\boldsymbol{x}_{12}) \,  \cI^{(2)}{}_{\cB_{2}}{}^{\cA'_{2}}(\boldsymbol{x}_{23}) \nonumber \\[-2mm]
	& \hspace{50mm} \times \cI^{(3)}{}_{\cA_{3}}{}^{\cA'_{3}}(\boldsymbol{x}_{13}) \, \cH_{\cA'_{1} \cA'_{2} \cA'_{3}}(\boldsymbol{X}_{12}, \Q_{12}) \, ,
\end{align}
where we have ignored any signs due to Grassmann parity (they can be included later). Before we can simplify the above equation, we must understand how the inversion tensor acts on $\cH(\boldsymbol{X},\Q)$. Now let:
\begin{align}
	\cH_{ \cA_{1} \cA_{2} \cA_{3} }(\boldsymbol{X},\Q) &= \boldsymbol{X}^{\D_{3} - \D_{3}- \D_{1}} \hat{\cH}_{\cA_{1} \cA_{2} \cA_{3}}(\boldsymbol{X}, \Q) \, ,
\end{align}
where $\hat{\cH}_{\cA_{1} \cA_{2} \cA_{3}}(\boldsymbol{X}, \Q)$ is homogeneous degree 0 in $(\boldsymbol{X}, \Q)$, i.e.
\begin{align}
	\hat{\cH}_{\cA_{1} \cA_{2} \cA_{3}}(\l^{2} \boldsymbol{X}, \l \Q) &= \hat{\cH}_{\cA_{1} \cA_{2} \cA_{3}}(\boldsymbol{X}, \Q) \, .
\end{align}
The tensor $\hat{\cH}_{\cA_{1} \cA_{2} \cA_{3}}(\boldsymbol{X}, \Q)$ can be constructed from totally symmetric, homogeneous degree 0 combinations of $\ve$, $\boldsymbol{X}$ and $\Q$, compatible with the set of indices $\cA_{1}, \cA_{2}, \cA_{3}$, hence, we consider the following objects:
\begin{align}
	\ve_{\a \b} \, , \hspace{5mm} \hat{\boldsymbol{X}}_{\a \b} \, , \hspace{5mm} \hat{\Q}_{\a} \, , \hspace{5mm} (\hat{\boldsymbol{X}} \cdot \hat{\Q})_{\a} =  \hat{\boldsymbol{X}}_{\a \b} \hat{\Q}^{\b} \, , \hspace{5mm} \boldsymbol{J} = \hat{\Q}^{2} \, .
\end{align}
Now to simplify \eqref{Ch05.1-Htilde and H relation}, consider
\begin{equation}
	\cI^{(1)}{}_{\cA_{1}}{}^{\cA'_{1}}(\boldsymbol{x}_{13}) \, \cI^{(2)}{}_{\cA_{2}}{}^{\cA'_{2}}(\boldsymbol{x}_{13}) \, \cI^{(3)}{}_{\cA_{3}}{}^{\cA'_{3}}(\boldsymbol{x}_{13}) \, \hat{\cH}_{\cA'_{1} \cA'_{2} \cA'_{3}}(\boldsymbol{X}_{12}, \Q_{12}) \, .
\end{equation}
Only combinations of the following fundamental products may appear in the result:
\begin{subequations}
	\begin{align}
		\cI_{\a}{}^{\a'}(\boldsymbol{x}_{13}) \, \cI_{\b}{}^{\b'}(\boldsymbol{x}_{13}) \, \ve_{\a' \b'} &= - \ve_{\a \b} \, , \\
		\cI_{\a}{}^{\a'}(\boldsymbol{x}_{13}) \, \cI_{\b}{}^{\b'}(\boldsymbol{x}_{13}) \, \hat{\boldsymbol{X}}_{12 \, \a' \b'} &= \hat{\boldsymbol{X}}{}^{I}_{23 \, \a \b} \, , \\
		\cI_{\a}{}^{\a'}(\boldsymbol{x}_{13}) \, \hat{\Q}_{12 \, \a'} &= \hat{\Q}^{I}_{23 \, \a} \, , \\
		\cI_{\a}{}^{\a'}(\boldsymbol{x}_{13}) \, (\hat{\boldsymbol{X}}_{12} \cdot \hat{\Q}_{12})_{\a'} &= - (\hat{\boldsymbol{X}}{}^{I}_{23} \cdot \hat{\Q}^{I}_{23})_{\a} \, ,
	\end{align}
\end{subequations}
where $\hat{\Q}^{I}_{ij}$ was defined in~\eqref{Ch05.1-zh3}.
For correlation functions involving the superconformal invariant, $\boldsymbol{J}$, we must note that $\boldsymbol{J}^{I} = (\hat{\Q}^{I})^2 = - \boldsymbol{J}$. These identities are consequences of \eqref{Ch05.1-Three-point building blocks 1a - properties 1}, \eqref{Ch05.1-Three-point building blocks 1a - properties 2}. If we now denote the above transformations by $\cI_{13}$, it acts on $\hat{\cH}(\boldsymbol{X}_{12},\Q_{12})$ as follows:
\begin{subequations}
	\begin{align} \label{Ch05.1-Inversion even objects}
		\hat{\boldsymbol{X}}_{12} \xrightarrow[]{\cI_{13}} \hat{\boldsymbol{X}}{}^{I}_{23} \, , \hspace{10mm} \hat{\Q}_{12} \xrightarrow[]{\cI_{13}} \hat{\Q}^{I}_{23} \, ,
	\end{align}
	\vspace{-10mm}
	\begin{align} \label{Ch05.1-Inversion odd objects}
		\ve \xrightarrow[]{\cI_{13}} -\ve  \, , \hspace{10mm} \hat{\boldsymbol{X}}_{12} \cdot \hat{\Q}_{12} \xrightarrow[]{\cI_{13}} - \hat{\boldsymbol{X}}{}^{I}_{23} \cdot \hat{\Q}^{I}_{23} \, ,\hspace{10mm} \boldsymbol{J} \xrightarrow[]{\cI_{13}} - \boldsymbol{J}^{I} \, .
	\end{align}
\end{subequations}
Hence, due to their transformation properties under $\cI$, the objects \eqref{Ch05.1-Inversion even objects} are classified as ``parity-even" as they are invariant under $\cI$, while the objects \eqref{Ch05.1-Inversion odd objects} are classified as ``parity-odd", as they are pseudo-invariant under $\cI$. At this point it is convenient to partition our solution into ``even" and ``odd" sectors as follows:
\begin{equation}
	\cH_{\cA_{1} \cA_{2}  \cA_{3} }(\boldsymbol{X}, \Q) = \cH^{(+)}_{\cA_{1} \cA_{2}  \cA_{3} }(\boldsymbol{X}, \Q) + \cH^{(-)}_{\cA_{1} \cA_{2}  \cA_{3} }(\boldsymbol{X}, \Q) \, , 
\end{equation}
where $\cH^{(+)}$ contains all structures that are invariant under $\cI$, and $\cH^{(-)}$ contains all structures that are pseudo-invariant under $\cI$. With this choice of convention, as a consequence of \eqref{Ch05.1-Three-point building blocks 1a - properties 1}, \eqref{Ch05.1-Three-point building blocks 1a - properties 2}, the following relation holds:
\begin{align} \label{Ch05.1-Hc and H relation}
	\hat{\cH}^{I \, (\pm)}_{\cA_{1} \cA_{2} \cA_{3}}(\boldsymbol{X}_{23}, \Q_{23}) &= \pm \, \cI^{(1)}{}_{\cA_{1}}{}^{\cA'_{1}}(\boldsymbol{x}_{13}) \, \cI^{(2)}{}_{\cA_{2}}{}^{\cA'_{2}}(\boldsymbol{x}_{13}) \nonumber \\
	& \hspace{20mm} \times \cI^{(3)}{}_{\cA_{3}}{}^{\cA'_{3}}(\boldsymbol{x}_{13}) \, \hat{\cH}^{(\pm)}_{\cA'_{1} \cA'_{2} \cA'_{3}}(\boldsymbol{X}_{12}, \Q_{12}) \, ,
\end{align}
where $\hat{\cH}^{I \, (\pm)}_{\cA_{1} \cA_{2} \cA_{3}}(\boldsymbol{X}, \Q) = \hat{\cH}^{(\pm)}_{\cA_{1} \cA_{2} \cA_{3}}(\boldsymbol{X}^{I}, \Q^{I})$. A result analogous to \eqref{Ch05.1-Inversion even objects}, \eqref{Ch05.1-Inversion odd objects} that follows from the properties of the inversion tensor acting on $(\boldsymbol{X}, \Q)$ is
\begin{subequations}
	\begin{align} \label{Ch05.1-Inverison even objects - X}
		\hat{\boldsymbol{X}} \xrightarrow[]{\cI_{\boldsymbol{X}}}   \hat{\boldsymbol{X}}{}^{I} \, , \hspace{10mm} \hat{\Q} \xrightarrow[]{\cI_{\boldsymbol{X}}} \hat{\Q}^{I} \, , 
	\end{align}
	\vspace{-10mm}
	\begin{align} \label{Ch05.1-Inverison odd objects - X}
		\ve \xrightarrow[]{\cI_{\boldsymbol{X}}} -\ve  \, , \hspace{10mm} \hat{\boldsymbol{X}} \cdot \hat{\Q} \xrightarrow[]{\cI_{\boldsymbol{X}}} - \hat{\boldsymbol{X}}{}^{I} \cdot \hat{\Q}^{I}\, ,\hspace{10mm} \boldsymbol{J} \xrightarrow[]{\cI_{\boldsymbol{X}}} - \boldsymbol{J}^{I} \, .
	\end{align}
\end{subequations}
Hence, we also have the general formula
\begin{align} \label{Ch05.1-H inversion}
	\cH^{I \, (\pm)}_{\cA_{1} \cA_{2} \cA_{3}}(\boldsymbol{X}, \Q) = \pm \, \cI^{(1)}{}_{\cA_{1}}{}^{\cA'_{1}}(\boldsymbol{X}) \, \cI^{(2)}{}_{\cA_{2}}{}^{\cA'_{2}}(\boldsymbol{X}) \, \cI^{(3)}{}_{\cA_{3}}{}^{\cA'_{3}}(\boldsymbol{X}) \, \cH^{(\pm)}_{\cA'_{1} \cA'_{2} \cA'_{3}}(\boldsymbol{X}, \Q) \, ,
\end{align}
which is generally more simple to compute. After substituting \eqref{Ch05.1-Hc and H relation} into \eqref{Ch05.1-Htilde and H relation}, we obtain the following relation between $\cH$ and $\tilde{\cH}$:
\begin{equation} \label{Ch05.1-Htilde and Hc relation}
	\tilde{\cH}^{(\pm)}_{\cA_{1} \cA_{2} \cA_{3} }(\boldsymbol{X},\Q) = (\boldsymbol{X}^{2})^{\D_{1} - \D_{3}} \, \cI^{(2)}{}_{\cA_{2}}{}^{\cA'_{2}}(\boldsymbol{X}) \, \cH^{I \, (\pm)}_{\cA_{1} \cA'_{2} \cA_{3}}(\boldsymbol{X}, \Q) \, . 
\end{equation}
Again, it is now apparent that $\cI$ acts as an intertwining operator between the various representations of the correlation function. Once $\tilde{\cH}$ is obtained we can then impose conservation on $\Pi$ as if it were located at the ``first point'', using identities analogous to \eqref{Ch05.1-Three-point building blocks 1c - differential identities 3}, \eqref{Ch05.1-Three-point building blocks 1c - differential identities 4}.  

If we now consider the correlation function of three conserved primary superfields $\mathbf{J}^{}_{\a(I)}$, $\mathbf{J}'_{\b(J)}$, $\mathbf{J}''_{\g(K)}$, where $I=2s_{1}$, $J=2s_{2}$, $K=2s_{3}$, then the general ansatz is
\begin{align} \label{Ch05.1-Conserved correlator ansatz}
	\langle \, \mathbf{J}^{}_{\a(I)}(z_{1}) \, \mathbf{J}'_{\b(J)}(z_{2}) \, \mathbf{J}''_{\g(K)}(z_{3}) \rangle = \frac{ \cI_{\a(I)}{}^{\a'(I)}(\boldsymbol{x}_{13}) \,  \cI_{\b(J)}{}^{\b'(J)}(\boldsymbol{x}_{23}) }{(\boldsymbol{x}_{13}^{2})^{\D_{1}} (\boldsymbol{x}_{23}^{2})^{\D_{2}}}
	\; \cH_{\a'(I) \b'(J) \g(K)}(\boldsymbol{X}_{12}, \Q_{12}) \, ,
\end{align} 
where $\D_{i} = s_{i} + 1$. The constraints on $\cH$ are then as follows:
\begin{enumerate}
	\item[\textbf{(i)}] {\bf Homogeneity:}
	\begin{equation}
		\cH_{\a(I) \b(J) \g(K)}(\l^{2} \boldsymbol{X}, \l \Q) = (\l^{2})^{\D_{3} - \D_{2} - \D_{1}} \, \cH_{\a(I) \b(J) \g(K)}(\boldsymbol{X}, \Q) \, ,
	\end{equation}
	It is often convenient to introduce $\hat{\cH}_{\a(I) \b(J) \g(K)}(\boldsymbol{X}, \Q)$, such that
	\begin{align}
		\cH_{\a(I) \b(J) \g(K)}(\boldsymbol{X},\Q) &= \boldsymbol{X}^{\D_{3} - \D_{3}- \D_{1}} \hat{\cH}_{\a(I) \b(J) \g(K)}(\boldsymbol{X}, \Q) \, ,
	\end{align}
	where $\hat{\cH}_{\a(I) \b(J) \g(K)}(\boldsymbol{X}, \Q)$ is homogeneous degree 0 in $(\boldsymbol{X}, \Q)$, i.e.
	\begin{align}
		\hat{\cH}_{\a(I) \b(J) \g(K)}(\l^{2} \boldsymbol{X}, \l \Q) &= \hat{\cH}_{\a(I) \b(J) \g(K)}(\boldsymbol{X}, \Q) \, .
	\end{align}
	
	\item[\textbf{(ii)}] {\bf Differential constraints:} \\
	After application of the identities \eqref{Ch05.1-Three-point building blocks 1c - differential identities 3}, \eqref{Ch05.1-Three-point building blocks 1c - differential identities 4} we obtain the following constraints:
	\begin{subequations} \label{Ch05.1-Conservation on H - tensor formalism}
		\begin{align}
			\text{Conservation at $z_{1}$:} && \cD^{\a} \cH_{\a \a(I - 1) \b(J) \g(K)}(\boldsymbol{X}, \Q) &= 0 \, , \\
			\text{Conservation at $z_{2}$:} && \cQ^{\b} \cH_{\a(I) \b \b(J-1) \g(K)}(\boldsymbol{X}, \Q) &= 0 \, , \\
			\text{Conservation at $z_{3}$:} && \cQ^{\g} \tilde{\cH}_{\a(I) \b(J) \g \g(K-1)  }(\boldsymbol{X}, \Q) &= 0 \, ,
		\end{align}
	\end{subequations}
	where
	\begin{equation}
		\tilde{\cH}^{(\pm)}_{\a(I) \b(J) \g(K) }(\boldsymbol{X}, \Q) = (\boldsymbol{X}^{2})^{\D_{1} - \D_{3}} \, \cI_{\b(J)}{}^{\b'(J)}(\boldsymbol{X}) \, \cH^{I \, (\pm)}_{\a(I) \b'(J) \g(K)}(\boldsymbol{X}, \Q) \, . 
	\end{equation}

	\item[\textbf{(iii)}] {\bf Point-switch symmetries:} \\
	If the fields $\mathbf{J}$ and $\mathbf{J}'$ coincide, then we obtain the following point-switch identity
	\begin{equation}
		\cH_{\a(I) \b(I) \g(K)}(\boldsymbol{X}, \Q) = (-1)^{\e(\mathbf{J})} \cH_{\b(I) \a(I) \g(K)}(-\boldsymbol{X}^{\text{T}}, -\Q) \, ,
	\end{equation}
	where $\e(\mathbf{J})$ is the Grassmann parity of $\mathbf{J}$. Likewise, if the fields $\mathbf{J}$ and $\mathbf{J}''$ coincide, then we obtain the constraint
	\begin{equation}
		\tilde{\cH}_{\a(I) \b(J) \g(I) }(\boldsymbol{X}, \Q) = (-1)^{\e(\mathbf{J})} \cH_{\g(I) \b(J) \a(I)}(-\boldsymbol{X}^{\text{T}}, -\Q) \, .
	\end{equation}
\end{enumerate}
In practice, imposing these constraints on correlation functions involving higher-spin supercurrents quickly becomes unwieldy using the tensor formalism, particularly due to the sheer number of possible tensor structures for a given set of superspins. It is therefore advantageous to use auxiliary spinors analagous to the 3D CFT case; recalling that $I = 2s_{1}$, $J = 2s_{2}$, $K = 2s_{3} $, first we define:
\begin{subequations}
	\begin{align}
		\mathbf{J}^{}_{s_{1}}(z_{1}; u) & = \mathbf{J}_{\a(I)}(z_{1}) \, \boldsymbol{u}^{\a(I)} \, , & \mathbf{J}'_{s_{2}}(z_{2}; v) &= \mathbf{J}_{\b(J)}(z_{2}) \, \boldsymbol{v}^{\a(J)} \, ,
	\end{align}
	\vspace{-10mm}
	\begin{align}
		\mathbf{J}''_{s_{3}}(z_{3}; w) &= \mathbf{J}_{\g(K)}(z_{3}) \, \boldsymbol{w}^{\g(K)} \, .
	\end{align}
\end{subequations}
The general ansatz for the three-point function is as follows:
\begin{align}
	\langle \, \mathbf{J}^{}_{s_{1}}(z_{1}; u) \, \mathbf{J}'_{s_{2}}(z_{2}; v) \, \mathbf{J}''_{s_{3}}(z_{3}; w) \rangle = \frac{ \cI^{(I)}(\boldsymbol{x}_{13}; u, \tilde{u}) \,  \cI^{(J)}(\boldsymbol{x}_{23}; v, \tilde{v}) }{(\boldsymbol{x}_{13}^{2})^{\D_{1}} (\boldsymbol{x}_{23}^{2})^{\D_{2}}}
	\; \cH(\boldsymbol{X}_{12}, \Q_{12}; \tilde{u},\tilde{v},w) \, ,
\end{align} 
where the generating polynomial $\cH(\boldsymbol{X},\Q; u,v,w)$ is defined as
\begin{align} \label{Ch05.1-H - generating polynomial}
	\cH(\boldsymbol{X}, \Q; u,v,w) = \cH_{\a(I) \b(J) \g(K)}(\boldsymbol{X}, \Q)  \boldsymbol{u}^{\a(I)}  \boldsymbol{v}^{\b(J)}  \boldsymbol{w}^{\g(K)} \, ,
\end{align}
and
\begin{equation}
	\cI^{(s)}(\boldsymbol{x}; u,\tilde{u}) \equiv \cI^{(s)}_{\boldsymbol{x}}(u,\tilde{u}) = \boldsymbol{u}^{\a(s)} \cI_{\a(s)}{}^{\a'(s)}(\boldsymbol{x}) \, \frac{\pa}{\pa \tilde{\boldsymbol{u}}^{\a'(s)}} \, ,
\end{equation}
is the inversion operator acting on polynomials degree $s$ in $\tilde{u}$, and $\D_{i} = s_{i} + 1$.
After converting the constraints summarised in the previous subsection into the auxiliary spinor formalism, we obtain:
\begin{enumerate}
	\item[\textbf{(i)}] {\bf Homogeneity:}
	\begin{equation}
		\cH(\l^{2} \boldsymbol{X}, \l \Q ; u(I), v(J), w(K)) = (\l^{2})^{\D_{3} - \D_{2} - \D_{1}} \, \cH(\boldsymbol{X}, \Q; u(I), v(J), w(K)) \, ,
	\end{equation}
	where we have used the notation $u(I)$, $v(J)$, $w(K)$ to keep track of the homogeneity of the auxiliary spinors $u$, $v$ and $w$.
	\item[\textbf{(ii)}] {\bf Differential constraints:} \\
	First, define the following three differential operators:
	\begin{align}
		D_{1} = \cD^{\a} \frac{\pa}{\pa u^{\a}} \, , && D_{2} = \cQ^{\a} \frac{\pa}{\pa v^{\a}} \, , && D_{3} = \cQ^{\a} \frac{\pa}{\pa w^{\a}} \, .
	\end{align}
	Conservation on all three points may be imposed using the following constraints:
	\begin{subequations} \label{Ch05.1-Conservation equations}
		\begin{align}
			\text{Conservation at $z_{1}$:} && D_{1} \, \cH(\boldsymbol{X}, \Q; u(I), v(J), w(K)) &= 0 \, , \\[1mm]
			\text{Conservation at $z_{2}$:} && D_{2} \, \cH(\boldsymbol{X}, \Q; u(I), v(J), w(K)) &= 0 \, , \\[1mm]
			\text{Conservation at $z_{3}$:} && D_{3} \, \tilde{\cH}(\boldsymbol{X}, \Q; u(I), v(J), w(K)) &= 0 \, ,
		\end{align}
	\end{subequations}
	where, in the auxiliary spinor formalism, $\tilde{\cH} = \tilde{\cH}^{(+)} + \tilde{\cH}^{(-)}$ is computed as follows:
	\begin{equation}
		\tilde{\cH}^{(\pm)}(\boldsymbol{X}, \Q; u(I), v(J), w(K) ) =  (\boldsymbol{X}^{2})^{\D_{1} - \D_{3}} \cI^{(J)}_{\boldsymbol{X}}(v,\tilde{v}) \, \cH^{I \, (\pm)}(\boldsymbol{X}, \Q; u(I), \tilde{v}(J), w(K)) \, , 
	\end{equation}
	where $\cI^{(s)}_{\boldsymbol{X}}(v,\tilde{v}) \equiv \cI^{(s)}(\boldsymbol{X}; v,\tilde{v})$.
	\item[\textbf{(iii)}] {\bf Point switch symmetries:} \\
	If the fields $\mathbf{J}$ and $\mathbf{J}'$ coincide (hence $I = J$), then we obtain the following point-switch constraint
	\begin{equation} \label{Ch05.1-Point switch A}
		\cH(\boldsymbol{X}, \Q; u(I), v(I), w(K)) = (-1)^{\e(\mathbf{J})} \cH(- \boldsymbol{X}^{\text{T}}, - \Q; v(I), u(I), w(K)) \, ,
	\end{equation}
	where, again, $\e(\mathbf{J})$ is the Grassmann parity of $\mathbf{J}$. Similarly, if the fields $\mathbf{J}$ and $\mathbf{J}''$ coincide (hence $I = K$) then we obtain the constraint
	\begin{equation} \label{Ch05.1-Point switch B}
		\tilde{\cH}(\boldsymbol{X}, \Q; u(I), v(J), w(I)) = (-1)^{\e(\mathbf{J})} \cH(- \boldsymbol{X}^{\text{T}}, - \Q; w(I), v(J), u(I)) \, .
	\end{equation}
\end{enumerate}
To find an explicit solution for the polynomial \eqref{Ch05.1-H - generating polynomial}, one must now consider all possible scalar combinations of $\boldsymbol{X}$, $\Q$, $\ve$, $u$, $v$ and $w$ with the appropriate homogeneity. Hence, let us introduce the following structures: \\[2mm]
\textbf{Bosonic:}
\begin{subequations} \label{Ch05.1-Basis scalar structures 1}
	\begin{align}
		P_{1} &= \ve_{\a \b} v^{\a} w^{\b} \, , & P_{2} &= \ve_{\a \b} w^{\a} u^{\b} \, , & P_{3} &= \ve_{\a \b} u^{\a} v^{\b} \, , \\
		\mathbb{Q}_{1} &= \hat{\boldsymbol{X}}_{\a \b} v^{\a} w^{\b} \, , & \mathbb{Q}_{2} &= \hat{\boldsymbol{X}}_{\a \b} w^{\a} u^{\b} \, , & \mathbb{Q}_{3} &= \hat{\boldsymbol{X}}_{\a \b} u^{\a} v^{\b} \, , \\
		\mathbb{Z}_{1} &= \hat{\boldsymbol{X}}_{\a \b} u^{\a} u^{\b} \, , & \mathbb{Z}_{2} &= \hat{\boldsymbol{X}}_{\a \b} v^{\a} v^{\b} \, , & \mathbb{Z}_{3} &= \hat{\boldsymbol{X}}_{\a \b} w^{\a} w^{\b} \, .
	\end{align}
\end{subequations}
\textbf{Fermionic:}
\begin{subequations} \label{Ch05.1-Basis scalar structures 2}
	\begin{align}
		R_{1} &= \ve_{\a \b} u^{\a} \hat{\Q}^{\b} \, , & R_{2} &= \ve_{\a \b} v^{\a} \hat{\Q}^{\b} \, , & R_{3} &= \ve_{\a \b} w^{\a} \hat{\Q}^{\b} \, , \\
		\mathbb{S}_{1} &= \hat{\boldsymbol{X}}_{\a \b} u^{\a} \hat{\Q}^{\b} \, , & \mathbb{S}_{2} &= \hat{\boldsymbol{X}}_{\a \b} v^{\a} \hat{\Q}^{\b} \, , & \mathbb{S}_{3} &= \hat{\boldsymbol{X}}_{\a \b} w^{\a} \hat{\Q}^{\b} \, .
	\end{align}
\end{subequations}
A general solution for $\cH(\boldsymbol{X}, \Q)$ is comprised of all possible combinations of $P_{i}, \mathbb{Q}_{i}, \mathbb{Z}_{i}, R_{i}, \mathbb{S}_{i}$ and $\boldsymbol{J}$ which possess the correct homogeneity in $u$, $v$ and $w$. Comparing with \eqref{Ch05.1-Inversion even objects}, \eqref{Ch05.1-Inversion odd objects}, we can identify the objects $P_{i}$, $\mathbb{S}_{i}$ and $\boldsymbol{J}$ as being ``parity-odd" due to their transformation properties under inversions. 

For the subsequent analysis of conserved three-point functions, due to the property \eqref{Ch05.1-Three-point building blocks 1a - properties 3}, and the fact that in $\cN=1$ theories $\Q^{3} = 0 \implies \boldsymbol{X}^{2} = X^{2}$, it is generally more convenient to construct the polynomial in terms of the symmetric spin-tensor, $X_{\a \b}$, rather than $\boldsymbol{X}_{\a \b}$, resulting in the polynomial $\cH(X, \Q)$. Hence, we expand $\mathbb{Q}_{i}, \mathbb{Z}_{i}, \mathbb{S}_{i}$ as follows:
\begin{align}
	\mathbb{Q}_{i} = Q_{i} - \frac{\text{i}}{2} \, P_{i} \, \boldsymbol{J} \, , && \mathbb{Z}_{i} &= Z_{i} \, ,
	&& \mathbb{S}_{i} = S_{i} \, ,
\end{align}
where we have defined
\begin{subequations}
	\begin{align}
		Q_{1} &= \hat{X}_{\a \b} v^{\a} w^{\b} \, , & Q_{2} &= \hat{X}_{\a \b} w^{\a} u^{\b} \, , & Q_{3} &= \hat{X}_{\a \b} u^{\a} v^{\b} \, , \\
		Z_{1} &= \hat{X}_{\a \b} u^{\a} u^{\b} \, , & Z_{2} &= \hat{X}_{\a \b} v^{\a} v^{\b} \, , & Z_{3} &= \hat{X}_{\a \b} w^{\a} w^{\b} \, , \\
		S_{1} &= \hat{X}_{\a \b} u^{\a} \hat{\Q}^{\b} \, , & S_{2} &= \hat{X}_{\a \b} v^{\a} \hat{\Q}^{\b} \, , & S_{3} &= \hat{X}_{\a \b} w^{\a} \hat{\Q}^{\b} \, .
	\end{align}
\end{subequations}
The polynomial $\cH(X, \Q)$ is now constructed from all possible combinations of $P_{i}$, $Q_{i}$, $Z_{i}$, $R_{i}$, $S_{i}$ 
and $\boldsymbol{J}$. Once a general solution for $\cH(X, \Q)$ is obtained, one can convert back to ``covariant form", $\cH(\boldsymbol{X}, \Q)$, by making the replacements
\begin{align}
	Q_{i} \rightarrow \mathbb{Q}_{i} + \frac{\text{i}}{2} \, P_{i} \, \boldsymbol{J} \, , && Z_{i} \rightarrow \mathbb{Z}_{i} \, ,
	&& S_{i} \rightarrow \mathbb{S}_{i} \, .
\end{align}

\subsubsection{Generating function formalism}\label{Ch05-subsection5.2.2.2}
In general, it is a non-trivial technical problem to come up with an exhaustive list of possible solutions for $\cH(X,\Q;u,v,w)$ for a given set of superspins, however, this process can be simplified by introducing generating functions for the polynomial $\cH(X,\Q; u, v, w)$. First we introduce the function $\cF(X)$, defined as follows:
\begin{align} \label{Ch05.1-Generating function 1}
	\cF(X) &= X^{\d} P_{1}^{k_{1}} P_{2}^{k_{2}} P_{3}^{k_{3}} Q_{1}^{l_{1}} Q_{2}^{l_{2}} Q_{3}^{l_{3}} Z_{1}^{m_{1}} Z_{2}^{m_{2}} Z_{3}^{m_{3}}
\end{align}
where, typically, $\d = \D_{3} - \D_{2} - \D_{1}$. The generating functions for Grassmann-even and Grassmann-odd correlators in $\cN=1$ theories are then defined as follows:
\begin{align} \label{Ch05.1-Generating function 2}
	\cG(X,\Q \, | \, \G) &= \begin{cases} 
		\cF(X) \, \boldsymbol{J}^{\s} \, , & \text{Bosonic} \\
		\cF(X) \, R_{1}^{p_{1}} R_{2}^{p_{2}} R_{3}^{p_{3}} S_{1}^{q_{1}} S_{2}^{q_{2}} S_{3}^{q_{3}} \, , & \text{Fermionic}
	\end{cases}
\end{align}
Here the non-negative integers, $ \G = \{ k_{i}, l_{i}, m_{i}, p_{i}, q_{i}, \s\}$, $i=1,2,3$, are  constrained; for overall bosonic correlation functions they are solutions to the following linear system
\begin{subequations} \label{Ch05.1-Diophantine equations 1}
	\begin{align}
		k_{2} + k_{3} + l_{2} + l_{3} + 2m_{1} &= I \, , \\
		k_{1} + k_{3} + l_{1} + l_{3} + 2m_{2} &= J \, , \\
		k_{1} + k_{2} + l_{1} + l_{2} + 2m_{3} &= K \, ,
	\end{align}
\end{subequations}
with $\s = 0,1$. Likewise, for overall fermionic correlation functions, the integers $\G$ are solutions to the following system
\begin{subequations} \label{Ch05.1-Diophantine equations 2}
	\begin{align}
		k_{2} + k_{3} + l_{2} + l_{3} + 2m_{1} + p_{1} + q_{1} &= I \, , \\
		k_{1} + k_{3} + l_{1} + l_{3} + 2m_{2} + p_{2} + q_{2} &= J \, , \\
		k_{1} + k_{2} + l_{1} + l_{2} + 2m_{3} + p_{3} + q_{3} &= K \, , \\
		p_{1} + p_{2} + p_{3} + q_{1} + q_{2} + q_{3} &= 1 \, ,
	\end{align}
\end{subequations}
where $I = 2s_{1}$, $J = 2s_{2}$, $K = 2s_{3}$ specify the spin-structure of the correlation function. These equations are obtained by comparing the homogeneity of the auxiliary spinors $u$, $v$, $w$ in the generating functions \eqref{Ch05.1-Generating function 2}, against the index structure of the tensor $\cH$. The solutions correspond to a linearly dependent basis of structures in which the polynomial $\cH$ can be decomposed. Using \textit{Mathematica} it is straightforward to generate all possible solutions to \eqref{Ch05.1-Diophantine equations 1}, \eqref{Ch05.1-Diophantine equations 2} for fixed values of the superspins. 

Now let us assume there exists a finite number of solutions $\G_{i}$, $i = 1, ..., N$ to \eqref{Ch05.1-Diophantine equations 1}, \eqref{Ch05.1-Diophantine equations 2} for a given choice of $I,J,K$. The set of solutions $\G = \{ \G_{i} \}$ may be partitioned into ``even" and ``odd" sets $\G^{+}$ and $\G^{-}$ respectively by counting the number of pseudo-invariant basis structures present in a particular solution. Therefore we define:
\begin{align}
	\G^{+} = \G|_{ \, k_{1} + k_{2} + k_{3} + q_{1} + q_{2} + q_{3} + \s \, ( \hspace{-1mm}\bmod 2 ) = 0} \, , && \G^{-} = \G|_{ \, k_{1} + k_{2} + k_{3} + q_{1} + q_{2} + q_{3} + \s\, ( \hspace{-1mm} \bmod 2 ) = 1} \, .
\end{align}
Hence, the even solutions are those such that $k_{1} + k_{2} + k_{3} + q_{1} + q_{2} + q_{3} + \s = \text{even}$ (i.e contains an even number of parity-odd building blocks), while the odd solutions are those such that $k_{1} + k_{2} + k_{3} + q_{1} + q_{2} + q_{3} + \s= \text{odd}$ (contains an odd number of parity-odd building blocks). Let $|\G^{+}| = N^{+}$ and $|\G^{-}| = N^{-}$, with $N = N^{+} + N^{-}$, then the most general ansatz for the polynomial $\cH$ in \eqref{Ch05.1-H - generating polynomial} is as follows:
\begin{equation}  \label{Ch05.1-H decomposition}
	\cH(X, \Q; u, v, w) = \cH^{(+)}(X, \Q; u, v, w) + \cH^{(-)}(X, \Q; u, v, w) \, ,
\end{equation}
where
\begin{subequations}
	\begin{align}
		\cH^{(+)}(X, \Q; u, v, w) &= \sum_{i=1}^{N^{+}} A_{i} \, \cG(X, \Q \, | \,\G^{+}_{i}) \, , \\
		\cH^{(-)}(X, \Q; u, v, w) &= \sum_{i=1}^{N^{-}} B_{i} \, \cG(X, \Q \, | \, \G^{-}_{i}) \, ,
	\end{align}
\end{subequations}
and $A_{i}$ and $B_{i}$ are real constants. Using this method one can generate all the possible structures for a given set of superspins $(s_{1}, s_{2}, s_{3} )$, however, at this stage we must recall that the solutions generated using this approach are linearly dependent. To form a linearly independent set of solutions we must systematically take into account the following non-linear relations between the primitive structures: 
\begin{subequations}
	\begin{align} \label{Ch05.1-Linear dependence 1}
		Z_{2} Z_{3} + P_{1}^{2} - Q_{1}^{2} &= 0 \, , \\
		Z_{1} Z_{3} + P_{2}^{2} - Q_{2}^{2} &= 0 \, , \\
		Z_{1} Z_{2} + P_{3}^{2} - Q_{3}^{2} &= 0 \, ,
	\end{align}
\end{subequations}
\vspace{-8mm}
\begin{subequations}
	\begin{align} \label{Ch05.1-Linear dependence 2}
		P_{1} Z_{1} + P_{2} Q_{3} + P_{3} Q_{2} &= 0 \, , & Q_{1} Z_{1} - Q_{2} Q_{3} - P_{2} P_{3} &= 0 \, , \\
		P_{2} Z_{2} + P_{1} Q_{3} + P_{3} Q_{1} &= 0 \, , & Q_{2} Z_{2} - Q_{1} Q_{3} - P_{1} P_{3} &= 0 \, , \\
		P_{3} Z_{3} + P_{1} Q_{2} + P_{2} Q_{1} &= 0 \, , & Q_{3} Z_{3} - Q_{1} Q_{2} - P_{1} P_{2} &= 0 \, .
	\end{align}
\end{subequations}
These allow elimination of the combinations $Z_{i} Z_{j}$, $Z_{i} P_{i}$, $Z_{i} Q_{i}$. There is also another relation involving triple products:
\begin{align} \label{Ch05.1-Linear dependence 3}
	P_{1} P_{2} P_{3} + P_{1} Q_{2} Q_{3} + P_{2} Q_{1} Q_{3} + P_{3} Q_{1} Q_{2} &= 0 \, ,
\end{align}
which allows elimination of $P_{1} P_{2} P_{3}$. The relations above are identical to those appearing in \eqref{Ch03-Linear dependence 1}-\eqref{Ch03-Linear dependence 4}, however, they must be supplemented by relations involving the fermionic structures:
\begin{subequations}
	\begin{align} \label{Ch05.1-Linear dependence 4}
		P_{1} R_{1} - Q_{2} S_{2} + Q_{3} S_{3} &= 0 \, , & P_{1} S_{1} - Q_{2} R_{2} + Q_{3} R_{3} &= 0 \, , \\
		P_{2} R_{2} - Q_{3} S_{3} + Q_{1} S_{1} &= 0 \, , & P_{2} S_{2} - Q_{3} R_{3} + Q_{1} R_{1} &= 0 \, , \\
		P_{3} R_{3} - Q_{1} S_{1} + Q_{2} S_{2} &= 0 \, , & P_{3} S_{3} - Q_{1} R_{1} + Q_{2} R_{2} &= 0 \, ,
	\end{align}
\end{subequations}
\vspace{-8mm}
\begin{subequations}
	\begin{align} \label{Ch05.1-Linear dependence 5}
		Z_{1} R_{2} - Q_{3} R_{1} + P_{3} S_{1} &= 0 \, , & Z_{2} R_{1} - Q_{3} R_{2} - P_{3} S_{2} &= 0 \, , \\
		Z_{2} R_{3} - Q_{1} R_{2} + P_{1} S_{2} &= 0 \, , & Z_{3} R_{2} - Q_{1} R_{3} - P_{1} S_{3} &= 0 \, , \\
		Z_{3} R_{1} - Q_{2} R_{3} + P_{2} S_{3} &= 0 \, , & Z_{1} R_{3} - Q_{2} R_{1} - P_{2} S_{1} &= 0 \, ,
	\end{align}
\end{subequations}
\vspace{-8mm}
\begin{subequations}
	\begin{align} \label{Ch05.1-Linear dependence 6}
		Z_{1} S_{2} - Q_{3} S_{1} + P_{3} R_{1} &= 0 \, , & Z_{2} S_{1} - Q_{3} S_{2} - P_{3} R_{2} &= 0 \, , \\
		Z_{2} S_{3} - Q_{1} S_{2} + P_{1} R_{2} &= 0 \, , & Z_{3} S_{2} - Q_{1} S_{3} - P_{1} R_{3} &= 0 \, , \\
		Z_{3} S_{1} - Q_{2} S_{3} + P_{2} R_{3} &= 0 \, , & Z_{1} S_{3} - Q_{2} S_{1} - P_{2} R_{1} &= 0 \, .
	\end{align}
\end{subequations}
These allow for elimination of the products $P_{i} R_{i}$, $P_{i} S_{i}$, $Z_{i} R_{j}$, $Z_{i} S_{j}$. As a consequence of \eqref{Ch05.1-Linear dependence 4}, the following also hold:
\begin{subequations}
	\begin{align}
		P_{1} R_{1} + P_{2} R_{2} + P_{3} R_{3} &= 0 \, , \\
		P_{1} S_{1} + P_{2} S_{2} + P_{3} S_{3} &= 0 \, .	
	\end{align}
\end{subequations}
Applying the above relations to a set of linearly dependent polynomial structures significantly reduces the number of structures to consider for a given three-point function, since we are now restricted to only linearly independent contributions. This process is relatively straightforward to implement using Mathematica's pattern matching functions.

Now that we have taken care of linear-dependence, it now remains to impose conservation on all three points in addition to the various point-switch symmetries; introducing the objects $P_{i}, Q_{i}, Z_{i}, R_{i}, S_{i}$ proves to streamline this analysis significantly. First let us consider conservation;
to impose conservation on $z_{1}$, (for either sector) we compute
\begin{align}
	D_{1} \cH(X, \Q; u,v,w) &= D_{1} \Bigg\{ \sum_{i=1}^{N} c_{i} \, \cG(X, \Q \, | \, \G_{i}) \Bigg\} \nonumber \\
	&= \sum_{i=1}^{N} c_{i} \, D_{1} \cG(X, \Q \, | \, \G_{i}) \, .
\end{align}
We then solve for the coefficient $c_{i}$ such that the result vanishes. For imposing superfield conservation equations on three-point correlation functions, the following identities are useful:
\begin{subequations} \label{App5A-Derivative identities}
	\begin{align}
		\cD^{\a} Q_{1} &= \frac{\text{i}}{X^{1/2}} \Big\{ v^{\a} R_{3} + w^{\a} R_{2} - Q_{1} \, 	(\hat{X}\cdot\hat{\Q})^{\a} \Big\} \, , \\
		\cD^{\a} Q_{2} &= \frac{\text{i}}{X^{1/2}} \Big\{ u^{\a} R_{3} + w^{\a} R_{1} - Q_{2} \, 	(\hat{X}\cdot\hat{\Q})^{\a} \Big\} \, , \\
		\cD^{\a} Q_{3} &= \frac{\text{i}}{X^{1/2}} \Big\{ u^{\a} R_{2} + v^{\a} R_{1} - Q_{3} \, 	(\hat{X}\cdot\hat{\Q})^{\a} \Big\} \, ,
	\end{align}
	\vspace{-5mm}
	\begin{align}
		\cD^{\a} Z_{1} &= \frac{\text{i}}{X^{1/2}} \Big\{ 2 u^{\a} R_{1} - Z_{1} \, (\hat{X}\cdot\hat{\Q})^{\a} \Big\} \, , \\
		\cD^{\a} Z_{2} &= \frac{\text{i}}{X^{1/2}} \Big\{ 2 v^{\a} R_{2} - Z_{2} \, (\hat{X}\cdot\hat{\Q})^{\a} \Big\} \, , \\
		\cD^{\a} Z_{3} &= \frac{\text{i}}{X^{1/2}} \Big\{ 2 w^{\a} R_{3} - Z_{3} \, (\hat{X}\cdot\hat{\Q})^{\a} \Big\} \, ,
	\end{align}
	\vspace{-5mm}
	\begin{align}
		\cD^{\a} R_{1} &= \frac{1}{X^{1/2}} \Big\{ - u^{\a} - \frac{\text{i}}{4} (\hat{X} \cdot u)^{\a} \boldsymbol{J} \Big\} \, , \\
		\cD^{\a} R_{2} &= \frac{1}{X^{1/2}} \Big\{ - v^{\a} - \frac{\text{i}}{4} (\hat{X} \cdot v)^{\a} \boldsymbol{J} \Big\} \, , \\
		\cD^{\a} R_{3} &= \frac{1}{X^{1/2}} \Big\{ - w^{\a} - \frac{\text{i}}{4} (\hat{X} \cdot w)^{\a} \boldsymbol{J} \Big\} \, ,
	\end{align}
	\vspace{-5mm}
	\begin{align}
		\cD^{\a} S_{1} &= \frac{1}{X^{1/2}} \Big\{  (\hat{X} \cdot 	u)^{\a} - \frac{3\text{i}}{4} \, u^{\a} \boldsymbol{J} \Big\} \, , \\
		\cD^{\a} S_{2} &= \frac{1}{X^{1/2}} \Big\{  (\hat{X} \cdot 	v)^{\a} - \frac{3\text{i}}{4} \, v^{\a} \boldsymbol{J} \Big\} \, , \\
		\cD^{\a} S_{3} &= \frac{1}{X^{1/2}} \Big\{ (\hat{X} \cdot 	w)^{\a} - \frac{3\text{i}}{4} \, w^{\a} \boldsymbol{J} \Big\} \, .
	\end{align}
\end{subequations}
The same approach applies for conservation on $z_{2}$, and similar identities may be derived for the action of $\cQ^{\a}$ on the basis structures.

As concerns imposing conservation on $z_{3}$, we must first obtain an explicit expression for $\tilde{\cH}(\boldsymbol{X},\Q)$ in terms of $\cH(\boldsymbol{X},\Q)$, that is, we must compute (e.g. for the even sector)
\begin{equation} \label{Ch05.1-H tilde}
	\tilde{\cH}(\boldsymbol{X}, \Q; u(I), v(J), w(K) ) = (\boldsymbol{X}^{2})^{\D_{1} - \D_{3}} \cI^{(J)}_{\boldsymbol{X}}(v,\tilde{v}) \, \cH^{I}(\boldsymbol{X}, \Q; u(I), \tilde{v}(J), w(K)) \, .
\end{equation}
Recall that any solution for $\cH(\boldsymbol{X}, \Q)$ can be written in terms of the structures \eqref{Ch05.1-Basis scalar structures 1}, \eqref{Ch05.1-Basis scalar structures 2}; given the transformation properties \eqref{Ch05.1-Hc and H relation}, and \eqref{Ch05.1-Htilde and Hc relation}, the computation of $\cH^{I}(\boldsymbol{X}, \Q)$ from $\cH(\boldsymbol{X}, \Q)$ is equivalent to the following replacements:
\begin{subequations} \label{Ch05.1-Inversion transformation 1}
	\begin{align} 
		P_{1} &\rightarrow - P_{1} \, , & P_{2} &\rightarrow -P_{2} \, , & P_{3} &\rightarrow -P_{3} \, , \\
		R_{1} &\rightarrow - \mathbb{S}_{1} \, , & R_{2} &\rightarrow - \mathbb{S}_{2} \, , & R_{3} &\rightarrow - \mathbb{S}_{3} \, , \\
		\mathbb{S}_{1} &\rightarrow R_{1} \, , & \mathbb{S}_{2} &\rightarrow R_{2} \, , & \mathbb{S}_{3} &\rightarrow R_{3} \, .
	\end{align}
\end{subequations}
Now to compute $\tilde{\cH}(\boldsymbol{X},\Q)$ from $\cH^{I}(\boldsymbol{X}, \Q)$, we make use of the fact that $P_{1}$, $P_{3}$, $\mathbb{Q}_{1}$, $\mathbb{Q}_{3}$, $\mathbb{Z}_{2}$, $R_{2}$, and $\mathbb{S}_{2}$ are the only objects with $\tilde{v}$ dependence, and apply the identities
\begin{subequations} \label{Ch05.1-Inversion transformation 2}
	\begin{align}
		\cI_{\boldsymbol{X}}(v,\tilde{v}) \, P_{1} &= - \mathbb{Q}_{1} \, , & \cI_{\boldsymbol{X}}(v,\tilde{v}) \, P_{3} &= \mathbb{Q}_{3} + \text{i} P_{3} \boldsymbol{J} \, , \\
		\cI_{\boldsymbol{X}}(v,\tilde{v}) \, \mathbb{Q}_{1} &= - P_{1} + \text{i} \, \mathbb{Q}_{1} \boldsymbol{J} \, , & \cI_{\boldsymbol{X}}(v,\tilde{v}) \, \mathbb{Q}_{3} &= P_{3} \, , \\
		\cI_{\boldsymbol{X}}(v,\tilde{v}) \, R_{2} &= -\mathbb{S}_{2} \, ,  & \cI_{\boldsymbol{X}}(v,\tilde{v}) \, \mathbb{S}_{2} &= - R_{2} \, ,
	\end{align}
	\vspace{-12mm}
	\begin{align}
		\cI^{(2)}_{\boldsymbol{X}}(v,\tilde{v}) \, \mathbb{Z}_{2} = - \mathbb{Z}_{2} \,.
	\end{align}
\end{subequations}
Hence, given a solution for the polynomial $\cH(\boldsymbol{X}, \Q)$, the computation of $\tilde{\cH}(\boldsymbol{X}, \Q)$ is now equivalent to the following replacements of the basis structures \eqref{Ch05.1-Basis scalar structures 1}, \eqref{Ch05.1-Basis scalar structures 2}:
\begin{subequations} \label{Ch05.1-Htilde structure replacements}
	\begin{align} 
		P_{1} &\rightarrow \mathbb{Q}_{1} \, , & P_{2} &\rightarrow - P_{2} \, , & P_{3} &\rightarrow - \mathbb{Q}_{3} - \text{i} P_{3} \boldsymbol{J} \, , \\
		\mathbb{Q}_{1} &\rightarrow - P_{1} + \text{i} \, \mathbb{Q}_{1} \boldsymbol{J} \, , & \mathbb{Q}_{2} &\rightarrow \mathbb{Q}_{2} \, , & \mathbb{Q}_{3} &\rightarrow P_{3} \, , \\
		\mathbb{Z}_{1} &\rightarrow \mathbb{Z}_{1} \, , & \mathbb{Z}_{2} &\rightarrow - \mathbb{Z}_{2} \, , & \mathbb{Z}_{3} &\rightarrow \mathbb{Z}_{3} \, \\
		R_{1} &\rightarrow - \mathbb{S}_{1} \, , & R_{2} &\rightarrow R_{2} \, , & R_{3} &\rightarrow - \mathbb{S}_{3} \, , \\
		\mathbb{S}_{1} &\rightarrow R_{1} \, , & \mathbb{S}_{2} &\rightarrow - \mathbb{S}_{2} \, , & \mathbb{S}_{3} &\rightarrow R_{3} \, .
	\end{align}
\end{subequations}
These rules are obtained by combining \eqref{Ch05.1-Inversion transformation 1}, \eqref{Ch05.1-Inversion transformation 2}. Conservation on $z_{3}$ can now be imposed using the operator $D_{3}$.

It now remains to find out how point-switch symmetries act on the basis structures; this analysis is more simple when working with $\cH(X,\Q)$, instead of $\cH(\boldsymbol{X},\Q)$. For permutation of superspace points $z_{1}$ and $z_{2}$, we have $X \rightarrow - X$, $\Q \rightarrow - \Q$, $u \leftrightarrow v$. This results in the following replacement rules for the basis objects \eqref{Ch05.1-Basis scalar structures 1}, \eqref{Ch05.1-Basis scalar structures 2}:
\begin{subequations} \label{Ch05.1-Point switch A - basis}
	\begin{align} 
		P_{1} &\rightarrow - P_{2} \, , & P_{2} &\rightarrow - P_{1} \, , & P_{3} &\rightarrow - P_{3} \, , \\
		Q_{1} &\rightarrow - Q_{2} \, , & Q_{2} &\rightarrow - Q_{1} \, , & Q_{3} &\rightarrow - Q_{3} \, , \\
		Z_{1} &\rightarrow - Z_{2} \, , & Z_{2} &\rightarrow - Z_{1} \, , & Z_{3} &\rightarrow - Z_{3} \, , \\
		R_{1} &\rightarrow - R_{2} \, , & R_{2} &\rightarrow - R_{1} \, , & R_{3} &\rightarrow - R_{3} \, , \\
		S_{1} &\rightarrow S_{2} \, , & S_{2} &\rightarrow S_{1} \, , & S_{3} &\rightarrow S_{3} \, .
	\end{align}
\end{subequations}
Likewise, for permutation of superspace points $z_{1}$ and $z_{3}$ we have  $X \rightarrow - X$, $\Q \rightarrow - \Q$, $u \leftrightarrow w$, resulting in the following replacements:
\begin{subequations} \label{Ch05.1-Point switch B - basis}
	\begin{align} 
		P_{1} &\rightarrow - P_{3} \, , & P_{2} &\rightarrow - P_{2} \, , & P_{3} &\rightarrow - P_{1} \, , \\
		Q_{1} &\rightarrow - Q_{3} \, , & Q_{2} &\rightarrow - Q_{2} \, , & Q_{3} &\rightarrow - Q_{1} \, , \\
		Z_{1} &\rightarrow - Z_{3} \, , & Z_{2} &\rightarrow - Z_{2} \, , & Z_{3} &\rightarrow - Z_{1} \, , \\
		R_{1} &\rightarrow - R_{3} \, , & R_{2} &\rightarrow - R_{2} \, , & R_{3} &\rightarrow - R_{1} \, , \\
		S_{1} &\rightarrow S_{3} \, , & S_{2} &\rightarrow S_{2} \, , & S_{3} &\rightarrow S_{1} \, .
	\end{align}
\end{subequations}
We have now developed all the formalism necessary to analyse the structure of three-point correlation functions in 3D $\cN=1$ SCFT. In the remaining sections of this chapter we will analyse the three-point functions of conserved higher-spin supercurrents (for both integer and half-integer superspin) using the following method:
\begin{enumerate}
	\item For a given set of superspins, we construct all possible (linearly dependent) structures for the polynomial $\cH(X, \Q; u,v,w)$, which is governed by the solutions to \eqref{Ch05.1-Diophantine equations 1}, \eqref{Ch05.1-Diophantine equations 2}. The solutions are sorted into even and odd sectors.
	\item We systematically apply the linear dependence relations \eqref{Ch05.1-Linear dependence 1}, \eqref{Ch05.1-Linear dependence 2}, \eqref{Ch05.1-Linear dependence 3}, \eqref{Ch05.1-Linear dependence 4}, \eqref{Ch05.1-Linear dependence 5}, \eqref{Ch05.1-Linear dependence 6} to the set of all polynomial structures. The process is identical to the one outlined in the relevant sections of Chapter \ref{Chapter3}. 
	\item Using the method outlined in subsection \ref{Ch05-subsection5.2.2.2}, we impose the superfield conservation equations on the correlation function, resulting in the differential contraints \eqref{Ch05.1-Conservation equations} on $\cH$. The result of each computation is a large polynomial in the basis structures \eqref{Ch05.1-Basis scalar structures 1}, \eqref{Ch05.1-Basis scalar structures 2}. The linear dependence relations are systematically applied to this polynomial again to ensure that it is composed of only linearly independent terms. The coefficients are read off the structures, resulting in algebraic constraint relations on the coefficients $A_{i}, B_{i}$. This process significantly reduces the number of structures in the three-point function.
	\item Once the general form of the polynomial $\cH(X,\Q; u,v,w)$ (associated with the conserved three-point function $\langle \mathbf{J}^{}_{s_{1}} \mathbf{J}'_{s_{2}} \mathbf{J}''_{s_{3}} \rangle$) is obtained for a given set of superspins $(s_{1},s_{2}, s_{3})$, we then impose any symmetries under permutation of superspace points, that is, \eqref{Ch05.1-Point switch A} and \eqref{Ch05.1-Point switch B} (if applicable). In certain cases, imposing these constraints can eliminate the remaining structures. The solution is then converted into covariant form $\cH(\boldsymbol{X},\Q; u,v,w)$.
\end{enumerate}
The computations are carried out analytically with the use of Mathematica and the Grassmann package. By using pattern matching functions, the calculations are carried out purely amongst the basis structures \eqref{Ch05.1-Basis scalar structures 1}, \eqref{Ch05.1-Basis scalar structures 2}, and as a result we do not have to fix superspace points to certain values in any of the calculations (compared to say the approach of \cite{Nizami:2013tpa}). Instead the only fixed parameters are the superspins of the fields. Due to computational limitations we could carry out the analysis up to $s_{i} = 20$ (some steps of the calculations involve millions of terms), however, with more optimisation and sufficient computational resources this approach should hold for arbitrary superspins. 
Since there are an enormous number of possible three-point functions with $s_{i} \leq 20$, we present the final results for $\cH(\boldsymbol{X},\Q; u,v,w)$ for some low superspin examples, as the solutions and coefficient constraints quickly become cumbersome to present for the higher superspin cases. We are primarily interested in counting the number of independent polynomial structures after imposing all the constraints. 

\section{Three-point functions of conserved supercurrents}\label{Ch05-section5.3}

In the next subsections we analyse the structure of three-point correlation functions involving conserved higher-spin supercurrents. As a test of our approach we begin with an analysis of three-point functions involving currents with low superspins, such as the supercurrent and flavour current multiplets. Some of these results were first obtained using the tensor formalism in \cite{Buchbinder:2015qsa,Buchbinder:2021gwu,Buchbinder:2021qlb}.

\subsection{Supercurrent and flavour current correlators}\label{Ch05-subsection5.3.1}

The fundamental conserved supercurrents in 3D $\cN=1$ superconformal field theories are the supercurrent and flavour current multiplets. The supercurrent multiplet is described by the spin-tensor superfield, $\mathbf{J}_{\a(3)}(z)$, with scale dimension $\Delta_{J} = 5/2$. It satisfies $D^{\a_{1}} \mathbf{J}_{\a_{1} \a_{2} \a_{3}}(z) = 0$ and contains the energy-momentum tensor, $T_{\a(4)}(x) = D_{(\a_{1}} \mathbf{J}_{\a_{2} \a_{3} \a_{4})}(z) |_{\q = 0} $, and the supersymmetry current, $Q_{\a(3)}(x) = \mathbf{J}_{\a(3)}(z) |_{\q = 0} $, as its independent component fields. Likewise, the flavour current multiplet is described by a spinor superfield, $\mathbf{L}_{\a}(z)$, with scale dimension $\Delta_{L} = 3/2$. It satisfies the superfield conservation equation $D^{\a} \mathbf{L}_{\a}(z) = 0$, and contains a conserved vector current $V_{\a(2)}(x) = D_{(\a_{1}} \mathbf{L}_{\a_{2})}(z) |_{\q = 0} $. Three-point functions of these supercurrents were originally studied in \cite{Buchbinder:2015qsa,Buchbinder:2021gwu}, here we present the solutions for them using our formalism. The possible three-point functions involving the supercurrent and flavour current multiplets are: 
\begin{align} \label{Ch05.1-Low-superspin component correlators}
	\langle \mathbf{L}_{\a}(z_{1}) \, \mathbf{L}_{\b}(z_{2}) \, \mathbf{L}_{\g}(z_{3}) \rangle \, , &&  \langle \mathbf{L}_{\a}(z_{1}) \, \mathbf{L}_{\b}(z_{2}) \, \mathbf{J}_{\g(3)}(z_{3}) \rangle \, , \\
	\langle \mathbf{J}_{\a(3)}(z_{1}) \, \mathbf{J}_{\b(3)}(z_{2}) \, \mathbf{L}_{\a}(z_{3}) \rangle \, , &&  \langle \mathbf{J}_{\a(3)}(z_{1}) \, \mathbf{J}_{\b(3)}(z_{2}) \, \mathbf{J}_{\g(3)}(z_{3}) \rangle \, .
\end{align}
We note that in all cases the correlation functions are overall Grassmann-odd, hence, it's expected that each of them are fixed up to a single parity-even solution after imposing conservation on all three points. The analysis of these three-point functions is relatively straightforward using our computational approach.

\noindent\textbf{Correlation function} $\langle \mathbf{L} \mathbf{L} \mathbf{L} \rangle$\textbf{:}

Let us first consider $\langle \mathbf{L} \mathbf{L} \mathbf{L} \rangle$; within the framework of our formalism we study the three-point function $\langle \mathbf{J}^{}_{1/2} \mathbf{J}'_{1/2} \mathbf{J}''_{1/2} \rangle$. The general ansatz for this correlation function, according to \eqref{Ch05.1-Conserved correlator ansatz} is
\begin{align}
	\langle \mathbf{J}^{}_{\a}(z_{1}) \, \mathbf{J}'_{\b}(z_{2}) \, \mathbf{J}''_{\g}(z_{3}) \rangle = \frac{ \cI_{\a}{}^{\a'}(\boldsymbol{x}_{13}) \,  \cI_{\b}{}^{\b'}(\boldsymbol{x}_{23}) }{(\boldsymbol{x}_{13}^{2})^{3/2} (\boldsymbol{x}_{23}^{2})^{3/2}}
	\; \cH_{\a' \b' \g}(\boldsymbol{X}_{12}, \Q_{12}) \, .
\end{align} 
Using the formalism outlined in Subsection \ref{Ch05-subsection5.2.2.2}, all information about this correlation function is encoded in the following polynomial:
\begin{align}
	\cH(\boldsymbol{X}, \Q; u(1), v(1), w(1)) = \cH_{ \a \b \g }(\boldsymbol{X}, \Q) \, \boldsymbol{u}^{\a}  \boldsymbol{v}^{\b}  \boldsymbol{w}^{\g} \, .
\end{align}
Using Mathematica we solve \eqref{Ch05.1-Diophantine equations 2} for the chosen spins and substitute each solution into the generating function \eqref{Ch05.1-Generating function 2}. This provides us with the following list of linearly dependent polynomial structures for the polynomial $\cH(X,\Q;u,v,w)$ in the even and odd sectors respectively:
\begin{subequations}
	\begin{align}
		\textbf{Even:}& \hspace{5mm} \{Q_3 R_3,Q_2 R_2,Q_1 R_1,P_3 S_3,P_2 S_2,P_1 S_1 \} \\
		\textbf{Odd:}& \hspace{5mm} \{Q_3 S_3,Q_2 S_2,Q_1 S_1,P_3 R_3,P_2 R_2,P_1 R_1 \}
	\end{align}
\end{subequations}
After systematic application of the linear dependence relations \eqref{Ch05.1-Linear dependence 1}-\eqref{Ch05.1-Linear dependence 6} we obtain the following linearly independent sets:
\begin{subequations}
	\begin{align}
		\textbf{Even:}& \hspace{5mm} \{Q_1 R_1,Q_2 R_2,Q_3 R_3 \} \\
		\textbf{Odd:}& \hspace{5mm} \{Q_1 S_1,Q_2 S_2,Q_3 S_3 \}
	\end{align}
\end{subequations}
Next, we impose conservation on all three points. As result we obtain a linear system in the coefficients $A_{i}$ and $B_{i}$ which may be solved computationally. After solving this system of equations we obtain the following explicit solution for $\cH(\boldsymbol{X}, \Q; u,v,w)$ (i.e. in covariant form):
\begin{subequations}
	\begin{align}
		\textbf{Even:}& \hspace{5mm} \frac{A_1}{\boldsymbol{X}^{3/2}} \Big(\mathbb{Q}_1 R_1+\mathbb{Q}_2 R_2-3 \mathbb{Q}_3 R_3 \Big) \\
		\textbf{Odd:}& \hspace{5mm} 0
	\end{align}
\end{subequations}
Hence, the three-point function is fixed up to a single parity-even polynomial structure. 
After imposing symmetries under permutation of spacetime points, 
e.g. $\mathbf{J}=\mathbf{J}'=\mathbf{J}''$, the remaining structure vanishes.

This vanishing result is not surprising because it corresponds to the contribution proportional to the symmetric invariant tensor of the flavour symmetry group. 
In four dimensions this contribution is related to the chiral anomaly, which does not exist in three dimensions.
However, in the case where the flavour currents possess flavour indices, the correlator $\langle \mathbf{J}_{1/2} \, \mathbf{J}_{1/2} \, \mathbf{J}_{1/2} \rangle$ has a non-vanishing contribution (corresponding to the structure above) proportional to the totally antisymmetric structure constant. The general form of three-point functions of flavour current multiplets were also computed in~\cite{Buchbinder:2015qsa, Buchbinder:2021gwu}, where it was shown that supersymmetry relates the free parity-even coefficients in the corresponding non-supersymmetric three-point function of vector currents.

\vspace{2mm}

\noindent
\textbf{Correlation function} $\langle \mathbf{L} \mathbf{L} \mathbf{J} \rangle$\textbf{:}

The next example to consider is the mixed correlator $\langle \mathbf{L} \mathbf{L} \mathbf{J} \rangle$; to study this case we may examine the correlation function $\langle \mathbf{J}^{}_{1/2} \mathbf{J}'_{1/2} \mathbf{J}''_{3/2} \rangle$. Using the general formula, the ansatz for this three-point function is
\begin{align}
	\langle \mathbf{J}^{}_{\a}(z_{1}) \, \mathbf{J}'_{\b}(z_{2}) \, \mathbf{J}''_{\g(3)}(z_{3}) \rangle = \frac{ \cI_{\a}{}^{\a'}(\boldsymbol{x}_{13}) \,  \cI_{\b}{}^{\b'}(\boldsymbol{x}_{23}) }{(\boldsymbol{x}_{13}^{2})^{3/2} (\boldsymbol{x}_{23}^{2})^{3/2}}
	\; \cH_{\a' \b' \g(3)}(\boldsymbol{X}_{12}, \Q_{12}) \, .
\end{align} 
All information about this correlation function is encoded in the following polynomial:
\begin{align}
	\cH(\boldsymbol{X}, \Q; u(1), v(1), w(3)) = \cH_{ \a \b \g(3) }(\boldsymbol{X}, \Q) \, \boldsymbol{u}^{\a}  \boldsymbol{v}^{\b} \boldsymbol{w}^{\g(3)} \, .
\end{align}
After solving \eqref{Ch05.1-Diophantine equations 2}, we obtain the following list of polynomial structures for $\cH(X,\Q;u,v,w)$ in the even and odd sectors respectively:
\begin{subequations}
	\begin{align}
		\textbf{Even:}& \hspace{5mm} \{Q_3 R_3 Z_3,Q_2 R_2 Z_3,Q_1 R_1 Z_3,Q_1 Q_2 R_3,P_3 S_3 Z_3, \nonumber \\
		& \hspace{20mm} P_2 S_2 Z_3,P_2 Q_1 S_3,P_1 S_1 Z_3,P_1 Q_2 S_3,P_1 P_2 R_3 \} \\
		\textbf{Odd:}& \hspace{5mm} \{Q_3 S_3 Z_3,Q_2 S_2 Z_3,Q_1 S_1 Z_3,Q_1 Q_2 S_3,P_3 R_3 Z_3, \nonumber \\
		&\hspace{20mm} P_2 R_2 Z_3,P_2 Q_1 R_3,P_1 R_1 Z_3,P_1 Q_2
		R_3,P_1 P_2 S_3 \}
	\end{align}
\end{subequations}
After systematic application of the linear dependence relations \eqref{Ch05.1-Linear dependence 1}-\eqref{Ch05.1-Linear dependence 6} we obtain the following linearly independent sets:
\begin{subequations}
	\begin{align}
		\textbf{Even:}& \hspace{5mm} \{P_1 P_2 R_3,Q_1 Q_2 R_3,P_2 Q_1 S_3,P_1 Q_2 S_3 \} \\
		\textbf{Odd:}& \hspace{5mm} \{P_2 Q_1 R_3,P_1 Q_2 R_3,P_1 P_2 S_3,Q_1 Q_2 S_3 \}
	\end{align}
\end{subequations}
Next, we impose conservation on all three points and obtain the following solution for $\cH(\boldsymbol{X}, \Q; u,v,w)$ after solving the linear system of equations in the coefficients $A_{i}$ and $B_{i}$:
\begin{subequations}
	\begin{align}
		\textbf{Even:}& \hspace{5mm} \frac{A_1}{\boldsymbol{X}^{1/2}} \Big(\sfrac{1}{5} P_2 \mathbb{Q}_1 \mathbb{S}_3-\sfrac{1}{5} P_1 \mathbb{Q}_2
		\mathbb{S}_3+P_1 P_2 R_3+\sfrac{1}{5} \mathbb{Q}_1 \mathbb{Q}_2 R_3\Big) \\
		\textbf{Odd:}& \hspace{5mm} 0
	\end{align}
\end{subequations}
Hence, after conservation, the three-point function is fixed up to a single even structure. This structure is also compatible with the symmetry $\mathbf{J}=\mathbf{J}'$, therefore $\langle \mathbf{L} \mathbf{L} \mathbf{J} \rangle$ is fixed up to a single structure. \\[5mm]
\noindent
\textbf{Correlation function} $\langle \mathbf{J} \mathbf{J} \mathbf{L} \rangle$\textbf{:}

The next example to consider is the mixed correlator $\langle \mathbf{J} \mathbf{J} \mathbf{J} \rangle$; to study this case we may examine the correlation function $\langle \mathbf{J}^{}_{3/2} \mathbf{J}'_{3/2} \mathbf{J}''_{1/2} \rangle$. Using the general formula, the ansatz for this three-point function is
\begin{align}
	\langle \mathbf{J}^{}_{\a(3)}(z_{1}) \, \mathbf{J}'_{\b(3)}(z_{2}) \, \mathbf{J}''_{\g}(z_{3}) \rangle = \frac{ \cI_{\a(3)}{}^{\a'(3)}(\boldsymbol{x}_{13}) \,  \cI_{\b(3)}{}^{\b'(3)}(\boldsymbol{x}_{23}) }{(\boldsymbol{x}_{13}^{2})^{5/2} (\boldsymbol{x}_{23}^{2})^{5/2}}
	\; \cH_{\a'(3) \b'(3) \g}(\boldsymbol{X}_{12}, \Q_{12}) \, .
\end{align} 
Using the formalism outlined in Subsection \ref{Ch05-subsection5.2.2.2}, all information about this correlation function is encoded in the following polynomial:
\begin{align}
	\cH(\boldsymbol{X}, \Q; u(3), v(3), w(1)) = \cH_{ \a(3) \b(3) \g }(\boldsymbol{X}, \Q) \, \boldsymbol{u}^{\a(3)}  \boldsymbol{v}^{\b(3)} \boldsymbol{w}^{\g} \, .
\end{align}
After solving \eqref{Ch05.1-Diophantine equations 2} and systematically applying the linear dependence relations \eqref{Ch05.1-Linear dependence 1}-\eqref{Ch05.1-Linear dependence 6}, we obtain the following linearly independent sets we obtain the following list of (linearly dependent) polynomial structures in the parity-even and parity-odd sectors respectively:
\begin{subequations}
	\begin{align}
		\textbf{Even:}& \hspace{5mm} \{P_3^2 Q_1 R_1,Q_1 Q_3^2 R_1,P_3^2 Q_2 R_2,Q_2 Q_3^2 R_2,Q_3^3 R_3,P_3 Q_1 Q_3 S_1,P_3 Q_2 Q_3 S_2 \} \\
		\textbf{Odd:}& \hspace{5mm} \{P_3 Q_1 Q_3 R_1,P_3 Q_2 Q_3 R_2,P_3^2 Q_1 S_1,Q_1 Q_3^2 S_1,P_3^2 Q_2 S_2,Q_2 Q_3^2 S_2,Q_3^3 S_3 \}
	\end{align}
\end{subequations}
Next, we impose conservation on all three points and obtain the following explicit solutions for $\cH(\boldsymbol{X}, \Q; u,v,w)$:
\begin{subequations}
	\begin{align}
		\textbf{Even:}& \hspace{5mm} \frac{A_1}{\boldsymbol{X}^{7/2}} \Big(P_3 \mathbb{Q}_1 \mathbb{Q}_3 \mathbb{S}_1-P_3 \mathbb{Q}_2 \mathbb{Q}_3
		\mathbb{S}_2-\sfrac{1}{5} P_3^2 \mathbb{Q}_1 R_1 \nonumber \\
		& \hspace{15mm} -\sfrac{1}{5} P_3^2 \mathbb{Q}_2 R_2-\sfrac{7}{3} \mathbb{Q}_3^3
		R_3+\mathbb{Q}_1 \mathbb{Q}_3^2 R_1+\mathbb{Q}_2 \mathbb{Q}_3^2 R_2 \Big) \\
		\textbf{Odd:}& \hspace{5mm} 0
	\end{align}
\end{subequations}
Hence, after imposing conservation on all three points, the three-point function is fixed up to a single even structure. This structure is not compatible with the symmetry property $\mathbf{J} = \mathbf{J}'$, hence, $\langle \mathbf{J} \mathbf{J} \mathbf{L} \rangle = 0$. \\[5mm]
\noindent
\textbf{Correlation function} $\langle \mathbf{J} \mathbf{J} \mathbf{J} \rangle$\textbf{:}

The last example to consider is the three-point function of the supercurrent, $\langle \mathbf{J} \mathbf{J} \mathbf{J} \rangle$. To study it we may examine the correlation function $\langle \mathbf{J}^{}_{3/2} \mathbf{J}'_{3/2} \mathbf{J}''_{3/2} \rangle$. Using the general formula, the ansatz for this three-point function is
\begin{align}
	\langle \mathbf{J}^{}_{\a(3)}(z_{1}) \, \mathbf{J}'_{\b(3)}(z_{2}) \, \mathbf{J}''_{\g(3)}(z_{3}) \rangle = \frac{ \cI_{\a(3)}{}^{\a'(3)}(\boldsymbol{x}_{13}) \,  \cI_{\b(3)}{}^{\b'(3)}(\boldsymbol{x}_{23}) }{(\boldsymbol{x}_{13}^{2})^{5/2} (\boldsymbol{x}_{23}^{2})^{5/2}}
	\; \cH_{\a'(3) \b'(3) \g(3)}(\boldsymbol{X}_{12}, \Q_{12}) \, .
\end{align} 
We now encode the structure of this correlation function in the following polynomial:
\begin{align}
	\cH(\boldsymbol{X}, \Q; u(3), v(3), w(3)) = \cH_{ \a(3) \b(3) \g(3) }(\boldsymbol{X}, \Q) \, \boldsymbol{u}^{\a(3)}  \boldsymbol{v}^{\b(3)} \boldsymbol{w}^{\g(3)} \, .
\end{align}
In this case there are a vast number of linearly dependent structures to consider and the list is too large to present. However, after application of the linear dependence relations \eqref{Ch05.1-Linear dependence 1}-\eqref{Ch05.1-Linear dependence 6} we obtain the following linearly independent structures:
\begin{subequations}
	\begin{align}
		\textbf{Even:}& \hspace{5mm} \{P_2 P_3 Q_1^2 R_1,Q_1^2 Q_2 Q_3 R_1,P_1 P_3 Q_2^2 R_2,Q_1 Q_2^2 Q_3 R_2, \nonumber \\
		&\hspace{10mm} P_1 P_2 Q_3^2 R_3,Q_1 Q_2 Q_3^2 R_3, P_3 Q_1^2 Q_2 S_1, P_2 Q_1^2 Q_3 S_1, \nonumber \\
		&\hspace{15mm} P_3 Q_1 Q_2^2 S_2,P_1 Q_2^2 Q_3 S_2,P_2 Q_1 Q_3^2 S_3,P_1 Q_2 Q_3^2
		S_3 \} \\
		\textbf{Odd:}& \hspace{5mm} \{P_3 Q_1^2 Q_2 R_1,P_2 Q_1^2 Q_3 R_1,P_3 Q_1 Q_2^2 R_2,P_1 Q_2^2 Q_3 R_2, \nonumber \\
		& \hspace{10mm} P_2 Q_1 Q_3^2 R_3, P_1 Q_2 Q_3^2 R_3,P_2 P_3 Q_1^2 S_1,Q_1^2 Q_2 Q_3 S_1, \nonumber \\
		&\hspace{15mm} P_1 P_3 Q_2^2 S_2,Q_1 Q_2^2 Q_3 S_2,P_1 P_2 Q_3^2 S_3,Q_1 Q_2 Q_3^2 S_3 \}
	\end{align}
\end{subequations}
%
Next, we impose conservation on all three points and obtain the following explicit solution for $\cH(\boldsymbol{X}, \Q; u,v,w)$:
\begin{subequations}
	\begin{align}
		\textbf{Even:}& \hspace{5mm} \frac{A_1}{\boldsymbol{X}^{5/2}} \Big(\sfrac{3}{11} P_3 \mathbb{Q}_2 \mathbb{Q}_1^2 \mathbb{S}_1+\sfrac{7}{11} P_2 \mathbb{Q}_3
		\mathbb{Q}_1^2 \mathbb{S}_1-\sfrac{3}{11} P_3 \mathbb{Q}_2^2 \mathbb{Q}_1 \mathbb{S}_2+\sfrac{7}{11} P_2
		\mathbb{Q}_3^2 \mathbb{Q}_1 \mathbb{S}_3 \nonumber \\
		&\hspace{15mm} -\sfrac{7}{11} P_1 \mathbb{Q}_2^2 \mathbb{Q}_3
		\mathbb{S}_2-\sfrac{7}{11} P_1 \mathbb{Q}_2 \mathbb{Q}_3^2 \mathbb{S}_3-\sfrac{1}{11} P_2 P_3 \mathbb{Q}_1^2
		R_1-\sfrac{1}{11} P_1 P_3 \mathbb{Q}_2^2 R_2 \nonumber \\
		& \hspace{18mm} +\sfrac{21}{11} P_1 P_2 \mathbb{Q}_3^2 R_3+\mathbb{Q}_2 \mathbb{Q}_3
		\mathbb{Q}_1^2 R_1-\sfrac{21}{11} \mathbb{Q}_2 \mathbb{Q}_3^2 \mathbb{Q}_1 R_3+\mathbb{Q}_2^2 \mathbb{Q}_3
		\mathbb{Q}_1 R_2 \Big) \\
		\textbf{Odd:}& \hspace{5mm} 0
	\end{align}
\end{subequations}
Hence, the three-point function $\langle \mathbf{J}^{}_{3/2} \mathbf{J}'_{3/2} \mathbf{J}''_{3/2} \rangle$ is fixed up to a single parity-even structure. The solution is also compatible with the symmetry property $\mathbf{J}=\mathbf{J}'=\mathbf{J}''$, and so the three-point function of the supercurrent, $\langle \mathbf{J} \mathbf{J} \mathbf{J} \rangle$, is fixed up to a single parity-even structure. In terms of the number of independent structures, these results are consistent with \cite{Buchbinder:2015qsa}.



\subsection{General structure of \texorpdfstring{$\langle \mathbf{J}^{}_{s_{1}} \mathbf{J}'_{s_{2}} \mathbf{J}''_{s_{3}} \rangle$
	}{< J J' J'' >}}
\label{Ch05-subsection5.3.2}


We performed a comprehensive analysis of the general structure of the three-point correlation function 
$\langle \mathbf{J}^{}_{s_{1}} \mathbf{J}'_{s_{2}} \mathbf{J}''_{s_{3}} \rangle$ using our computational approach. 
Due to computational limitations we were able to carry out this analysis for $s_{i} \leq 20$, however, the pattern in the solutions is very clear and we propose that the results 
stated in this section hold for arbitrary superspins. We also want to emphasise that for given $(s_1, s_2, s_3)$ our method produces a result which can be presented in an explicit form even for relatively
high superspins (see examples below). With a sufficiently powerful computer one could extend our results to larger values of $s_i$.

Based on our analysis we found that the general structure of the three-point correlation function 
$\langle \mathbf{J}^{}_{s_{1}} \mathbf{J}'_{s_{2}} \mathbf{J}''_{s_{3}} \rangle$ is constrained to the following form:
%
\begin{equation}
	\langle \mathbf{J}^{}_{s_{1}} \mathbf{J}'_{s_{2}} \mathbf{J}''_{s_{3}} \rangle = a \, \langle \mathbf{J}^{}_{s_{1}} \mathbf{J}'_{s_{2}} \mathbf{J}''_{s_{3}} \rangle_{E} + b \, \langle \mathbf{J}^{}_{s_{1}} \mathbf{J}'_{s_{2}} \mathbf{J}''_{s_{3}} \rangle_{O} \, .
\end{equation}
One of our main conclusions is that the parity-odd structure, $\langle  \mathbf{J}^{}_{s_{1}}  \mathbf{J}'_{s_{2}}  \mathbf{J}''_{s_{3}} \rangle_{O}$ does not appear in correlators that are overall 
Grassmann-odd (or fermionic). The existence of the odd solution in the Grassmann-even (bosonic) correlators depend on the following superspin triangle inequalities:
\begin{align} \label{Ch05.1-Triangle inequalities}
	s_{1} &\leq s_{2} + s_{3} \, , & s_{2} &\leq s_{1} + s_{3} \, , & s_{3} &\leq s_{1} + s_{2} \, .
\end{align}
When the triangle inequalities are simultaneously satisfied, there is one even solution and one odd solution, however, if any of the above inequalities are not satisfied then the odd solution is incompatible 
with current conservation.
Further, if any of the $ \mathbf{J}$, $ \mathbf{J}'$, $ \mathbf{J}''$ coincide then the resulting point-switch symmetries can kill off the remaining structures. 

Before we discuss in more detail Grassmann-even and Grassmann-odd correlators let us make some general comments. 
In particular, we observe that if the triangle inequalities are simultaneously satisfied, each polynomial structure in the three-point functions can be written as a product of at most 5 of the $P_{i}$, $Q_{i}$, with the $Z_{i}$ 
completely eliminated. Another useful observation is that, analogous to the non-supersymmetric case, the triangle inequalities can be encoded in a discriminant, $\s$, which we define as follows:
\begin{align} \label{Ch05.1-Discriminant}
	\s(s_{1}, s_{2}, s_{3}) = q_{1} q_{2} q_{3} \, , \hspace{10mm} q_{i} = s_{i} - s_{j} - s_{k} - 1 \, ,
\end{align}
where $(i,j,k)$ is a cyclic permutation of $(1,2,3)$. For $\s(s_{1}, s_{2}, s_{3}) < 0$, there is one even solution and one odd solution, while for $\s(s_{1}, s_{2}, s_{3}) \geq 0$ there is a single even solution. 
Also recall that the correlation function can be encoded in a tensor $\cH$, which is a function of two three-point covariants, $X$ and $\Q$. There are three different (equivalent) representations of a given three-point function, 
call them $\cH^{(i)}$, where the superscript $i$ denotes which point we set to act as the ``third point" in the ansatz \eqref{Ch05.1-H ansatz}. As shown in subsection \ref{Ch05-subsection5.2.2.1}, 
the representations are related by the intertwining operator, $\cI$. Since the dimensions of the conserved supercurrents $\Delta_i$ are related to the superspins as $\Delta_i= s_i+1$ 
it follows that each $\cH^{(i)}$ is homogeneous of degree $q_{i}$. Then it follows that the odd structure survives if and only if $\forall i$, $q_{i} < 0$. In other words, 
each $\cH^{(i)}$ must be a rational function of $X$ and $\Q$ with homogeneity $q_{i} < 0$.



\subsubsection{Grassmann-even correlators}\label{Ch05-subsection5.3.2.1}

The complete classification of results for Grassmann-even conserved three-point functions, including cases where there is a point-switch symmetry, is as follows:
\begin{itemize}
	\item In all cases we have examined ($s_{i} \leq 20$) there is one even solution and one odd solution, however, the odd solution vanishes if the superspin triangle inequalities are not satisfied. 
	\item $\langle \mathbf{J}^{}_{s_{1}} \mathbf{J}^{}_{s_{1}} \mathbf{J}'_{s_{2}} \rangle$: 
	Note that in this case $s_2$ must  be an integer. 
	For $s_{2}$ even, the solutions survive the point-switch symmetry for arbitrary $s_{1}$ (integer or half-integer). For $s_{2}$ odd the point-switch symmetry is not satisfied and the three-point function vanishes. 
	\item $\langle \mathbf{J}_{s} \, \mathbf{J}_{s} \, \mathbf{J}_{s} \rangle$: in this case $s$ is restricted to integer values. For $s$ even the solutions are compatible with the point-switch symmetries. 
\end{itemize}
The number of linearly independent structures grows rapidly with the superspins, and we present the results for some low superspin cases after imposing conservation on all three points in Appendix \ref{Appendix5B}.

\subsubsection{Grassmann-odd correlators}\label{Ch05-subsection5.3.2.2}

The classification of results for Grassmann-odd three-point functions, including cases where there is a point-switch symmetry, is as follows:
\begin{itemize}
	\item In all cases we have examined ($s_{i} \leq 20$), the three-point functions are fixed up to a single parity-even solution after conservation on all three points. In general, any parity-odd structures are incompatible with conservation. Supported by the observations from our computational approach, in Section \ref{Ch05-subsection5.4.1} we prove that the parity-odd structure cannot exist in these three-point functions for arbitrary spins.
	\item $\langle \mathbf{J}^{}_{s_{1}} \mathbf{J}^{}_{s_{1}} \mathbf{J}'_{s_{2}} \rangle$: 
	Note that in this case $s_2$ must be half-integer.
	For $s_{1} \neq s_{2}$, the classification is as follows:
	\begin{itemize}
		\item Let $s_{2} = 2k+\tfrac{1}{2}$, $k \in \mathbb{Z}_{\geq 0}$; for arbitrary $s_{1}$ (integer or half-integer) the point-switch symmetry is not satisfied and therefore the three-point function vanishes in general.
		\item Let $s_{2} = 2k+\tfrac{3}{2}$, $k \in \mathbb{Z}_{\geq 0}$; for arbitrary $s_{1}$ (integer or half-integer) the point-switch symmetry is satisfied and therefore the three-point function is fixed up to a single parity-even structure.
	\end{itemize}
	
	\item $\langle \mathbf{J}_{s} \, \mathbf{J}_{s} \, \mathbf{J}_{s} \rangle$: for $s = 2k + \tfrac{3}{2}$, $k \in \mathbb{Z}_{\geq 0}$ the solution is compatible with the point-switch symmetry. For $s = 2k+\tfrac{1}{2}$, $k \in \mathbb{Z}_{\geq 0}$ the three-point function vanishes. 
	
\end{itemize}
We present the results for some low superspin cases after imposing conservation on all three points in Appendix \ref{Appendix5B}.

\subsection{Three-point functions of scalar superfields}\label{Ch05-subsection5.3.3}

For completeness, in this section we analyse three-point correlation functions involving scalar superfields and conserved supercurrents. Some of the three-point functions contain parity-odd solutions, with their existence depending on both triangle inequalities and the weights of the scalars. We found that the following general results hold:
\begin{subequations}
	\begin{align}
		\langle \cO \, \cO' \, \mathbf{J}_{s} \rangle &= a \, \langle \cO \, \cO' \, \mathbf{J}_{s} \rangle_{E} \, , \\[2mm]
		\langle \mathbf{J}^{}_{s_{1}} \mathbf{J}'_{s_{2}} \, \cO \rangle &= a \, \langle \mathbf{J}^{}_{s_{1}} \mathbf{J}'_{s_{2}} \, \cO \rangle_{E} + b \, \langle \mathbf{J}^{}_{s_{1}} \mathbf{J}'_{s_{2}} \, \cO \rangle_{O} \, .
	\end{align}
\end{subequations}
The correlation functions are analysed using the same methods as in the previous sections; the full classification of results (for cases where there is a point-switch symmetry), is summarised below:
\begin{itemize}
	\item $\langle \cO \, \cO' \, \mathbf{J}_{s} \rangle$: in general there are solutions only for $\D_{\cO} = \D_{\cO'}$. For the Grassmann-even case the solution satisfies the point-switch symmetry $\cO = \cO'$ only for even $s$. For the Grassmann-odd case the solution satisfies the point-switch symmetry only for $s = 2k+\tfrac{3}{2}$, $k \in \mathbb{Z}_{\geq 0}$. The general solutions (for $\cH$) are:
	\begin{subequations}
	\begin{align}
		\text{integer}-s:& \hspace{10mm} a \, \mathbb{Z}_{3}^{s} \, \boldsymbol{X}^{s - 2 \D +1 } \, , \\
		\text{half-integer}-s:& \hspace{10mm} a \, R_{3} \, \mathbb{Z}_{3}^{s-1/2} \, \boldsymbol{X}^{s - 2 \D  + 1 } \, ,
	\end{align}
	\end{subequations}
	where $a$ is a free coefficient.
	
	\item $\langle \mathbf{J}^{}_{s_{1}} \mathbf{J}'_{s_{2}} \, \cO \rangle$: for $s_{1} \neq s_{2}$, there is a single even solution for $\Delta_{\cO}=1$, otherwise the three-point function vanishes. For $s_{1} = s_{2}$ there is one even and one odd solution and the point-switch symmetries are satisfied.
\end{itemize}
Some explicit solutions for the above cases are presented in Appendix \ref{Appendix5C}.

\section{Analytic construction of Grassmann-odd three-point functions}\label{Ch05-section5.4}

In the previous section it was demonstrated that the parity-odd part of Grassmann-odd three-point functions appears to vanish, while the parity-even part is shown to be unique in all cases examined.
In the next sections, we will demonstrate that conservation on $z_{1}$ and $z_{2}$ is sufficient to constrain the structure of the three-point function to a unique parity-even solution, while the parity-odd solution must vanish. There are only two possibilities for Grassmann-odd three-point functions in superspace (up to permutations of the fields):
\begin{equation}
	\langle \mathbf{J}^{}_{F} \mathbf{J}'_{F} \mathbf{J}''_{F} \rangle \, , \hspace{20mm} \langle \mathbf{J}^{}_{F} \mathbf{J}'_{B} \mathbf{J}''_{B} \rangle \, ,
\end{equation}
where ``$B$" represents a Grassmann-even (bosonic) field, and ``$F$" represents a Grassmann-odd (fermionic) field. Each of these correlation functions require separate analysis, however, they have a similar underlying structure.


\subsection{Grassmann-odd three-point functions -- \texorpdfstring{$\langle \mathbf{J}^{}_{F} \mathbf{J}'_{F} \mathbf{J}''_{F} \rangle$
	}{< F F F >}}\label{Ch05-subsection5.4.1}


\subsubsection{Method of irreducible decomposition}\label{Ch05-subsubsection5.4.1.1}

First let us analyse the case $\langle \mathbf{J}^{}_{F} \mathbf{J}'_{F} \mathbf{J}''_{F} \rangle$; we consider three Grassmann-odd currents: 
$\mathbf{J}^{}_{\a(2A+1)}$, $\mathbf{J}'_{\a(2B+1)}$, $\mathbf{J}''_{\g(2C+1)}$, where $A,B,C$ are positive integers. Therefore, the superfields $\mathbf{J}$, $\mathbf{J}'$, $\mathbf{J}''$ 
are of superspin 
$s_{1} = A+ \tfrac{1}{2}$, $s_{2} = B+ \tfrac{1}{2}$, $s_{3} = C+ \tfrac{1}{2}$ respectively. Using the formalism above, all information about the correlation function
\begin{equation}
	\langle \, \mathbf{J}^{}_{\a(2A+1)}(z_{1}) \, \mathbf{J}'_{\b(2B+1)}(z_{2}) \, \mathbf{J}''_{\g(2C+1)}(z_{3}) \rangle \, ,
\end{equation}
is encoded in a homogeneous tensor field $\cH_{\a(2A+1) \b(2B+1) \g(2C+1)}(\boldsymbol{X}, \Q)$, which is a function of a single superspace 
variable $\cZ = ( \boldsymbol{X} , \Q)$ and satisfies the scaling property
\begin{equation}
	\cH_{\a(2A+1) \b(2B+1) \g(2C+1)}( \l^{2} \boldsymbol{X}, \l \Q) = (\l^{2})^{C-A-B-\tfrac{3}{2}}\cH_{\a(2A+1) \b(2B+1) \g(2C+1)}(\boldsymbol{X}, \Q) \, .
\end{equation}
To simplify the problem, for each set of totally symmetric spinor indices (the $\a$'s, $\b$'s and $\g$'s respectively), we convert pairs of them into vector indices as follows:
\begin{align}
	\cH_{\a(2A+1) \b(2B+1) \g(2C+1)}(\boldsymbol{X}, \Q) &\equiv \cH_{\a \a(2A), \b \b(2B), \g \g(2C)}(\boldsymbol{X}, \Q) \nonumber \\
	&= (\g^{m_{1}})_{\a_{1} \a_{2}} ... (\g^{m_{A}})_{\a_{2A-1} \a_{2A}} \nonumber \\
	& \times (\g^{n_{1}})_{\b_{1} \b_{2}} ... (\g^{n_{B}})_{\b_{2B-1} \b_{2B}} \nonumber \\
	& \times (\g^{k_{1}})_{\g_{1} \g_{2}} ... (\g^{k_{C}})_{\g_{2C-1} \g_{2C}} \nonumber \\
	& \times \cH_{\a \b \g, m_{1} ... m_{A}, n_{1} ... n_{B}, k_{1} ... k_{C} }(\boldsymbol{X}, \Q) \, .
\end{align}
The equality above holds only if and only if $\cH_{\a \b \g, m_{1} ... m_{A} n_{1} ... n_{B} k_{1} ... k_{C} }(\boldsymbol{X}, \Q)$ is totally symmetric
\begin{equation}
	\cH_{\a \b \g, m_{1} ... m_{A} n_{1} ... n_{B} k_{1} ... k_{C} }(\boldsymbol{X}, \Q) = \cH_{\a \b \g, (m_{1} ... m_{A}) (n_{1} ... n_{B}) (k_{1} ... k_{C}) }(\boldsymbol{X}, \Q) \, ,
\end{equation}
and traceless in each group of vector indices, i.e. $\forall i,j$
\begin{subequations} \label{Ch05.2-FFF - traceless}
	\begin{align}
		\eta^{m_{i} m_{j}} \cH_{\a \b \g, m_{1} ... m_{i} m_{j} ... m_{A}, n_{1} ... n_{B}, k_{1} ... k_{C} }(\boldsymbol{X}, \Q) &= 0 \, ,
		\\
		\eta^{n_{i} n_{j}} \cH_{\a \b \g, m_{1} ... m_{A}, n_{1} ... n_{i} n_{j} ... n_{B}, k_{1} ... k_{C} }(\boldsymbol{X}, \Q) &= 0 \, , \\
		\eta^{k_{i} k_{j}} \cH_{\a \b \g, m_{1} ... m_{A}, n_{1} ... n_{B}, k_{1} ... k_{i} k_{j} ... k_{C} }(\boldsymbol{X}, \Q) &= 0 \, .
	\end{align}
\end{subequations}
It is also required that $\cH$ is subject to the $\g$-trace constraints
\begin{subequations} \label{Ch05.2-FFF - Gamma trace}
	\begin{align}
		(\g^{m_{1}})_{\s}{}^{\a} \cH_{\a \b \g, m_{1} ... m_{A}, n_{1} ... n_{B}, k_{1} ... k_{C} }(\boldsymbol{X}, \Q) &= 0 \, , \label{Ch05.2-FFF - Gamma trace 1} \\
		(\g^{n_{1}})_{\s}{}^{\b} \cH_{\a \b \g, m_{1} ... m_{A}, n_{1} ... n_{B}, k_{1} ... k_{C} }(\boldsymbol{X}, \Q) &= 0 \, , \label{Ch05.2-FFF - Gamma trace 2} \\
		(\g^{k_{1}})_{\s}{}^{\g} \cH_{\a \b \g, m_{1} ... m_{A}, n_{1} ... n_{B}, k_{1} ... k_{C} }(\boldsymbol{X}, \Q) &= 0 \, . \label{Ch05.2-FFF - Gamma trace 3}
	\end{align}
\end{subequations}
Now since $\cH$ is Grassmann-odd, it must be linear in $\Q$, and, using the property \eqref{Ch05.1-Three-point building blocks 1a - properties 3}, we decompose $\cH$ as follows (raising all indices for convenience):
\begin{subequations} \label{Ch05.2-FFF - irreducible decomposition}
	\begin{equation}
		\cH^{\a \b \g, m(A) n(B) k(C) }(\boldsymbol{X}, \Q) = \sum_{i=1}^{4} \cH_{i}^{\a \b \g, m(A) n(B) k(C) }(X, \Q)\,,
	\end{equation}
	\vspace{-3mm}
	\begin{align}
		\cH_{1}^{\a \b \g, m(A) n(B) k(C) }(X, \Q) &= \ve^{\a \b} \Q^{\g} A^{m(A) n(B) k(C)}(X) \, , \\
		\cH_{2}^{\a \b \g, m(A) n(B) k(C) }(X, \Q) &= \ve^{\a \b} (\gamma_{r})^{\g}{}_{\d} \Q^{\d} B^{r, m(A) n(B) k(C)}(X) \, , \\
		\cH_{3}^{\a \b \g, m(A) n(B) k(C) }(X, \Q) &= (\g_{p})^{\a \b} \Q^{\g} C^{p, m(A) n(B) k(C)}(X) \, , \\
		\cH_{4}^{\a \b \g, m(A) n(B) k(C) }(X, \Q) &= (\g_{p})^{\a \b} (\gamma_{r})^{\g}{}_{\d} \Q^{\d} D^{pr, m(A) n(B) k(C)}(X) \, .
	\end{align}
\end{subequations}
Here it is convenient to view the contributions $\cH_{i}$ as functions of the symmetric tensor $X$ (which is equivalent to a three-dimensional vector) rather than of $\boldsymbol{X}$. 
In fact, since $\cH$ is linear in $\Q$ and $\Q^3=0$ we have $\cH (\boldsymbol{X}, \Q)= \cH (X, \Q)$. 
The tensors $A,B,C,D$ are constrained by conservation equations and any algebraic symmetry properties which $\cH$ possesses. In particular, the equations for conservation on $z_{1}$ and $z_{2}$ in  \eqref{Ch05.1-Conservation on H - tensor formalism} are now equivalent to the following constraints on $\cH$ with vector indices
\begin{subequations}
	\begin{align} \label{Ch05.2-FFF - conservation equations}
		\cD_{\a} \cH^{\a \b \g, m(A) n(B) k(C) }(\boldsymbol{X}, \Q) = 0 \, , \\
		\cQ_{\b} \cH^{\a \b \g, m(A) n(B) k(C) }(\boldsymbol{X}, \Q) = 0 \, .
	\end{align}
\end{subequations}
We also need consider the constraint for conservation at $z_{3}$, however, this is technically challenging to impose using this analytic approach and is more appropriately handed by the computational approach outlined in the previous sections. Since $\cH$ is linear in $\Q$, the conservation conditions \eqref{Ch05.2-FFF - conservation equations} split up into constraints $O(\Q^{0})$
\begin{subequations}
	\begin{align} 
		\frac{\pa}{\pa \Q^{\a}} \cH^{\a \b \g, m(A) n(B) k(C) }(\boldsymbol{X}, \Q) &= 0  \, , \label{Ch05.2-FFF - conservation equations O(0) - 1} \\
		\frac{\pa}{\pa \Q^{\b}} \cH^{\a \b \g, m(A) n(B) k(C) }(\boldsymbol{X}, \Q) &= 0 \, , \label{Ch05.2-FFF - conservation equations O(0) - 2}
	\end{align}
\end{subequations}
and $O(\Q^{2})$
\begin{subequations}
	\begin{align} 
		(\g^{m})_{\a \d} \Q^{\d} \frac{\pa}{\pa X^{m}} \cH^{\a \b \g, m(A) n(B) k(C) }(\boldsymbol{X}, \Q) &= 0 \, , \label{Ch05.2-FFF - conservation equations O(2) - 1} \\
		(\g^{m})_{\b \d} \Q^{\d} \frac{\pa}{\pa X^{m}} \cH^{\a \b \g, m(A) n(B) k(C) }(\boldsymbol{X}, \Q) &= 0 \, . \label{Ch05.2-FFF - conservation equations O(2) - 2}
	\end{align}
\end{subequations}
Using the irreducible decomposition \eqref{Ch05.2-FFF - irreducible decomposition}, equation \eqref{Ch05.2-FFF - conservation equations O(0) - 1} results in the algebraic relations
\begin{subequations}
	\begin{align} \label{Ch05.2-O(0) constraints - 1}
		A^{m(A) n(B) k(C)} + \eta_{pr} D^{pr, m(A) n(B) k(C) } &= 0 \, , \\
		B^{q,m(A) n(B) k(C)} + C^{q, m(A) n(B) k(C)} - \e^{q}{}_{pr} D^{pr, m(A) n(B) k(C)} &= 0 \, ,
	\end{align}
\end{subequations}
while \eqref{Ch05.2-FFF - conservation equations O(0) - 2} gives
\begin{subequations}
	\begin{align} \label{Ch05.2-O(0) constraints - 2}
		- A^{m(A) n(B) k(C)} + \eta_{pr} D^{pr, m(A) n(B) k(C) } &= 0 \, , \\
		- B^{q,m(A) n(B) k(C)} + C^{q, m(A) n(B) k(C)} - \e^{q}{}_{pr} D^{pr, m(A) n(B) k(C)} &= 0 \, .
	\end{align}
\end{subequations}
Hence, equations \eqref{Ch05.2-O(0) constraints - 1}, \eqref{Ch05.2-O(0) constraints - 2} together result in $A = B = 0$, while $C$ and $D$ satisfy:
\begin{subequations} \label{Ch05.2-C and D constraints}
	\begin{align} 
		\eta_{pr} D^{pr, m(A) n(B) k(C) } &= 0 \, , \\
		C^{q, m(A) n(B) k(C)} - \e^{q}{}_{pr} D^{pr, m(A) n(B) k(C)} &= 0 \, .
	\end{align}
\end{subequations}
%
Next we consider the relations arising from the conservation equations at $O(\Q^{2})$. 
Using the decomposition \eqref{Ch05.2-FFF - irreducible decomposition}, from a straightforward computation we obtain
\begin{subequations}
	\begin{align} \label{Ch05.2-O(2) constraints}
		\pa_{t} \big( \e^{t}{}_{pr} D^{pr,m(A) n(B) k(C)} + C^{t,m(A) n(B) k(C)} \big) &= 0 \, , \\
		\pa_{t} \big( - \e^{tq}{}_{p} C^{p,m(A) n(B) k(C)} + D^{qt,m(A) n(B) k(C)} + D^{tq,m(A) n(B) k(C)}\big) &= 0 \, .
	\end{align}
\end{subequations}
After substituting the algebraic relations \eqref{Ch05.2-C and D constraints} into \eqref{Ch05.2-O(2) constraints}, we obtain
\begin{align} \label{Ch05.2-C and D conservation equations}
	\pa_{p} C^{p,m(A) n(B) k(C)} = 0 \, , \hspace{10mm} \pa_{p} D^{pr, m(A) n(B) k(C)} = 0 \, .
\end{align}
We now must impose the $\g$-trace conditions, starting with \eqref{Ch05.2-FFF - Gamma trace 1} and \eqref{Ch05.2-FFF - Gamma trace 2}. Making use of the decomposition \eqref{Ch05.2-FFF - irreducible decomposition}, equation \eqref{Ch05.2-FFF - Gamma trace 1} results in the algebraic constraints
\begin{subequations}
	\begin{align}
		\eta_{mp} C^{p, m m(A-1) n(B) k(C)} &= 0 \, , & \e_{qmp} C^{p, m m(A-1) n(B) k(C)} &= 0 \, , \\
		\eta_{mp} D^{pr, m m(A-1) n(B) k(C)} &= 0 \, , & \e_{qmp} D^{pr, m m(A-1) n(B) k(C)} &= 0 \, , 
	\end{align}
\end{subequations}
while from \eqref{Ch05.2-FFF - Gamma trace 2} we find
\begin{subequations}
	\begin{align}
		\eta_{np} C^{p, m(A) n n(B-1) k(C)} &= 0 \, , & \e_{qnp} C^{p, m(A) n n(B-1) k(C)} &= 0 \, , \\
		\eta_{np} D^{pr, m(A) n n(B-1) k(C)} &= 0 \, , & \e_{qnp} D^{pr, m(A) n n(B-1) k(C)} &= 0 \, .
	\end{align}
\end{subequations}
Altogether these relations imply that both $C$ and $D$ are symmetric and traceless in the indices $p, m_{1}, ..., m_{A}, n_{1}, ..., n_{B}$, i.e.,
\begin{equation}
	C^{p, m(A) n(B) k(C)} \equiv C^{(p m(A) n(B)) k(C)} \, , \hspace{5mm} D^{pr, m(A) n(B) k(C)} \equiv D^{(p m(A) n(B)), r k(C)} \, .
\end{equation}
Next, from the $\g_{k}$-trace constraint \eqref{Ch05.2-FFF - Gamma trace 3}, we obtain the algebraic relations
\begin{subequations}  \label{Ch05.2-C and D constraints 2} 
	\begin{align}
		\eta_{kr} D^{pr, m(A) n(B) k k(C-1) } = 0 \, , \\
		C^{p, m(A) n(B) k k(C-1)} + \e^{k}{}_{rs} D^{pr, m(A) n(B) s k(C-1)} = 0 \, .
	\end{align}
\end{subequations}
To make use of these relations, we decompose $D$ into symmetric and anti-symmetric parts on the indices $r, k_{1}, ..., k_{C}$ as follows:
\begin{align} \label{Ch05.2-D decomposition}
	D^{(p m(A) n(B)), r ( k_{1} ... k_{C} ) } = D_{S}^{ (p m(A) n(B)), ( r k_{1} ... k_{C} ) } + \sum_{i=1}^{C} \e^{r k_{i}}{}_{t} D_{A}^{ (p m(A) n(B)), ( t k_{1} ... \hat{k}_{i} ... k_{C} ) } \, ,
\end{align}
where 
the notation $\hat{k}_{i}$ denotes removal of the index $k_{i}$ from $D_{A}$. After substituting this decomposition into \eqref{Ch05.2-C and D constraints 2}, we obtain:
\begin{subequations}
	\begin{equation}
		\eta_{k r}  D_{S}^{ (p m(A) n(B)), ( r k k(C-1) ) } = 0 \, ,
	\end{equation}
	\begin{equation}
		D_{A}^{  (p m(A) n(B)), ( r k(C-1) ) } = \frac{1}{C+1} C^{ (p m(A) n(B)), ( r k(C-1) ) } \, .
	\end{equation}
\end{subequations}
We see that the tensor $D_A$ is fully determined in terms of $C$. 
To continue we now substitute \eqref{Ch05.2-D decomposition} into \eqref{Ch05.2-C and D constraints} to obtain equations relating $C$ and $D_{S}$; we obtain:
\begin{subequations} \label{Ch05.2-C and D algebraic relations}
	\begin{align}
		&\eta_{pr} D_{S}^{(p m(A) n(B)),(r k(C))} + \frac{1}{C+1} \sum_{i=1}^{C} \e^{k_{i}}{}_{p r} C^{ (p m(A) n(B)), ( r k_{1} ... \hat{k}_{i} ... k_{C} ) } = 0 \, , \\
		&\e^{p}{}_{qr} D_{S}^{(q m(A) n(B)), (r k(C))} - C^{(p m(A) n(B)), (r k(C))} \\ 
		&+ \frac{1}{C+1} \sum_{i=1}^{C} \Big\{ \eta^{p k_{i}} \eta_{qr} C^{(q m(A) n(B)), (r k_{1} ... \hat{k}_{i} ... k(C)) } - C^{(k_{i} m(A) n(B)),(p k_{1} ... \hat{k}_{i} ... k(C))} \Big\} = 0 \, . \nonumber
	\end{align}
\end{subequations}
Further, the conservation equations \eqref{Ch05.2-C and D conservation equations} are now equivalent to
\begin{equation}
	\pa_{p} C^{(pm(A) n(B)), k(C)} = 0 \, , \hspace{10mm} \pa_{p} D_{S}^{(p m(A) n(B)),(rk(C))} = 0 \, .
	\label{Ch05.2-e4}
\end{equation}
Hence, finding the solution for the three-point function $\langle \mathbf{J}^{}_{F} \mathbf{J}'_{F} \mathbf{J}''_{F} \rangle$ is now equivalent to finding two transverse tensors $C$ and $D_{S}$, which are related by the algebraic constraints \eqref{Ch05.2-C and D algebraic relations}. It may be checked for $A=B=C=1$ that the constraints reproduce those of the supercurrent three-point 
function found in~\cite{Buchbinder:2015qsa}. 

Let us now briefly comment on the analysis of three-point functions involving flavour currents, i.e. when $A,B,C = 0$. In these cases one can simply ignore the relevant tracelessness and $\g$-trace conditions \eqref{Ch05.2-FFF - traceless}, \eqref{Ch05.2-FFF - Gamma trace} respectively and omit the appropriate groups of tensor indices. The analysis of the conservation equations proves to be essentially the same and we will not elaborate further on these cases.


\subsubsection{Conservation equations}\label{Ch05-subsubsection5.4.1.2}

Let us now summarise the constraint analysis in the previous subsection in a way that makes the symmetries more apparent. For the $\langle \mathbf{J}^{}_{F} \mathbf{J}'_{F} \mathbf{J}''_{F} \rangle$ correlators we have two tensors; $C$ of rank $N+C+1$, and $D$ of rank $N+C+2$, where $N= A +B$, which possess the following symmetries:
\begin{align}
C^{(r_{1} ... r_{N+1})(k_{1} ... k_{C})} \, , \hspace{10mm} D_{S}^{(r_{1} ... r_{N+1})(k_{1} ... k_{C+1})} \, .
\end{align}
The tensors $C$ and $D_{S}$ are totally symmetric and traceless in the groups of indices $r$ and $k$ respectively. They also satisfy the conservation conditions
\begin{align} \label{Ch05.2-C and D conservation equations - more symmetric}
\pa_{r_{1}} C^{(r_{1} ... r_{N+1})(k_{1} ... k_{C})} = 0 \, , \hspace{5mm} \pa_{r_{1}} D_{S}^{(r_{1} ... r_{N+1})(k_{1} ... k_{C+1})} = 0 \, ,
\end{align}
and the algebraic relations
\begin{subequations} \label{Ch05.2-C and D algebraic relations - more symmetric}
\begin{align}
	&\eta_{r_{1} k_{1}} D_{S}^{(r_{1} ... r_{N+1})(k_{1} ... k_{C+1})} + \frac{1}{C+1} \sum_{i=2}^{C+1} \e^{k_{i}}{}_{r_{1} k_{1}} C^{ (r_{1} ... r_{N+1})( k_{1} k_{2} ... \hat{k}_{i} ... k_{C+1} ) } = 0 \, , \\
	&\e^{p}{}_{r_{1} k_{1}} D_{S}^{(r_{1} ... r_{N+1})(k_{1} ... k_{C+1})} - C^{(p r_{2} ... r_{N+1})(k_{2} ... k_{C+1})} \\ 
	&+ \frac{1}{C+1} \sum_{i=2}^{C+1} \Big\{ \eta^{p k_{i}} \eta_{r_{1} k_{1}} C^{(r_{1} ... r_{N+1})(k_{1} k_{2} ... \hat{k}_{i} ... k_{C+1}) } - C^{(k_{i} r_{2} ... r_{N+1})(p k_{2} ... \hat{k}_{i} ... k_{C+1})} \Big\} = 0 \, . \nonumber
\end{align}
\end{subequations}
The ``full" tensor $D$, present in the decomposition \eqref{Ch05.2-FFF - irreducible decomposition}, is constructed from $C$ and $D_{S}$ as follows:
\begin{align} \label{Ch05.2-D decomposition - more symmetric}
D^{q, (r_{1} ... r_{N+1})( k_{1} ... k_{C} ) } = D_{S}^{ (r_{1} ... r_{N+1})( q k_{1} ... k_{C} ) } + \frac{1}{C+1} \sum_{i=1}^{C} \e^{q k_{i}}{}_{t} C^{ (r_{1} ... r_{N+1})( t k_{1} ... \hat{k}_{i} ... k_{C} ) } \, .
\end{align}
As we will see later, the algebraic relations \eqref{Ch05.2-C and D algebraic relations - more symmetric} are sufficient to determine $D$ completely in terms of $C$. 
However, before we prove this it is prudent to analyse the conservation equation \eqref{Ch05.2-C and D conservation equations - more symmetric} for $C$.

Since we have identified the algebraic symmetries of $C$, it is convenient to convert $C$ back into spinor notation and contract it with commuting auxiliary spinors as follows:
\begin{align}
C(X; u(2N+2), v(2C)) = C_{\a(2N+2) \b(2C)}(X) \, u^{\a_{1}} \, ... \, u^{\a_{2N+2}} v^{\b_{1}} \, ... \, v^{\b_{2C}} \, .
\end{align}
We now introduce a basis of monomials out of which $C$ can be constructed. Adapting the results from the previous sections we use
\begin{align} \label{Ch05.2-Basis structures}
P_{3} &= \ve_{\a \b} u^{\a} v^{\b} \, , & Q_{3} &= \hat{X}_{\a \b} u^{\a} v^{\b} \, ,  & Z_{1} &= \hat{X}_{\a \b} u^{\a} u^{\b} \, , & Z_{2} &= \hat{X}_{\a \b} v^{\a} v^{\b} \, .
\end{align}
A general ansatz for $C(X;u,v)$ which is homogeneous degree $2(N+1)$ in $u$, $2C$ in $v$ and $C-N-2$ in $X$ is of the following form:
\begin{equation}
C(X; u,v ) = X^{C-N-2} \sum_{a,b} \a(a,b) P_{3}^{a} Q_{3}^{2(C-b) - a} Z_{1}^{b + N - C + 1} Z_{2}^{b} \, .
\end{equation}
However, there is linear dependence between the monomials \eqref{Ch05.2-Basis structures} of the form
\begin{equation}
Z_{1} Z_{2} = Q_{3}^{2} - P_{3}^{2} \, ,
\end{equation}
which allows for elimination of $Z_{2}$. Hence, the ansatz becomes:
\begin{equation} \label{Ch05.2-C polynomial ansatz}
C(X; u,v ) = X^{C-N-2} \sum_{k=0}^{2C} \a_{k} P_{3}^{k} Q_{3}^{2C - k} Z_{1}^{N-C+1} \, .
\end{equation}
This expansion is valid for $C \leq N +1$, which
is always fulfilled for field configurations where the third superspin triangle inequality, $s_{3} \leq s_{1} + s_{2} $, is satisfied. 
Note that if $s_{3} > s_{1} + s_{2} $ it then follows that $s_{1} \leq s_{2} + s_{3}$ and $s_{2} \leq s_{1} + s_{3}$. That is, if one of the triangle inequalities is 
violated the remaining two are necessarily satisfied. It implies that one can always arrange the fields in the three-point function to obtain a configuration for which $s_{3} \leq s_{1} + s_{2} $, 
or equivalently, 
$C \leq N < N +1$. We will assume that we have performed such an arrangement and use eq.~\eqref{Ch05.2-C polynomial ansatz}. 

Requiring that the three-point function is conserved at $z_{1}$ and $z_{2}$ is now tantamount to imposing
\begin{equation}
\frac{\pa}{\pa u^{\a}} \frac{\pa}{\pa u^{\b}} \frac{\pa}{\pa X_{ \a \b}} C(X; u,v ) = 0 \, .
\end{equation}
By acting with this operator on the ansatz \eqref{Ch05.2-C polynomial ansatz}, we obtain
\begin{align}
X^{C-N-3} \sum_{k=0}^{2C} \a_{k} P_{3}^{k-2} Q_{3}^{2C - 2 - k} Z_{1}^{N-C} \Big\{ \s_{1}(k) P_{3}^{2} Q_{3}^{2} + \s_{2}(k) Q_{3}^{4} + \s_{3}(k) P_{3}^{4} \Big\} = 0 \, ,
\end{align}
where
\begin{subequations}
\begin{align}
	\s_{1}(k) &= - 2 k^3 + 4 C k^2 + 2 k (1 + C + 2 N (2 + N)) - 2 C (3 + 4 N (2 + N)) \, , \\
	\s_{2}(k) &= - k(k-1) (2 N - k + 3)  \, , \\
	\s_{3}(k) &= (2 C - k - 1) (2 C - k) (2 N + k + 3) \, .
\end{align}
\end{subequations}
The sum above may now be split up into three contributions so that the coefficients may be easily read off
\begin{align}
\sum_{k=0}^{2C} \a_{k} \s_{1}(k) P_{3}^{k} Q_{3}^{2C - k} + \sum_{k=0}^{2C-2} \a_{k+2} \s_{2}(k+2) P_{3}^{k} Q_{3}^{2C - k} + \sum_{k=2}^{2C} \a_{k-2} \s_{3}(k-2) P_{3}^{k} Q_{3}^{2C - k} = 0\, .
\end{align}
Hence, we obtain the following linear system:
\begin{subequations} \label{Ch05.2-Linear homogeneous equations}
\begin{align}
	\a_{k} \s_{1}(k) + \a_{k+2} \s_{2}(k+2) + \a_{k-2} \s_{3}(k-2) = 0 \, , \hspace{10mm} 2 \leq k \leq 2C-2 \, ,
\end{align}
\vspace{-12mm}
\begin{align}
	\a_{0} \s_{1}(0) + \a_{2} \s_{2}(2) &= 0 \, , & \a_{2C} \s_{1}(2C) + \a_{2C-2} \s_{3}(2C-2) &= 0 \, , \\[1mm]
	\a_{1} \s_{1}(1) + \a_{3} \s_{2}(3) &= 0 \, , & \a_{2C-1} \s_{1}(2C-1) + \a_{2C-3} \s_{3}(2C-3) &= 0 \, .
\end{align}
\end{subequations}
It must be noted that the equations above for the variables $\a_{k}$ split into independent $C+1$ and $C$ dimensional systems of linear homogeneous equations corresponding to the parity-even and parity-odd sectors respectively. The terms for which $k$ is even are denoted parity-even, while the terms for which $k$ is odd are denoted parity-odd, so that
\begin{equation}
C(X;u,v) = C_{E}(X;u,v) + C_{O}(X;u,v) \, ,
\end{equation}
where
\begin{subequations}
\begin{align}
	C_{E}(X; u,v ) &= X^{C-N-2} \sum_{k=0}^{C} \a_{2k} P_{3}^{2k} Q_{3}^{2(C - k)} Z_{1}^{N-C+1} \, , \label{Ch05.2-zh7} \\
	C_{O}(X; u,v ) &= X^{C-N-2} \sum_{k=1}^{C} \a_{2k-1} P_{3}^{2k-1} Q_{3}^{2(C - k) + 1} Z_{1}^{N-C+1} \, .
\end{align}
\end{subequations}
Indeed, this convention is consistent with \eqref{Ch05.1-H inversion}. Hence, in the linear homogeneous system \eqref{Ch05.2-Linear homogeneous equations}, we define the parity-even coefficients, $a_{k}$, $b_{k}$, $c_{k}$, as
\begin{subequations}
\begin{align}
	a_{k} &= \s_{1}(2k-2) \, , & &1\leq k \leq C+1 \, ,\\
	b_{k} &= \s_{2}(2k) \, , & &1\leq k \leq C \, , \\
	c_{k} &= \s_{3}(2k-2) \, , & &1\leq k \leq C \, ,
\end{align}
\end{subequations}
and the parity-odd coefficients, $\tilde{a}_{k}$, $\tilde{b}_{k}$, $\tilde{c}_{k}$ as
\begin{subequations}
\begin{align}
	\tilde{a}_{k} &= \s_{1}(2k-1) \, , & &1\leq k \leq C \, , \\
	\tilde{b}_{k} &= \s_{2}(2k+1) \, , & &1\leq k \leq C-1 \, , \\
	\tilde{c}_{k} &= \s_{3}(2k-1) \, , & &1\leq k \leq C-1 \, .
\end{align}
\end{subequations}
With the above definitions, the linear homogeneous equations \eqref{Ch05.2-Linear homogeneous equations} split into two independent systems which can be written in the form 
$\mathbf{M} \vec{\a} = \mathbf{0}$. More explicitly
\begin{subequations} \label{Ch05.2-zh100}
\begin{align}
	\textbf{Even:}& &
	\mathbf{M}_{E} &= 
	\setlength{\arraycolsep}{5pt}
	\begin{bmatrix}
		\; a_{1} & b_{1} & 0 & 0 & \dots & 0 \\
		\; c_{1} & a_{2} & b_{2} & 0 & \dots & 0 \\
		\; 0 & c_{2} & a_{3} & b_{3} & \dots & 0 \\
		\; \vdots & \vdots & \ddots & \ddots & \ddots & \vdots \\
		\; 0 & \dots & \dots & c_{C-1} & a_{C} & b_{C} \\
		\; 0 & 0 & \dots & 0 & c_{C} & a_{C+1}
	\end{bmatrix} , &
	\vec{\a}_{E} &=
	\begin{bmatrix}
		\a_{0} \\
		\a_{2} \\
		\a_{4} \\
		\vdots \\
		\a_{2C-2} \\
		\a_{2C} 
	\end{bmatrix} , \label{Ch05.2-Parity even homogeneous system}\\[5mm]
	\textbf{Odd:}& &
	\mathbf{M}_{O} &= 
	\setlength{\arraycolsep}{5pt}
	\begin{bmatrix}
		\; \tilde{a}_{1} & \tilde{b}_{1} & 0 & 0 & \dots & 0 \\
		\; \tilde{c}_{1} & \tilde{a}_{2} & \tilde{b}_{2} & 0 & \dots & 0 \\
		\; 0 & \tilde{c}_{2} & \tilde{a}_{3} & \tilde{b}_{3} & \dots & 0 \\
		\; \vdots & \vdots & \ddots & \ddots & \ddots & \vdots \\
		\; 0 & \dots & \dots & \tilde{c}_{C-2} & \tilde{a}_{C-1} & \tilde{b}_{C-1} \\
		\; 0 & 0 & \dots & 0 & \tilde{c}_{C-1} & \tilde{a}_{C}
	\end{bmatrix} , &
	\vec{\a}_{O} &=
	\begin{bmatrix}
		\a_{1} \\
		\a_{3} \\
		\a_{5} \\
		\vdots \\
		\a_{2C-3} \\
		\a_{2C-1} 
	\end{bmatrix} , \label{Ch05.2-Parity odd homogeneous system}
\end{align}
\end{subequations}
where $\mathbf{M}_{E}$, $\mathbf{M}_{O}$ are square matrices of dimension $C+1$, $C$ respectively. The system of equations \eqref{Ch05.2-Parity even homogeneous system} is associated with the solution $C_{E}(X;u,v)$, while the system \eqref{Ch05.2-Parity odd homogeneous system} is associated with the solution $C_{O}(X;u,v)$. The question now is whether there exists explicit solutions to these linear homogeneous systems for arbitrary $A,B,C$. 

The matrices of the form~\eqref{Ch05.2-zh100} are referred to as tri-diagonal matrices, and before we continue with the analysis let us comment on some of their features.
The sufficient conditions under which a tri-diagonal matrix is invertible for arbitrary sequences $a_{k}, b_{k}, c_{k}$ has been discussed in e.g. \cite{BRUGNANO1992131,YHuang_1997}. 
Now consider the determinant of a tri-diagonal matrix
\begin{align}
\D_{k} = 
\setlength{\arraycolsep}{5pt}
\begin{vmatrix}
	\; a_{1} & b_{1} & 0 & 0 & \dots & 0 \\
	\; c_{1} & a_{2} & b_{2} & 0 & \dots & 0 \\
	\; 0 & c_{2} & a_{3} & b_{3} & \dots & 0 \\
	\; \vdots & \vdots & \ddots & \ddots & \ddots & \vdots \\
	\; 0 & \dots & \dots & c_{k-2} & a_{k-1} & b_{k-1} \; \\
	\; 0 & 0 & \dots & 0 & c_{k-1} & a_{k}
\end{vmatrix} \, .
\end{align}
One of its most important properties is that it satisfies the recurrence relation
\begin{equation} \label{Ch05.2-Continuant equation}
\D_{k} = a_{k} \D_{k-1} - b_{k-1} c_{k-1} \D_{k-2} \, , \hspace{10mm} \D_{0} = 1 \, , \hspace{4mm} \D_{-1} = 0 \, ,
\end{equation}
which may be seen by performing a Laplace expansion on the last row. The sequence $\D_{k}$ is often called the ``generalised continuant" with respect to the sequences $a_{k}, b_{k}, c_{k}$. There are methods to compute this determinant in closed form for simple cases, for example if the tri-diagonal matrix under consideration is ``Toeplitz" 
(i.e. if the sequences are constant $a_{k} = a$, $b_{k} = b$, $c_{k} = c$). 
For Toeplitz tri-diagonal matrices one obtains
%
\begin{align}
\D_{k} = \frac{1}{\sqrt{a^{2} - 4 b c}} \, \Bigg\{ \bigg(\frac{a+\sqrt{a^{2} - 4 b c}}{2} \bigg)^{k+1} - \: \bigg(\frac{a - \sqrt{a^{2} - 4 b c}}{2}\bigg)^{k+1} \Bigg\} \, ,
\end{align}
for $a^{2} - 4 b c \neq 0$, 
while for $a^{2} - 4 b c = 0$ we obtain $\D_{k} = (k+1) (\frac{a}{2})^{k}$. For general sequences $a_{k}, b_{k}, c_{k}$ there is no straightforward approach to compute $\D_{k}$ and it must computed recursively using \eqref{Ch05.2-Continuant equation}. Another important feature of tri-diagonal matrices is that in general their nullity (co-rank) is either $0$ or $1$,
which implies that any system of linear homogeneous equations with a tri-diagonal matrix has at most one non-trivial solution.

In the next subsections, we study the continuants of $\mathbf{M}_{E},\mathbf{M}_{O}$ for the homogeneous systems \eqref{Ch05.2-Parity even homogeneous system}, \eqref{Ch05.2-Parity odd homogeneous system}, and obtain their explicit form for arbitrary $A,B,C$. This determines whether $\mathbf{M}_{E},\mathbf{M}_{O}$ are invertible, and the number of solutions for the homogeneous systems \eqref{Ch05.2-Parity even homogeneous system}, \eqref{Ch05.2-Parity odd homogeneous system}. 


\subsubsection*{Parity-odd case}\label{Ch05.2-subsection3.3}

First we will analyse the system of equations for the parity-odd sector. 
Note that if $\det[ \mathbf{M}_{O} ] \neq 0$ then the system of equations 
$\mathbf{M}_{O} \vec{\a}_{O} = \mathbf{0}$ admits only the trivial solution. 
Let us denote $\mathbf{M}_{O} := \mathbf{M}_{O}^{(C)}$ to indicate that the dimension of the tri-diagonal matrix in~\eqref{Ch05.2-Parity odd homogeneous system} is $C \times C$.
We also introduce the $ k \times k$ continuant $\tilde{\D}_{k}^{(C)}$, where $1  \leq k \leq C$. It satisfies the continuant equation~\eqref{Ch05.2-Continuant equation} with
\begin{subequations} \label{Ch05.2-Parity odd coefficients}
\begin{align}
	\tilde{a}_{k} &= 4 C (-1 + k (4k-3) - 2 N (N+2) ) \nonumber\\
	& \hspace{10mm} - 4 (2 k - 1) (2 k (k - 1) - N (N + 2))  \, , & &1\leq k \leq C \, , \\
	\tilde{b}_{k} &= - 4 k (1 + 2 k) (N - k+1) \, , & &1\leq k \leq C-1 \, , \\
	\tilde{c}_{k} &= 4 (1 + 2 C - 2 k) (C - k) (1 + k + N) \, , & &1\leq k \leq C-1 \, .
\end{align}
\end{subequations}
We also have $\tilde{\D}_{C}^{(C)} = \det[ \mathbf{M}_{O}^{(C)} ]$. Below we present some examples of the matrix $\mathbf{M}_{O}$ and the determinant 
$\tilde{\D}_{C}^{(C)}$ for arbitrary $N=A+B$ and fixed $C$:
\begin{subequations} \label{Ch05.2-Parity odd matrix examples}
\begin{align}
	\mathbf{M}_{O}^{(1)} = \left[ \,
	-4 N (N+2) \, \right] \, ,
\end{align}
\begin{align}
	\mathbf{M}_{O}^{(2)} = 
	\left[
	\begin{array}{cc}
		-12 N (N+2) & -12 N \\
		12 (N+2) & 24-4 N (N+2) \\
	\end{array}
	\right] ,
\end{align}
\begin{align}
	\mathbf{M}_{O}^{(3)} = 
	\left[
	\begin{array}{ccc}
		-20 N (N+2) & -12 N & 0 \\
		40 (N+2) & 60-12 N (N+2) & -40 (N-1) \\
		0 & 12 (N+3) & 72-4 N (N+2) \\
	\end{array}
	\right] , 
\end{align}
\begin{align}
	\mathbf{M}_{O}^{(4)} = 
	\left[
	\scalemath{0.9}{\begin{array}{cccc}
			-28 N (N+2) & -12 N & 0 & 0 \\
			84 (N+2) & 96-20 N (N+2) & -40 (N-1) & 0 \\
			0 & 40 (N+3) & 176-12 N (N+2) & -84 (N-2) \\
			0 & 0 & 12 (N+4) & 144-4 N (N+2) \\
	\end{array}}
	\right] \,.
\end{align}
\end{subequations}
The corresponding determinants are:
\begin{subequations}
\begin{align}
	\tilde{\D}_{1}^{(1)} &= -4 N (N+2) \, ,\\
	\tilde{\D}_{2}^{(2)} &= 48 (N-1) N (N+2) (N+3) \, , \\
	\tilde{\D}_{3}^{(3)} &= -960 (N-2) (N-1) N (N+2) (N+3) (N+4) \, , \\
	\tilde{\D}_{4}^{(4)} &= 26880 (N-3) (N-2) (N-1) N (N+2) (N+3) (N+4) (N+5) \, . 
\end{align}
\end{subequations}
Indeed the matrices above are invertible as $\det[\mathbf{M}_{O}] \neq 0$, therefore we have the solution $\vec{\a}_{O} = \mathbf{0}$. The determinant can be efficiently computed using the recursion formula \eqref{Ch05.2-Continuant equation}, and the pattern appears holds for large integers $N,C$. Using Mathematica we explicitly computed $\tilde{\D}_{C}^{(C)}$ for arbitrary $N$, up to $C=500$. In all cases we found it is non-trivial.

However, the continuant $\tilde{\D}_{k}^{(C)}$ can also be obtained explicitly for all $1 \leq k \leq C$, and arbitrary $C$. First, one can show that $\tilde{\D}_{k}^{(C)}$ satisfies the following recurrence relation:
\begin{align} \label{Ch05.2-Trick recursion - odd sector}
\tilde{\D}_{k}^{(C)} = \tilde{\g}_{k-1}^{(C)} \tilde{\D}_{k-1}^{(C)} \, , \hspace{5mm} \tilde{\D}_{1}^{(C)} = -4(2C-1) N (2+N) \, , \hspace{5mm} 1 \leq k \leq C \, ,
\end{align}
where $\tilde{\g}^{(C)}_{k}$ is given by 
%
\begin{align} \label{Ch05.2-e1-Odd}
\tilde{\g}^{(C)}_{k} = - 4 (N-k) (2 + k + N ) (-1 + 2C - 2k ) \, .
\end{align}
To see this, consider the combination $\tilde{\D}_{k+2}^{(C)} - \tilde{a}_{k+2} \tilde{\D}_{k+1}^{(C)} + \tilde{b}_{k+1} \tilde{c}_{k+1} \tilde{\D}_{k}^{(C)}$. 
Using eqs. \eqref{Ch05.2-Trick recursion - even sector}, \eqref{Ch05.2-e1-Odd}, this combination becomes $( \tilde{\g}_{k+1} \tilde{\g}_{k} - \tilde{a}_{k+2} \tilde{\g}_{k}  + \tilde{b}_{k+1} \tilde{c}_{k+1} ) \tilde{\D}^{(C)}_{k}$. 
However it may be shown using eqs.~\eqref{Ch05.2-Parity odd coefficients} and~\eqref{Ch05.2-e1-Odd} that 
\begin{equation}
\label{Ch05.2-e2}
\tilde{\g}_{k+1} \tilde{\g}_{k} - \tilde{a}_{k+2} \tilde{\g}_{k}  + \tilde{b}_{k+1} \tilde{c}_{k+1} = 0 \, ,
\end{equation}
for arbitrary $k, N, C$, which implies that the recurrence relation~\eqref{Ch05.2-Continuant equation} is indeed satisfied. We find the following general solution for \eqref{Ch05.2-Trick recursion - odd sector}:\footnote{Recall that we have arranged the operators in the three-point function so that $s_3 \leq s_1+ s_2$ which implies $C \leq N$.}
\begin{align} \label{Ch05.2-Parity-odd continuant}
\tilde{\D}_{k}^{(C)} = 2^{3k} \, \frac{\Gamma(\tfrac{1}{2} - C + k) }{\Gamma(\tfrac{1}{2}-C) } \frac{(N+k+1)!}{ (N+1) (N-k)!} \, , \hspace{5mm} 1 \leq k \leq C \, .
\end{align}
One can check (using e.g. Mathematica) that it solves the continuant equation~\eqref{Ch05.2-Continuant equation}. Recalling that $\det[ \mathbf{M}_{O}^{(C)} ] = \tilde{\D}_{C}^{(C)}$, it then follows that 
\begin{align} \label{Ch05.2-Odd determinant}
\det[ \mathbf{M}_{O}^{(C)} ] = \frac{2^{3C} \sqrt{\pi }}{\Gamma (\tfrac{1}{2}-C)} \frac{ (N+C+1)!}{ (N+1) (N-C)!} \, , \hspace{5mm} C \leq N \, ,
\end{align}
%
which is always non-trivial. This implies that $\mathbf{M}_{O}$ is of full rank and, hence, the system of equations \eqref{Ch05.2-Parity odd homogeneous system} admits only the trivial solution $\vec{\a}_{O} = \mathbf{0}$. 
Therefore, recalling that $C_{O}$ is the parity-odd solution corresponding to the system of equations \eqref{Ch05.2-Parity odd homogeneous system}, from the analysis above we have shown 
that $C_{O}(X;u,v) = 0$ for arbitrary $A, B, C$. 

Now we will show that the tensor $D_{S}$ associated with $C_{O}$, which are related by the algebraic relations \eqref{Ch05.2-C and D algebraic relations - more symmetric}, also vanishes. 
Since we have shown that $C_{O}=0$ in general, eq.~\eqref{Ch05.2-C and D algebraic relations - more symmetric} implies
\begin{subequations}
\begin{align}
	\eta_{r_{1} k_{1}} D_{S}^{(r_{1} ... r_{N+1})(k_{1} ... k_{C+1})} &= 0 \, , \\
	\e^{p}{}_{r_{1} k_{1}} D_{S}^{(r_{1} ... r_{N+1})(k_{1} ... k_{C+1})} &= 0 \, .
\end{align}
\end{subequations}
Hence, $D_{S}$ is totally symmetric and traceless in all tensor indices
\begin{equation}
D_{S}^{(r_{1} ... r_{N+1})(k_{1} ... k_{C+1})} \equiv D_{S}^{(r_{1} ... r_{N+1} k_{1} ... k_{C+1})} \, .
\end{equation}
We can now construct a solution for $D_{S}$ using auxiliary spinors as follows:
\begin{align}
D_{S}(X; u(2N+2C+4) ) = D_{S \, \a(2N+2C+4)}(X) \, u^{\a_{1}} \, ... \, u^{\a_{2N+2C+4}} \, .
\end{align}
Since $D_{S}$ is transverse, the polynomial $D_{S}(X;u)$ must satisfy the conservation equation
\begin{equation} \label{Ch05.2-D conservation equation}
\frac{\pa}{\pa u^{\a}} \frac{\pa}{\pa u^{\b}} \frac{\pa}{\pa X_{ \a \b}} D_{S}(X; u) = 0 \, .
\end{equation}
The only possible structure (up to a constant coefficient) for $D_{S}(X;u)$ is of the form
\begin{equation}
D_{S}(X; u(2N+2C+4) ) =X^{C-N-2} Z_{1}^{N+C+2} \, .
\end{equation}
Explicit computation of \eqref{Ch05.2-D conservation equation} gives
\begin{equation}
\frac{\pa}{\pa u^{\a}} \frac{\pa}{\pa u^{\b}} \frac{\pa}{\pa X_{ \a \b}} D_{S}(X; u) = -4(N+C+2)^{2}(1+2C) \, X^{C-N-3} Z_{1}^{N+C+1} \, ,
\end{equation}
which is always non-zero. Therefore $D_{S} = 0$ for the parity-odd sector. Hence, for Grassmann-odd three-point functions of the 
form $\langle \mathbf{J}^{}_{F} \mathbf{J}'_{F} \mathbf{J}''_{F} \rangle$, there is no parity-odd solution for arbitrary superspins.

Let us point out that we did not need to use the constraint arising from conservation on the third point. Imposing the conservation equations at the first two points was sufficient to prove vanishing of the parity-odd contribution. Note, however, that we assume the operator inserted at the third point is a conserved supercurrent, i.e., we used the fact that its dimension is $s_3+1$, saturating the unitarity bound.


\subsubsection*{Parity-even case}\label{Ch05.2-subsection3.4}

For the parity even case, we follow the same approach. Let us denote $\mathbf{M}_{E} := \mathbf{M}_{E}^{(C)}$ (recall that the dimension of the matrix $\mathbf{M}_{E}^{(C)}$ 
in~\eqref{Ch05.2-Parity even homogeneous system} is now $(C+1) \times (C+1)$) and consider the continuant $\D_{k}^{(C)}$, $ 1 \leq k \leq C+1$, with 
$\D_{C+1}^{(C)} = \det[ \mathbf{M}_{E}^{(C)} ]$. The continuant satisfies eq.~\eqref{Ch05.2-Continuant equation}, where 
%
\begin{subequations} \label{Ch05.2-Parity even coefficients}
\begin{align}
	a_{k} &= 2 C (3 + 2 k (4 k - 7) - 4 N (N + 2)) \nonumber\\
	& \hspace{10mm} - 
	4 (k-1) (3 + 4 k (k - 2) - 2 N (N + 2))  \, , & &1\leq k \leq C + 1 \, , \\
	b_{k} &= - 2 k (2k-1) (3 + 2N - 2k) \, , & &1\leq k \leq C \, , \\
	c_{k} &= 2 (1 + 2 C - 2 k) (1 + C - k) (1 + 2 k + 2 N) \, , & &1\leq k \leq C \, .
\end{align}
\end{subequations}
%
Below are some examples of the matrix $\mathbf{M}_{E}$ and the determinant $\D_{C+1}^{(C)}$ for fixed $C$:
\begin{subequations} \label{Ch05.2-Parity even matrix examples}
\begin{align}
	\mathbf{M}_{E}^{(N,1)}= \left[ 
	\begin{array}{cc}
		-2 (2 N+1) (2 N+3) & -2 (2 N+1) \\
		2 (2 N+3) & 2 \\
	\end{array} \right]  ,
\end{align}
\begin{align}
	\mathbf{M}_{E}^{(N,2)} = 
	\left[
	\begin{array}{ccc}
		-4 (2 N+1) (2 N+3) & -2 (2 N+1) & 0 \\
		12 (2 N+3) & -8 \left(N^2+2 N-2\right) & -12 (2 N-1) \\
		0 & 2 (2 N+5) & 12 \\
	\end{array}
	\right] ,
\end{align}
\begin{align}
	\mathbf{M}_{E}^{(N,3)} = 
	\left[
	\scalemath{0.8}{\begin{array}{cccc}
			-6 (2 N+1) (2 N+3) & -2 (2 N+1) & 0 & 0 \\
			30 (2 N+3) & -2 \left(8 N^2+16 N-15\right) & -12 (2 N-1) & 0 \\
			0 & 12 (2 N+5) & -2 \left(4 N^2+8 N-39\right) & -30 (2 N-3) \\
			0 & 0 & 2 (2 N+7) & 30 \\
	\end{array}}
	\right] , 
\end{align}
\end{subequations}
Contrary to the parity-odd case, all of the matrices above are singular, i.e. $\D_{2}^{(1)} = \D_{3}^{(2)} = \D_{4}^{(3)} = \D_{5}^{(4)} =  0$. 
For large integers we use the recursion formula \eqref{Ch05.2-Continuant equation} to analyse the general structure of $\D_{C+1}^{(C)}$. Analogous to the parity-odd case, we computed it for arbitrary $N$ up to $C = 500$ and found in all cases $\D_{C+1}^{(C)} = 0$. 
It then follows that the matrix $\mathbf{M}_{E}^{(C)}$ is of co-rank one and 
the solution, $\vec{\a}_{E}$, to $\mathbf{M}_{E}^{(N,C)} \vec{\a}_{E} = \mathbf{0}$, is fixed up to a single overall constant in all cases, and therefore $C_{E}$ is unique.

For the parity-even case it also turns out to be possible to solve for the continuants $\D_{k}^{(C)}$ for $ 1 \leq k \leq C+1$. One can show that  $\D_{k}^{(C)}$
satisfies the following recurrence relation:
\begin{align} \label{Ch05.2-Trick recursion - even sector}
\D_{k}^{(C)} = \g_{k-1}^{(C)} \D_{k-1}^{(C)} \, , \hspace{5mm} \D_{1}^{(C)} = -2C (1+2 N) (3+2 N) \, , \hspace{5mm} 1 \leq k \leq C + 1 \, ,
\end{align}
where $\g^{(C)}_{k}$ is given by 
%
\begin{align} \label{Ch05.2-e1}
\g^{(C)}_{k} = - 2 (C-k) (1 + 2 N - 2k) (3 + 2N + 2k ) \, .
\end{align}
%
To show this, consider again the combination $\D_{k+2}^{(C)} - a_{k+2} \D_{k+1}^{(C)} + b_{k+1} c_{k+1} \D_{k}^{(C)}$. 
Using eqs. \eqref{Ch05.2-Trick recursion - even sector}, \eqref{Ch05.2-e1}, we obtain $( \g_{k+1} \g_{k} - a_{k+2} \g_{k}  + b_{k+1} c_{k+1} ) \D^{(C)}_{k}$. 
It is then simple to show using eqs.~\eqref{Ch05.2-Parity even coefficients} and~\eqref{Ch05.2-e1} that this combination vanishes for arbitrary $k, N, C$, which implies that the recurrence relation~\eqref{Ch05.2-Continuant equation} is indeed satisfied. 
Analogous to the parity-odd case, it is possible to find an explicit solution for the recurrence relation~\eqref{Ch05.2-Trick recursion - even sector}, and we find the following general formula for the continuant:
\begin{align} \label{Ch05.2-Even continuant}
\D_{k}^{(C)} =  (-1)^{N+1} \frac{ 2^{3k} C! }{ \pi \, (C-k)! } \, \Gamma(\tfrac{1}{2} - N + k ) \, \Gamma( \tfrac{3}{2} + N + k )  \, , \hspace{5mm} 1 \leq k \leq C \,.
\end{align}
%
For $k = C+1$, we have $\Delta_{C+1}^{(C)} = \g_{C}^{(C)} \Delta_{C}^{(C)}$. However, we note that $\g_{C}^{(C)} = 0$, which implies 
$\Delta_{C+1}^{(C)} = \det[ \mathbf{M}_{E}^{(C)} ] = 0$, for arbitrary $N, C$. Hence, we have shown that the matrix $\mathbf{M}_{E}$ is always of co-rank one and
the system~\eqref{Ch05.2-Parity even homogeneous system} has a unique non-trivial solution.\footnote{Alternatively, note that \eqref{Ch05.2-Even continuant} implies that the largest non-trivial minor of $\mathbf{M}_{E}$ is of dimension $C \times C$. Therefore, $\text{Rank}(\mathbf{M}_{E}) = C$, which implies $\text{Nullity}( \mathbf{M}_{E} ) = 1$.} This, in turn, implies that the tensor $C_{E}(X; u,v)$ is 
unique up to an overall coefficient. The explicit solution to the system $\mathbf{M} \vec{\a} = \mathbf{0}$ and, thus, for the tensor $C_E$ 
can also be found for arbitrary $N,C$. Indeed, by analysing the nullspace of $\mathbf{M}_{E}$ we obtain the following solution for the coefficients $\a_{2k}$ of \eqref{Ch05.2-zh7}:
\begin{align}
\a_{2k} = \frac{ (-1)^k 2^{2 k} \, \Gamma(C+1) \, \Gamma(k+N+\tfrac{3}{2}) }{ \Gamma(2k+1) \, \Gamma(C-k+1) \, \Gamma(N+\tfrac{3}{2}) } \, \a_{0} \, , \hspace{10mm} 1 \leq k \leq C \, .
\end{align}
This is also a solution to the parity-even part of the recurrence relations \eqref{Ch05.2-Linear homogeneous equations}, which can be explicitly checked. Hence, we have obtained a unique solution (up to an overall coefficient) for $C_{E}$ in explicit form for arbitrary superspins.


Now recall that the three-point functions under consideration are determined not just by the tensor $C^{(r_{1} ... r_{N+1})(k_{1} ... k_{C})}$  
but also by the tensor $D_{S}^{(r_{1} ... r_{N+1})(k_{1} ... k_{C+1})}$ which is transverse (see eq.~\eqref{Ch05.2-e4}) and related to $C^{(r_{1} ... r_{N+1})(k_{1} ... k_{C})}$  
by the algebraic relation~\eqref{Ch05.2-C and D algebraic relations}. Remarkably, it is possible to solve for $D_S$ in terms of $C$ using~\eqref{Ch05.2-C and D algebraic relations}. 
To show it let us begin by constructing irreducible decompositions for $C$ and $D$. Since we know that $C$ is parity-even, it cannot contain $\e$, and we use the decompositions
\begin{align} \label{Ch05.2-C irreducible decomposition}
C^{(r_{1} ... r_{N+1})(k_{1} ... k_{C})} = C_{1}^{(r_{1} ... r_{N+1} k_{1} ... k_{C})} + \sum_{i=1}^{N+1} \sum_{j=1}^{C} \eta^{r_{i} k_{j}} C_{2}^{(r_{1} ... \hat{r}_{i} ... r_{N+1} k_{1} ... \hat{k}_{j} ... k_{C})} \nonumber \\ 
+ \sum_{j > i=1}^{N+1} \eta^{r_{i} r_{j}} C_{3}^{(r_{1} ... \hat{r}_{i} \hat{r}_{j} ... r_{N+1} k_{1} ... k_{C})}
+ \sum_{j > i=1}^{C} \eta^{k_{i} k_{j}} C_{4}^{(r_{1} ... r_{N+1} k_{1} ... \hat{k}_{i} \hat{k}_{j} ... k_{C})} \, ,
\end{align}
where $C_{2}, C_{3}, C_{4}$ are the irreducible components of rank $N+C-1$ ($C_{4}$ exists only for $C > 1$). 
Requiring that the above ansatz is traceless in the appropriate groups of indices fixes $C_{3}$ and $C_{4}$ in terms of $C_{2}$ as follows (indices suppressed):
\begin{align}
C_{3} = - \frac{2C}{2N+1} \, C_{2} \, , \hspace{15mm} C_{4} = - \frac{2(N+1)}{2C-1} \, C_{2} \, .
\end{align}
Hence, $C$ is determined completely in terms of the totally symmetric and traceless tensors $C_{1}$ and $C_{2}$. Now let us construct an irreducible decomposition of $D_{S}$. 
Due to the algebraic relation \eqref{Ch05.2-C and D algebraic relations - more symmetric}, we know that $D_{S}$ must be linear in $\e$. 
The only way to construct $D_{S}$ such that it contains $\e$ is by using the following decomposition:
\begin{equation} \label{Ch05.2-D_{S} irreducible decomposition}
D_{S}^{(r_{1} ... r_{N+1})(k_{1} ... k_{C+1})} = \sum_{i=1}^{N+1} \sum_{j=1}^{C+1} \e^{q r_{i} k_{j}} T^{q, (r_{1} ... \hat{r}_{i} ... r_{N+1} k_{1} ... \hat{k}_{j} ... k_{C+1})} \, ,
\end{equation}
where the tensor $T$ (of rank $N+C+1$) is decomposed as follows:
\begin{align}
T^{q, (r_{1} ... r_{N})(k_{1} ... k_{C})} = T_{1}^{(r_{1} ... r_{N} k_{1} ... k_{C} q)} + \sum_{i=1}^{N} \sum_{j=1}^{C} \eta^{r_{i} k_{j}} T_{2}^{(r_{1} ... \hat{r}_{i} ... r_{N} k_{1} ... \hat{k}_{j} ... k_{C} q)} \nonumber \\ 
+ \sum_{j > i=1}^{N} \eta^{r_{i} r_{j}} T_{3}^{(r_{1} ... \hat{r}_{i} \hat{r}_{j} ... r_{N} k_{1} ... k_{C} q)}
+ \sum_{j > i=1}^{C} \eta^{k_{i} k_{j}} T_{4}^{(r_{1} ... r_{N} k_{1} ... \hat{k}_{i} \hat{k}_{j} ... k_{C} q)} \, .
\label{Ch05.2-zh3}
\end{align}
Here $T_1$ is the irreducible component of rank $N+C+1$ and $T_{2}, T_{3}, T_{4}$ are the irreducible components of rank $N+C-1$ (where $T_{4}$ exists only for $C > 1$). It should be noted that one could also consider contributions to $T$ proportionate to $\eta^{q r_{i}}$, $\eta^{q k_{j}}$, 
but such contributions will cancel when substituted into \eqref{Ch05.2-D_{S} irreducible decomposition} and, hence, they do not contribute to the irreducible decomposition of $D_S$.  
Requiring that $D$ is traceless on the appropriate groups of indices fixes $T_{3}$ and $T_{4}$ in terms of $T_{2}$ as follows (indices suppressed):
\begin{align}
T_{3} = - \frac{2C}{2N+1} \, T_{2} \, , \hspace{15mm} T_{4} = - \frac{2N}{2C-1} \, T_{2} \, .
\label{Ch05.2-zh4}
\end{align}
Hence, $D_{S}$ is described completely in terms of the totally symmetric and traceless tensors $T_{1}$ and $T_{2}$. If we now consider the algebraic relations \eqref{Ch05.2-C and D algebraic relations - more symmetric} and substitute in the above decompositions, after some tedious calculation one obtains:
\begin{align}
0 &= \eta_{r_{1} k_{1}} D_{S}^{(r_{1} ... r_{N+1})(k_{1} ... k_{C+1})} + \frac{1}{C+1} \sum_{i=2}^{C+1} \e^{k_{i}}{}_{r_{1} k_{1}} C^{ (r_{1} ... r_{N+1})( k_{1} k_{2} ... \hat{k}_{i} ... k_{C+1} ) } \nonumber \\
&= \sum_{i=2}^{N+1} \sum_{j=2}^{C+1} \e^{q r_{i} k_{j}} \Big\{ \t(N,C) \, T_{2}^{(\hat{r}_{1} ... \hat{r}_{i} ... r_{N+1} \hat{k}_{1} ... \hat{k}_{j} ... k_{C+1} q)}  \\
& \hspace{30mm} + \frac{1}{C+1} \bigg( 1 + \frac{2C}{2N + 1} \bigg) C_{2}^{(\hat{r}_{1} ... \hat{r}_{i} ... r_{N+1} \hat{k}_{1} ... \hat{k}_{j} ... k_{C+1} q)} \Big\}\,, \nonumber
\end{align}
where the constant $\t(N,C)$ is defined as follows:
\begin{equation}
\t(N,C) = \frac{ 6 C^2 + 2 N^2 + 3 C (1 + 4 N) - 5 N - 3 }{(2 C - 1) (1 + 2 N)} \, .
\end{equation}
Requiring that the above combination vanishes gives $T_{2} = \xi_{2} \, C_{2}$, where
\begin{equation}
\xi_{2} = - \frac{ (2 C - 1) (1 + 2 C + 2 N) }{ (1 + C) ( 6 C^2 + 2 N^2 + 3 C (1 + 4 N) - 5 N - 3 ) } \, .
\end{equation}
One can proceed in a similar way with the second algebraic relation in \eqref{Ch05.2-C and D algebraic relations - more symmetric}. After some calculation we found
that it relates $T_1$ and $C_1$ as  $T_{1} = \xi_{1} \, C_{1}$, where
\begin{equation}
\xi_{1} = - \frac{ 2 C + 1 }{ (C+1)(N+C+2)} \, .
\end{equation}
A similar calculation was performed in \cite{Buchbinder:2015qsa}, where it was shown that for $A = B = C = 1$, $\xi_{1} = - \frac{3}{10}$, $\xi_{2} = - \frac{1}{8}$, 
which is in full agreement with the expressions above. 

Finally, we need to show that $D_S$ found this way is transverse, i.e.
\begin{equation}
\pa_{p} D_{S}^{(p r_1 \dots r_N) (k_1 \dots k_{C+1})} = 0\,. 
\label{Ch05.2-e5}
\end{equation}
For this let us define the tensor 
\begin{equation}
E^{( r_1 \dots r_N) (k_1 \dots k_{C+1})}= \pa_{p} D_{S}^{(p r_1 \dots r_N) (k_1 \dots k_{C+1})} \, . 
\end{equation}
Our aim is to show that $E^{( r_1 \dots r_N) (k_1 \dots k_{C+1})}=0$. To find $E^{( r_1 \dots r_N) (k_1 \dots k_{C+1})}$ we contract $D_S$ in 
eqs.~\eqref{Ch05.2-D_{S} irreducible decomposition}, \eqref{Ch05.2-zh3}, \eqref{Ch05.2-zh4}  with the derivative. 
However, from the algebraic relations~\eqref{Ch05.2-C and D algebraic relations} and the fact that $C$ is transverse it follows that $E^{( r_1 \dots r_N) (k_1 \dots k_{C+1})}$ is totally symmetric and traceless:
\begin{equation}
E^{(r_{1} ... r_{N})(k_{1} ... k_{C+1})} \equiv E^{(r_{1} ... r_{N} k_{1} ... k_{C+1})}  \, .
\end{equation}
Using eqs.~\eqref{Ch05.2-D_{S} irreducible decomposition}, \eqref{Ch05.2-zh3}, \eqref{Ch05.2-zh4} we find that the totally symmetric and traceless contribution is given by 
\begin{align}
E^{(r_{2} ... r_{N+1} k_{1} ... k_{C+1})} &= -\sum_{i=2}^{N+1} \e_{ p q }{}^{ r_{i} } \pa^{p} T_{1}^{(q r_{2} ... \hat{r}_{i} ... r_{N+1} k_{1} ... k_{C+1} )} \nonumber  \\
& \hspace{15mm} -\sum_{j=1}^{C+1} \e_{ p q }{}^{ k_{j} } \pa^{p} T_{1}^{(q r_{2} ... r_{N+1} k_{1} ... \hat{k}_{j} ... k_{C+1})} \, .
\label{Ch05.2-zh6}
\end{align}
Since the tensor $T_1$ is proportional to $C_1$, it is constructed out of the vector $X^m$ and the Minkowski metric $\eta^{mn}$. It is not difficult to see that for
a tensor $T_{1}$ constructed out of $X^m$ and $\eta^{mn}$, the combination \eqref{Ch05.2-zh6} must vanish. Hence $E = 0$ and $D_{S}$ is transverse.

Let us summarise the results of this subsection. We have shown that the parity-even contribution is fixed up to an overall coefficient. Moreover, it can be explicitly found 
for arbitrary superspins by solving a linear homogeneous system of equations with the tri-diagonal matrix \eqref{Ch05.2-Parity even homogeneous system}. 
Once the solution for the parity-even coefficients $\alpha_{2k}$ is found, the tensor $C$ is obtained using eq.~\eqref{Ch05.2-zh7} and the tensor $D_S$ is obtained 
from $C$ as discussed above. Note that our analysis for the parity even solution is incomplete since we have not imposed the conservation condition at the third point. 
This is technically difficult to impose using the approach outlined in this section, and from this viewpoint the computational approach developed in Section \ref{Ch05-subsection5.2.2} of this thesis is superior. The analysis in this subsection assumes that this property continues to hold for arbitrary superspins.

\subsubsection{Point-switch symmetries}\label{Ch05-subsection5.4.1.3}

For the $\langle \mathbf{J}^{}_{F} \mathbf{J}'_{F} \mathbf{J}''_{F} \rangle$ three-point functions we can also examine the case where $\mathbf{J} = \mathbf{J}'$ for arbitrary superspins. We want to determine the conditions under which the parity-even solution satisfies the point-switch symmetry. If we consider the condition \eqref{Ch05.1-Point switch A} and the irreducible decomposition \eqref{Ch05.2-FFF - irreducible decomposition}, we obtain the following conditions on $C$ and $D$:
\begin{subequations}
\begin{align}
	C_{E}^{(r_{1} ... r_{N+1})(k_{1} ... k_{C})}(X) - C_{E}^{(r_{1} ... r_{N+1})(k_{1} ... k_{C})}(-X) &= 0 \, , \label{Ch05.2-Point switch 1 - C} \\
	D_{E}^{(r_{1} ... r_{N+1})(k_{1} ... k_{C+1})}(X) - D_{E}^{(r_{1} ... r_{N+1})(k_{1} ... k_{C+1})}(-X) &= 0 \, . \label{Ch05.2-Point switch 1 - D}
\end{align}
\end{subequations}
Let us consider \eqref{Ch05.2-Point switch 1 - C} first. Using auxiliary spinors, this condition may be written as
\begin{align} \label{Ch05.2-Point switch 1 - C in auxiliary spinors}
C_{E}( X; u,v ) - C_{E}(- X; u,v ) = 0 \, .
\end{align}
However, recall that $C_{E}(X; u,v)$ is of the form
\begin{align}
C_{E}(X; u,v ) &= X^{C-N-2} \sum_{k=0}^{C} \a_{2k} P_{3}^{2k} Q_{3}^{2(C - k)} Z_{1}^{N-C+1} \, .
\end{align}
For $\mathbf{J} = \mathbf{J}'$, we have $N = A + B = 2A$, and from \eqref{Ch05.2-Point switch 1 - C in auxiliary spinors} we obtain
\begin{equation}
\sum_{k=0}^{C} (1 + (-1)^{C}) \, \a_{2k} P_{3}^{2k} Q_{3}^{2(C - k)} Z_{1}^{N-C+1} = 0 \, .
\end{equation}
Hence, the parity-even solution $C_{E}(X; u, v)$ satisfies the point-switch only for $C$ odd, i.e. for $s_{3} = 2k + \tfrac{3}{2}$, $k \in \mathbb{Z}_{\geq 0}$, 
which is consistent with the results of \cite{Buchbinder:2023fqv}. 

Now assume that $C_{E}$ satisfies the point-switch symmetry \eqref{Ch05.2-Point switch 1 - C}. 
We want to show that $D_{E}$, which is fully determined by $C_{E}$, satisfies \eqref{Ch05.2-Point switch 1 - D}. Since \eqref{Ch05.2-Point switch 1 - C} is satisfied, 
from \eqref{Ch05.2-C irreducible decomposition} we must have $C_{1}(X) = C_{1}(-X)$, $C_{2}(X) = C_{2}(-X)$ (indices suppressed). 
Now consider \eqref{Ch05.2-D decomposition - more symmetric} and the irreducible decomposition for $D_{S}$ given by \eqref{Ch05.2-D_{S} irreducible decomposition}. 
Since $T_{1} \propto C_{1}$, $T_{2} \propto C_{2}$, we have $T_{1}(X) = T_{1}(-X)$, $T_{2}(X) = T_{2}(-X)$. It is then easy to see by 
substituting \eqref{Ch05.2-D decomposition - more symmetric}, \eqref{Ch05.2-D_{S} irreducible decomposition} into \eqref{Ch05.2-Point switch 1 - D} that $D_{E}$ also 
satisfies the point-switch symmetry. Concerning point-switch symmetries that involve $\mathbf{J}''$, these are difficult to check using the approach 
outlined in this paper, as one must compute $\tilde{\cH}$ using \eqref{Ch05.1-H tilde} and check \eqref{Ch05.1-Point switch A}. The computational approach elucidated in section \ref{Ch05-subsection5.2.2.2} is more appropriate for this task.


\subsection{Grassmann-odd three-point functions -- \texorpdfstring{$\langle \mathbf{J}^{}_{F} \mathbf{J}'_{B} \mathbf{J}''_{B} \rangle$
}{< F B B >}}\label{Ch05-subsection5.4.2}

Let us now consider the case $\langle \mathbf{J}^{}_{F} \mathbf{J}'_{B} \mathbf{J}''_{B} \rangle$, which proves to be considerably simpler. 
Let us begin with making important comments on the arrangement of the operators in this three-point function. First, we arrange them in such a way that 
the operator at the third position is bosonic. Second, we arrange them in such a way that the third superspin satisfies the triangle inequality, that is  $s_3 \leq s_1+ s_2$. 
As was mentioned in the previous section if one of the triangle inequalities is violated the remaining two are necessarily satisfied. This means that we can always 
place a bosonic operator with the superspin satisfying the triangle inequality at the third position. 

We consider one Grassmann-odd current, $\mathbf{J}_{\a(2A+1)}$, of spin $s_{1} = A+ \tfrac{1}{2}$, and two Grassmann-even currents $\mathbf{J}'_{\a(2B)}$, $\mathbf{J}''_{\g(2C)}$, of spins $s_{2} = B$, $s_{3} = C$ respectively, where $A,B,C$ are positive integers. All information about the correlation function
\begin{equation}
\langle \, \mathbf{J}^{}_{\a(2A+1)}(z_{1}) \, \mathbf{J}'_{\b(2B)}(z_{2}) \, \mathbf{J}''_{\g(2C)}(z_{3}) \rangle \, ,
\end{equation}
is now encoded in a homogeneous tensor field $\cH_{\a(2A+1) \b(2B+1) \g(2C+1)}(\boldsymbol{X}, \Q)$, which satisfies the scaling property
\begin{equation}
\cH_{\a(2A+1) \b(2B) \g(2C)}( \l^{2} \boldsymbol{X}, \l \Q) = (\l^{2})^{C-A-B-\tfrac{3}{2}}\cH_{\a(2A+1) \b(2B) \g(2C)}(\boldsymbol{X}, \Q) \, .
\end{equation}
Analogous to the previous case, for each set of totally symmetric spinor indices (the $\a$'s, $\b$'s and $\g$'s respectively), we convert pairs of them into vector indices as follows:
\begin{align}
\cH_{\a(2A+1) \b(2B) \g(2C)}(\boldsymbol{X}, \Q) &\equiv \cH_{\a \a(2A), \b(2B), \g(2C)}(\boldsymbol{X}, \Q) \nonumber \\
&= (\g^{m_{1}})_{\a_{1} \a_{2}} ... (\g^{m_{A}})_{\a_{2A-1} \a_{2A}} \nonumber \\
& \times (\g^{n_{1}})_{\b_{1} \b_{2}} ... (\g^{n_{B}})_{\b_{2B-1} \b_{2B}} \nonumber \\
& \times (\g^{k_{1}})_{\g_{1} \g_{2}} ... (\g^{k_{C}})_{\g_{2C-1} \g_{2C}} \nonumber \\
& \times \cH_{\a, m_{1} ... m_{A}, n_{1} ... n_{B}, k_{1} ... k_{C} }(\boldsymbol{X}, \Q) \, .
\end{align}
Again, the equality above holds only if and only if $\cH_{\a, m_{1} ... m_{A} n_{1} ... n_{B} k_{1} ... k_{C} }(\boldsymbol{X}, \Q)$ is totally symmetric and traceless in each group of vector indices. It is also required that $\cH$ is subject to the $\g$-trace constraint
\begin{align}
(\g^{m_{1}})_{\s}{}^{\a} \cH_{\a, m_{1} ... m_{A}, n_{1} ... n_{B}, k_{1} ... k_{C} }(\boldsymbol{X}, \Q) = 0 \, . \label{Ch05.2-FBB - Gamma trace 1}
\end{align}
Indeed, since $\cH$ is Grassmann-odd it is linear in $\Q$, and we decompose $\cH$ it follows (again, raising all indices for convenience):
\begin{subequations} \label{Ch05.2-FBB - irreducible decomposition}
\begin{equation}
	\cH^{\a, m(A) n(B) k(C) }(\boldsymbol{X}, \Q) = \sum_{i=1}^{2} \cH_{i}^{\a, m(A) n(B) k(C) }(X, \Q) \, , 
\end{equation}
\vspace{-5mm}
\begin{align}
	\cH_{1}^{\a, m(A) n(B) k(C) }(X, \Q) &= \Q^{\a} A^{m(A) n(B) k(C)}(X) \, , \\
	\cH_{2}^{\a, m(A) n(B) k(C) }(X, \Q) &= (\gamma_{p})^{\a}{}_{\d} \Q^{\d} B^{p, m(A) n(B) k(C)}(X) \, .
\end{align}
\end{subequations}
In this case there are only two contributions to consider. The conservation equations \eqref{Ch05.1-Conservation equations} are now equivalent to the following constraints on $\cH$ with vector indices:
\begin{subequations}
\begin{align}
	\cD_{\a} \cH^{\a, m(A) n(B) k(C) }(\boldsymbol{X}, \Q) = 0 \, , \\
	(\gamma_{n})_{\s}{}^{\b} \cQ_{\b} \cH^{\a, m(A) n n(B-1) k(C) }(\boldsymbol{X}, \Q) = 0 \, .
\end{align}
\end{subequations}
They split up into constraints $O(\Q^{0})$
\begin{subequations}
\begin{align}
	\frac{\pa}{\pa \Q^{\a}} \cH^{\a, m(A) n(B) k(C) }(\boldsymbol{X}, \Q) &= 0 \, , \label{Ch05.2-FBB - conservation equations O(0) - 1} \\
	(\gamma_{n})_{\s}{}^{\b} \frac{\pa}{\pa \Q^{\b}} \cH^{\a, m(A) n n(B-1) k(C) }(\boldsymbol{X}, \Q) &= 0 \, , \label{Ch05.2-FBB - conservation equations O(0) - 2}
\end{align}
\end{subequations}
and $O(\Q^{2})$
\begin{subequations}
\begin{align}
	(\g^{m})_{\a \d} \Q^{\d} \frac{\pa}{\pa X^{m}} \cH^{\a, m(A) n(B) k(C) }(\boldsymbol{X}, \Q) &= 0 \, , \label{Ch05.2-FBB - conservation equations O(2) - 1} \\
	(\gamma_{n})_{\s}{}^{\b} (\g^{m})_{\b \d} \Q^{\d} \frac{\pa}{\pa X^{m}} \cH^{\a, m(A) n n(B-1) k(C) }(\boldsymbol{X}, \Q) &= 0 \, . \label{Ch05.2-FBB - conservation equations O(2) - 2}
\end{align}
\end{subequations}
Using the irreducible decomposition \eqref{Ch05.2-FBB - irreducible decomposition}, equation \eqref{Ch05.2-FBB - conservation equations O(0) - 1} immediately results in $A = 0$, while \eqref{Ch05.2-FBB - conservation equations O(0) - 2} gives
\begin{align} \label{Ch05.2-FBB - B constraints 1}
\eta_{pn} B^{p, m(A) n n(B-1) k(C) } &= 0 \, , & \e_{qpn} B^{p,m(A) n n(B-1) k(C)} &= 0 \, .
\end{align}
Next, after imposing the $\g$-trace condition \eqref{Ch05.2-FBB - Gamma trace 1}, we find that $B$ must satisfy
\begin{align} \label{Ch05.2-FBB - B constraints 2}
\eta_{pm} B^{p, m m(A-1) n(B) k(C) } &= 0 \, , & \e_{qpm} B^{p, m m(A-1) n(B) k(C)} &= 0 \, .
\end{align}
Altogether \eqref{Ch05.2-FBB - B constraints 1} and \eqref{Ch05.2-FBB - B constraints 2} imply that $B$ is symmetric and traceless in the indices $p$, $m_{1}, ..., m_{A}$, $n_{1}, ..., n_{B}$, i.e.
\begin{equation}
B^{p, m(A) n(B) k(C)} \equiv B^{(p m(A) n(B)), k(C)} \, .
\end{equation}
If we now consider the equations arising from conservation at $O(\Q^{2})$, a simple computation shows that $B$ must satisfy
\begin{align}
\pa_{p} B^{(p m(A) n(B)), k(C)} = 0 \, .
\end{align}
Therefore we need to construct a single transverse tensor $B$ of rank $A+B+C+1$. We see that the tensor $B$ has properties identical to the tensor $C$ from the previous 
section. Hence, the analysis becomes exactly the same as for  $C$ in the $\langle \mathbf{J}^{}_{F} \mathbf{J}'_{F} \mathbf{J}''_{F} \rangle$ case. 
Recalling the results from the previous section, we find that $\langle \mathbf{J}^{}_{F} \mathbf{J}'_{B} \mathbf{J}''_{B} \rangle$ has vanishing parity-odd contribution 
for all values of the superspins and a unique parity-even structure.\footnote{Here we have assumed that for our arrangement of the operators 
the conservation condition at the third point is automatically satisfied for all superspins. Within the framework of our computational approach this assumption is valid up to $s_i=20$.}
The explicit form of the parity-even solution can be found from the tri-diagonal system of linear equations just like in the previous section.

\section{Summary of results}\label{Ch05-section5.6}

In this chapter we embarked on the analysis of three-point functions of conserved currents in three-dimensional $\cN=1$ superconformal field theory, where a complete classification of their general structure is provided. In particular we demonstrated up to superspin-20 that the three-point functions which are Grassmann-even in 3D $\cN=1$ superspace are fixed up to one parity-even structure and one parity-odd structure, with the existence of the latter depending on whether the superspin triangle inequalities are satisfied. For the Grassmann-odd three-point functions we provided evidence that only a single parity-even structure is allowed, while the parity-odd structure vanishes. We then proved this assertion analytically for arbitrary superspins by using a method of irreducible decomposition similar to the approach outlined in \cite{Buchbinder:2015qsa}. In particular, we showed that the
construction of both the parity-even and parity-odd sector is governed by a
system of linear homogeneous equations with a tri-diagonal matrix. By computing the determinants of
the tridiagonal matrices for the parity-even and parity-odd sectors, in the former case
we prove that the matrix has co-rank one, and hence the parity-even solution is unique
for arbitrary superspins. In the latter case, we prove that the matrix is non-degenerate,
meaning that the parity-odd solution vanishes in general.

An explanation for the reduction in the number of parity-even structures as compared with the non-supersymmetric case is that supersymmetry relates the free coefficients in the corresponding non-supersymmetric three-point functions. This was shown in \cite{Buchbinder:2015qsa} for the three-point function of flavour currents. There are, however, many technical issues associated with carrying out these calculations for three-point functions of conserved currents for arbitrary superspins. See e.g. \cite{Buchbinder:2015qsa,Buchbinder:2015wia,Buchbinder:2021gwu} for examples of how the superspace reduction process is carried out for three-point functions in $\cN$-extended theories. 

We also point out that the results presented in this chapter have been independently reproduced in a recent work by Jain et. al. \cite{Jain:2023idr} which utilises the formalism of super spinor-helicity variables (see \cite{Jain:2020rmw,Jain:2021gwa,Jain:2021vrv,Jain:2021whr} for a similar approach to three-point functions in 3D CFT). Moreover, the results obtained are potentially of relevance to AdS/CFT duality. For example the duality between 3D $\cN=1$ SCFT and $\cN=1$ Vasiliev higher-spin theory on $\text{AdS}_{4}$ was investigated in \cite{Leigh:2003gk}. Once more general expressions are worked out, perhaps a holographic reconstruction procedure similar to that of \cite{Sleight:2016dba} could be applied to the supersymmetric case.

\begin{subappendices}
	\section{Examples: three-point functions of higher-spin supercurrents}\label{Appendix5B}
Below we present some results for three-point functions of conserved supercurrents. Due to the size of the solutions for increasing superspins, we consider only some low superspin cases after imposing conservation on all three points. To demonstrate the effectiveness of our computational approach we present the results as outputs from Mathematica. 

\subsection*{Grassmann-even correlators}

\noindent
\textbf{Correlation function} $\langle \mathbf{J}^{}_{1/2} \mathbf{J}'_{1/2} \mathbf{J}''_{1} \rangle$\textbf{:}
\begin{flalign*}
	\hspace{5mm} \includegraphics[width=0.9\textwidth]{1-1-2.pdf} &&
\end{flalign*} 

\noindent
\textbf{Correlation function} $\langle \mathbf{J}^{}_{1} \mathbf{J}'_{1} \mathbf{J}''_{1} \rangle$\textbf{:}
\begin{flalign*}
	\hspace{5mm} \includegraphics[width=0.9\textwidth]{2-2-2.pdf} &&
\end{flalign*}

\noindent
\textbf{Correlation function} $\langle \mathbf{J}^{}_{1/2} \mathbf{J}'_{1/2} \mathbf{J}''_{2} \rangle$\textbf{:}
\begin{flalign*}
	\hspace{5mm} \includegraphics[width=0.9\textwidth]{1-1-4.pdf} &&
\end{flalign*} 
This three-point function was initially studied in \cite{Nizami:2013tpa}, which asserted the existence of a parity-odd solution. However, it was proven later in \cite{Buchbinder:2021qlb} that such a structure cannot be consistent with the superfield conservation equations. We see that this structure also does not appear in our computational approach. This is a simple consequence of the fact that the superspin triangle inequalities are not satisfied for this three-point function.

\vspace{4mm}

\noindent
\textbf{Correlation function} $\langle \mathbf{J}^{}_{1/2} \mathbf{J}'_{1/2} \mathbf{J}''_{3} \rangle$\textbf{:}
\begin{flalign*}
	\hspace{5mm} \includegraphics[width=0.9\textwidth]{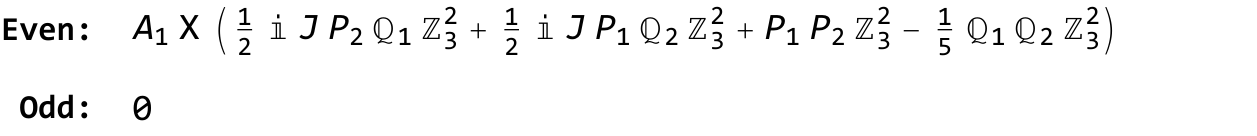} &&
\end{flalign*} 
This is another case where the superspin triangle inequalities are not satisfied, hence, the odd structure vanishes as expected.

\vspace{4mm}

\noindent
\textbf{Correlation function} $\langle \mathbf{J}^{}_{1/2} \mathbf{J}'_{3/2} \mathbf{J}''_{2} \rangle$\textbf{:}
\begin{flalign*}
	\hspace{5mm} \includegraphics[width=0.9\textwidth]{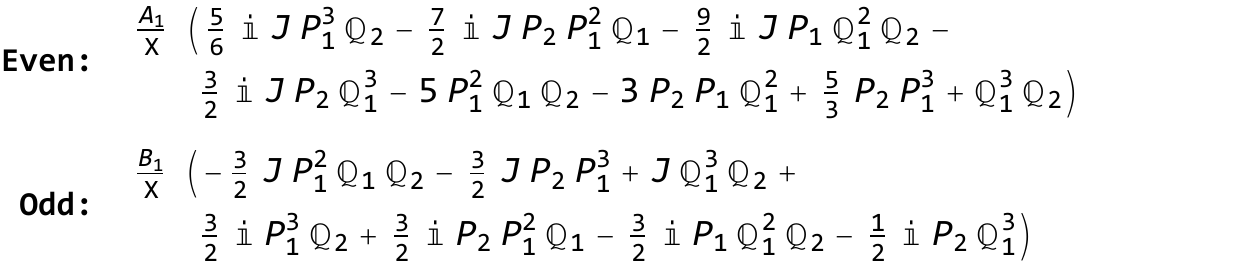} &&
\end{flalign*}

\noindent
\textbf{Correlation function} $\langle \mathbf{J}^{}_{3/2} \mathbf{J}'_{3/2} \mathbf{J}''_{2} \rangle$\textbf{:}
\begin{flalign*}
	\hspace{5mm} \includegraphics[width=0.9\textwidth]{3-3-4.pdf} &&
\end{flalign*}

\noindent
\textbf{Correlation function} $\langle \mathbf{J}^{}_{2} \mathbf{J}'_{2} \mathbf{J}''_{2} \rangle$\textbf{:}
\begin{flalign*}
	\hspace{5mm} \includegraphics[width=0.9\textwidth]{4-4-4.pdf} &&
\end{flalign*}
This three-point function was studied using the tensor formalism in \cite{Buchbinder:2021qlb}, where it was also shown that a parity-odd solution could arise in the three-point function. 

\vspace{5mm}

\noindent
\textbf{Correlation function} $\langle \mathbf{J}^{}_{1} \mathbf{J}'_{2} \mathbf{J}''_{4} \rangle$\textbf{:}
\begin{flalign*}
	\hspace{5mm} \includegraphics[width=0.9\textwidth]{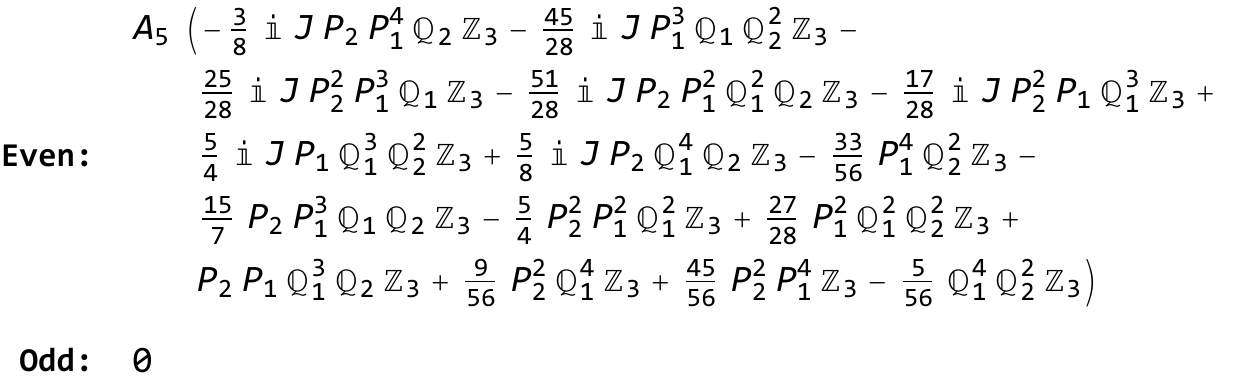} &&
\end{flalign*}
In this case we note that the superspin triangle inequalities are not satisfied and therefore the odd solution vanishes after current conservation.

\vspace{4mm}

\vfill



\vfill

\newpage

\noindent
\textbf{Correlation function} $\langle \mathbf{J}^{}_{4} \mathbf{J}'_{4} \mathbf{J}''_{4} \rangle$\textbf{:}
\begin{flalign*}
	\hspace{5mm} \includegraphics[width=0.9\textwidth]{8-8-8.pdf} &&
\end{flalign*}

\newpage

\subsection*{Grassmann-odd correlators}

We now present results for three-point functions of conserved currents which are Grassmann-odd in superspace. An important feature of all these three-point functions is that they do not possess a parity-violating contribution.

\vspace{2mm}

\noindent
\textbf{Correlation function} $\langle \mathbf{J}^{}_{1/2} \mathbf{J}'_{3/2} \mathbf{J}''_{5/2} \rangle$\textbf{:}
\begin{flalign*}
	\hspace{5mm} \includegraphics[width=0.9\textwidth]{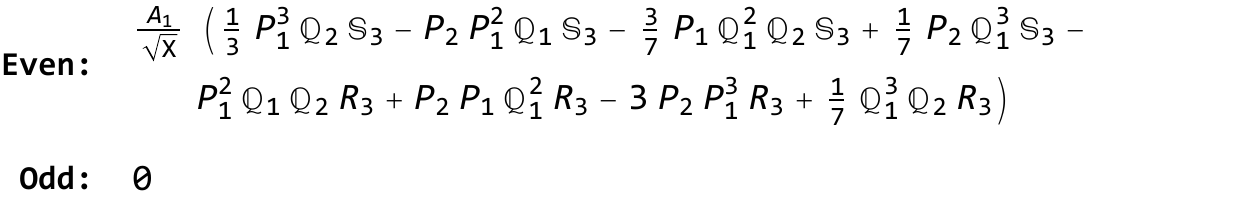} &&
\end{flalign*}

\noindent
\textbf{Correlation function} $\langle \mathbf{J}^{}_{2} \mathbf{J}'_{2} \mathbf{J}''_{1/2} \rangle$\textbf{:}
\begin{flalign*}
	\hspace{5mm} \includegraphics[width=0.9\textwidth]{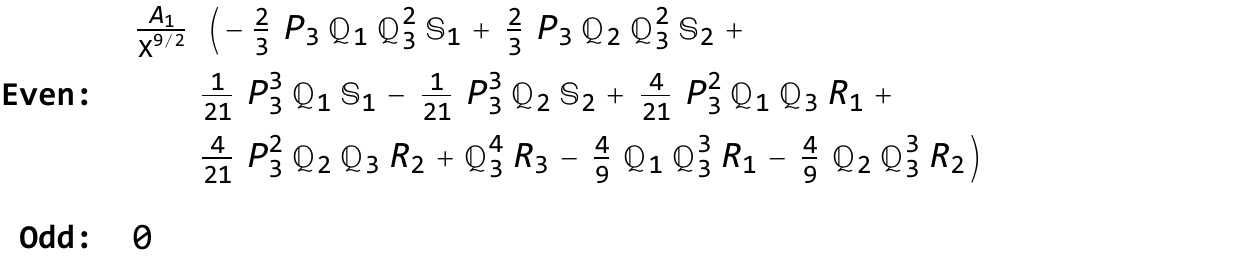} &&
\end{flalign*}
In this instance we note that the superspin triangle inequalities are not satisfied and therefore the odd solution vanishes after current conservation.

\vspace{2mm}

\noindent
\textbf{Correlation function} $\langle \mathbf{J}^{}_{2} \mathbf{J}'_{2} \mathbf{J}''_{3/2} \rangle$\textbf{:}
\begin{flalign*}
	\hspace{5mm} \includegraphics[width=0.9\textwidth]{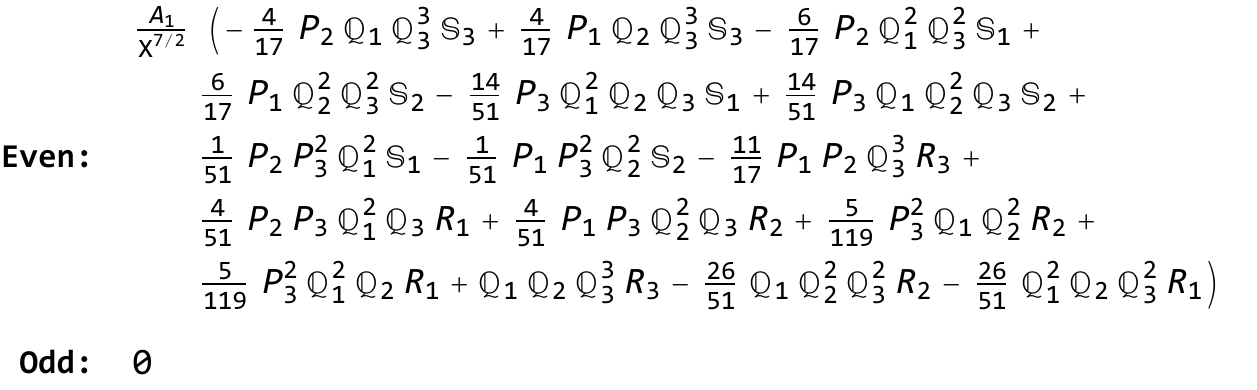} &&
\end{flalign*}

\noindent
\textbf{Correlation function} $\langle \mathbf{J}^{}_{3/2} \mathbf{J}'_{3/2} \mathbf{J}''_{5/2} \rangle$\textbf{:}
\begin{flalign*}
	\hspace{5mm} \includegraphics[width=0.9\textwidth]{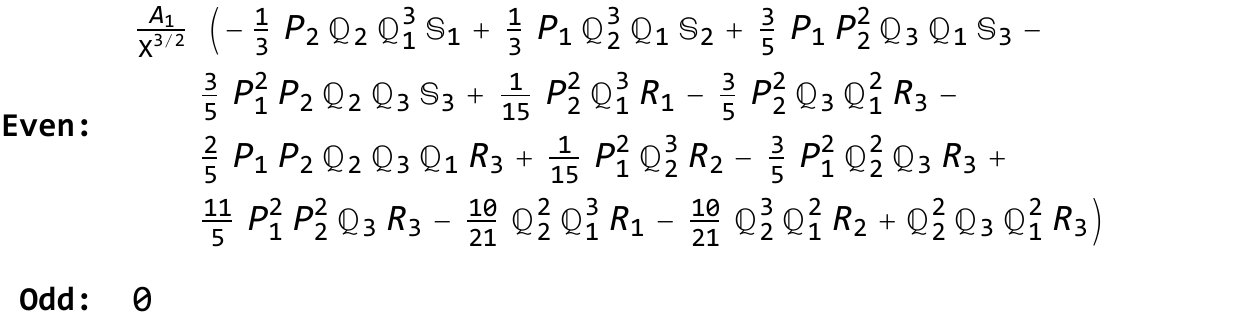} &&
\end{flalign*}

\noindent
\textbf{Correlation function} $\langle \mathbf{J}^{}_{3/2} \mathbf{J}'_{3/2} \mathbf{J}''_{7/2} \rangle$\textbf{:}
\begin{flalign*}
	\hspace{5mm} \includegraphics[width=0.9\textwidth]{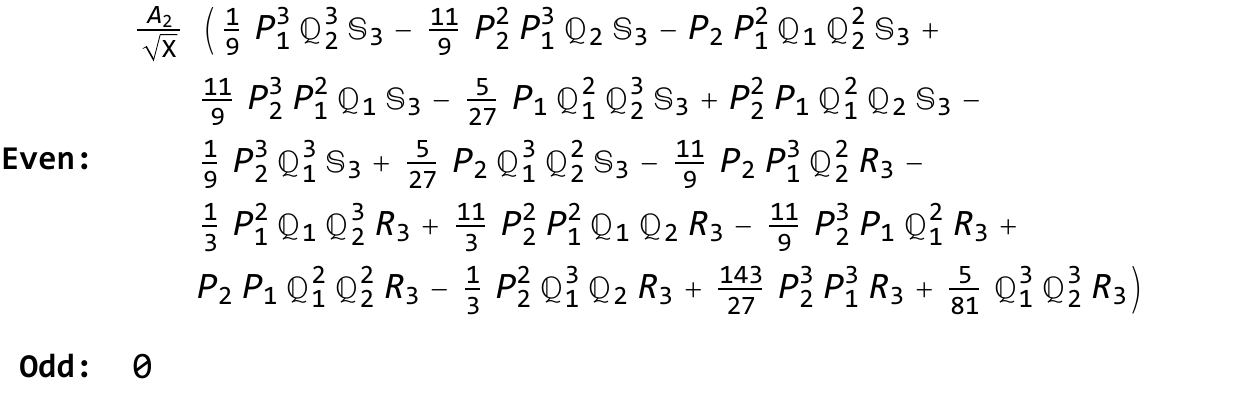} &&
\end{flalign*}

\noindent
\textbf{Correlation function} $\langle \mathbf{J}^{}_{2} \mathbf{J}'_{2} \mathbf{J}''_{7/2} \rangle$\textbf{:}
\begin{flalign*}
	\hspace{5mm} \includegraphics[width=0.9\textwidth]{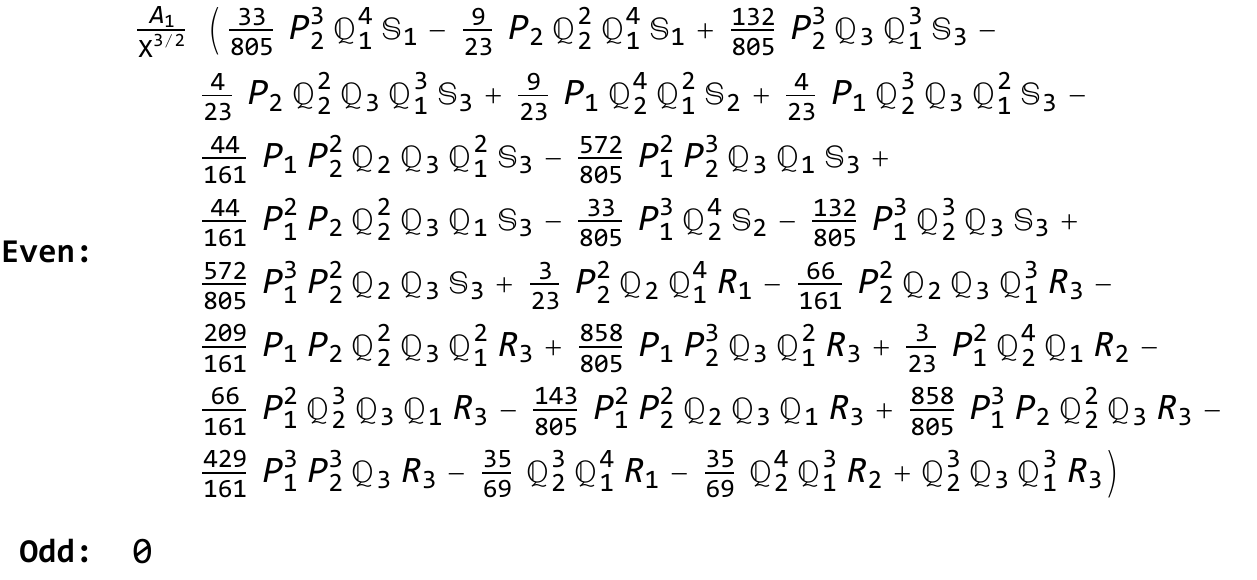} &&
\end{flalign*}

\noindent
\textbf{Correlation function} $\langle \mathbf{J}^{}_{5/2} \mathbf{J}'_{5/2} \mathbf{J}''_{5/2} \rangle$\textbf{:}
\begin{flalign*}
	\hspace{5mm} \includegraphics[width=0.9\textwidth]{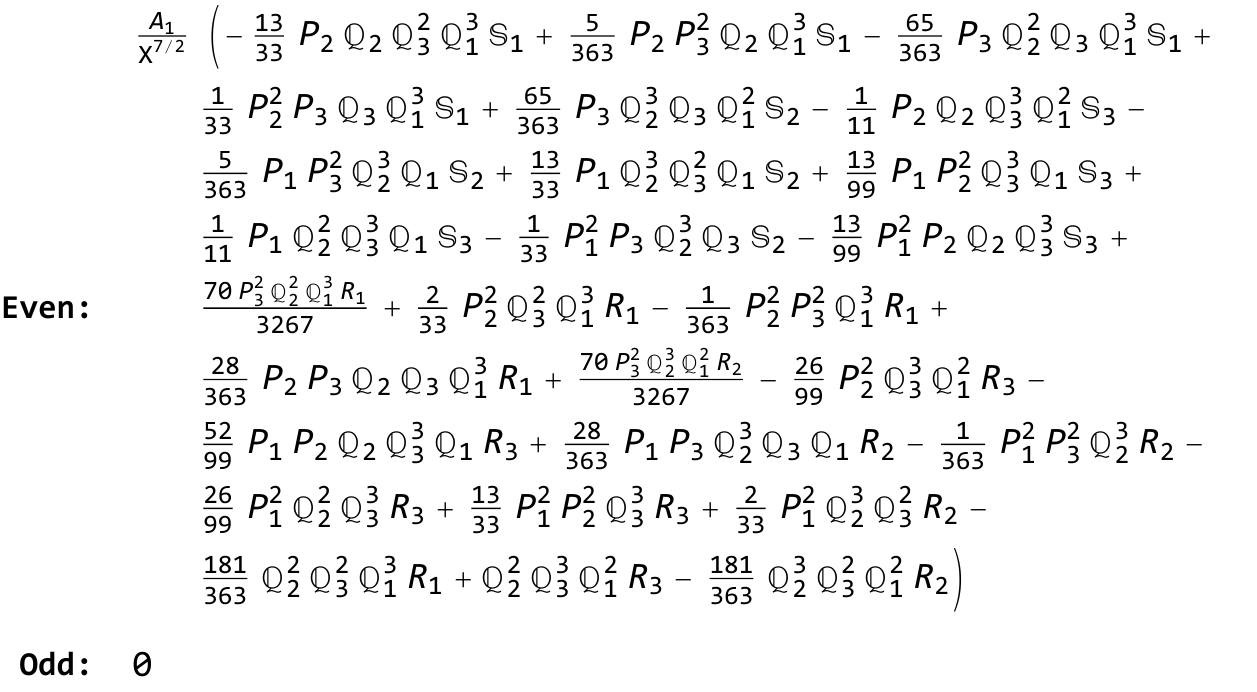} &&
\end{flalign*}

\vfill

\newpage

\noindent
\textbf{Correlation function} $\langle \mathbf{J}^{}_{7/2} \mathbf{J}'_{7/2} \mathbf{J}''_{7/2} \rangle$\textbf{:}
\begin{flalign*}
	\hspace{5mm} \includegraphics[width=0.9\textwidth]{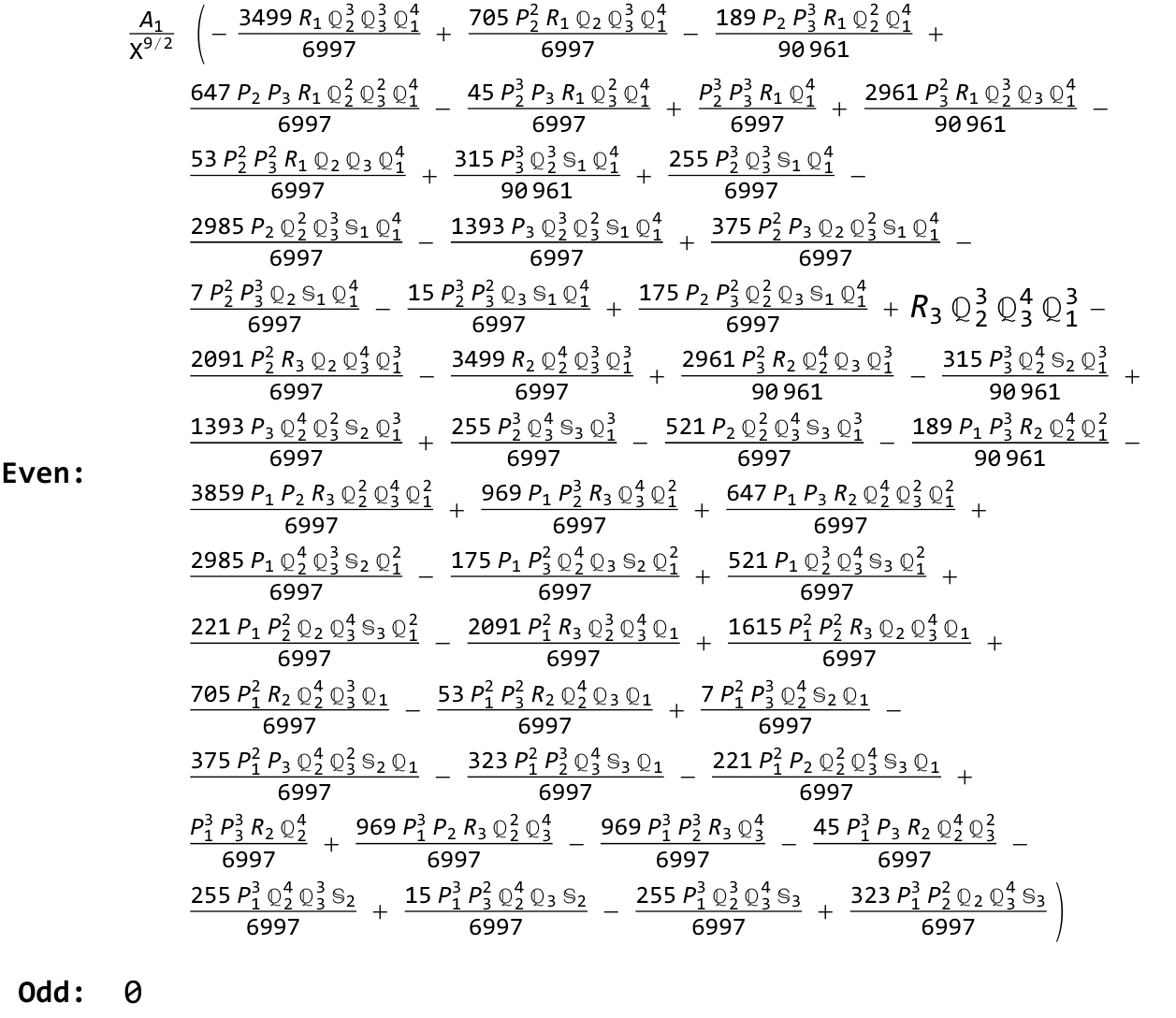} &&
\end{flalign*}

	\section{Examples: three-point functions involving scalar superfields} \label{Appendix5C}
\noindent
\textbf{Correlation function} $\langle \cO \, \cO' \, \mathbf{J}_{1/2} \rangle$\textbf{:}\\
For $\delta_{1} = \delta_{2} = \delta$, there is a single even solution compatible with conservation
\begin{flalign*}
	\hspace{5mm} \includegraphics[width=0.97\textwidth]{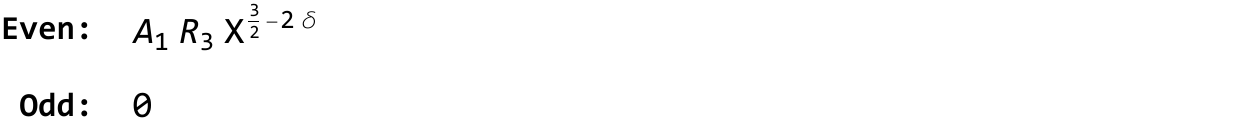} &&
\end{flalign*}

\noindent
\textbf{Correlation function} $\langle \cO \, \cO' \, \mathbf{J}_{1} \rangle$\textbf{:} \\
For $\delta_{1} = \delta_{2} = \delta$, there is a single even solution compatible with conservation
\begin{flalign*}
	\hspace{5mm} \includegraphics[width=0.97\textwidth]{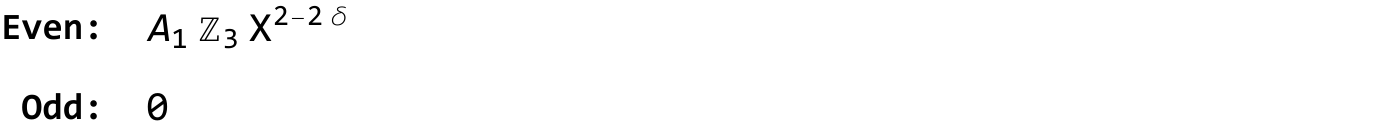} &&
\end{flalign*}

\noindent
\textbf{Correlation function} $\langle \cO \, \cO' \, \mathbf{J}_{3/2} \rangle$\textbf{:}\\
For $\delta_{1} = \delta_{2} = \delta$, there is a single even solution compatible with conservation
\begin{flalign*}
	\hspace{5mm} \includegraphics[width=0.97\textwidth]{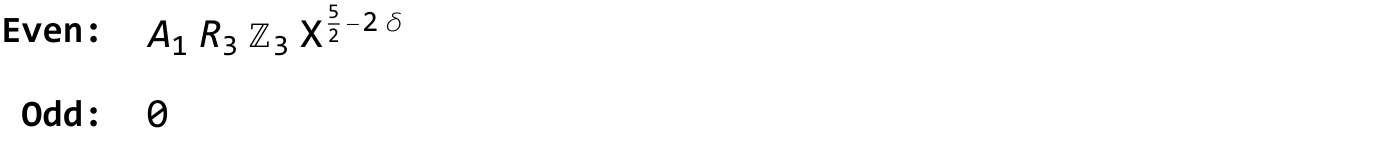} &&
\end{flalign*}

\noindent
\textbf{Correlation function} $\langle \cO \, \cO' \, \mathbf{J}_{2} \rangle$\textbf{:}\\
For $\delta_{1} = \delta_{2} = \delta$, there is a single even solution compatible with conservation.
\begin{flalign*}
	\hspace{5mm} \includegraphics[width=0.97\textwidth]{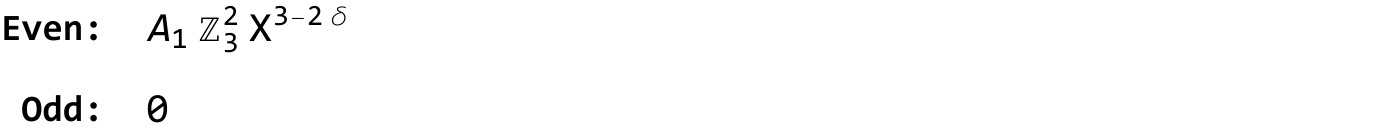} &&
\end{flalign*}
It is simple to see the generalisation to arbitrary integer or half-integer $s$.

\vspace{2mm}

\noindent
\textbf{Correlation function} $\langle \mathbf{J}^{}_{1/2} \mathbf{J}'_{1/2} \cO \rangle$\textbf{:}\\
In this case, the superspin triangle inequalities are satisfied and there is one even and one odd solution for arbitrary $\d$:
\begin{flalign*}
	\hspace{5mm} \includegraphics[width=0.97\textwidth]{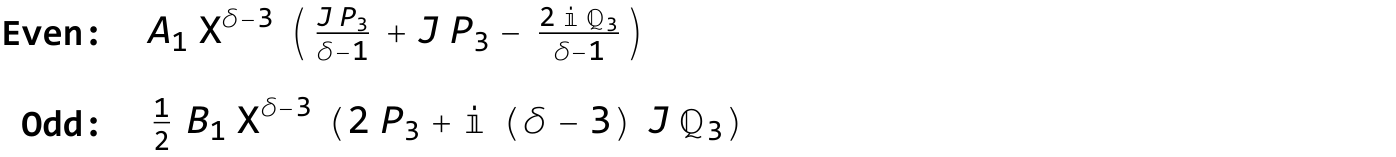} &&
\end{flalign*}

\noindent
\textbf{Correlation function} $\langle \mathbf{J}^{}_{1/2} \mathbf{J}'_{3/2} \cO \rangle$\textbf{:}\\
In this case there is a solution only for $\d = 1$:
\begin{flalign*}
	\hspace{5mm} \includegraphics[width=0.97\textwidth]{1-3-0-B.pdf} &&
\end{flalign*}

\noindent
\textbf{Correlation function} $\langle \mathbf{J}^{}_{3/2} \mathbf{J}'_{3/2} \cO \rangle$\textbf{:} \\
In this case, the superspin triangle inequalities are satisfied and there is one even and one odd solution for arbitrary $\d$:
\begin{flalign*}
	\hspace{5mm} \includegraphics[width=0.97\textwidth]{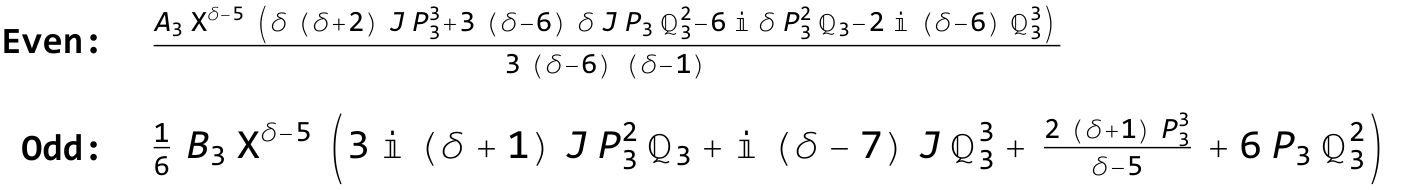} &&
\end{flalign*}

\noindent
\textbf{Correlation function} $\langle \mathbf{J}^{}_{1} \mathbf{J}'_{2} \cO \rangle$\textbf{:} \\
In this case there is a solution only for $\d = 1$:
\begin{flalign*}
	\hspace{5mm} \includegraphics[width=0.97\textwidth]{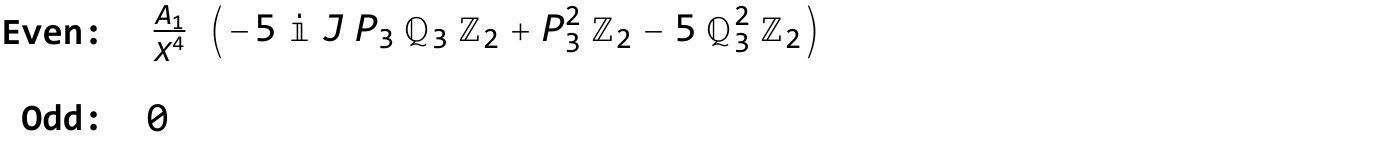} &&
\end{flalign*}
%


\end{subappendices}

\chapter{Conclusion}\label{Chapter6}

In this thesis we explored the implications of conformal and superconformal symmetry on the general structure of the three-point functions of conserved currents. In particular, we utilised the formalism of Osborn \& Petkou \cite{Osborn:1993cr} and specialised it to appropriate spinor representations in three and four spacetime dimensions. A key advantage of the approach in \cite{Osborn:1993cr} is that the general structure of the three-point function is completely encoded in a constrained homogeneous tensor which is a function of a single three-point conformally covariant building block, instead of all three spacetime points. This simplifies imposing constraints due to conservation equations and symmetries under permutations of spacetime points. To carry out the calculations for higher-spin currents, we developed the formalism further by augmenting it with auxiliary spinors. This drastically simplifies the computations, allowing us to determine a complete classification of the general structure for currents of integer or half-integer spin. This was achieved by automating all aspects of the calculations in \textit{Mathematica}. Then, by building on the covariant superspace formalisms developed by Park \cite{Park:1999cw} and later by Buchbinder, Kuzenko, Samsonov \cite{Buchbinder:2015qsa}, we repeated the analysis for three-point functions of conserved supercurrents in three-dimensional superconformal field theory. Compared with the non-supersymmetric case, we demonstrated that supersymmetry imposes further constraints on the structure of three-point functions of conserved currents.

Let us now recapitulate the results obtained in each chapter. First, in Chapter \ref{Chapter3} we systematically analysed the general structure of three-point functions of conserved currents of an arbitrary integer and half-integer spins in three spacetime dimensions. To undertake the analysis we generalised the construction of \cite{Osborn:1993cr} to a suitable spinor representation and augmented it with auxiliary spinors, which greatly simplifies the computations for three-point functions involving higher-spin currents. We demonstrated the effectiveness of the formalism by analysing in detail the three-point functions of the energy-momentum tensor and conserved vector currents, where we obtained the known results (in terms of the number of independent structures) \cite{Osborn:1993cr,Giombi:2011rz,Zhiboedov:2012bm,Giombi:2016zwa}. We then proposed a general classification for the structure of three-point functions of conserved currents for arbitrary spins by performing an explicit analysis for spins $s_{i} \leq 20$ (integer or half-integer). The computational approach is highly efficient and completely automated. In particular, we provided strong evidence that three-point functions of conserved currents are fixed up to two parity even structures and one parity-violating structure, with the existence of the latter structure subject to a set of triangle inequalities in the spins.

Next, in Chapter \ref{Chapter4} we carried out the analysis of three-point functions of conserved currents in four dimensional conformal field theory, where the results are quite different. We generalised the construction of \cite{Osborn:1993cr} to primary fields in an arbitrary Lorentz representation. For conserved vector-like currents in the $(s,s)$ representation, we demonstrated that the three-point functions of conserved currents are fixed up to $2 \min(s_{i}) + 1$ structures, as shown in \cite{Stanev:2012nq,Zhiboedov:2012bm} some time ago. We then extended these results by proposing a classification for the general structure of three-point function for currents of currents in an arbitrary Lorentz representation, where we obtained new results. We showed that for some three-point functions the number of structures deviates from $2 \min(s_{i}) + 1$. It remains an open problem to determine how these structures may arise in three-point functions, and whether they can be obtained from free field theories.

In Chapter \ref{Chapter5} we embarked on the analysis of three-point functions of conserved currents in three-dimensional $\cN=1$ superconformal field theory. In particular we demonstrated up to superspin-20 that in 3D $\cN=1$ superspace the Grassmann-even three-point functions are fixed up to one parity-even structure and one parity-odd structure, with the existence of the latter depending on whether the superspin triangle inequalities are satisfied. For the Grassmann-odd three-point functions we provided evidence that only a single parity-even structure is allowed, while the parity-odd structure vanishes. We then proved this for arbitrary superspins by using a method of irreducible decomposition similar to the approach outlined in \cite{Buchbinder:2015qsa}. In particular, we showed that the
construction of both the parity-even and parity-odd sector is governed by a
system of linear homogeneous equations with a tri-diagonal matrix. By computing the determinants of
the tridiagonal matrices for the parity-even and parity-odd sectors, in the former case
we prove that the matrix has co-rank one, hence, the parity-even solution is unique
for arbitrary superspins. In the latter case, we prove that the matrix is non-degenerate,
meaning that the parity-odd solution must vanish in general. By solving the linear homogeneous equations for the parity-even sector, we obtained an explicit solution for the parity-even structure for arbitrary superspins, which is a new result presented for the first time in this thesis. It is important to note that the Grassmann-odd three-point functions are those such as the supercurrent and flavour current three-point functions. 

An explanation for the reduction of the number of parity-even structures as compared with the non-supersymmetric case is that supersymmetry relates the free coefficients in the corresponding non-supersymmetric three-point functions. This was shown in \cite{Buchbinder:2015qsa} for the three-point function of flavour currents. Generalising such calculations to three-point functions of conserved higher-spin currents remains is a technical challenge, and such an analysis would have to be carried out on a case-by-case basis.

Let us now comment on some future directions for research. A noteworthy feature of the computational approach is that it allows for efficient computation of three-point functions where the symmetric tensor operators are \textit{not necessarily} conserved currents (and hence have arbitrary scale dimensions). We did not explicitly consider three-point correlation functions involving non-conserved tensor operators, but in principle this could be considered for future work. Another interesting open problem is to determine the generating functions which encode the three-point functions of conserved currents for \textit{all} spins in 3D CFT, analogous to the results given \cite{Giombi:2011rz,Maldacena:2011jn}. These generating functions have also been obtained using AdS/CFT techniques in \cite{Didenko:2012tv,Didenko:2013bj}. However, since we are utilising the formalism of Osborn and Petkou, the generating functions would be for the tensor $\cH$ instead of the entire three-point function. Attempts to construct such generating functions are mostly restricted to intelligent guesses, however, given the enormous number of results obtained in this thesis the task is simplified considerably. Once the generating functions are found, in principle it should not be too difficult to obtain the generating functions for the three-point functions of conserved supercurrents in 3D $\cN=1$ theories; their construction remains an open problem. The supersymmetric generating functions, when expanded in power series in the three-point covariant $\Theta$, would naturally contain the non-supersymmetric generating functions.

It would also be interesting to consider generalising the construction presented for 3D $\cN=1$ superconformal theories to 3D $\cN$-extended superconformal theories. Some preliminary results for the three-point functions of the supercurrent and flavour current multiplets in $\cN$-extended theories may be found in \cite{Buchbinder:2015qsa,Buchbinder:2015wia,Kuzenko:2016cmf,Buchbinder:2021gwu,Jain:2022izp}. On the computational side, some care will be required to handle computations involving the $R$-symmetry indices on the superspace variables and conserved supercurrents, however, most of the methods developed in this thesis should be directly applicable. One might also consider developing an analogous approach to compute three-point functions in 4D $\cN=1$ SCFT. The group theoretic formalism to analyse such three-point functions has been outlined in e.g. \cite{Park:1997bq,Osborn:1998qu,Buchbinder:2021izb,Buchbinder:2021kjk,Buchbinder:2022kmj}, however, a general classification of the structure of three-point functions is yet to be worked out. The index-free methods presented in this thesis will possibly allow for a complete classification of 4D $\cN=1$ three-point functions for arbitrary superspins (up to a suitably high computational bound). 

It's also possible that the computational methods developed for this project could be applied to three-point functions in (super)conformally flat spacetimes, such as $\cN$-extended AdS superspace, using the superembedding methods detailed in \cite{Kuzenko:2023yak} for $(p,q)$ AdS superspaces in three-dimensions, and in \cite{Koning:2023ruq} for $\cN$-extended AdS superspaces in four dimensions. In these works the two-point covariants were computed explicitly, however, the details concerning the construction of superconformally covariant three-point functions using such superembedding methods are yet to be worked out. Finally, we point out that another potential application of our results is to compute the cubic sector of induced superconformal higher-spin actions in three- and four-dimensional (super)conformal field theories, using the methods recently outlined in \cite{Kuzenko:2022qeq}. This sector is completely characterised by the three-point functions of primary conserved currents, and hence the results of this thesis are directly applicable. We leave the considerations above for future work.




\clearpage
\fancyhead[LE,LO]{Bibliography}
\printbibliography[heading=bibintoc,title={Bibliography}]




\end{document}